\documentclass[a4paper,fleqn]{cas-dc}

\usepackage[authoryear]{natbib}

\usepackage{tabularx}
\usepackage{algorithm}
\usepackage{booktabs}
\usepackage{algpseudocode}
\usepackage{subcaption}

\usepackage{xcolor}
\usepackage{booktabs}
\usepackage{multirow}
\usepackage{siunitx}
\usepackage{graphicx} 

\def\tsc#1{\csdef{#1}{\textsc{\lowercase{#1}}\xspace}}
\tsc{WGM}
\tsc{QE}

\begin{document}
\let\WriteBookmarks\relax
\def\floatpagepagefraction{1}
\def\textpagefraction{.001}

\shorttitle{}    

\shortauthors{}  

\title [mode = title]{LMFPPO-UBP: Local Mean Field Proximal Policy Optimization with Unbalanced Punishment for Spatial Public Goods Games}  



%

\author[1,2]{Jinshuo Yang}[style=chinese,orcid=0009-0008-2629-9406]


\ead{jsyang2002@163.com}

\tnotetext[1]{https://github.com/geek12138/LMFPPO-UBP}


\author[1,2]{Zhaoqilin Yang}[style=chinese,orcid=0000-0002-3676-4761]
\cormark[1]
\ead{zqlyang@gzu.edu.cn}

\author[1,2]{Wenjie Zhou}[style=chinese,orcid=0009-0007-9537-6770]

\ead{13158323712@163.com}

\author[3]{Xin Wang}[style=chinese,orcid=0009-0003-3551-7114]

\ead{xinwang2@bjtu.edu.cn}

\author[4,2]{Youliang Tian}[style=chinese,orcid=0000-0002-5974-1570]
\ead{yltian@gzu.edu.cn}

\affiliation[1]{organization={State Key Laboratory of Public Big Data, College of Computer Science and Technology},
	addressline={Guizhou University}, 
	city={Guiyang},
	postcode={550025}, 
	state={Guizhou},
	country={China}}

\affiliation[2]{organization={Institute of Cryptography and Data Security},
	addressline={Guizhou University}, 
	city={Guiyang},
	postcode={550025}, 
	state={Guizhou},
	country={China}}

\affiliation[3]{organization={School of Mathematics and Statistics},
	addressline={Beijing Jiaotong University}, 
	city={Beijing},
	postcode={100044}, 
	state={Beijing},
	country={China}}

\affiliation[4]{organization={State Key Laboratory of Public Big Data, College of Big Data and Information Engineering},
	addressline={Guizhou University}, 
	city={Guiyang},
	postcode={550025}, 
	state={Guizhou},
	country={China}}
	
	\cortext[cor1]{Corresponding author}


\begin{abstract}
Spatial public goods games are characterized by high-dimensional state spaces and localized externalities, which pose significant challenges for achieving stable and widespread cooperation. Traditional approaches often struggle to effectively capture neighborhood-level strategic interactions and dynamically align individual incentives with collective welfare. To resolve this issue, this paper introduces a novel intelligent decision-making framework called Local Mean-Field Proximal Policy Optimization with Unbalanced Punishment (LMFPPO-UBP). The conventional mean field concept is reformulated as a socio-statistical sensor embedded directly into the policy gradient space of deep reinforcement learning, allowing agents to adapt their strategies based on mesoscale neighborhood dynamics. Additionally, an unbalanced punishment mechanism is integrated to penalize defectors proportionally to the local density of cooperators, thereby reshaping the payoff structures without imposing direct costs on cooperative agents. Experimental results demonstrate that the LMFPPO-UBP promotes rapid and stable global cooperation even under low enhancement factors, consistently outperforming baseline methods such as Q-learning and Fermi update rules. Statistical analyses further validate the framework's effectiveness in lowering the cooperation threshold and achieving better coordinated outcomes.
\end{abstract}





\maketitle

\section{Introduction}
Distributed learning frameworks present fundamental coordination challenges in multi-agent environments. For instance, in urban traffic signal networks, each intersection operates as an autonomous agent that must optimize local flow while contributing to system-wide efficiency \citep{2025Asynchronous,2010Promotion}. The global objective is to minimize city-wide congestion. However, in the absence of central coordination, agents may prioritize local gains and effectively transfer congestion to neighboring nodes, thereby undermining collective efficiency. This conflict epitomizes the free-rider problem pervasive in distributed artificial intelligence and cyber-physical systems \citep{Hardin1968, Ledyard1995}. Similar dilemmas arise in domains such as distributed network resource allocation and smart energy grids, where agents decide whether to contribute to or draw from a shared pool \citep{2025Cooperation}. The decentralized nature and complex interdependencies in these systems necessitate a robust theoretical abstraction that captures the core tension between individual gain and collective welfare. These problems are formally modeled through Spatial Public Goods Games (SPGG) \citep{NowakMay1992, SzaboFath2007}. While spatial reciprocity can sustain cooperation \citep{Nowak2006}, its emergence and stability are governed by a critical threshold \citep{Lv2025, 2025Environmental}. Consequently, a significant gap persists between theoretical models and deployable algorithmic solutions. Although punishment mechanisms can promote cooperation \citep{FehrGachter2000, FehrGachter2002}, designing adaptive rules that are both responsive to local contexts and robust against exploitation remains a challenge. Fixed, exogenous rules fail to capture the essential co-evolution of agent behavior and institutional feedback \citep{HECKATHORN1989Collective}. Furthermore, classical multi-agent methods often struggle with convergence \citep{Watkins1992, 2021Multi}, and imitation-based rules provide only myopic guidance \citep{schuster_1983_replicator}.

The Mean-Field Proximal Policy Optimization (MFPPO) \citep{yang2020meanfieldmultiagentreinforcement} reduces multi-agent interactions by considering global population averages. In spatially structured environments like SPGG, however, global averages lack structural granularity as interactions are intrinsically localized. We therefore propose a Local Mean-Field (LMF) representation, wherein each agent’s perceptual input is derived strictly from its topological neighborhood. This ensures consistency with the spatial interaction topology, eliminates extraneous perceptual noise, and preserves scalability while respecting structural constraints. Building on this, we propose a Local Mean-Field Proximal Policy Optimization (LMFPPO) algorithm that embeds neighborhood-level strategic statistics into a policy gradient architecture. This enables agents to adapt to local cooperation pressure through a differentiable policy network, coupling individual learning with mesoscale social dynamics. To further shape the strategic landscape, we incorporate an Unbalanced Punishment (UBP) mechanism inspired by inequity aversion theory \citep{1999A}. Unlike conventional costly punishment, this mechanism penalizes defectors proportionally to the number of cooperating neighbors in their vicinity, translating local fairness concerns into algorithmic incentives without imposing costs on cooperators. Integrating the above components, this paper proposes LMFPPO-UBP.

LMFPPO-UBP establishes a principled computational framework for analyzing cooperation and designing algorithms within spatially structured multi-agent systems. This paper makes the following contributions:
\begin{itemize}
	\item To our knowledge, this is the first integration of MFPPO into the SPGG domain, providing a scalable multi-agent learning framework for spatially structured strategic interactions.
	\item We introduce a novel LMF design within the MFPPO architecture, aligning each agent's perceptual scope with its immediate topological neighborhood.
	\item We propose an UBP mechanism that penalizes defectors based on local
	cooperative density, transforming fairness concerns into a cost-free algorithmic incentive for cooperators.
\end{itemize}
\section{Related Works}
Previous research on SPGG has demonstrated that localized interactions and network structures can facilitate cooperation through mechanisms such as spatial reciprocity and cluster formation \citep{NowakMay1992,SzaboFath2007}. In this context, the critical enhancement factor acts as a pivotal threshold that governs the emergence and stability of cooperation \citep{2008Public,SzolnokiPerc2009}. More recent studies have integrated environmental feedback and stochastic interactions, illustrating their considerable influence on cooperation thresholds and phase behavior \citep{Lv2025,2025Environmental,Ling_2025}. Furthermore, environmental feedback and bottom-up reputation learning further bolster cooperation in spatial settings \citep{tang_2024_cooperative,shen_2022_high,Ren2025}.

Punishment has been widely investigated as a means to sustain cooperation in social dilemmas. Laboratory experiments reveal that individuals are willing to bear costs to penalize defectors, resulting in elevated levels of cooperation \citep{FehrGachter2000,FehrGachter2002}. Theoretical models have embedded punishment within evolutionary game dynamics, underscoring its efficacy while also acknowledging challenges such as the second-order free-rider problem \citep{Sigmund2001,Boyd2003,HECKATHORN1989Collective}. In spatial contexts, punishment interacts strongly with local structure. Recent work introduces layered or hybrid updating mechanisms that markedly reduce cooperation thresholds compared to static punishment models \citep{CSF2025Punishment}. Targeted punishment strategies, which prioritize defectors according to local impact or reputation, exhibit improved efficiency in deterring invasion dynamics  \citep{2022Cooperation,2020Blocking}. Probabilistic punishment combined with reputation mechanisms further enhances robustness in stochastic environments    \citep{2025Cooperation}. Regarding algorithmic approaches, imitation learning offers a tractable approach for modeling strategic adaptation in multi-agent systems. Imitation-based rules, while computationally feasible, provide a limited representation of long-term strategic adaptation \citep{schuster_1983_replicator,szabo_1998_evolutionary}. Such mechanisms often rely on local stochastic updates and fail to capture the complex, path-dependent dynamics that characterize spatially structured games.   Reinforcement learning offers a methodologically grounded approach for enabling adaptive decision-making in contexts of social dilemma. Classical value-based methods such as Q-learning have been applied to these settings but encounter difficulties with convergence and stability \citep{Watkins1992,2021Multi}. These methods typically struggle to represent the asymmetric strategic interactions and long-term dependencies inherent in partially observable spatial games.

Deep reinforcement learning (DRL) extends these approaches by leveraging function approximation to model adaptive behavior in high-dimensional state spaces. Mean-field approximations enable scalable modeling of large populations by reducing many-agent interactions to tractable two-agent problems \citep{yang2020meanfieldmultiagentreinforcement}. Policy optimization methods, most notably Proximal Policy Optimization (PPO), have demonstrated notable stability and scalability in non-stationary environments \citep{Schulman2017}. Contemporary PPO variants, including curriculum-enhanced and adversarial PPO, significantly improve the emergence of cooperation in spatial games \citep{Yang2025PPOACT}. Moreover, constrained PPO frameworks integrate collective welfare objectives directly into policy updates, allowing decentralized agents to balance individual and group incentives \citep{YANG_2025_116928}. Applications of DRL to public goods games continue to expand, revealing intricate behavioral transitions and path-dependent dynamics not captured by traditional evolutionary rules \citep{KangNeighbor2025,2025CooperationLi,Kulkarni2024}. 

Current research has started to merge punishment mechanisms with DRL. However, most existing frameworks still rely on externally defined punishment rules, which restricts the ability of agents to learn and adapt sanctioning strategies independently \citep{2021Modular}. This limitation impedes their applicability in complex adaptive systems where norms must emerge endogenously \citep{9684494}.

\section{Model}
\label{sec:model}
\subsection{SPGG}

This study investigates the SPGG conducted on a two-dimensional periodic lattice system with dimensions \(L\times L\). Within this system, each agent occupies a node on a grid. Agents interact according to a von Neumann neighborhood structure, meaning each node is connected to exactly \(k=4\)  adjacent neighbors \citep{2021Rationality}. Each agent simultaneously participates in \(G=5\) public goods game groups, centered on itself and its four immediate neighbors, forming a locally overlapping game network.

The strategy set is defined as \(\mathcal{S}=\{C,D\}\). In this notation, $C$ represents a cooperator who contributes one unit of resource to the common pool, while $D$ represents a defector who contributes no resource. For any given game group \(g\), the payoff function for agent \(i\) is expressed as:

\begin{equation}\label{eq:base}
	\Pi(s_{i}^{g})=\begin{cases}\dfrac{r\cdot N_{C}^{g}}{G}-1,&s_{i}^{g}=C\\[8pt]
		\dfrac{r\cdot N_{C}^{g}}{G},&s_{i}^{g}=D\end{cases}
\end{equation}
where, \(N_{C}^{g}=\sum_{j\in g}\mathbb{I}(s_{j}^{g}=C)\) represents the total number of cooperators in group \(g\). \(r>1\) is the enhancement factor of the public good, reflecting the amplifying effect of cooperative behavior on collective payoff.

Since each agent engages in multiple game groups, its total payoff is the summation of payoffs from all associated groups:

\begin{equation}\label{eq:total_base}
	\Pi_{i}=\sum_{g\in G_{i}}\Pi(s_{i}^{g})
\end{equation}
where \(G_{i}\) denotes the set of all game groups in which agent \(i\) participates.

\subsection{LMFPPO}

To effectively capture the localized nature of strategic interactions in spatial games, we propose a LMF approach. This formulation shifts the basis of decision-making from a global population average to the average strategy of each agent's immediate neighbors. This local average delivers a contextually relevant signal that is firmly grounded in the spatial structure of interactions. Formally, for agent \(i\), the mean action of its neighbors is defined as:

\begin{equation}\label{eq:mean_action}
	\mu_{i}=\frac{1}{k}\sum_{j\in\mathcal{N}(i)}s_{j}
\end{equation}
where \(\mathcal{N}(i)\) denotes the neighbor set of agent \(i\), and \(s_{j}\in\{0,1\}\) represents the strategy of neighbor. Here, 0 represents defection, and 1 represents cooperation. This LMF representation \(\mu_{i}\) compactly represents the strategic pressure from the immediate social environment. 
At each time step \(t\), the observation of agent \(i\) is encoded as a four-dimensional feature vector:

\begin{equation}\label{eq:observation} \mathbf{x}_{t}^{i}=\begin{bmatrix}s_{t}^{i},&n_{t}^{i},&g_{t},&\mu_{t}^{i}\end{bmatrix}\in\mathbb{R}^{4}
\end{equation}
where \(s_{t}^{i}\in\{0,1\}\) is the agent's own current strategy, \(n_{t}^{i}\in\{0,1,2,3,4\}\) is the count of cooperators in its von Neumann neighborhood. \(g_{t}\in[0,1]\) is the global cooperation frequency. The LMF representation \(\mu_{t}^{i}\in[0,1]\) is embedded into the agent's state representation and processed by the shared encoder to inform both the policy and value networks. It enables the agent to perceive local cooperative pressure, thereby guiding action selection and value estimation within the LMFPPO framework.

The objective function for the agent is formulated as:

\begin{equation}\label{eq:objective}
	\mathcal{L}^{\text{LMFPPO}}(\theta)=\mathcal{L}^{\text{CLIP}}(\theta)+\delta \mathcal{L}^{\text{VF}}(\theta)-\rho\mathcal{L}^{\text{ENT}}(\theta)
\end{equation}
where \(\mathcal{L}^{\text{CLIP}}(\theta)\) is the PPO clipped surrogate objective, which prevents excessively large policy updates by constraining the probability ratio \( r_t(\theta) = \pi_\theta(a_t \mid \mathbf{x}_t) / \pi_{\theta_\text{old}}(a_t \mid \mathbf{x}_t) \):

\begin{equation}\label{eq:clipped}
	\mathcal{L}^{\text{CLIP}}(\theta)=\mathbb{E}_{t}\left[\min\left(r_{t}(\theta)\hat{A}_{t},\operatorname{clip}(r_{t}(\theta),1-\epsilon,1+\epsilon)\hat{A}_{t}\right)\right]
\end{equation}\(\mathcal{L}^{\text{VF}}(\theta)\) is the value function loss, typically the mean-squared error between the predicted value and the estimated return:

\begin{equation}\label{eq:loss}
	\mathcal{L}^{\text{VF}}(\theta)=\mathbb{E}_{t}\left[(V_{\phi}(\mathbf{x}_{t})-R_{t})^{2}\right]
\end{equation}\(\mathcal{L}^{\text{ENT}}(\theta)\) is an entropy bonus term that encourages exploration:

\begin{equation}\label{eq:entropy_bonus}
	\mathcal{L}^{\text{ENT}}(\theta)=\mathbb{E}_{t}\left[-\sum_{a\in\mathcal{A}}\pi_{\theta}(a|\mathbf{x}_{t})\log\pi_{\theta}(a|\mathbf{x}_{t})\right]
\end{equation}with \(\delta\) and \(\rho\) as coefficients that balance the importance of the value function loss and the entropy regularization, respectively.

The advantage estimate \(\hat{A}_{t}\), which is crucial for updating the policy, is computed using generalized advantage estimation. This calculation is performed over trajectories that integrate both the LMF augmented state and the UBP reward:

\begin{equation}\label{eq:advantage_estimate}
	\hat{A}_{t}=\sum_{l=0}^{T-t-1}(\gamma\lambda)^{l}\delta_{t+l}
\end{equation}where

\begin{equation}\label{eq:reward}
	\delta_{t+l}=R_{t+l}+\gamma V(\mathbf{x}_{t+l+1})-V(\mathbf{x}_{t+l})
\end{equation}

\subsection{LMFPPO-UBP}

To promote the evolution of cooperation, we introduce a cost-free punishment mechanism grounded in the theory of unbalanced interactions. This theory posits that defectors create social imbalance by undermining collective welfare and should consequently incur penalties that do not impose additional costs on cooperative agents \citep{2014The,2000ERC,1997Altruistic}. Unlike reward shaping methods that alter internal preferences, our UBP mechanism is formulated as an institutional constraint. While this mechanism operates as an external rule within the public goods game, its theoretical justification rests on the endogenous loss of social standing experienced by defectors in structured societies. This approach reconciles systemic sanctions with the concept of institutional capital by ensuring that punishment does not represent a private expenditure. Drawing on inequity aversion theory \citep{1999A}, the UBP mechanism penalizes defectors in proportion to the contributions of their neighbors. The local density of cooperators functions as a proxy for the strength of social norms and facilitates the penalization of deviants without requiring explicit effort from individual agents. The punishment reward for agent $i$ is defined as follows:

\begin{equation}\label{eq:punishment}
	R^{i}_{\text{punish}}=-p\cdot\mathbb{I}(s_{i}=D)\cdot N^{\text{neigh}}_{C}(i)
\end{equation}where \(p > 0\) represents the punishment strength. The term \(N^{\text{neigh}}_{C}(i)\) denotes the count of cooperating neighbors within agent \(i\)’s von Neumann neighborhood. Additionally, \(\mathbb{I}(s_{i} = D)\) is an indicator function that takes the value 1 if agent \(i\) adopts the defector strategy and 0 otherwise.

The total reward for an agent is then the sum of its base game payoff and the punishment reward:

\begin{equation}\label{eq:toral_reward}
	R^{i}_{\text{total}}=\Pi_{i}+R^{i}_{\text{punish}}
\end{equation}

This mechanism selectively penalizes defectors in proportion to their local disruptive impact, specifically the number of cooperating neighbors they have. By doing so, it effectively increases the relative cost of defection within cooperative neighborhoods while leaving the payoff of cooperators unchanged.

\subsection{Actor–Critic Network}

Figure~\ref{fig:AC} shows the actor–critic
network architecture. This feature vector is processed by a shared encoder network with ReLU activations\citep{pmlr-v15-glorot11a}:
\begin{figure}[htbp!]
	\centering
	\includegraphics[width=\linewidth]{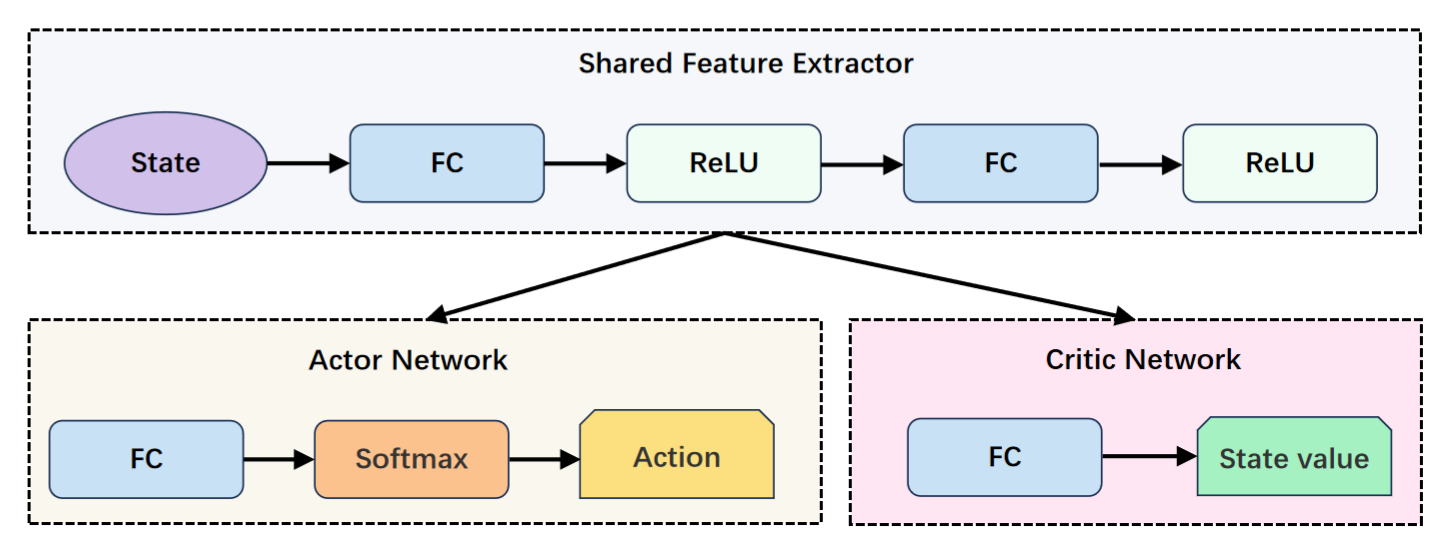}
	\caption{The architecture of actor-critic network.}
	\label{fig:AC}
\end{figure}

\begin{equation}\label{eq:RELU}
	\mathbf{h}_{t}^{i}=\sigma(\mathbf{W}_{2}\cdot\sigma(\mathbf{W}_{1}\mathbf{x}_{t}^{i}+\mathbf{b}_{1})+\mathbf{b}_{2})
\end{equation}
where \(\mathbf{x}_t^i\) is the agent's observation defined in Eq.~\ref{eq:observation}, \(\sigma(\cdot)\) denotes the ReLU activation function, \(\mathbf{W}_{1},\mathbf{W}_{2}\) are weight matrices, and \(\mathbf{b}_{1},\mathbf{b}_{2}\) are bias terms. The actor network outputs a categorical distribution over actions via a softmax layer:

\begin{equation}\label{eq:softmax}
	\pi_{\theta}(a_{t}^{i}|\mathbf{x}_{t}^{i})=\text{softmax}(\mathbf{W}_{a}\mathbf{h}_{t}^{i}+\mathbf{b}_{a})
\end{equation}

The critic network estimates the state value:

\begin{equation}\label{eq:critic_network}
	V_{\phi}(\mathbf{x}_{t}^{i})=\mathbf{w}_{v}^{\top}\mathbf{h}_{t}^{i}+b_{v}
\end{equation}

The proposed LMFPPO‑UBP framework is described as Algorithm~\ref{alg:LMFPPO-ubp}, which integrates LMF perception with an UBP mechanism in a spatially structured environment.
\begin{algorithm}[t]
	\caption{LMFPPO-UBP Framework for SPGG}
	\label{alg:LMFPPO-ubp}
	\begin{algorithmic}[1]
		\State \textbf{Initialize:} Policy network \(\pi_{\theta}\), value network \(V_{\phi}\), punishment strength \(p\), hyperparameters \(\delta, \rho, \gamma, \lambda, \epsilon\).
		\For{epoch = 1 to \(T\)}
		\For{each agent \(i\) in the grid}
		\State Observe state \(\mathbf{x}_t^i = [s_t^i, n_t^i, g_t, \mu_t^i]\)(Eq.~\ref{eq:observation})
		\State Sample action \(a_t^i \sim \pi_\theta(\cdot | \mathbf{x}_t^i)\)
		\State Execute \(a_t^i\), observe next state \(\mathbf{x}_{t+1}^i\)
		\State Calculate base reward \(\Pi_i\) (Eq.~\ref{eq:total_base})  
		\State Calculate punishment reward \(R_{\text{punish}}^i\) (Eq.~\ref{eq:punishment})
		\State Compute total reward \(R_{\text{total}}^i = \Pi_i + R_{\text{punish}}^i\) (Eq.~\ref{eq:toral_reward})
		\EndFor
		\State Compute advantages \(\hat{A}_t\) and returns \(R_{\mathrm{total}}^{i}\)
		\For{iteration = 1 to \(K\) (LMFPPO epochs)}
		\State Update \(\theta\) by maximizing \(\mathcal{L}^{\text{LMFPPO}}(\theta)\) (Eq.~\ref{eq:objective})
		\State Update \(\phi\) by minimizing \(\mathcal{L}^{\text{VF}}(\theta)\) (Eq.~\ref{eq:loss})
		\EndFor
		\EndFor
	\end{algorithmic}
\end{algorithm}
\section{Experimental Results}
\label{sec:exp}
\subsection{Experimental Setup}

All models were implemented on a periodic spatial grid of dimensions $L \times L = 200 \times 200$. The PPO and LMFPPO algorithms employed a learning rate $\alpha = 1 \times 10^{-3}$, a discount factor $\gamma = 0.99$, a clipping threshold $\epsilon = 0.2$, and a generalized advantage estimation parameter $\lambda = 0.95$. An entropy regularization coefficient $\rho = 0.001$. Training proceeded for up to $T = 1000$ iterations using the Adam optimizer with StepLR learning rate scheduling\citep{2014Adam}. The punishment strength parameter $p$ in the UBP mechanism was set to a baseline value of $0.5$.

\subsection{LMFPPO Hyperparameter Sensitivity Analysis}

\begin{figure}[htbp!]
	\centering
	\includegraphics[width=\linewidth]{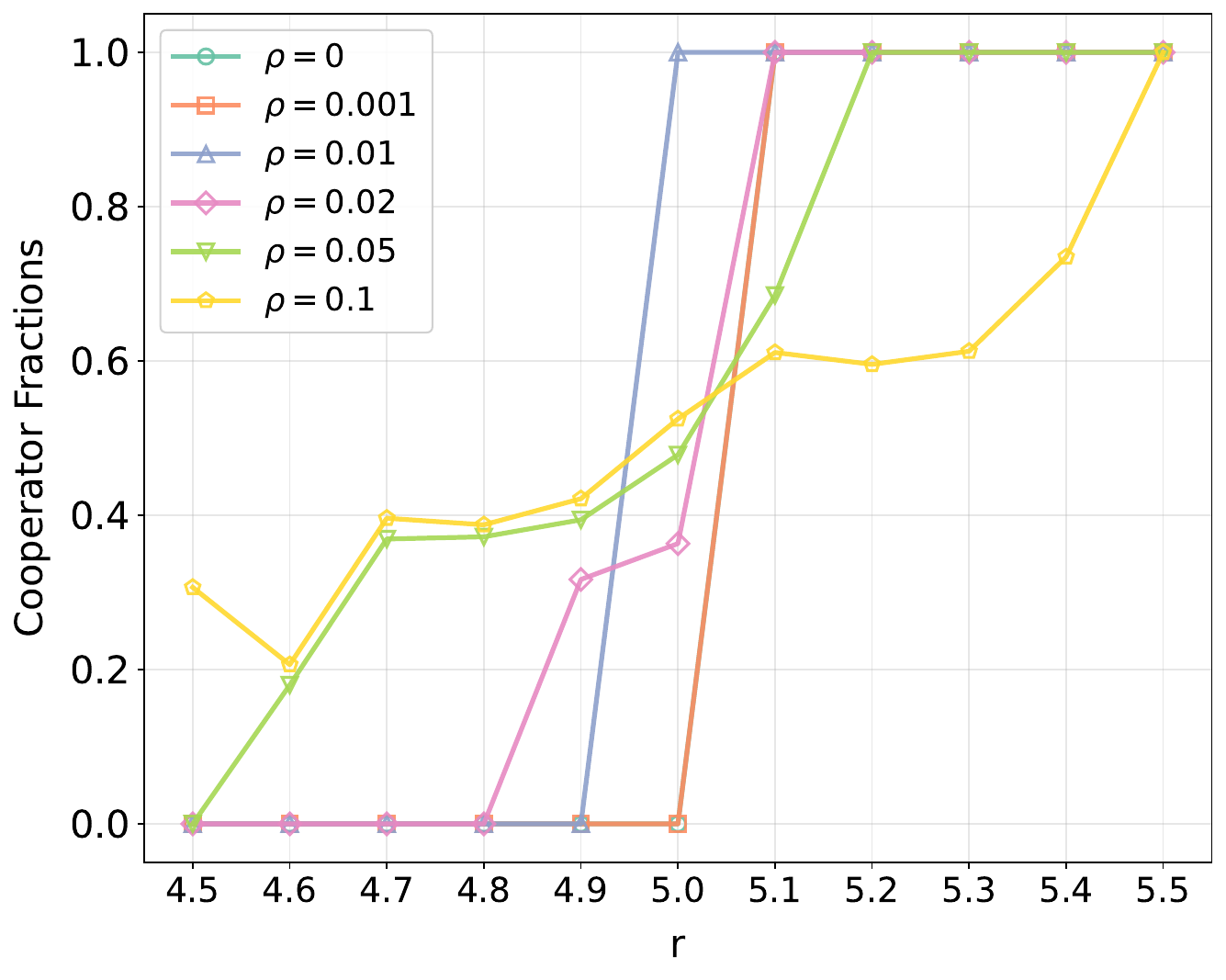}
	\caption{The entropy regularization coefficient $\rho$ significantly affects LMFPPO performance. Empirically, 
		$\rho$=0.01 is optimal. Higher $\rho$ destabilizes gradient trajectories and value function estimation, impairing convergence.}
	\label{fig:rho_sensitivity}
\end{figure}

To determine the optimal entropy weight $\rho$, a sensitivity analysis was conducted. This hyperparameter balances exploration and exploitation, with lower values favoring exploitation of known strategies and higher values encouraging exploration. The goal was to identify a value promoting cooperative evolution without disrupting learning. A systematic hyperparameter sensitivity analysis was performed to evaluate the entropy weight $\rho$. As shown in  Figure~\ref{fig:rho_sensitivity}, six distinct values $\rho = 0$, $0.001$, $0.01$, $0.02$, $0.05$, and $0.1$ were tested across a range of enhancement factors $r$ from $4.5$ to $5.5$. Each configuration was executed multiple times to establish statistical reliability.

Experimental results indicated a non-monotonic effect of $\rho$ near the critical point $r=5.0$. Theoretically, smaller values like $0.001$ should perform better by focusing on exploitation. However, $\rho=0.01$ achieved full cooperation at $r=5.0$, whereas $\rho=0.001$ led to complete defection. This occurs because near the critical threshold, cooperative and defective strategies exist in an unstable equilibrium, making the system highly sensitive to perturbations. Moderate exploration with $\rho=0.01$ helps escape local optima and discover cooperative equilibria, while overly conservative exploration with $\rho=0.001$ traps the system in defective states. Additionally, increasing $\rho$ shifts the cooperation threshold leftward. For example, with $\rho=0.05$, cooperation emerges at $r=4.6$, and with $\rho=0.1$, it sustains at $r=4.5$. This suggests moderate exploration facilitates cooperation under stricter conditions. Given the discrete strategy space and threshold nature of SPGG, moderate exploration proves crucial. Therefore, $\rho=0.01$ was selected for subsequent experiments, as it demonstrated stable performance at the critical threshold $r=5.0$ and maintained good performance across a broader range of $r$ values.

\subsection{LMFPPO-UBP Hyperparameter Sensitivity Analysis}
\begin{figure}[htbp!]
	\centering
	\includegraphics[width=\linewidth]{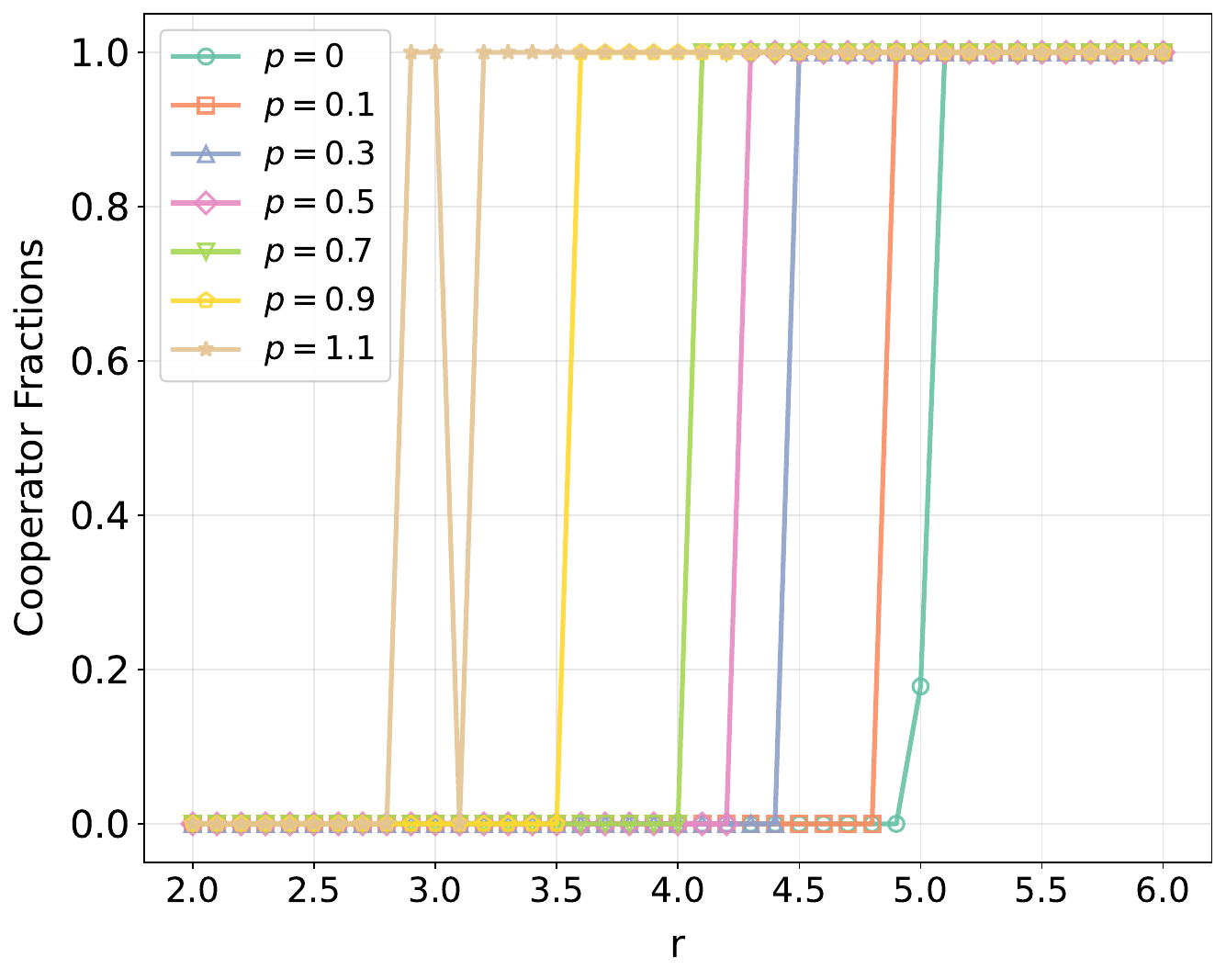}
	\caption{Impact of punishment strength $p$ on LMFPPO-UBP.}
	\label{fig:punishment_sensitivity}
\end{figure}

Following the hyperparameter sensitivity analysis for the LMFPPO algorithm, the entropy weight $\rho$ was set to $0.01$. A systematic sensitivity study of the punishment intensity parameter $p$ was then performed under the LMFPPO-UBP framework, as shown in Figure~\ref{fig:punishment_sensitivity}, to assess how punishment affects cooperative evolution. Seven punishment strengths were evaluated, ranging from $p = 0$ to $p = 1.1$, across enhancement factors $r$ between $2.0$ and $6.0$. 

Introducing punishment significantly improved cooperation at lower $r$ values. Without punishment, cooperation began only at $r \geq 5.0$, reaching just average cooperator fraction $17.81\%$ at $r = 5.0$. As $p$ increased, the cooperation threshold shifted leftward. For instance, $p = 0.1$ lowered the threshold to $r = 4.9$, while $p = 0.3$ reduced it to $r = 4.5$. At $p = 0.7$, the threshold dropped to $r = 4.1$, and full cooperation was achieved at $r = 3.6$ with $p = 0.9$. Moderate punishment levels between $p = 0.3$ and $p = 0.7$ produced stable and smooth cooperative transitions, indicating that well-calibrated punishment can foster cooperation without causing excessive behavioral fluctuations. Conversely, overly high punishment such as $p = 1.1$ led to irregular system behavior. Although full cooperation appeared at $r = 3.0$, it collapsed completely at $r = 3.1$ before recovering at higher $r$ values. This suggests that excessive punishment distorts payoff structures and can trap the system in local optima. Based on these findings, $p = 0.5$ was chosen as the default punishment intensity. This value balances cooperation promotion, lowering the threshold from $r = 5.0$ to $r = 4.3$, while maintaining system stability and avoiding anomalies caused by excessive punishment. This choice aligns with imbalance theory, where moderate punishment effectively curbs defection without disrupting the system's dynamic equilibrium.

\subsection{Statistical analysis of LMFPPO-UBP}

\begin{figure*}[h]
	\centering
	\begin{minipage}{\linewidth}
		\centering
		\includegraphics[width=0.7\linewidth]{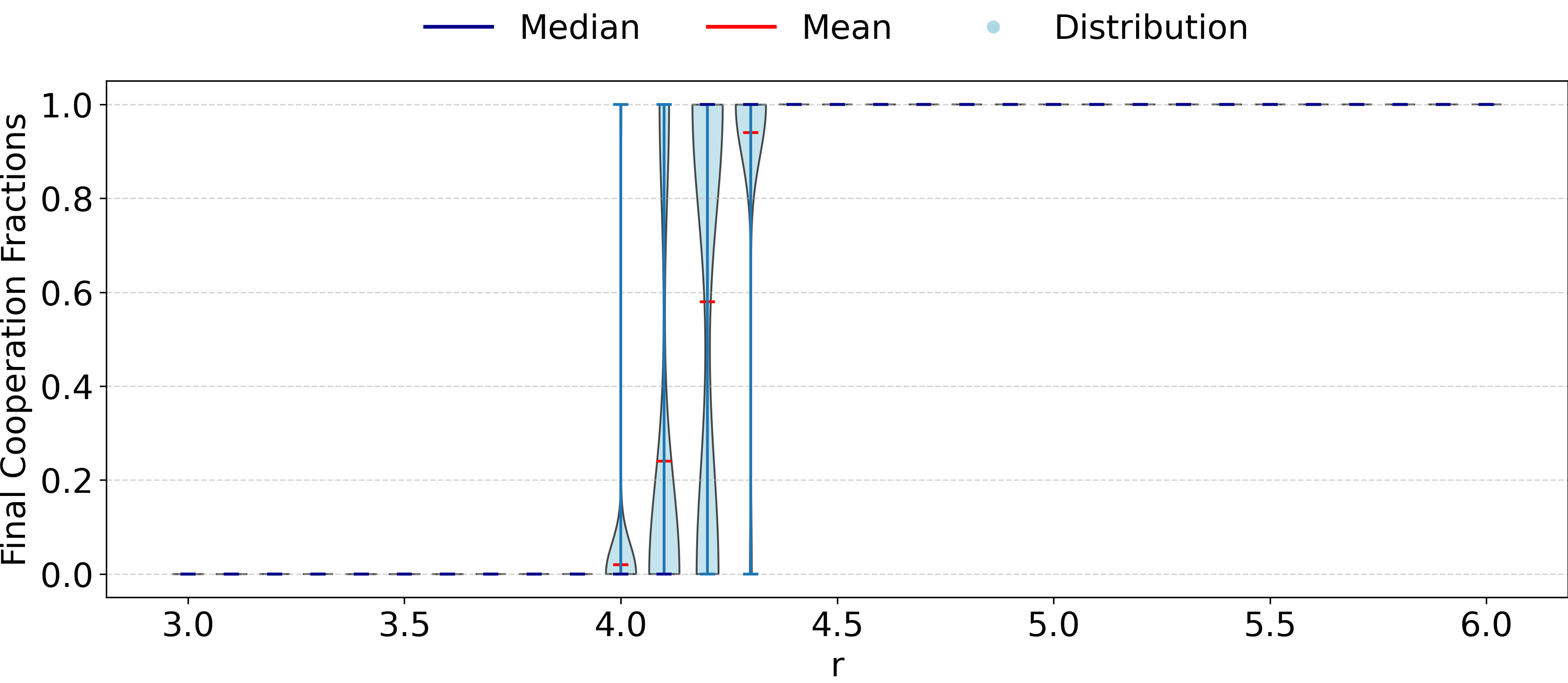}\\
		{\footnotesize (a) LMFPPO-UBP}
	\end{minipage}
	\\[3mm]
	\begin{minipage}{\linewidth}
		\centering
		\includegraphics[width=0.7\linewidth]{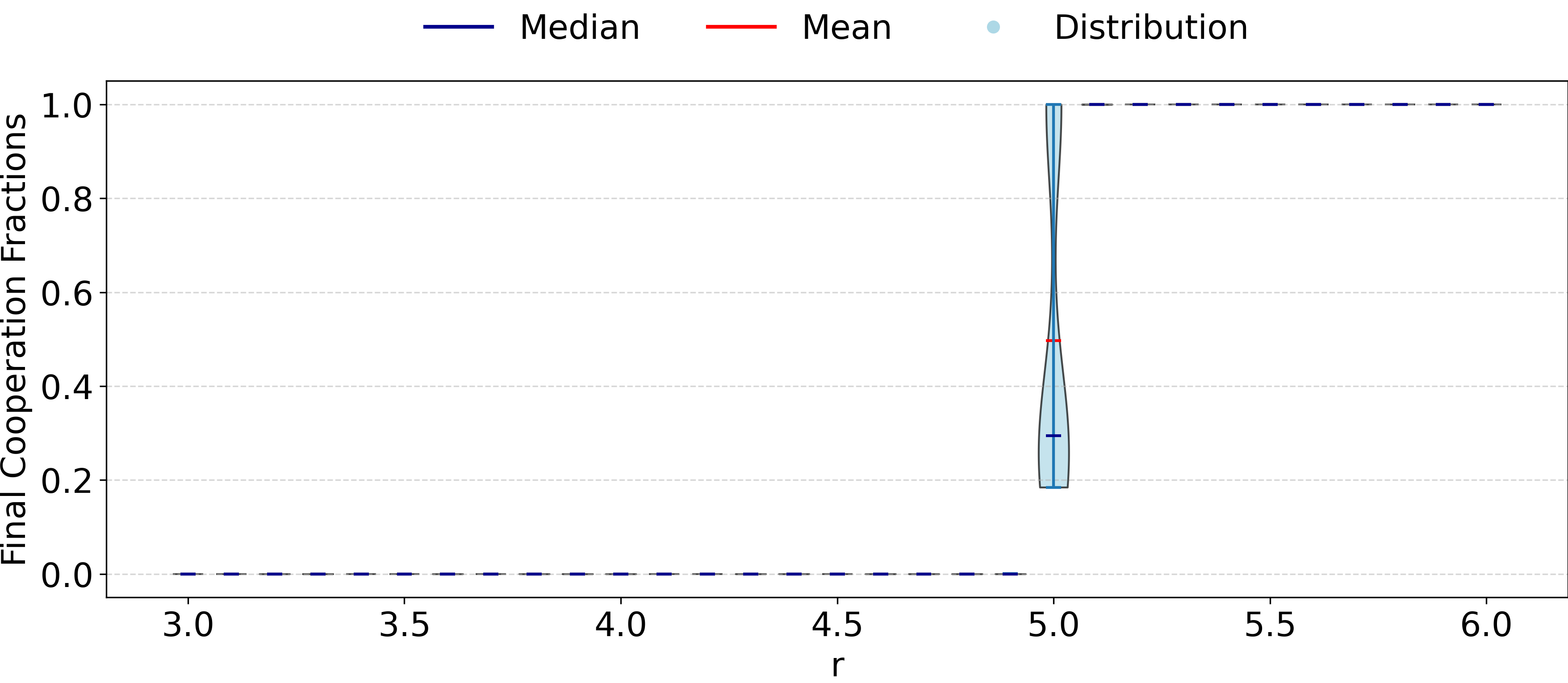}\\
		{\footnotesize (b) LMFPPO}
	\end{minipage}
	\begin{minipage}{\linewidth}
		\centering
		\includegraphics[width=0.7\linewidth]{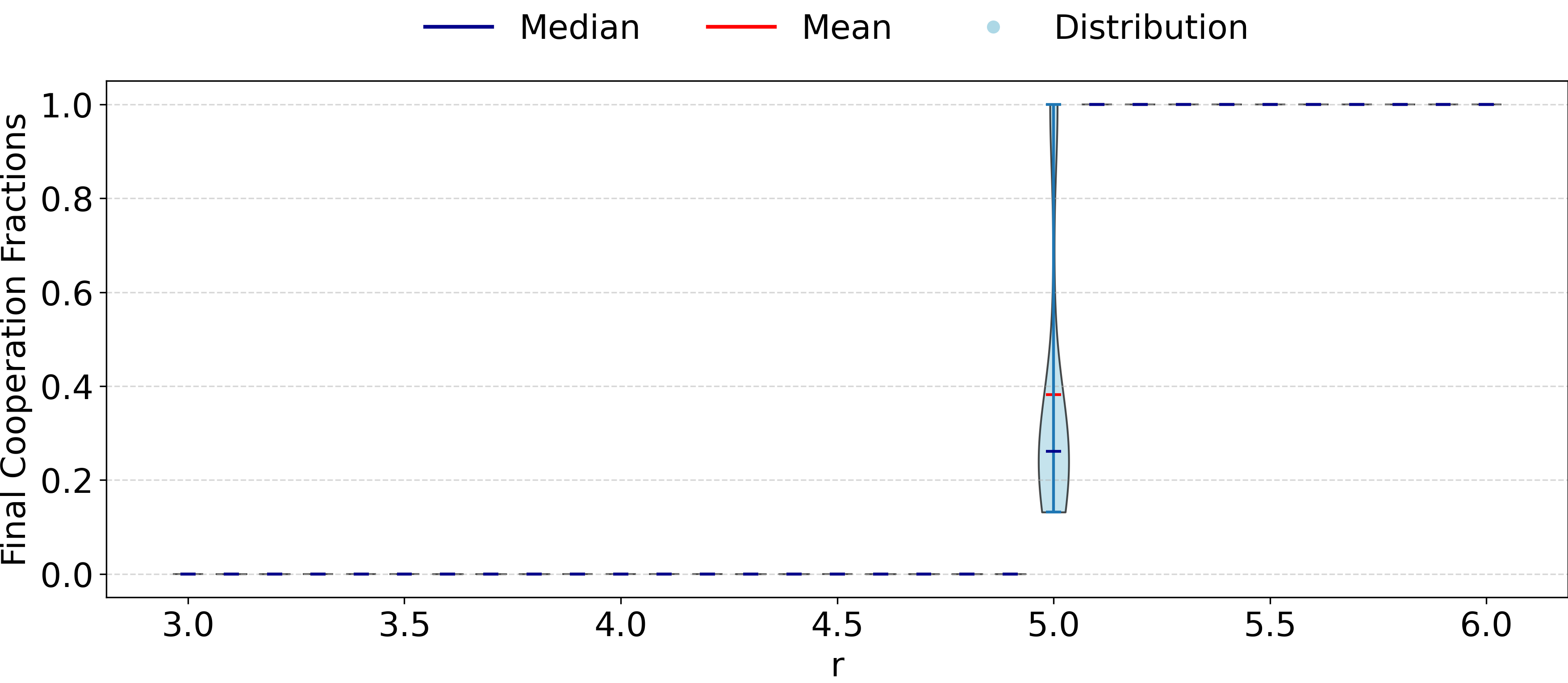}\\
		{\footnotesize (c) PPO}
	\end{minipage}
	\caption{Violin plots of final cooperation fractions from 50 trials for three algorithms. LMFPPO-UBP shows a sharp, left-shifted transition, while LMFPPO and PPO exhibit transitions only near $r=5.0$.}
	\label{fig:LMFPPO_UBP_vio}
\end{figure*}

\begin{table*}[h]
	\centering
	\footnotesize
	\caption{$95\%$ confidence intervals comparison for cooperation fractions}
	\label{tab:CI_comparison}
	\resizebox{\textwidth}{!}{ 
		\begin{tabular}{@{}c*{7}{S[table-format=1.2]@{\,--\,}S[table-format=1.2]}@{}}
			\toprule
			r & \multicolumn{2}{c}{3.5} & \multicolumn{2}{c}{3.6} & \multicolumn{2}{c}{3.7} & \multicolumn{2}{c}{3.8} & \multicolumn{2}{c}{3.9} & \multicolumn{2}{c}{4.0} & \multicolumn{2}{c}{4.1} \\
			\cmidrule(lr){2-3} \cmidrule(lr){4-5} \cmidrule(lr){6-7} \cmidrule(lr){8-9} \cmidrule(lr){10-11} \cmidrule(lr){12-13} \cmidrule(lr){14-15}
			LMFPPO-UBP & 0.00 & 0.00 & 0.00 & 0.00 & NaN & NaN & NaN & NaN & NaN & NaN & 0.00 & 0.06 &  0.12 & 0.36 \\
			LMFPPO & NaN & NaN & 0.00 & 0.00 & NaN & NaN  & 0.00 & 0.00  & 0.00 & 0.00 & 0.00 & 0.00  & NaN & NaN \\
			PPO & NaN & NaN & 0.00 & 0.00  & 0.00 & 0.00  & 0.00 & 0.00 & NaN & NaN & NaN & NaN & NaN & NaN \\
			\midrule
			
			r & \multicolumn{2}{c}{4.2} & \multicolumn{2}{c}{4.3} & \multicolumn{2}{c}{4.4} & \multicolumn{2}{c}{4.5} & \multicolumn{2}{c}{4.6} & \multicolumn{2}{c}{4.7} & \multicolumn{2}{c}{4.8} \\
			\cmidrule(lr){2-3} \cmidrule(lr){4-5} \cmidrule(lr){6-7} \cmidrule(lr){8-9} \cmidrule(lr){10-11} \cmidrule(lr){12-13} \cmidrule(lr){14-15}
			LMFPPO-UBP &  0.44 & 0.72 &  0.87 & 1.00 &  NaN & NaN &  NaN & NaN & NaN & NaN & NaN & NaN & NaN & NaN \\
			LMFPPO &NaN & NaN  & 0.00 & 0.00  & 0.00 & 0.00 & NaN & NaN & 0.00 & 0.00  & 0.00 & 0.00  & 0.00 & 0.00 \\
			PPO & NaN & NaN & 0.00 & 0.00  & 0.00 & 0.00  & NaN & NaN & 0.00 & 0.00 & 0.00 & 0.00 & 0.00 & 0.00 \\
			\midrule
			
			r & \multicolumn{2}{c}{4.9} & \multicolumn{2}{c}{5.0} & \multicolumn{2}{c}{5.1} & \multicolumn{2}{c}{5.2} & \multicolumn{2}{c}{5.3} & \multicolumn{2}{c}{5.4} & \multicolumn{2}{c}{5.5} \\
			\cmidrule(lr){2-3} \cmidrule(lr){4-5} \cmidrule(lr){6-7} \cmidrule(lr){8-9} \cmidrule(lr){10-11} \cmidrule(lr){12-13} \cmidrule(lr){14-15}
			LMFPPO-UBP & NaN & NaN & NaN & NaN & NaN & NaN & NaN & NaN & NaN & NaN& NaN & NaN & NaN & NaN \\
			LMFPPO & 0.00 & 0.00 & 0.40 & 0.60  & 1.00 & 1.00  & NaN & NaN & NaN & NaN & NaN & NaN & NaN & NaN \\
			PPO & 0.00 & 0.00 & 0.30 & 0.47  & NaN & NaN  & NaN & NaN & NaN & NaN & NaN & NaN & NaN & NaN \\
			\bottomrule
		\end{tabular}
	}
\end{table*}

\begin{figure}[h]
	\begin{minipage}{\linewidth}
		\centering
		\includegraphics[width=\linewidth]{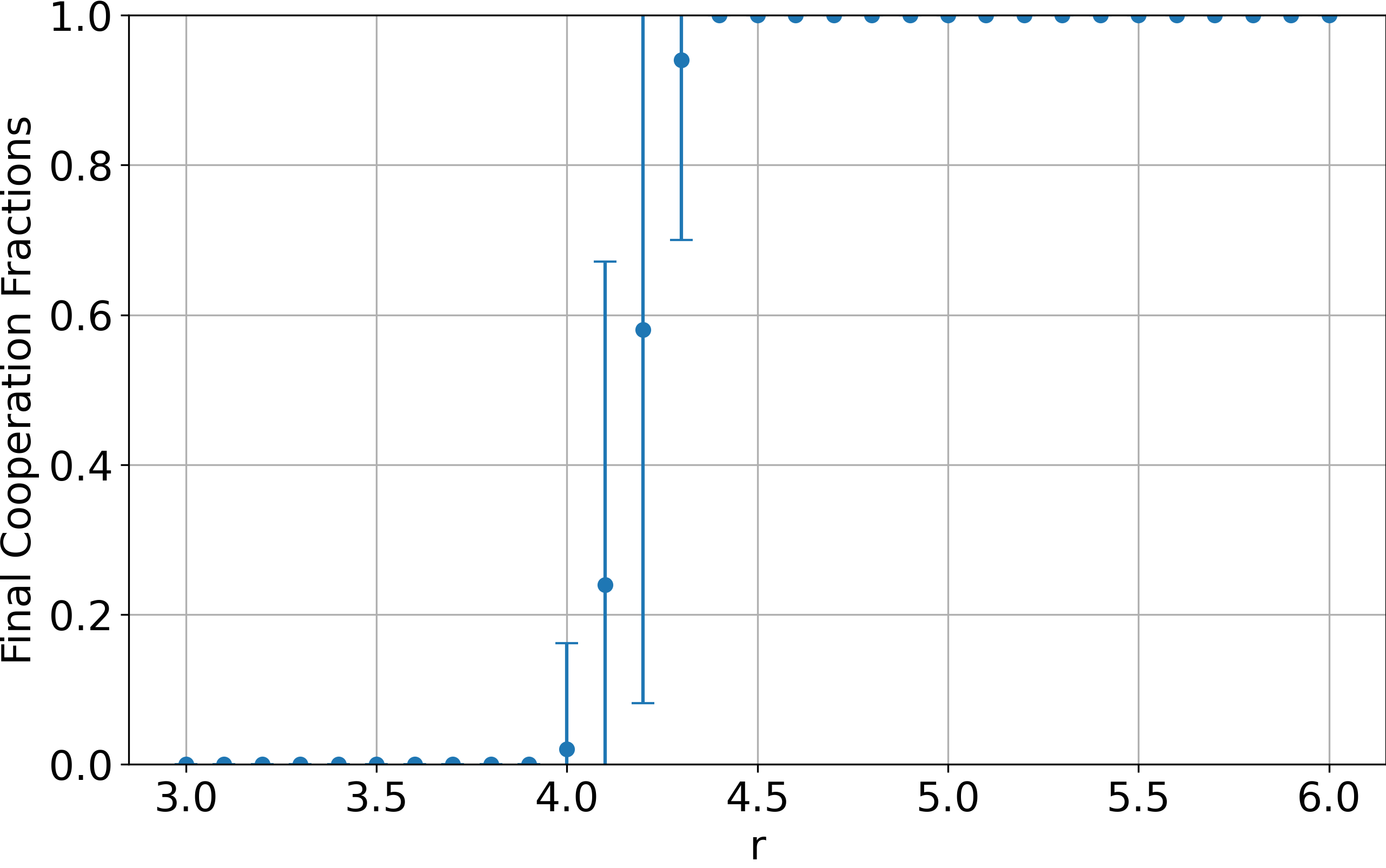}\\
		{\footnotesize (a) LMFPPO-UBP}
	\end{minipage}
	\\[3mm]
	\begin{minipage}{\linewidth}
		\centering
		\includegraphics[width=\linewidth]{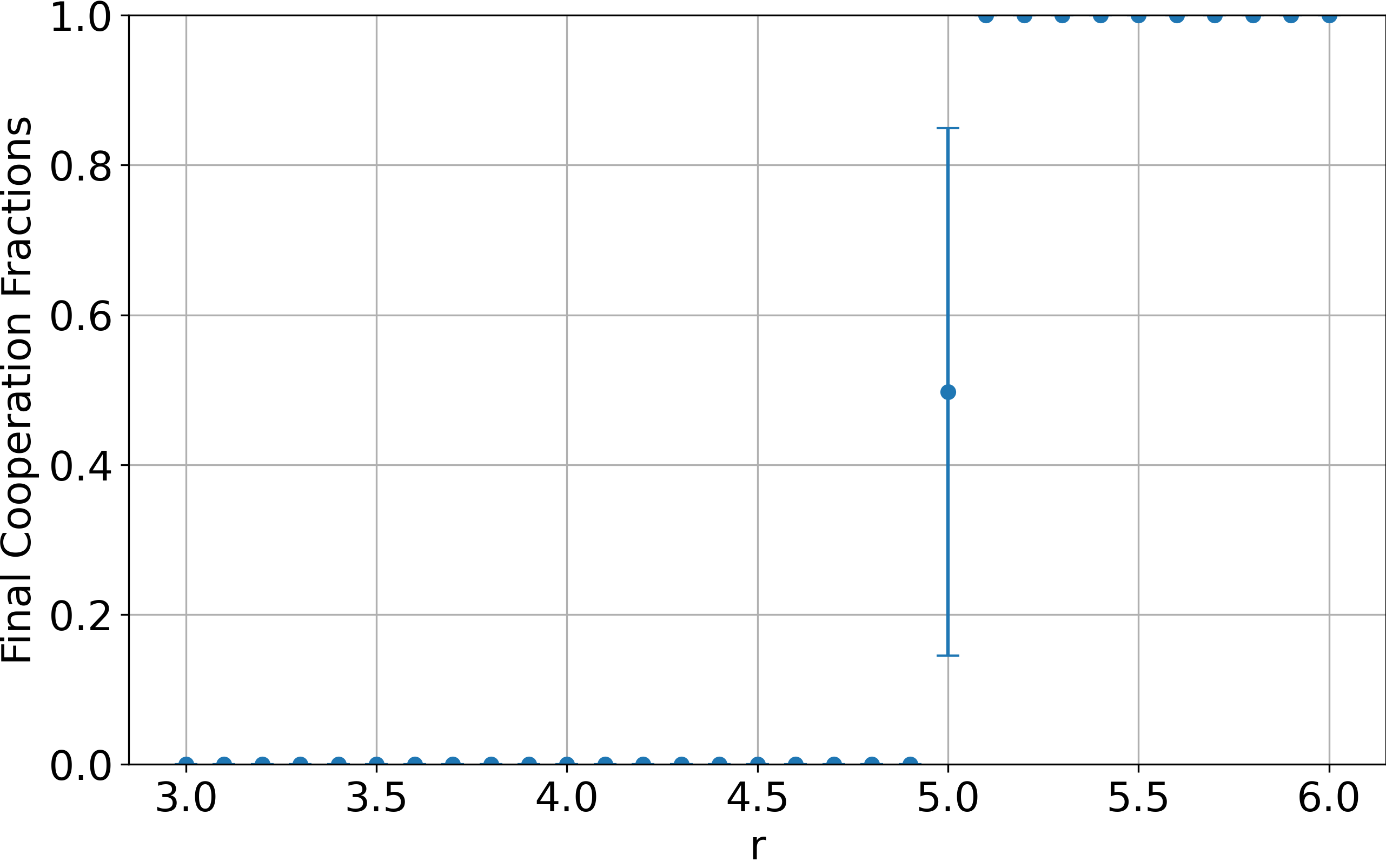}\\
		{\footnotesize (b) LMFPPO}
	\end{minipage}
	\\[3mm]
	\begin{minipage}{\linewidth}
		\centering
		\includegraphics[width=\linewidth]{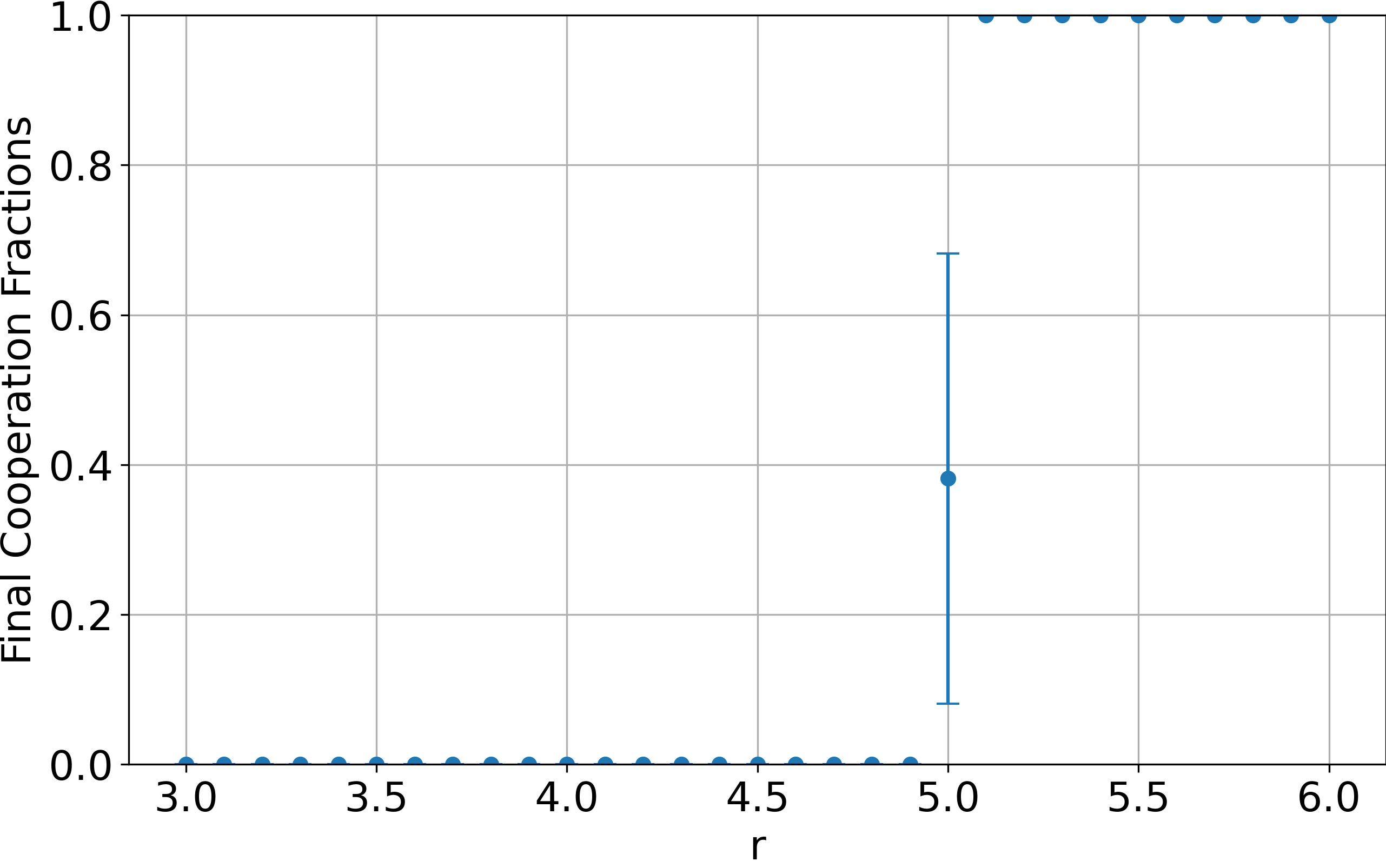}\\
		{\footnotesize (c) PPO}
	\end{minipage}
	\caption{Comparison of cooperation fractions from 50 trials for (a) LMFPPO-UBP, (b) LMFPPO, and (c) PPO. Error bars indicate means and standard deviations. LMFPPO-UBP shows a sharp transition at $r\in[4.0,4.3]$, whereas LMFPPO and PPO exhibit delayed, variable transitions only near $r=5.0$.}
	\label{fig:LMFPPO_UBP_r_stat_err}
\end{figure}

To rigorously evaluate the performance and robustness of the proposed LMFPPO-UBP framework, a comprehensive statistical analysis was conducted across 50 independent trials under varying enhancement factors \( r \). The analysis employs violin plots, 95\% confidence intervals, and error bar comparisons to illustrate the distribution and reliability of final cooperation fractions.

Figure~\ref{fig:LMFPPO_UBP_vio} presents violin plots for LMFPPO-UBP, LMFPPO, and PPO. The plot for LMFPPO-UBP demonstrates a sharp, first-order-like transition from full defection to full cooperation. This transition occurs within a narrow interval between \( r = 4.0 \) and \( r = 4.3 \). In contrast, the transitions for LMFPPO and PPO are significantly delayed and more variable, occurring only near the theoretical threshold of \( r = 5.0 \). The compact and left-shifted distribution for LMFPPO-UBP visually confirms its effectiveness in lowering the critical enhancement factor required for cooperation. The numerical stability of these results is further substantiated by 95\% confidence intervals, as summarized in Table~\ref{tab:CI_comparison}. The confidence intervals denoted as 'NaN' in the table correspond to cases where agents achieved absolute consensus through total defection or full cooperation. In these scenarios, the absence of variance makes the calculation of standard confidence intervals mathematically inapplicable. For LMFPPO-UBP, the confidence intervals for the cooperation fraction remain at zero below \( r = 4.0 \). They then expand through a transitional phase between \( r = 4.1 \) and \( r = 4.3 \), before consistently reaching unity for \( r \geq 4.4 \). This pattern indicates a well-defined and reproducible phase transition. Both LMFPPO and PPO maintain a cooperation fraction of zero until \( r \) approaches 5.0, where their confidence intervals abruptly shift to one, reflecting the inherent instability and higher threshold associated with pure social learning mechanisms. Error bar comparisons, shown in Figure~\ref{fig:LMFPPO_UBP_r_stat_err}, reinforce these findings. LMFPPO-UBP exhibits a rapid and decisive increase in mean cooperation, accompanied by minimal variance once the threshold is crossed. LMFPPO shows a comparable but delayed transition at \( r = 5.0 \), yet with greater variability in outcomes, particularly near the critical point. PPO demonstrates a similar delayed transition but with the highest variance among the three algorithms, indicating less consistent convergence.

A key observation is the comparative performance between LMFPPO and PPO at the critical threshold. While both algorithms undergo a cooperative phase transition near \( r = 5.0 \), LMFPPO achieves a substantially higher mean cooperation fraction. This indicates that the LMF perception provides a more effective signal for exploiting cooperative opportunities under marginal conditions. Notably, this performance gain is accompanied by a broader distribution of outcomes, as reflected in a larger variance compared to PPO. The localised learning process in LMFPPO yields a higher average cooperation rate but also introduces greater outcome variability. This reflects the system's heightened sensitivity to stochastic exploration and local strategic configurations near the critical phase transition point.
In summary, the statistical analysis conclusively demonstrates that the LMFPPO-UBP framework induces a sharp, low-threshold transition to cooperation with high reliability. It also confirms that the LMF component within LMFPPO provides a measurable performance improvement over the global perspective of PPO, yielding higher and more stable cooperation at the theoretical critical point.

\subsection{Comparative Analysis of Algorithms}
\begin{figure*}[htbp!]
	\begin{minipage}{\linewidth}
		\begin{minipage}{0.24\linewidth}
			\centering
			\includegraphics[width=\linewidth]{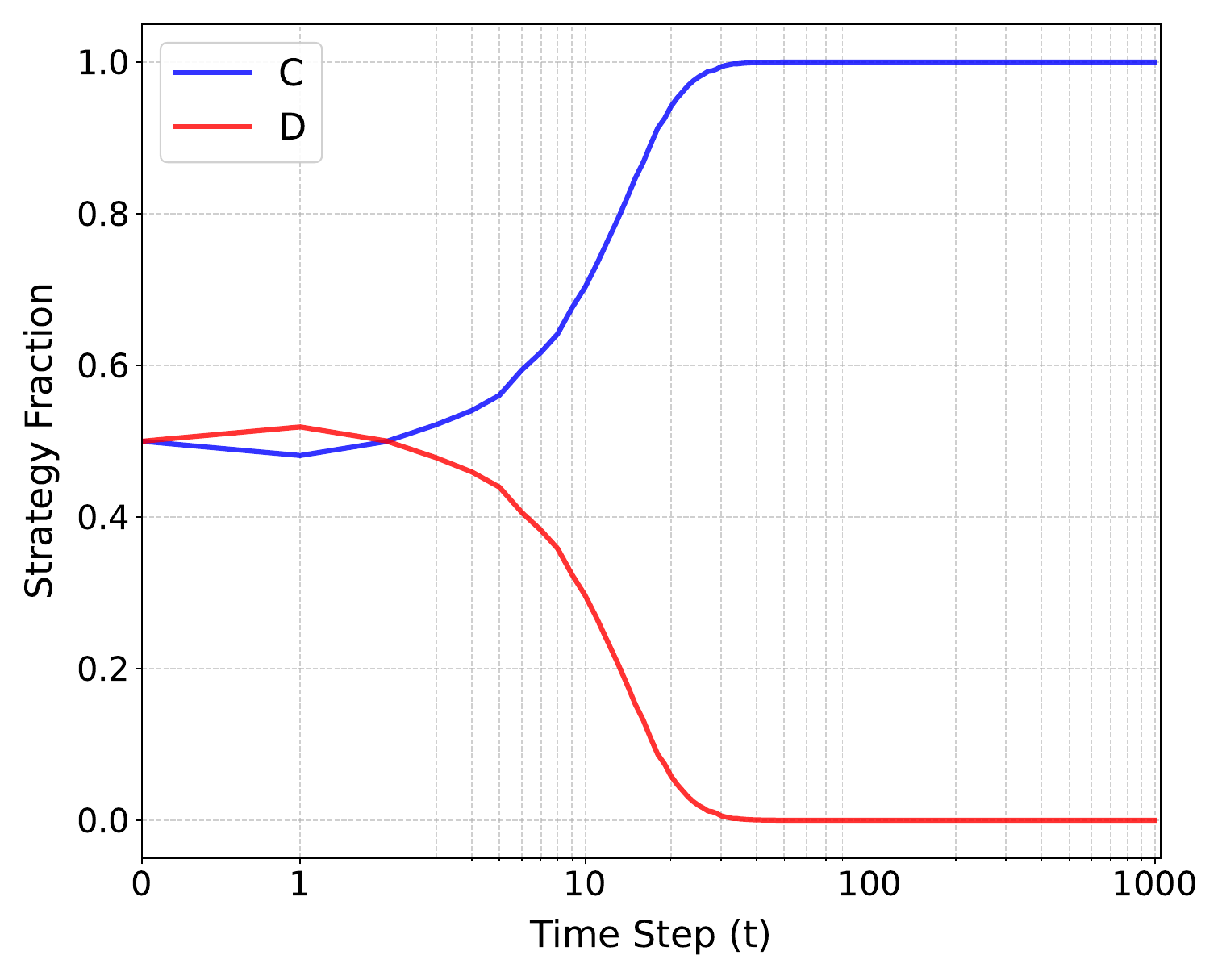}\\
		\end{minipage}
		\begin{minipage}{0.14\linewidth}
			\centering
			\includegraphics[width=\linewidth]{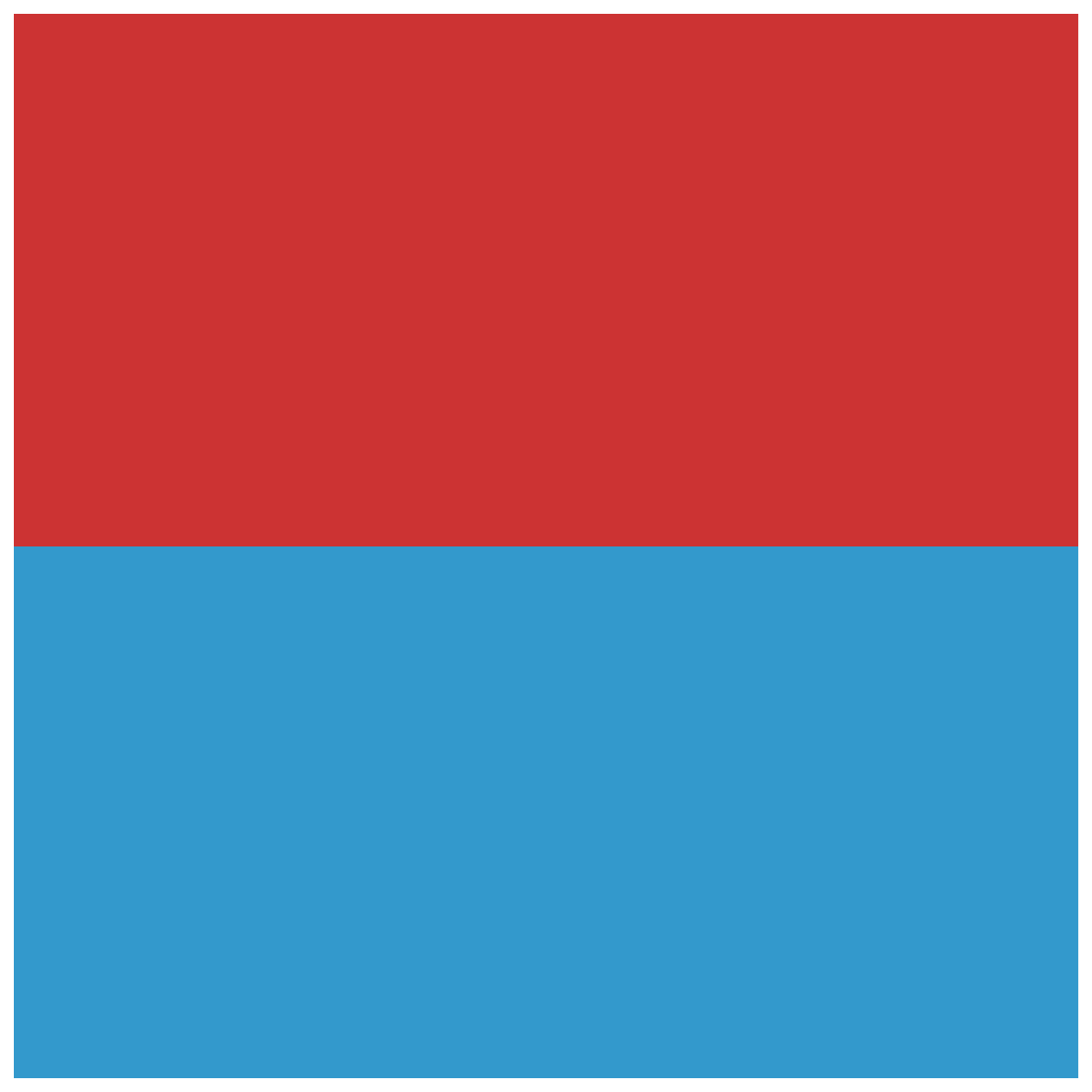}\\
			{\footnotesize t=0}
		\end{minipage}
		\begin{minipage}{0.14\linewidth}
			\centering
			\includegraphics[width=\linewidth]{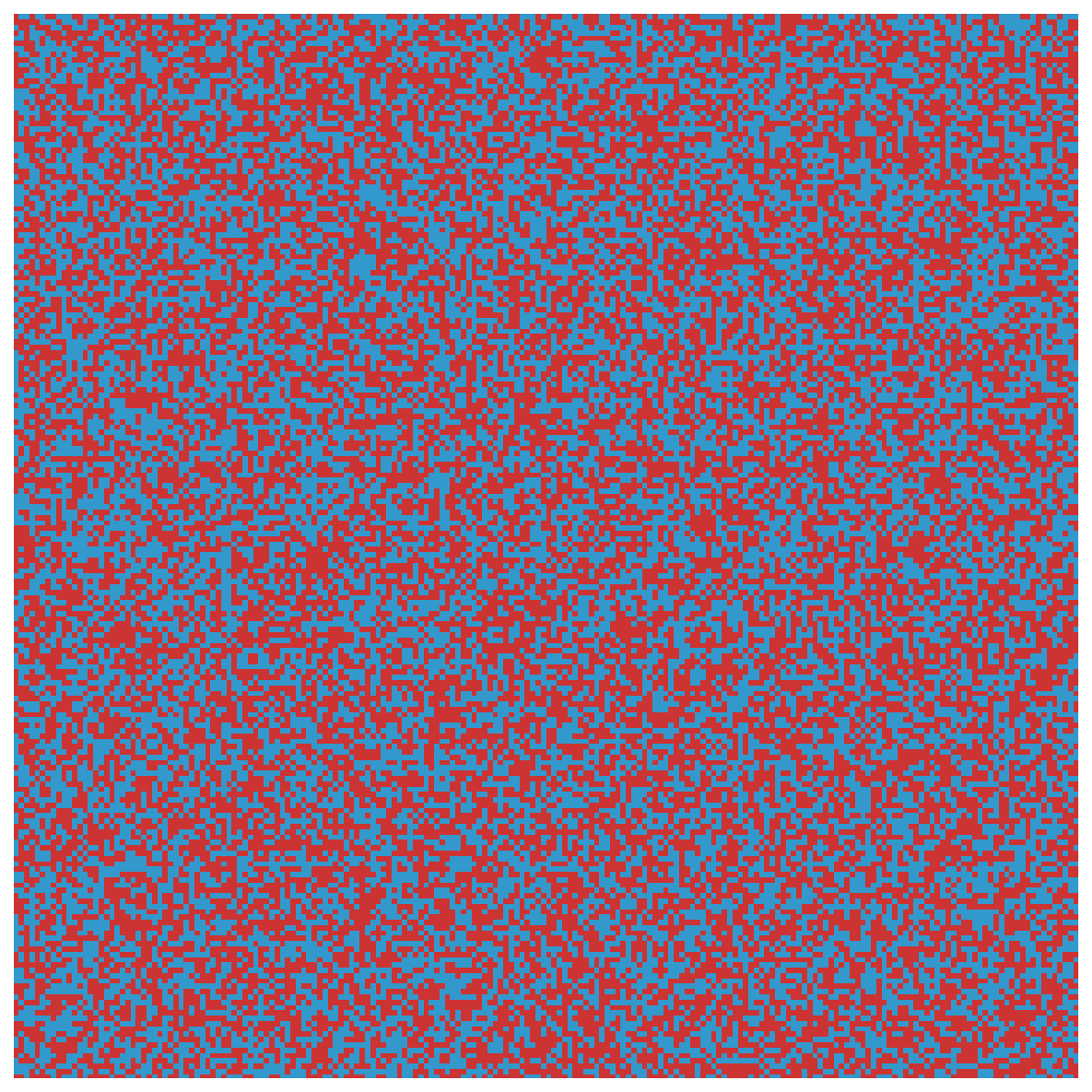}\\
			{\footnotesize t=1}
		\end{minipage}
		\begin{minipage}{0.14\linewidth}
			\centering
			\includegraphics[width=\linewidth]{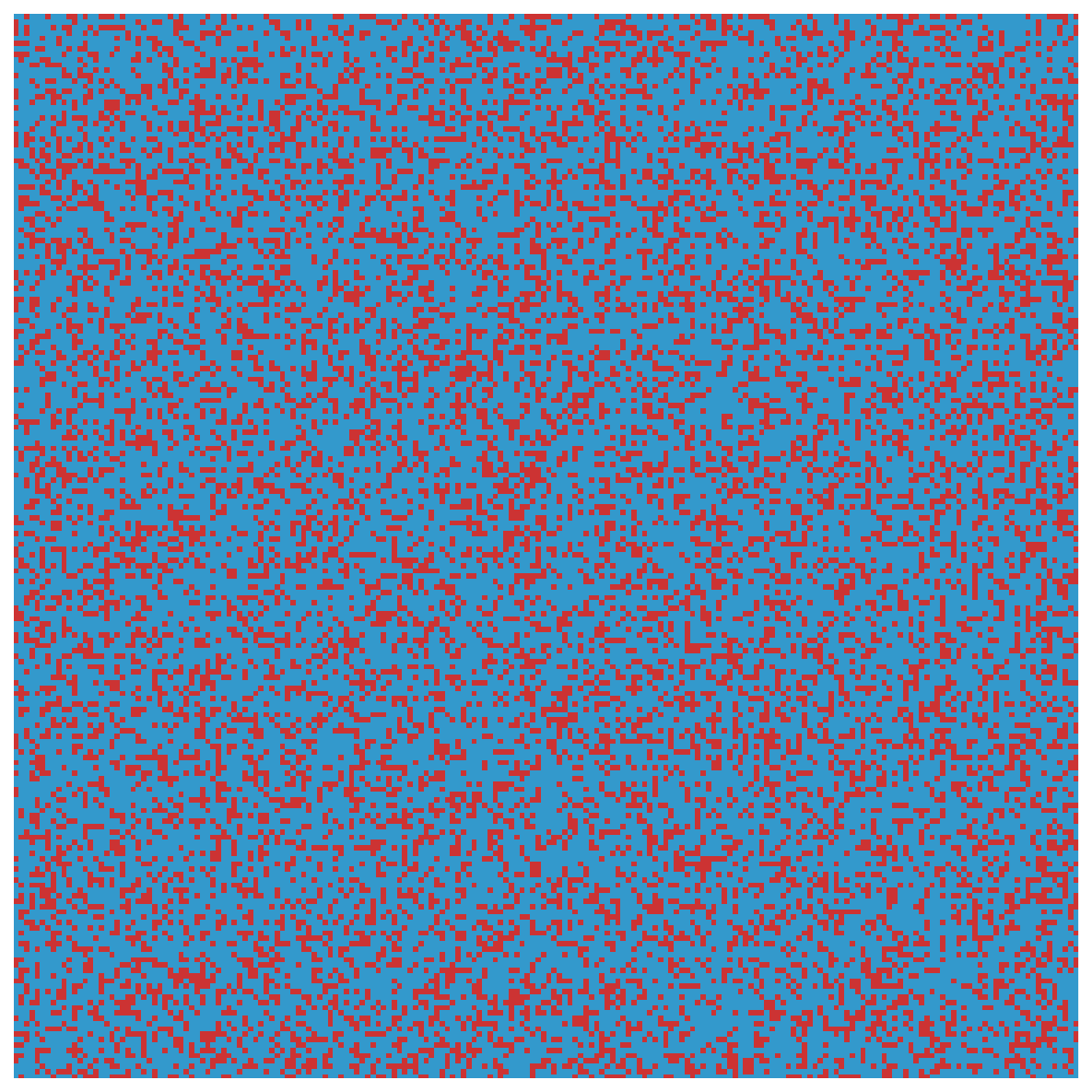}\\
			{\footnotesize t=10}
		\end{minipage}
		\begin{minipage}{0.14\linewidth}
			\centering
			\includegraphics[width=\linewidth]{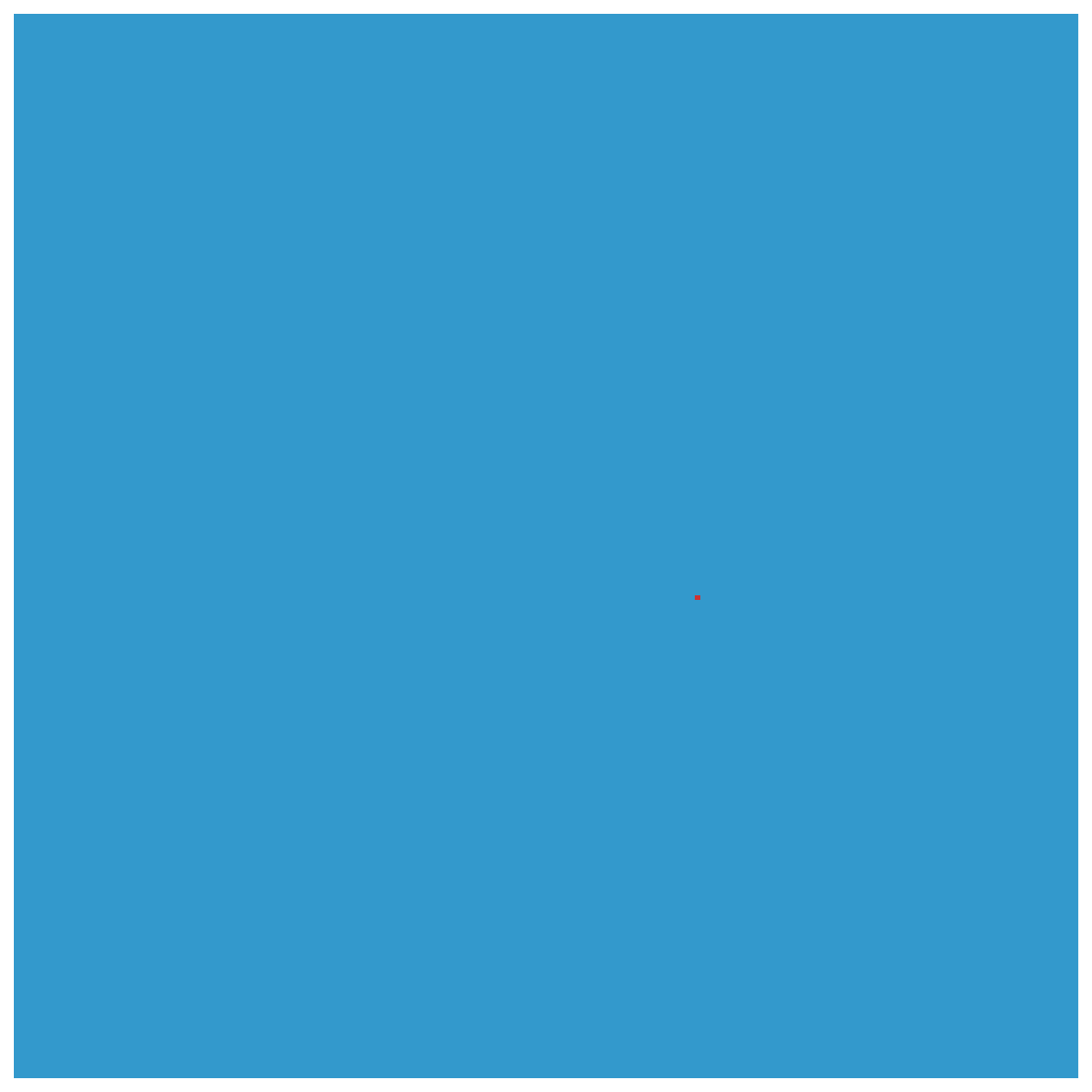}\\
			{\footnotesize t=100}
		\end{minipage}
		\begin{minipage}{0.14\linewidth}
			\centering
			\includegraphics[width=\linewidth]{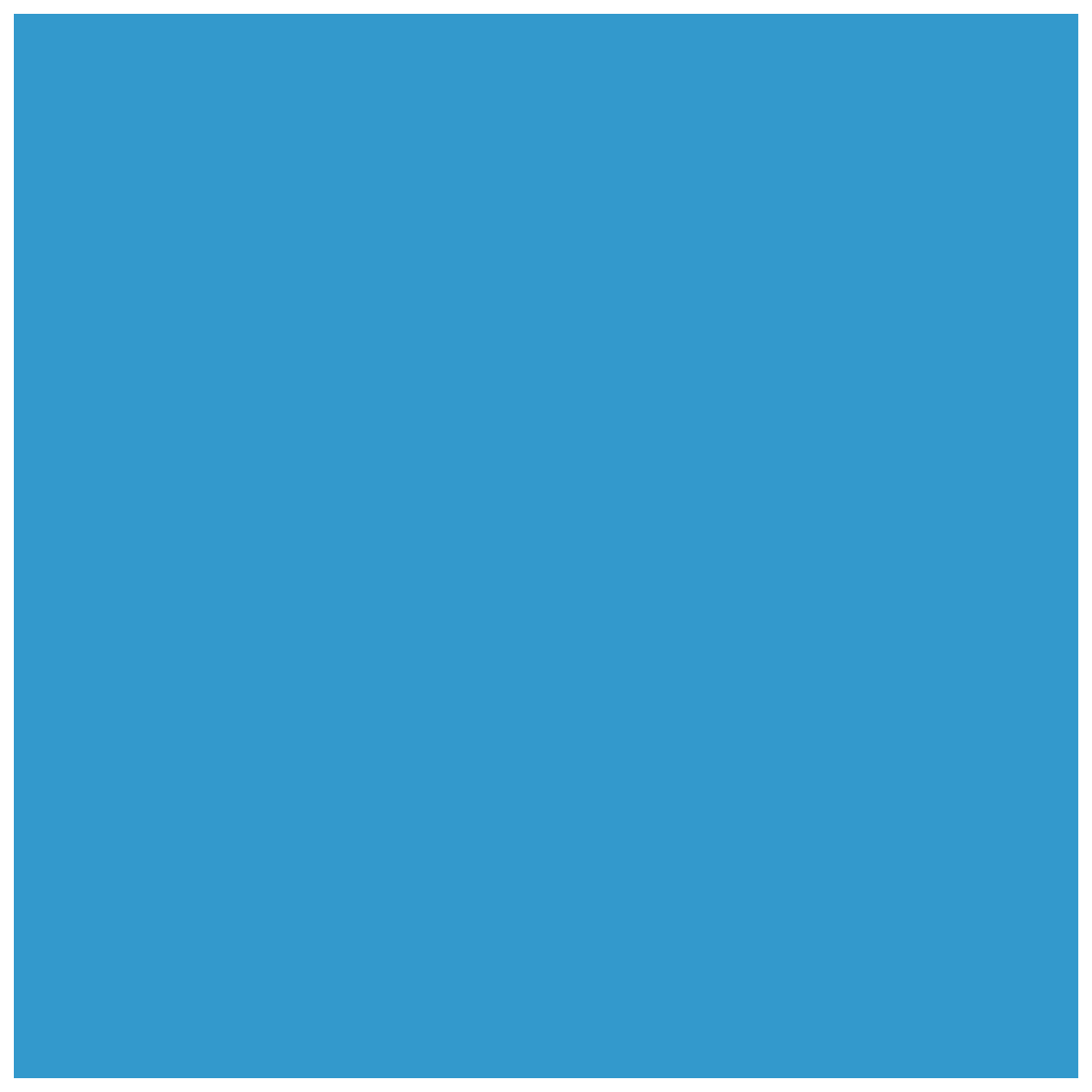}\\
			{\footnotesize t=1000}
		\end{minipage}
		\vspace{2mm}
		\\
		\centering
		{\footnotesize (a) LMFPPO-UBP}
	\end{minipage}
	\begin{minipage}{\linewidth}
		\begin{minipage}{0.24\linewidth}
			\centering
			\includegraphics[width=\linewidth]{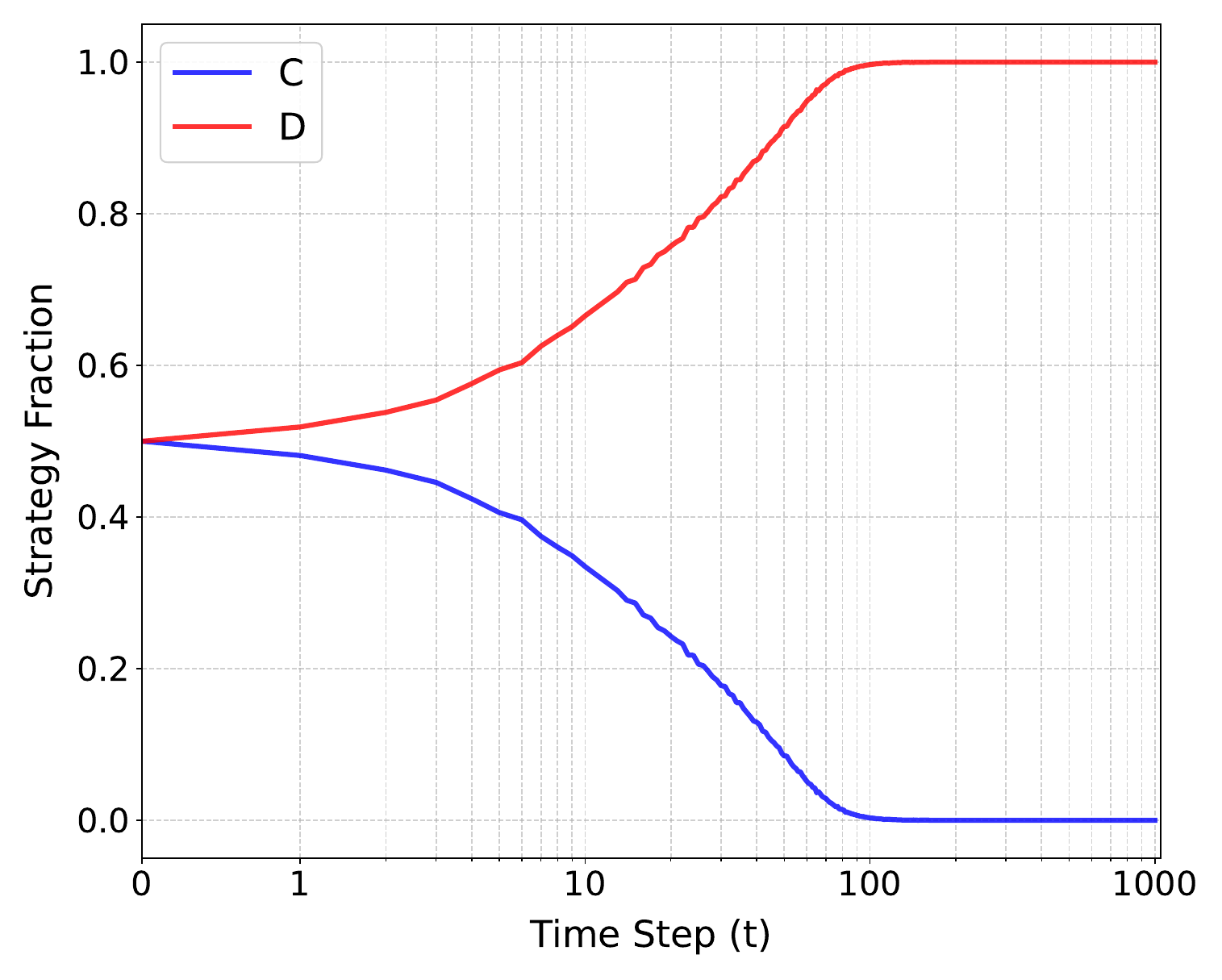}\\
		\end{minipage}
		\begin{minipage}{0.14\linewidth}
			\centering
			\includegraphics[width=\linewidth]{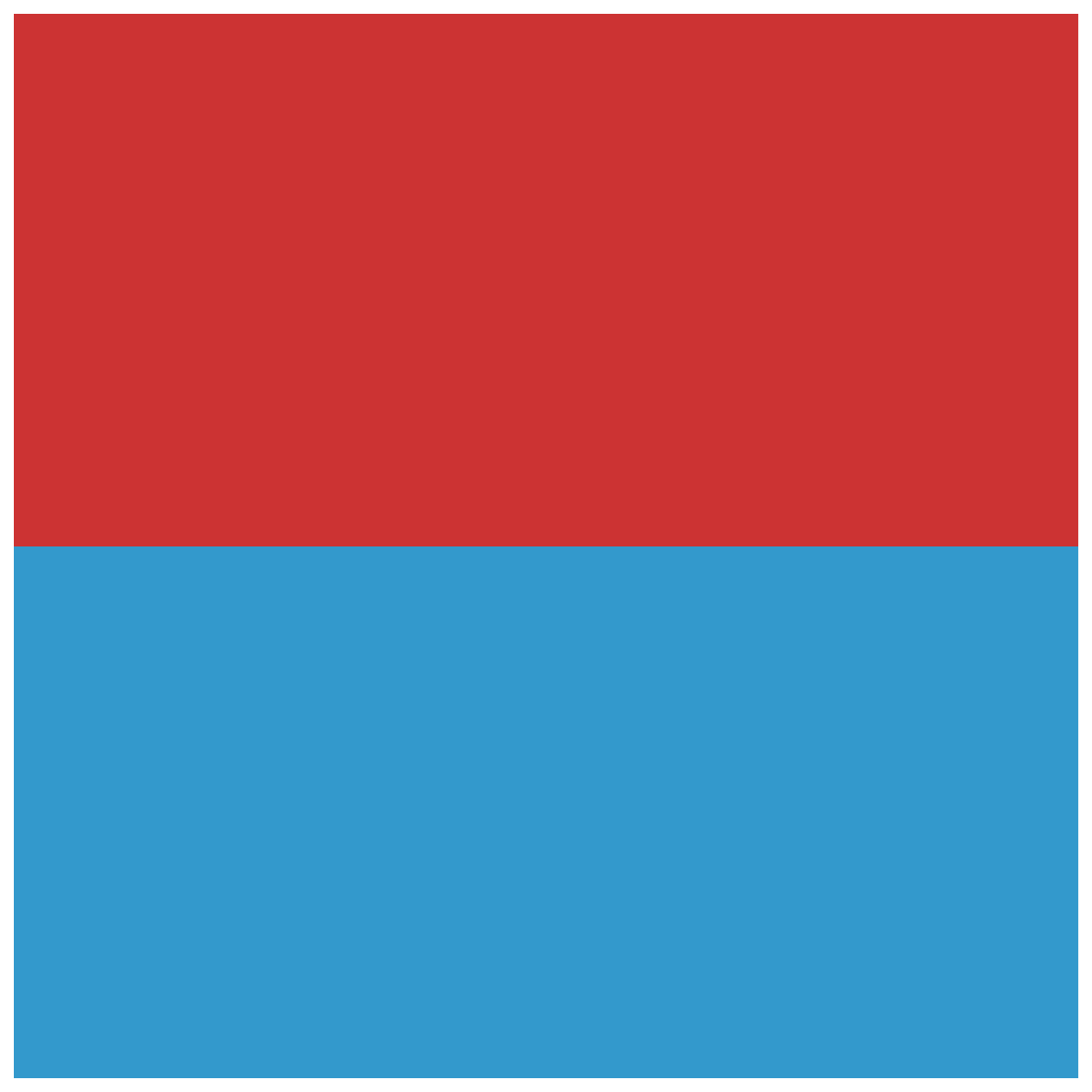}\\
			{\footnotesize t=0}
		\end{minipage}
		\begin{minipage}{0.14\linewidth}
			\centering
			\includegraphics[width=\linewidth]{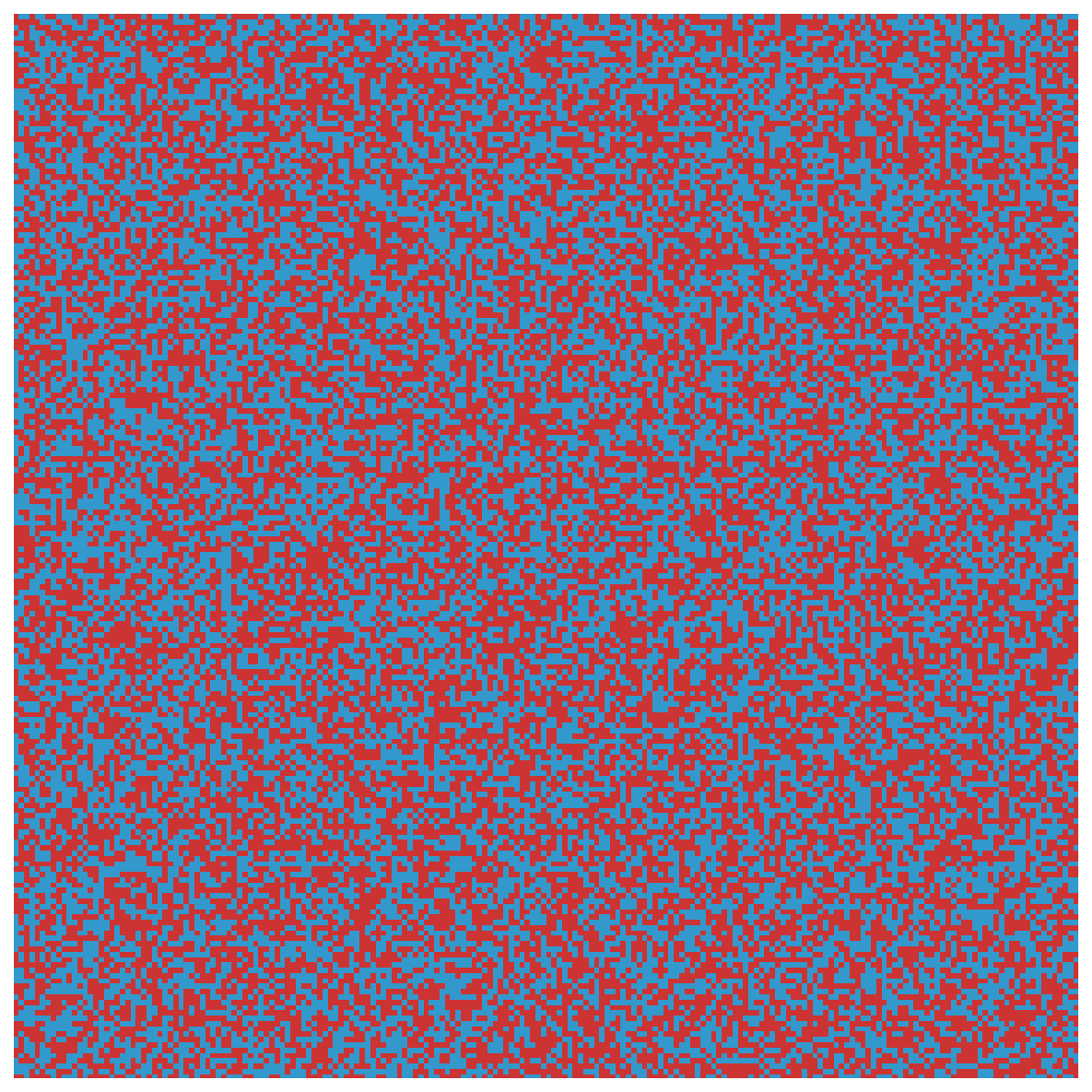}\\
			{\footnotesize t=1}
		\end{minipage}
		\begin{minipage}{0.14\linewidth}
			\centering
			\includegraphics[width=\linewidth]{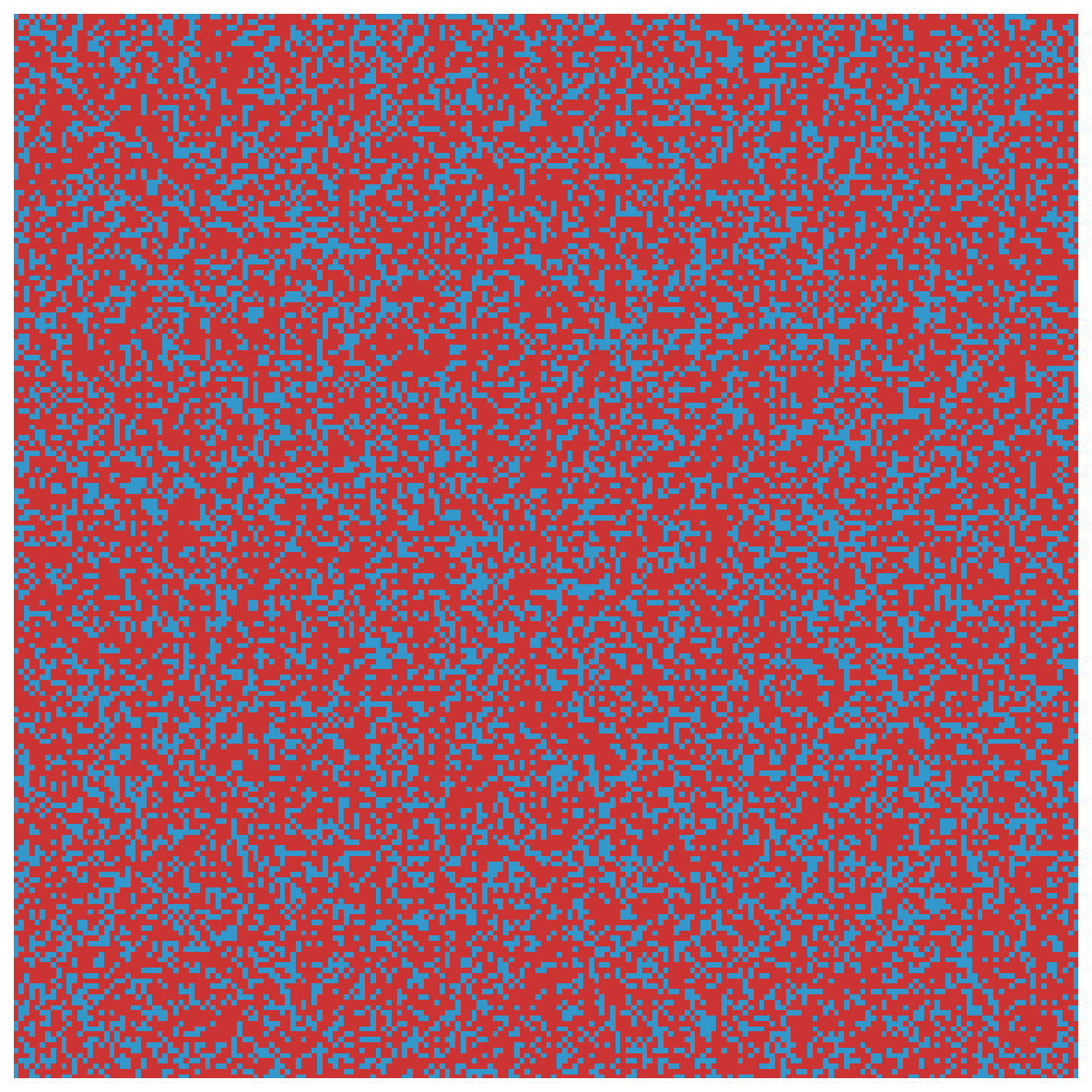}\\
			{\footnotesize t=10}
		\end{minipage}
		\begin{minipage}{0.14\linewidth}
			\centering
			\includegraphics[width=\linewidth]{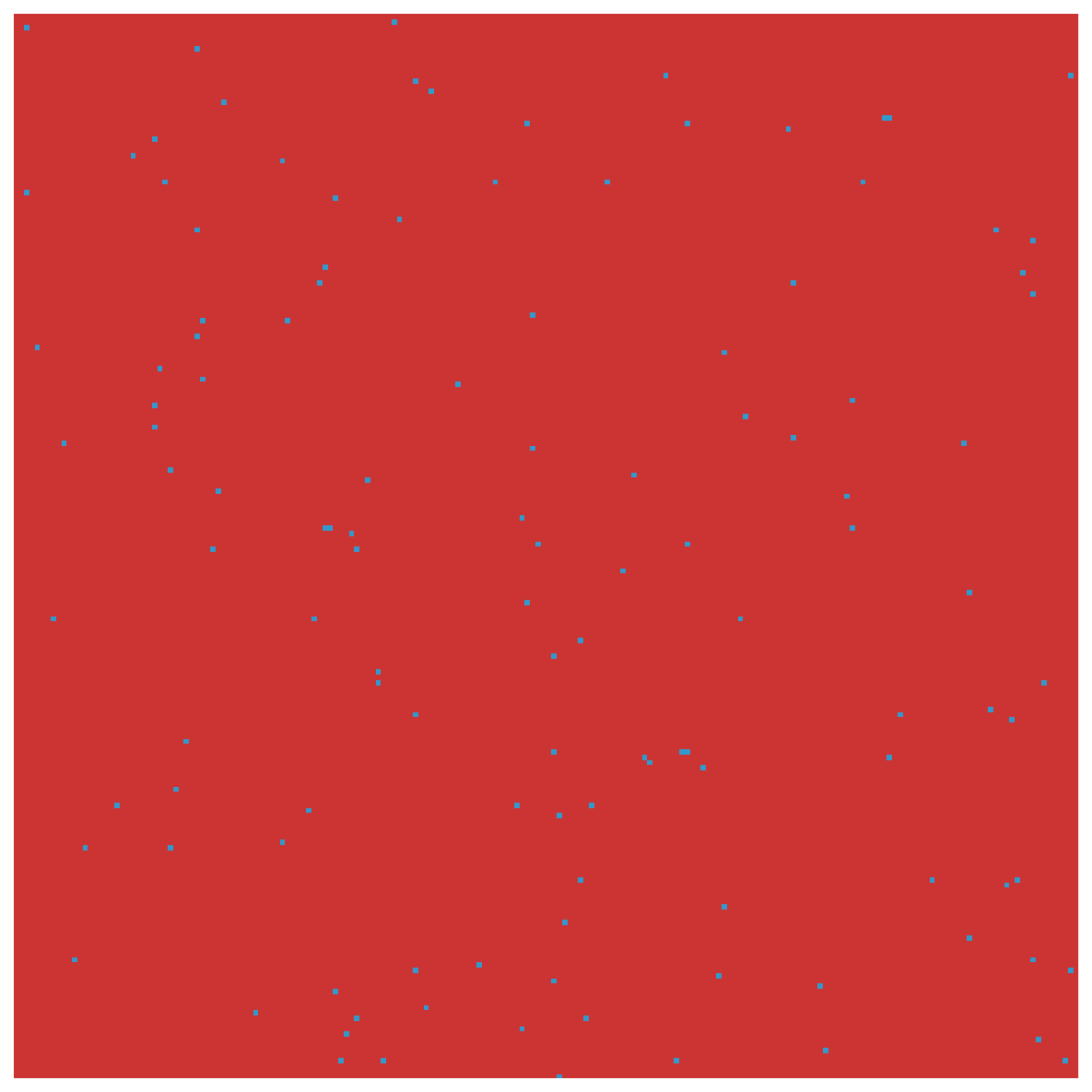}\\
			{\footnotesize t=100}
		\end{minipage}
		\begin{minipage}{0.14\linewidth}
			\centering
			\includegraphics[width=\linewidth]{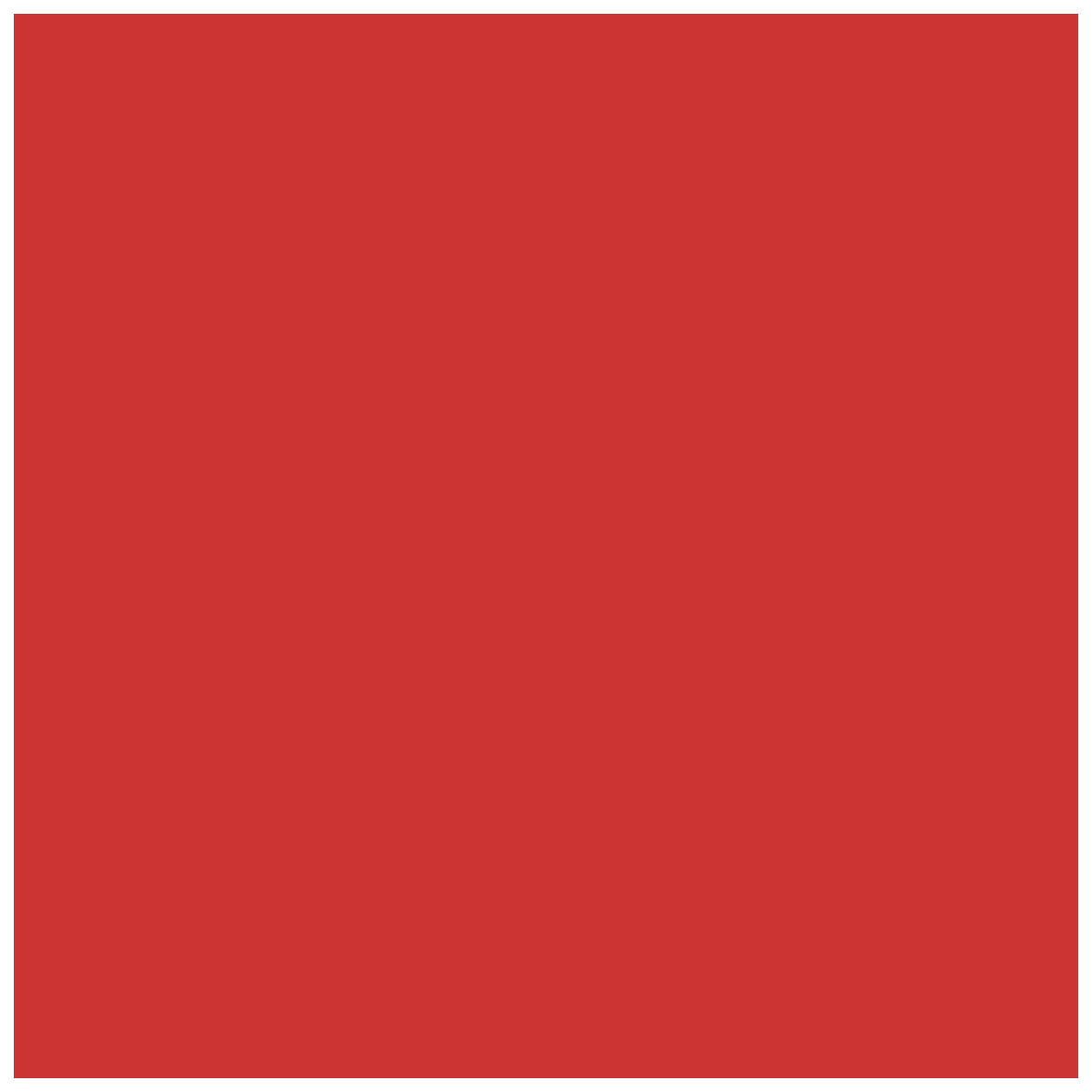}\\
			{\footnotesize t=1000}
		\end{minipage}
		\vspace{2mm}
		\\
		\centering
		{\footnotesize (b) LMFPPO}
	\end{minipage}
	\begin{minipage}{\linewidth}
		\begin{minipage}{0.24\linewidth}
			\centering
			\includegraphics[width=\linewidth]{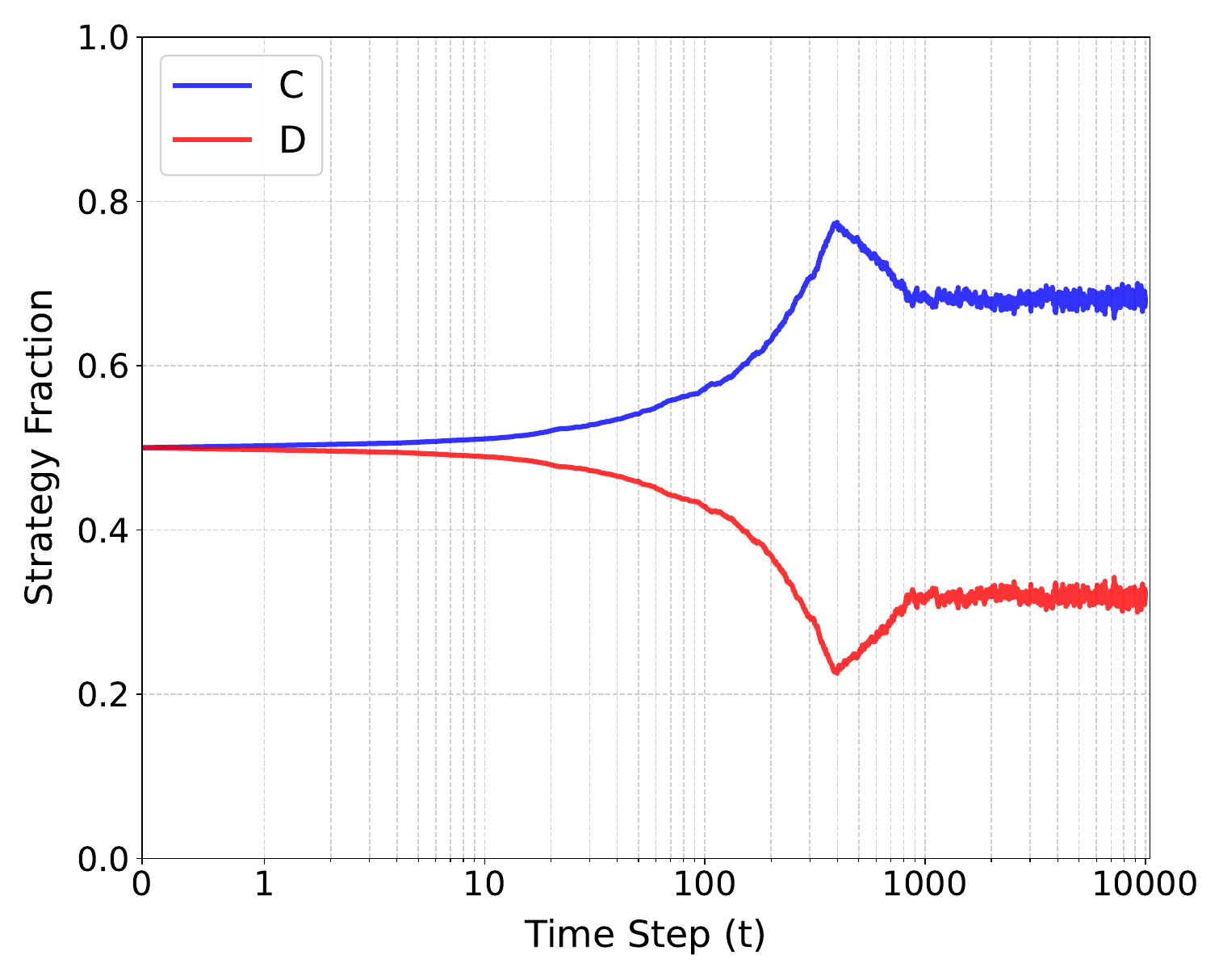}\\
		\end{minipage}
		\begin{minipage}{0.14\linewidth}
			\centering
			\includegraphics[width=\linewidth]{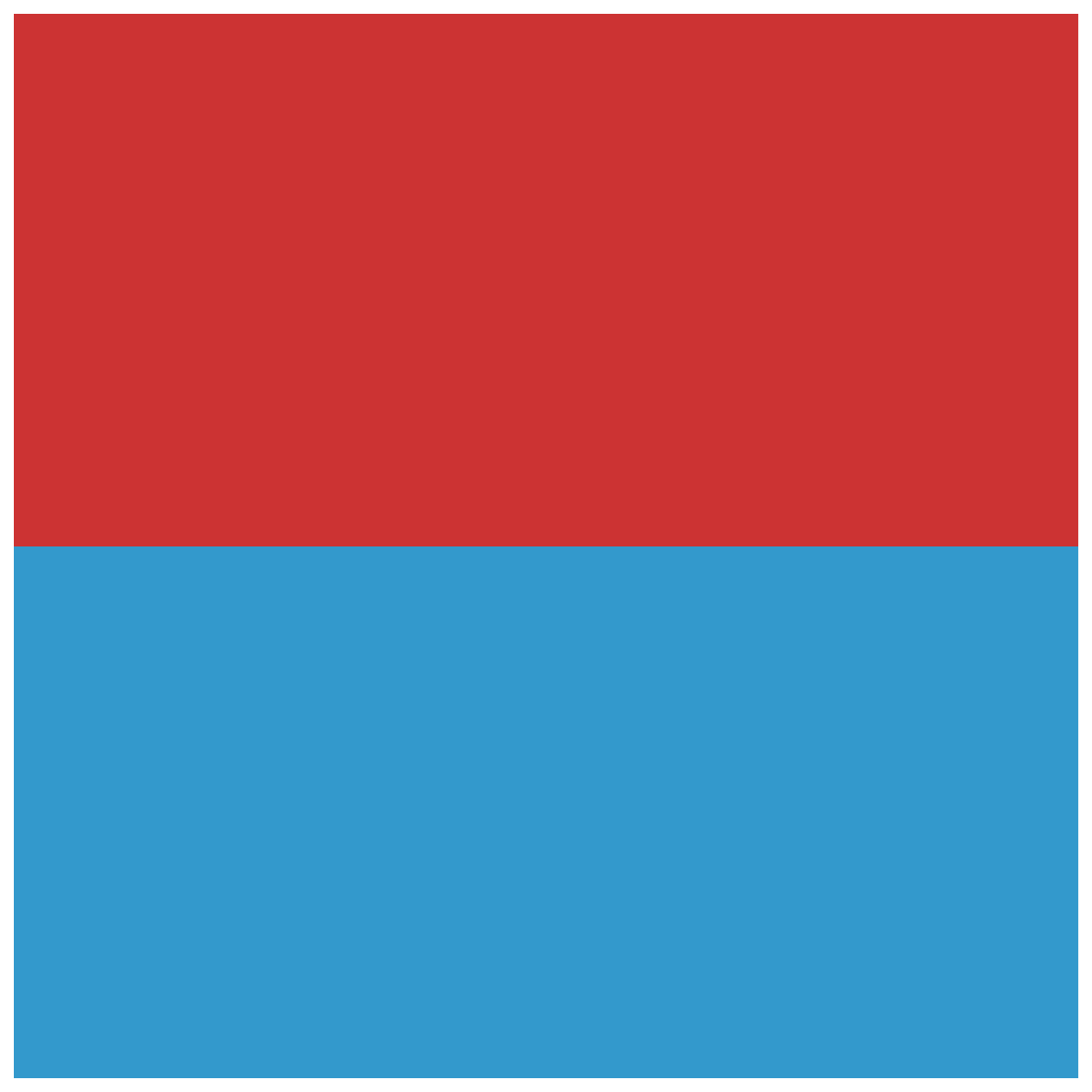}\\
			{\footnotesize t=0}
		\end{minipage}
		\begin{minipage}{0.14\linewidth}
			\centering
			\includegraphics[width=\linewidth]{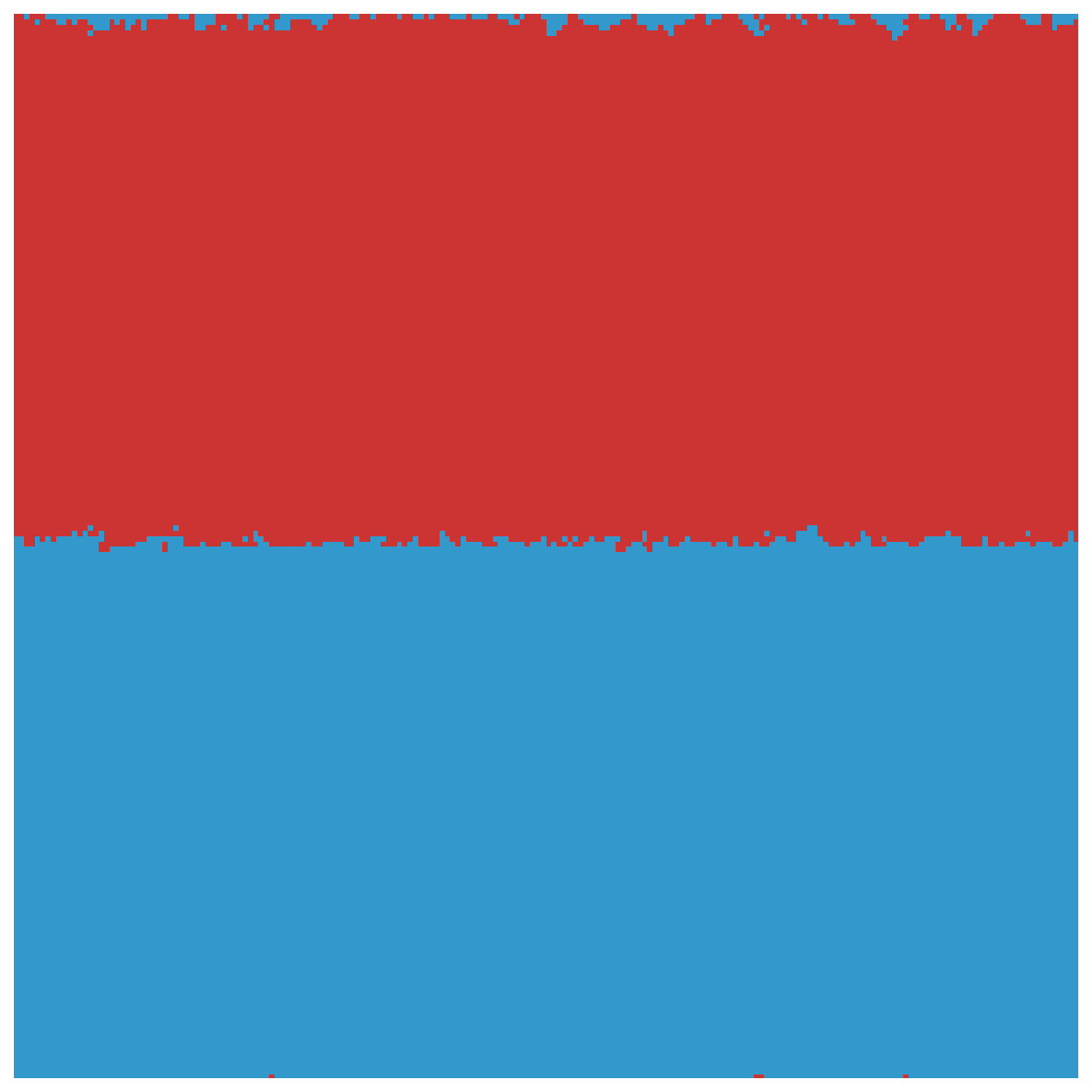}\\
			{\footnotesize t=10}
		\end{minipage}
		\begin{minipage}{0.14\linewidth}
			\centering
			\includegraphics[width=\linewidth]{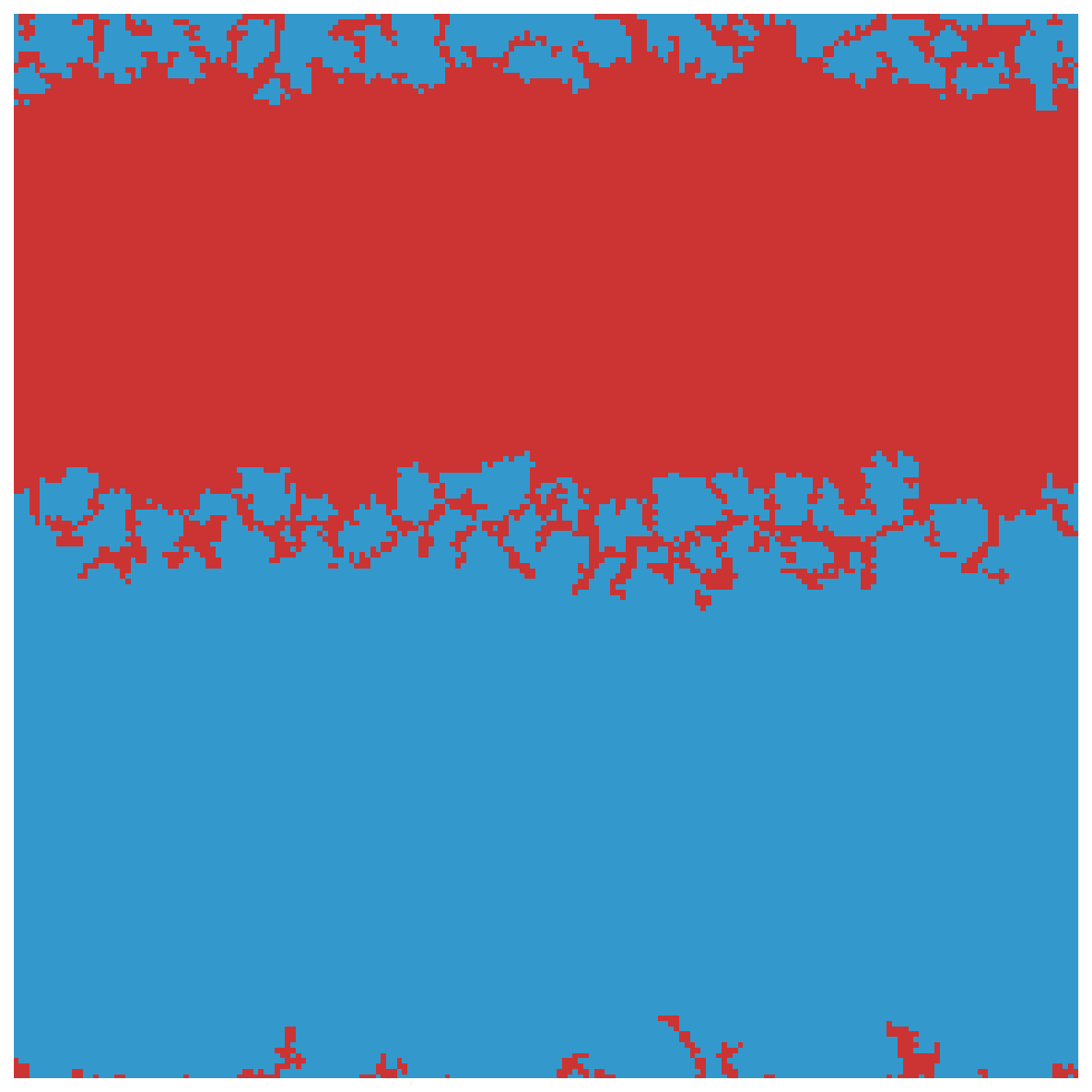}\\
			{\footnotesize t=100}
		\end{minipage}
		\begin{minipage}{0.14\linewidth}
			\centering
			\includegraphics[width=\linewidth]{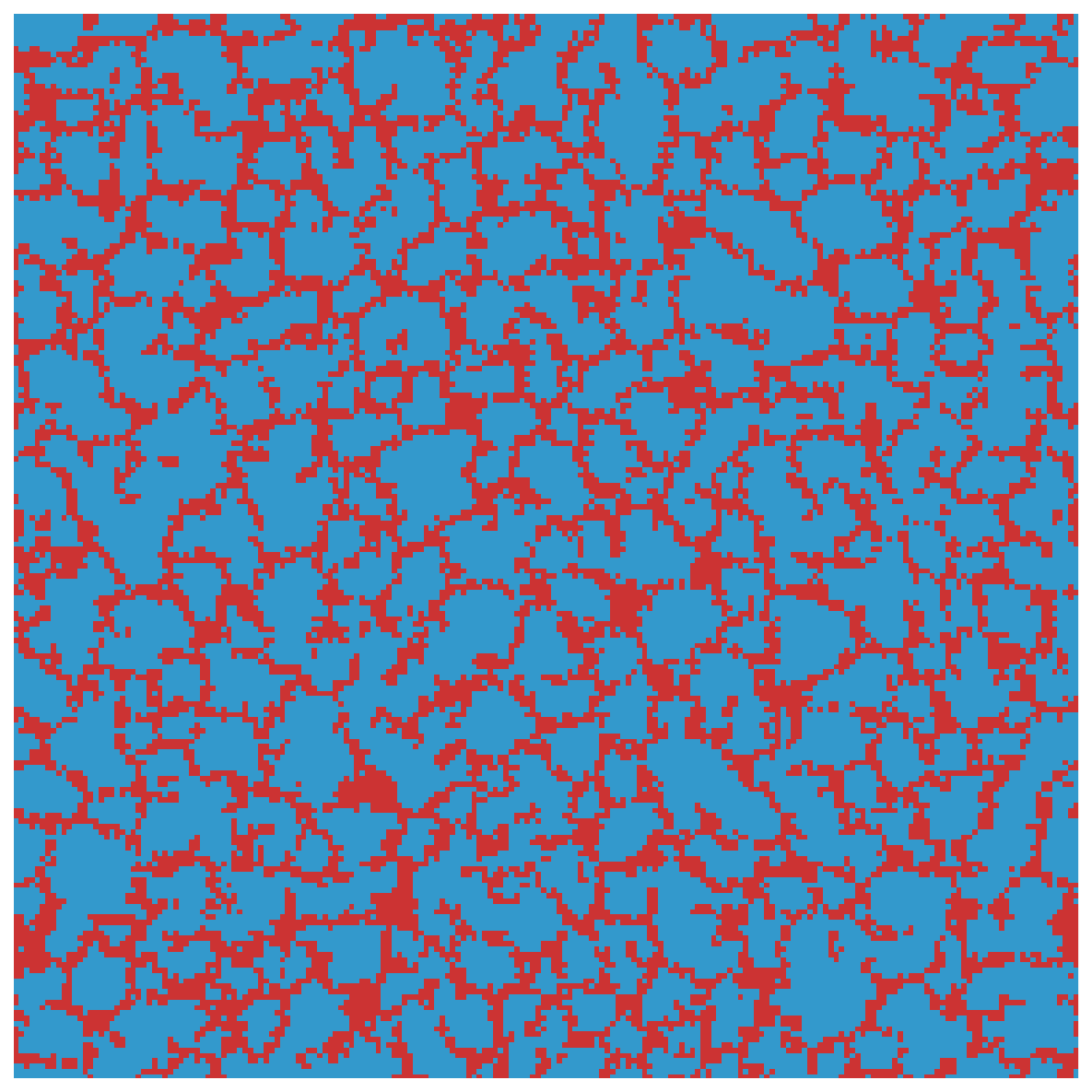}\\
			{\footnotesize t=1000}
		\end{minipage}
		\begin{minipage}{0.14\linewidth}
			\centering
			\includegraphics[width=\linewidth]{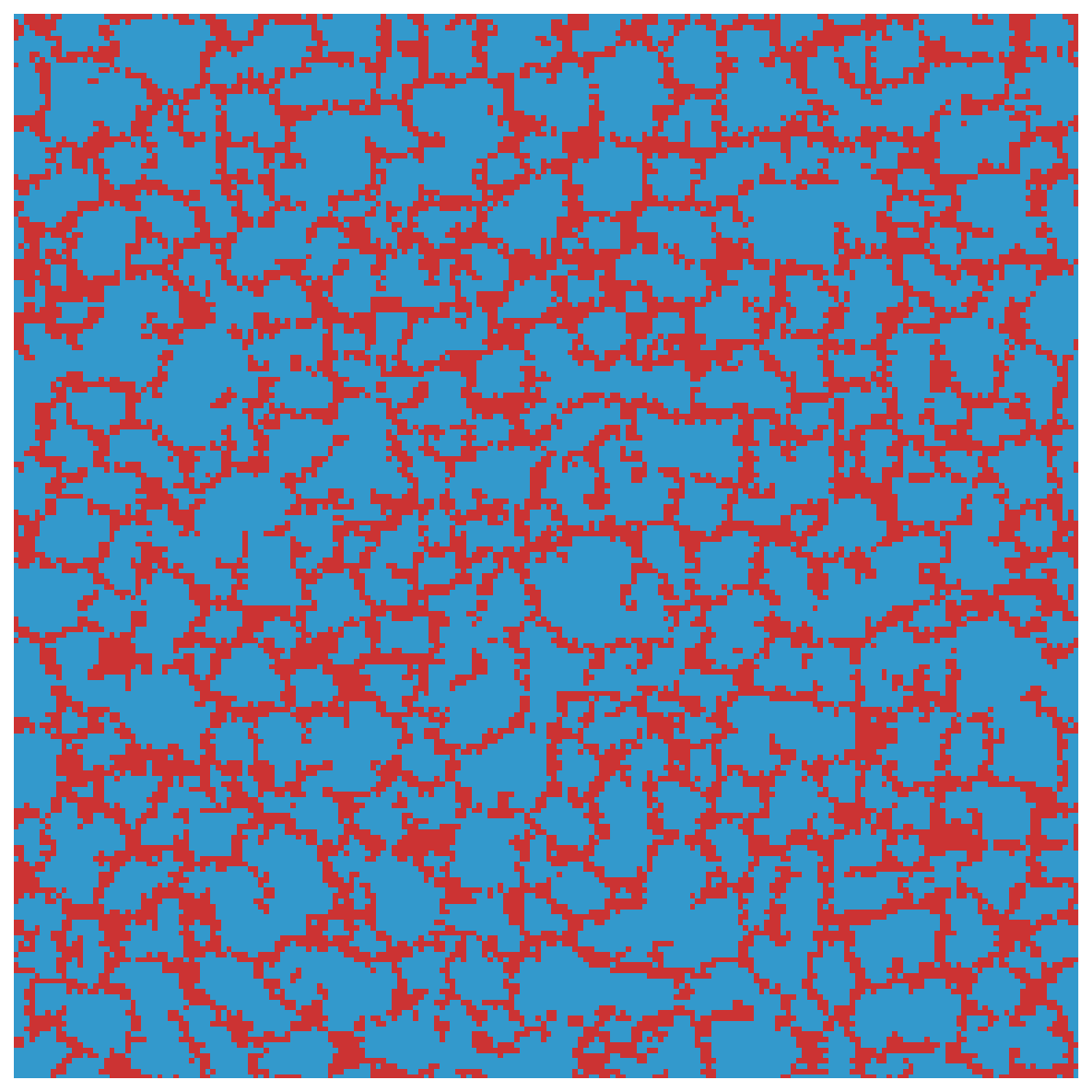}\\
			{\footnotesize t=10000}
		\end{minipage}
		\vspace{2mm}
		\\
		\centering
		{\footnotesize (c) Fermi update rule}
	\end{minipage}
	\begin{minipage}{\linewidth}
		\begin{minipage}{0.24\linewidth}
			\centering
			\includegraphics[width=\linewidth]{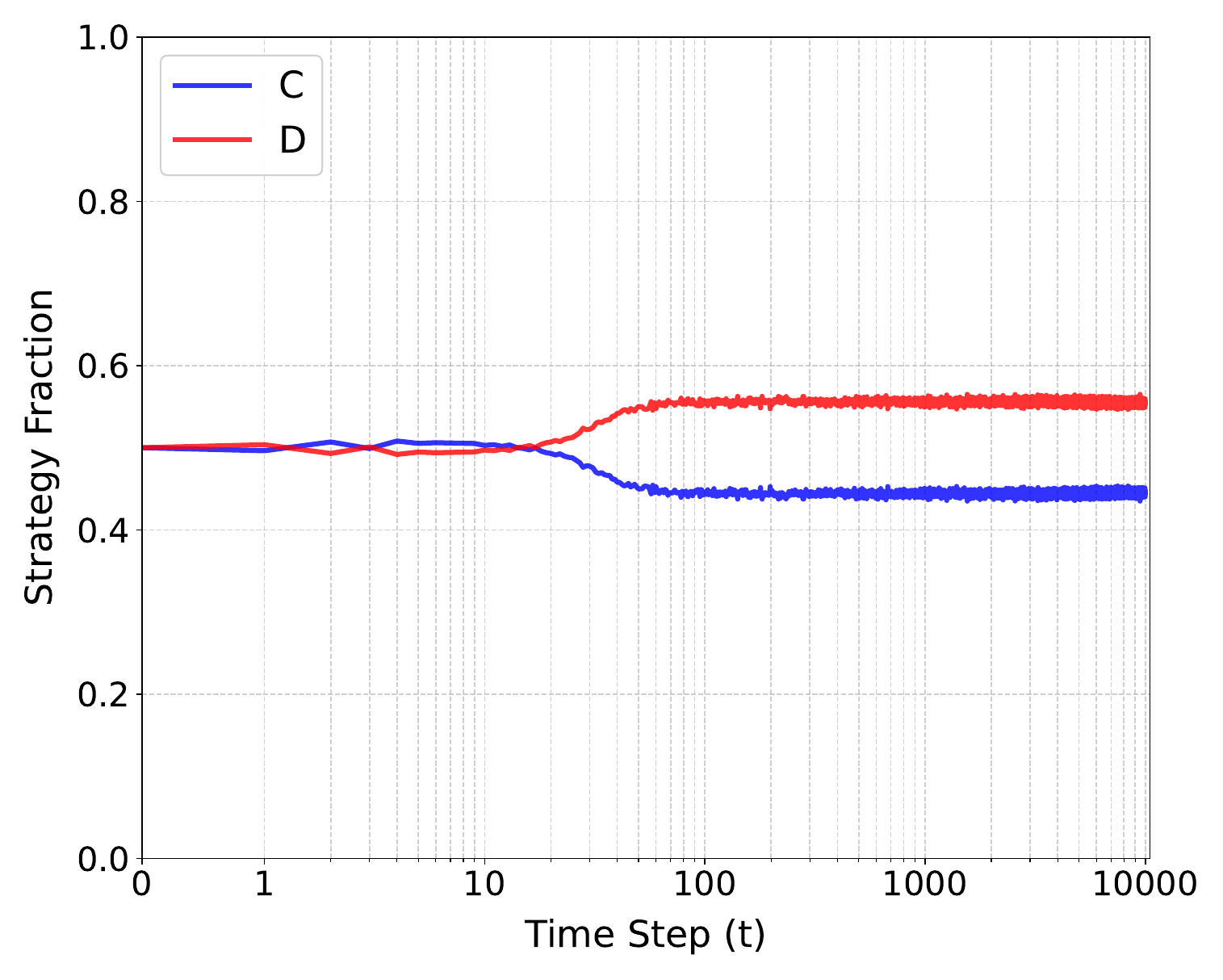}\\
		\end{minipage}
		\begin{minipage}{0.14\linewidth}
			\centering
			\includegraphics[width=\linewidth]{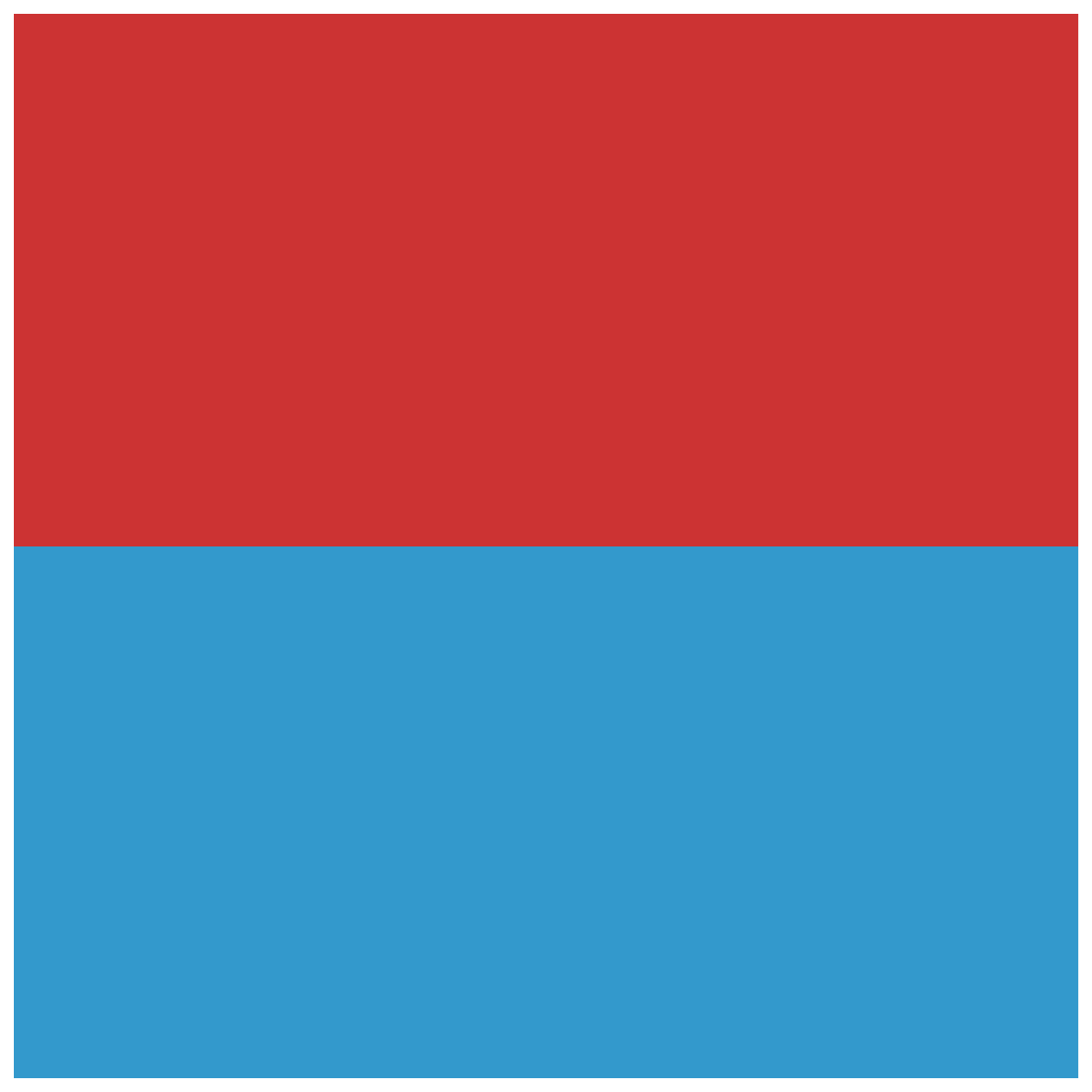}\\
			{\footnotesize t=0}
		\end{minipage}
		\begin{minipage}{0.14\linewidth}
			\centering
			\includegraphics[width=\linewidth]{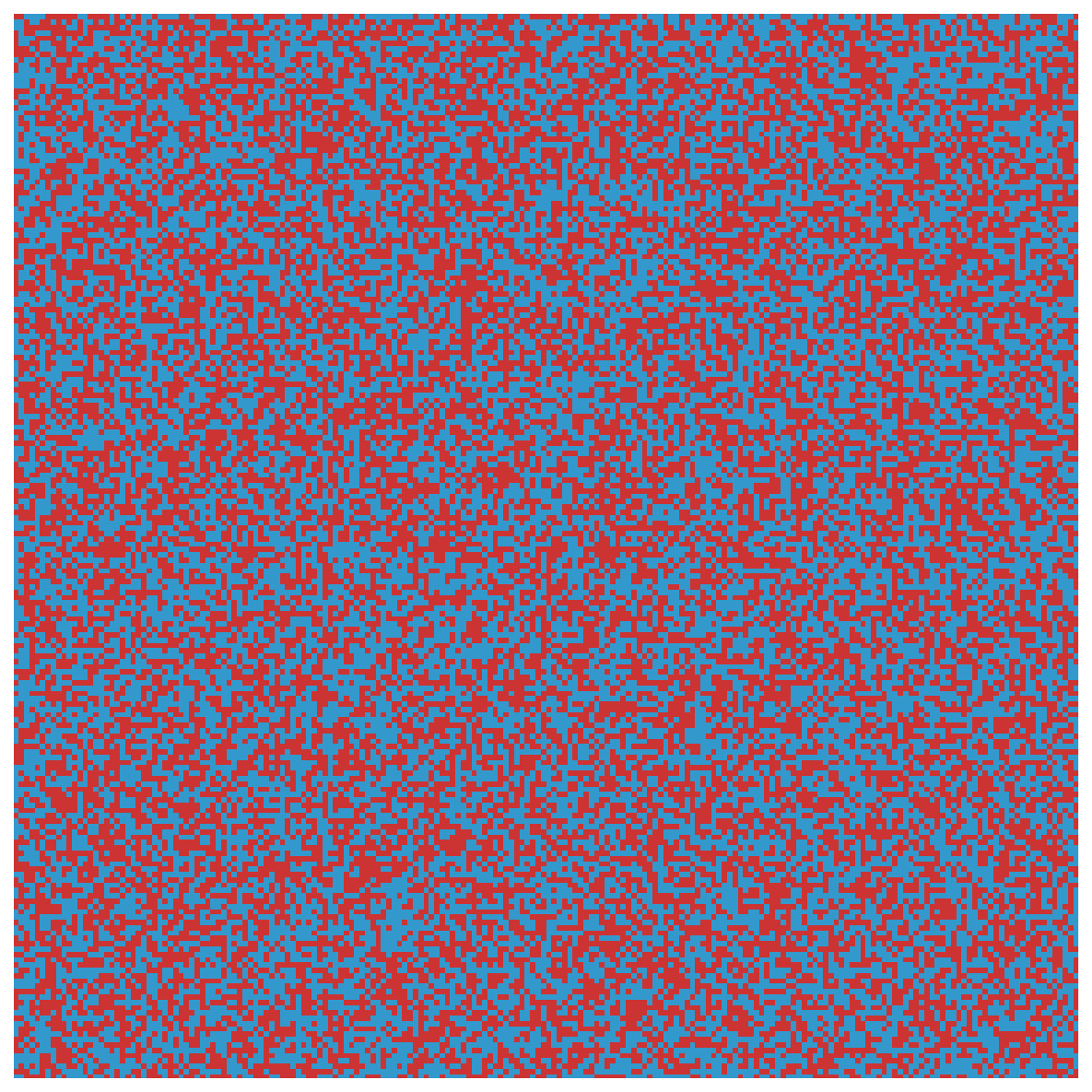}\\
			{\footnotesize t=10}
		\end{minipage}
		\begin{minipage}{0.14\linewidth}
			\centering
			\includegraphics[width=\linewidth]{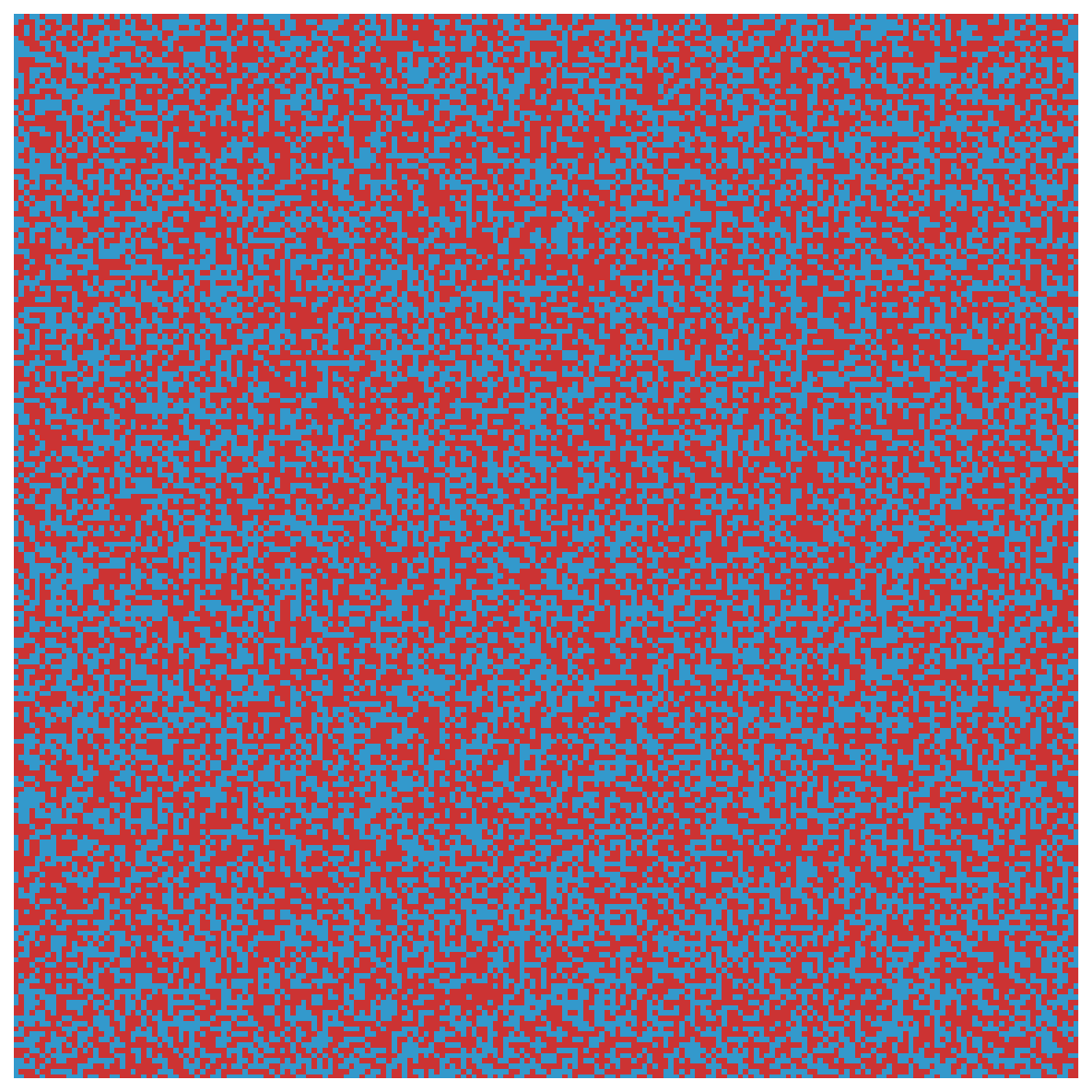}\\
			{\footnotesize t=100}
		\end{minipage}
		\begin{minipage}{0.14\linewidth}
			\centering
			\includegraphics[width=\linewidth]{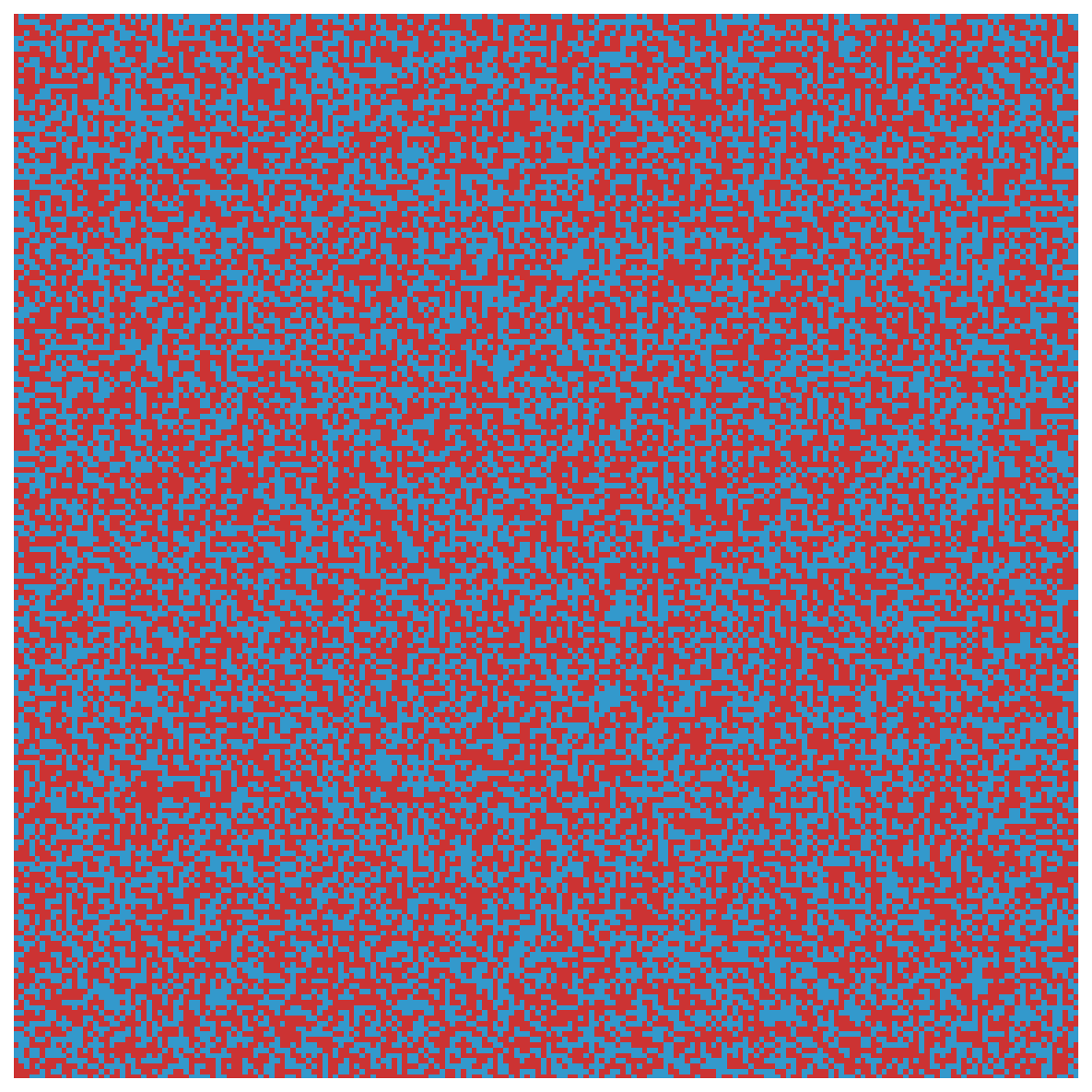}\\
			{\footnotesize t=1000}
		\end{minipage}
		\begin{minipage}{0.14\linewidth}
			\centering
			\includegraphics[width=\linewidth]{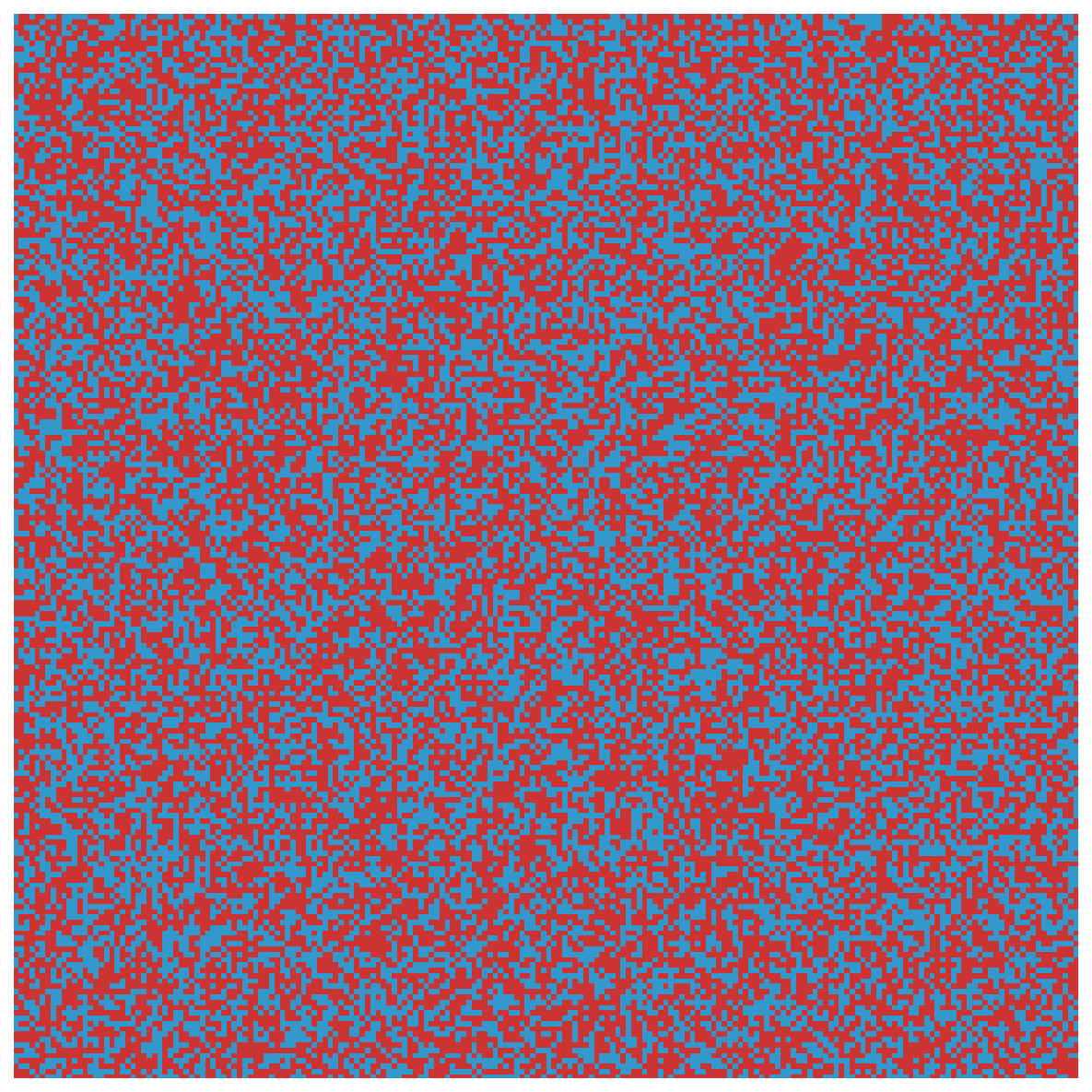}\\
			{\footnotesize t=10000}
		\end{minipage}
		\vspace{2mm}
		\\
		\centering
		{\footnotesize (d) Q-learning}
	\end{minipage}
	\caption{Spatial-temporal evolution of strategies under four algorithms at \( r = 4.3 \). 
		Rows (a) LMFPPO-UBP and (b) LMFPPO show snapshots at \( t = 0, 1, 10, 100, 1000 \),
		whereas (c) Fermi update rule and (d) Q-learning show snapshots at \( t = 0, 10, 100, 1000, 10000 \). 
		The cooperator fraction \( f_C \) and defector fraction \( f_D \) are labeled in each panel. 
		LMFPPO-UBP achieves rapid global cooperation, while LMFPPO converges to full defection. 
		Fermi update rule exhibits intermediate cooperation with spatial clustering, and Q-learning attains a suboptimal cooperation level.}
	\label{fig:algorithm_comparison}
\end{figure*}
A comprehensive comparative analysis was conducted to evaluate the performance of the LMFPPO-UBP algorithm. The algorithm was compared against three representative baseline methods LMFPPO, the Fermi update rule function based strategy update rule, and Q-learning. This comparison is visually presented in Figure~\ref{fig:algorithm_comparison}. All experiments were conducted under identical initial conditions. The enhancement factor was fixed at $r = 4.3$ and the lattice size was set to $L = 200$. The initial strategy configuration partitioned the grid spatially, with defectors $D$ occupying the upper half and cooperators $C$ occupying the lower half. The evolutionary dynamics were visualized through spatial strategy snapshots and the temporal evolution of the global cooperation fraction. For LMFPPO-UBP and LMFPPO, snapshots were captured at iterations $t = 0$, $1$, $10$, $100$, and $1000$. For the Fermi update rule and Q learning, snapshots were taken at $t = 0$, $10$, $100$, $1000$, and $10000$. Each subplot is annotated with the corresponding cooperator fraction $f_C$ and defector fraction $f_D$.

The results demonstrate that LMFPPO, augmented with the UBP mechanism, exhibits a remarkable capability for promoting cooperation and achieving rapid convergence. The algorithm accomplishes thorough strategic mixing and reaches global cooperation $f_C = 1.0$ within approximately 20 iterations, thereafter maintaining stability. In contrast, the LMFPPO algorithm without punishment also induces rapid initial mixing but, lacking additional suppression against defectors, the system converges to full defection $f_C = 0.0$ by $t \approx 100$. This suggests that at $r = 4.3$, a value below the traditional cooperation threshold, learning from local social information alone is insufficient to sustain cooperative outcomes.

Compared to classical update rules, the Fermi update rule imitation dynamics exhibit slow and unstable cooperative evolution. The cooperation fraction rises to nearly $0.8$ by $t \approx 400$, then declines sharply and eventually fluctuates around $f_C \approx 0.68$, accompanied by pronounced spatial clustering. This pattern highlights the limitation of local stochastic imitation in achieving global coordination. Meanwhile, Q learning attains a moderate cooperation level of approximately $f_C \approx 0.45$, yet convergence is slow and the outcome remains suboptimal. This indicates that traditional value-based methods struggle to capture the asymmetric strategic interactions and long term dependencies inherent in partially observable spatial games.	

\subsection{Algorithm performance evaluation under varying enhancement factors r}
\begin{figure*}[htbp!]
	\begin{minipage}{0.48\linewidth}
		\centering
		\includegraphics[width=\linewidth]{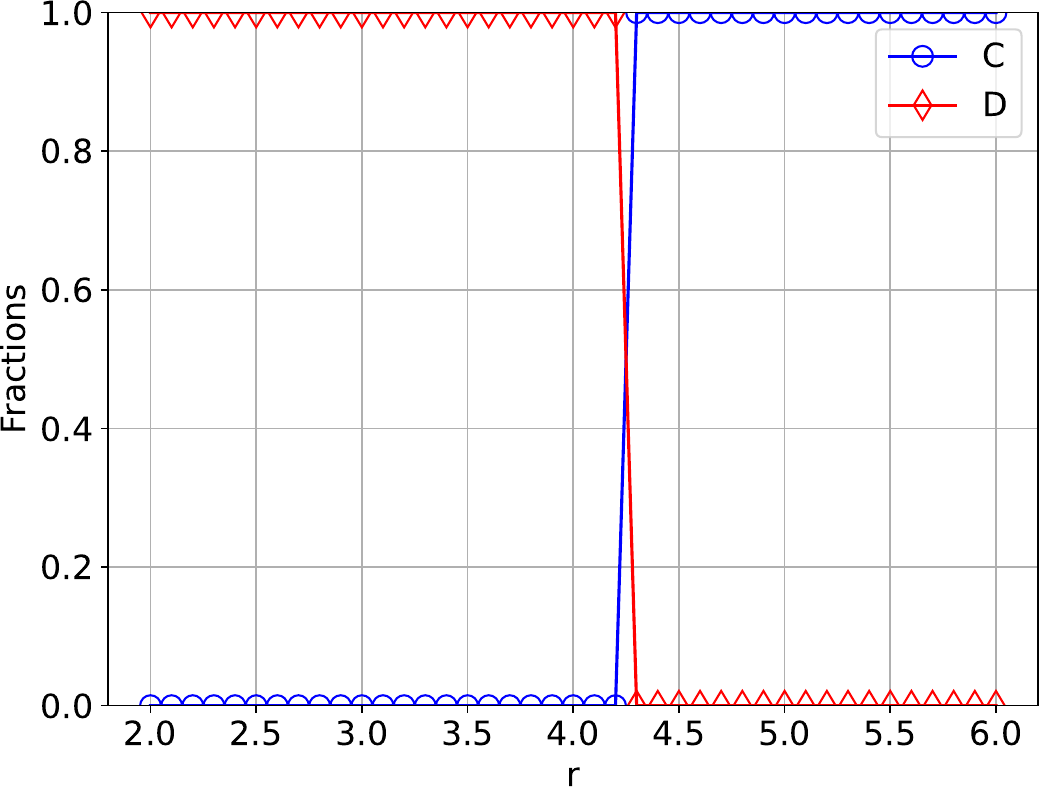}\\
		{\footnotesize (a) LMFPPO-UBP}
	\end{minipage}
	\begin{minipage}{0.48\linewidth}
		\centering
		\includegraphics[width=\linewidth]{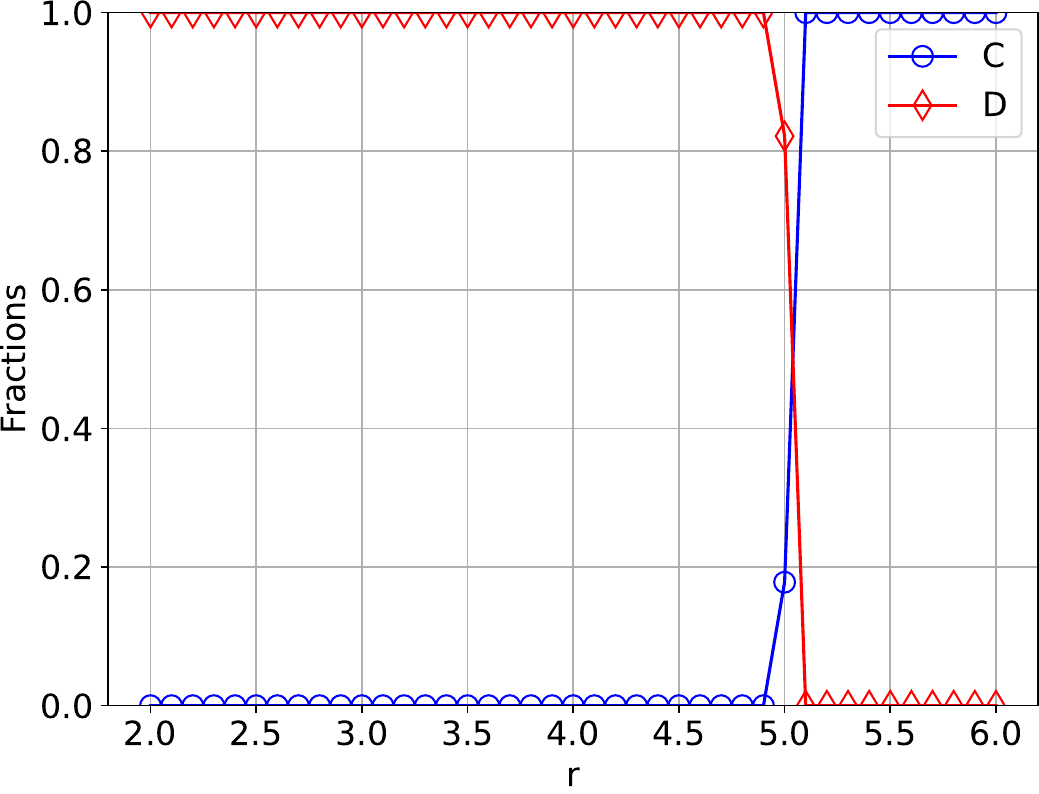}\\
		{\footnotesize (b) LMFPPO}
	\end{minipage}
	\\
	[3mm]
	\begin{minipage}{0.48\linewidth}
		\centering
		\includegraphics[width=\linewidth]{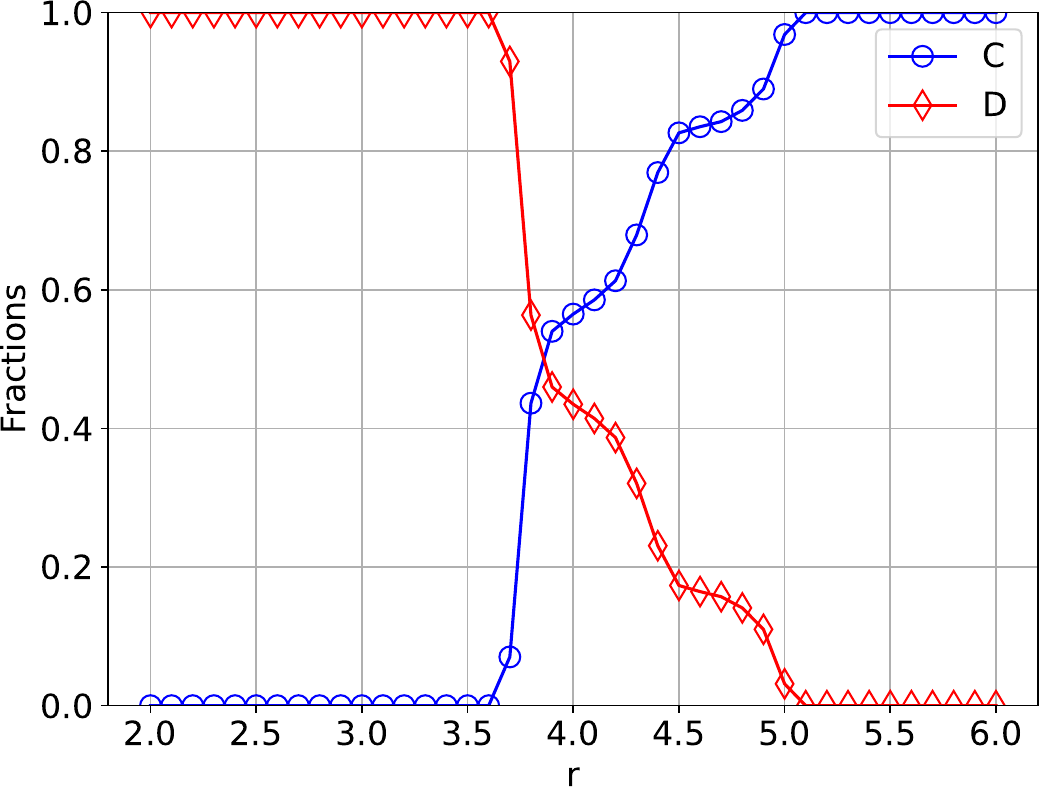}\\
		{\footnotesize (c) Fermi update rule}
	\end{minipage}	
	\begin{minipage}{0.48\linewidth}
		\centering
		\includegraphics[width=\linewidth]{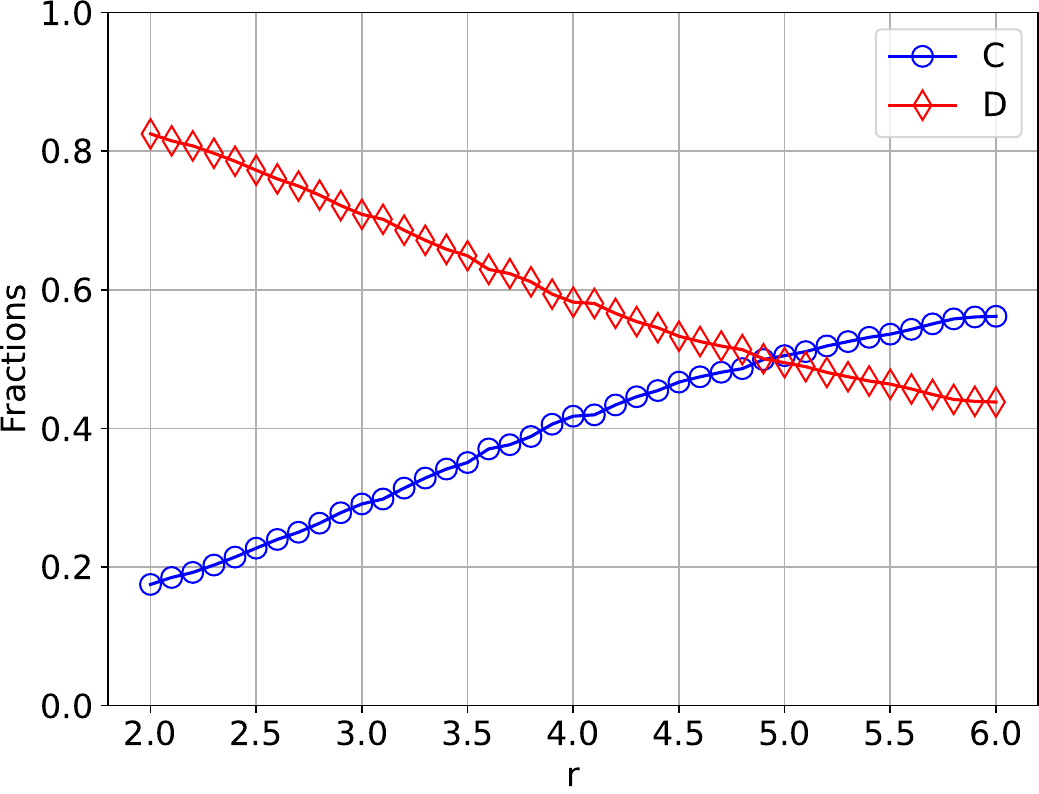}\\
		{\footnotesize (d) Q-learning}
	\end{minipage}	
	\caption{Cooperation fraction as a function of enhancement factor $r$ for four algorithms. LMFPPO-UBP achieves a sharp threshold transition to full cooperation at $r \geq 4.3$, while baseline algorithms (LMFPPO, Fermi update rule and Q-learning) require higher $r$ values for substantial cooperation. LMFPPO-UBP effectively lowers the critical enhancement factor for cooperation emergence.}
	\label{fig:algorithm_performance_evaluation}
\end{figure*}
A series of comparative experiments was carried out among four algorithms LMFPPO‑UBP, LMFPPO, the Fermi update rule, and Q‑learning. The enhancement factor $r$ was varied across the interval from 2.0 to 6.0. The results reveal distinct phase‑transition behaviors governed by the underlying structural incentives of SPGG, as shown in Figure~\ref{fig:algorithm_performance_evaluation}.

LMFPPO‑UBP demonstrates a sharp, first‑order‑like threshold response. It sustains full defection for $r < 4.3$ and achieves full cooperation thereafter. This behavior confirms the theoretical insight that cooperation can be stabilized by reshaping local incentive structures. The synergistic integration of LMF perception and the UBP mechanism directly reshapes the payoff landscape for defectors. This structural modification effectively reduces the critical enhancement factor required for the emergence of cooperation. In contrast, the LMFPPO algorithm, which lacks the punishment mechanism, only achieves substantial cooperation when the enhancement factor satisfies $r \geq 5.1$. This outcome underscores the inherent limitation of relying solely on pure social learning. While local perception provides strategic context, it is insufficient to overcome the fundamental free‑rider incentive when collective returns are low. The Fermi update rule exhibits a gradual and hysteresis‑prone increase in cooperation. This behavior illustrates the inherent inefficiency and instability of decentralized stochastic imitation as a mechanism for achieving global coordination. Q‑learning produces a smooth but persistently suboptimal level of cooperation across the parameter range. This outcome highlights the fundamental difficulty faced by traditional value‑based methods. Such approaches struggle to represent and exploit the complex interactions in these games. These include asymmetric strategic interactions and long-term dependencies, both of which characterize partially observable spatial games.

Collectively, these findings articulate a core principle for governing cooperation in spatial public goods dilemmas. Effective algorithmic design must jointly address three critical components. First, it requires accurate localized perception of the social environment \citep{2021Cooperation}. Second, it necessitates the adaptive alignment of individual incentives with collective welfare through structured mechanisms \citep{2025The}. Third, it must mitigate the strategic asymmetries that arise from spatial interactions \citep{2025Enhancing}. The LMFPPO‑UBP framework offers a concrete instantiation of this design principle. It bridges multi‑agent reinforcement learning with theories of institutional design to enable robust cooperation.

\subsection{LMFPPO-UBP with half-and-half initialization}

\begin{figure*}[htbp!]
	\begin{minipage}{0.45\linewidth}
		\begin{minipage}{\linewidth}
			\centering
			\includegraphics[width=\linewidth]{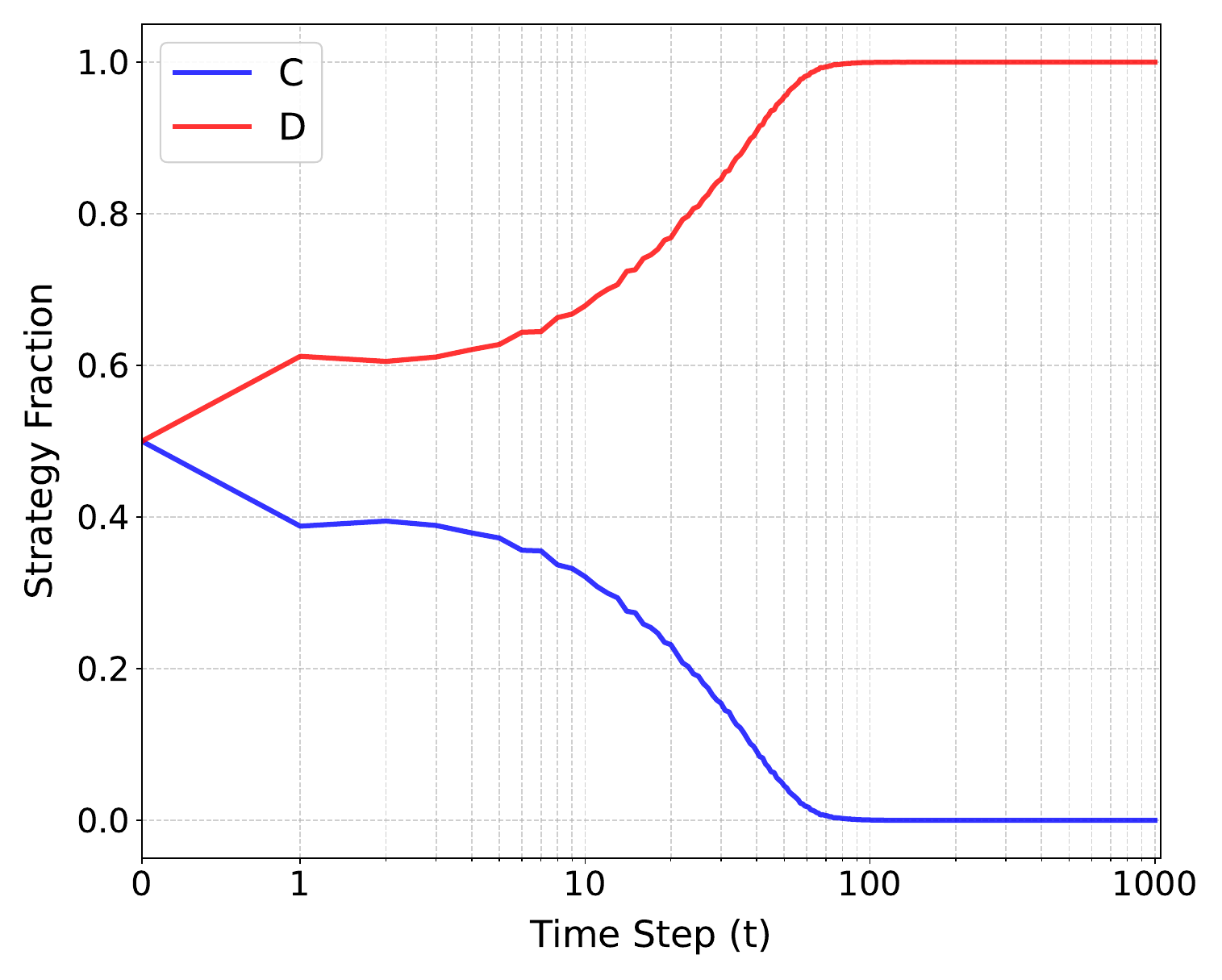}\\
		\end{minipage}
		\vspace{2mm}
		\\
		\begin{minipage}{0.188\linewidth}
			\centering
			\includegraphics[width=\linewidth]{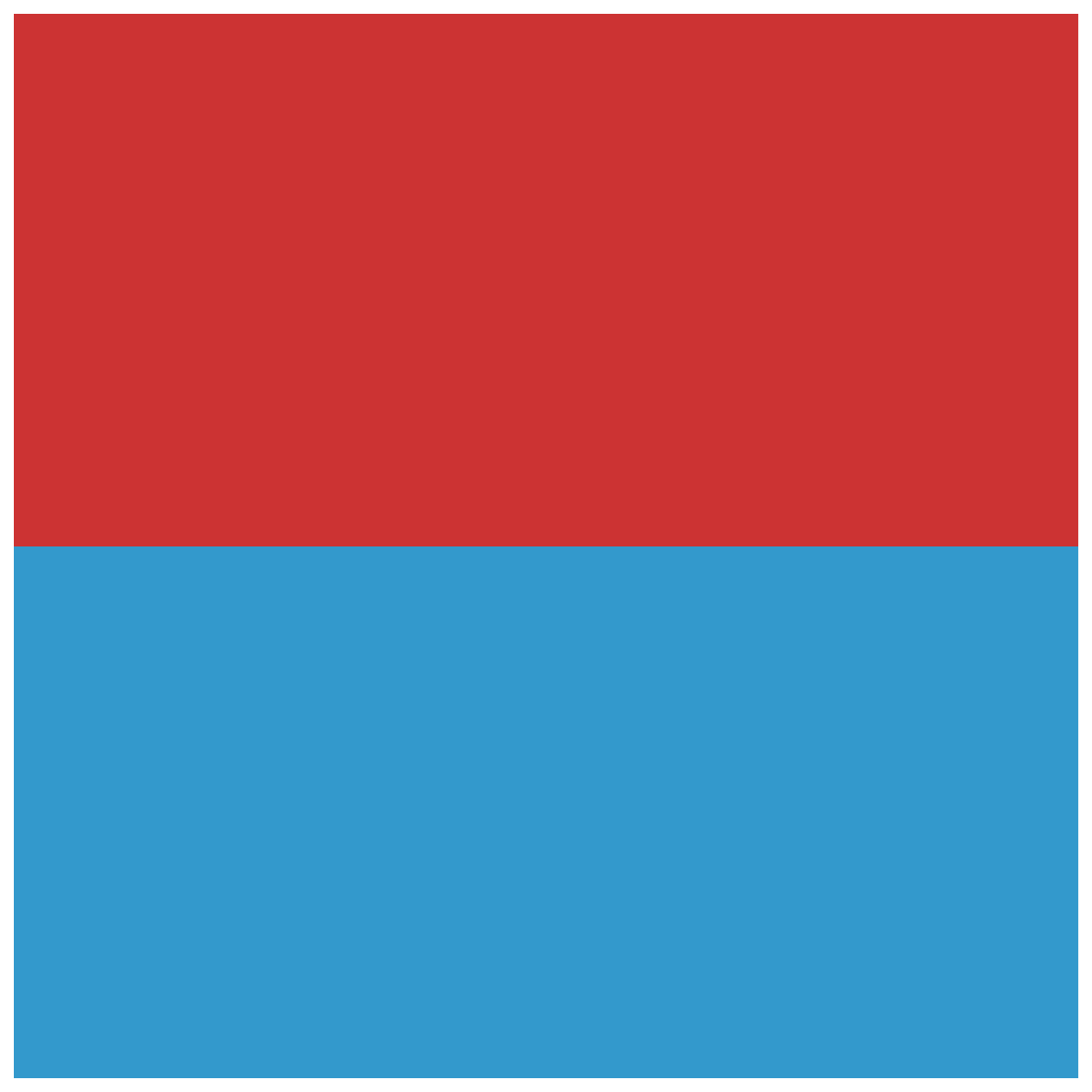}\\
			{\footnotesize t=0}
		\end{minipage}
		\begin{minipage}{0.188\linewidth}
			\centering
			\includegraphics[width=\linewidth]{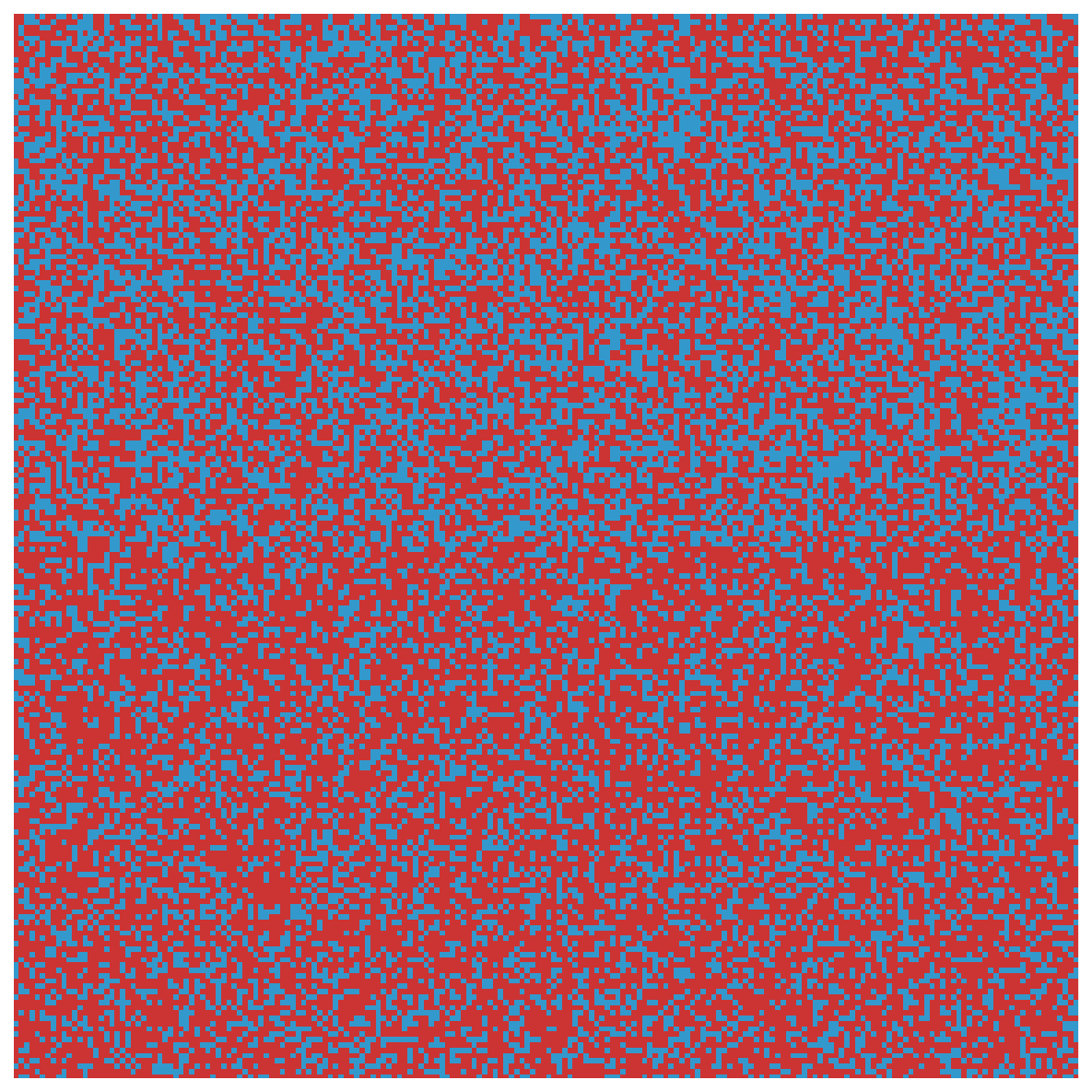}\\
			{\footnotesize t=1}
		\end{minipage}
		\begin{minipage}{0.188\linewidth}
			\centering
			\includegraphics[width=\linewidth]{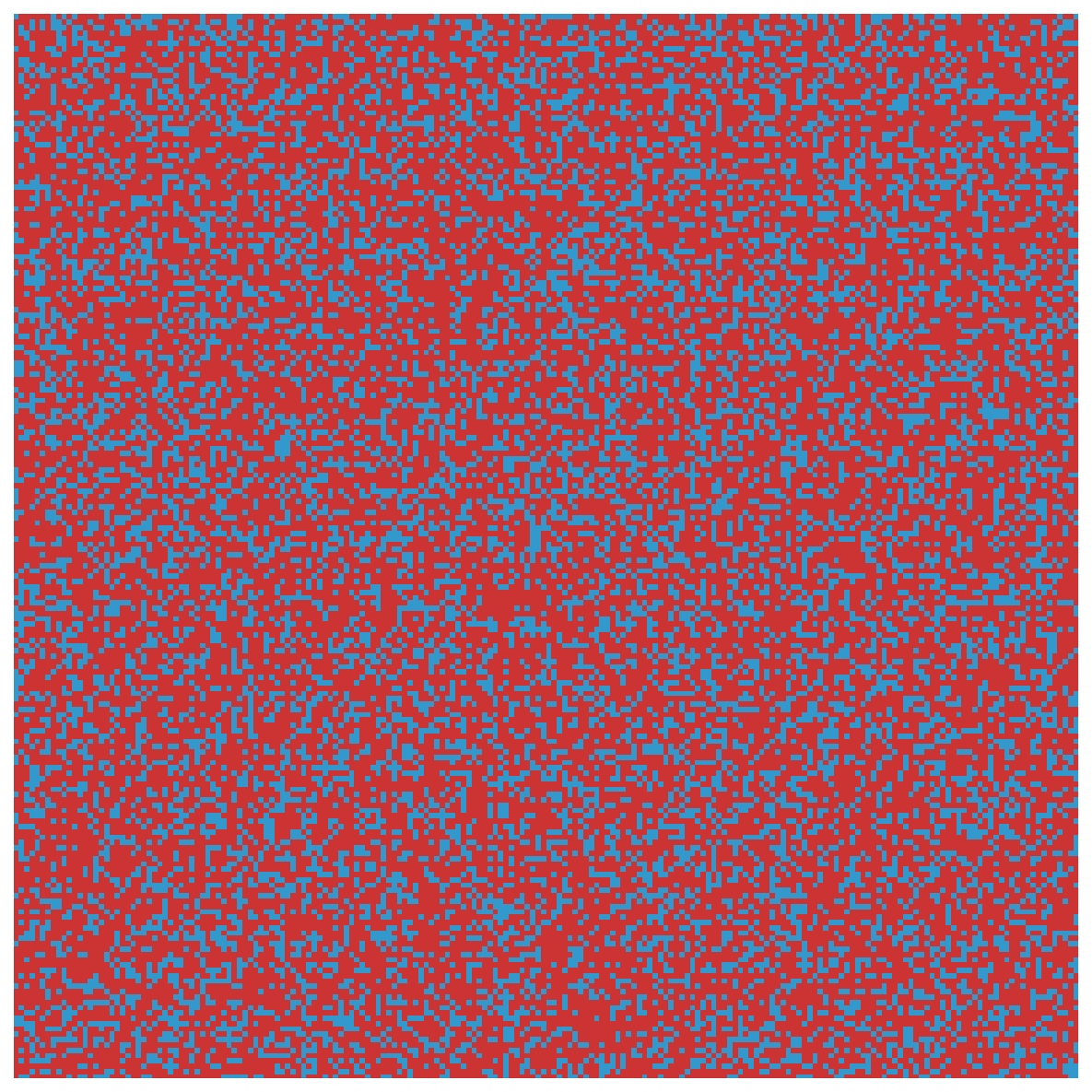}\\
			{\footnotesize t=10}
		\end{minipage}
		\begin{minipage}{0.188\linewidth}
			\centering
			\includegraphics[width=\linewidth]{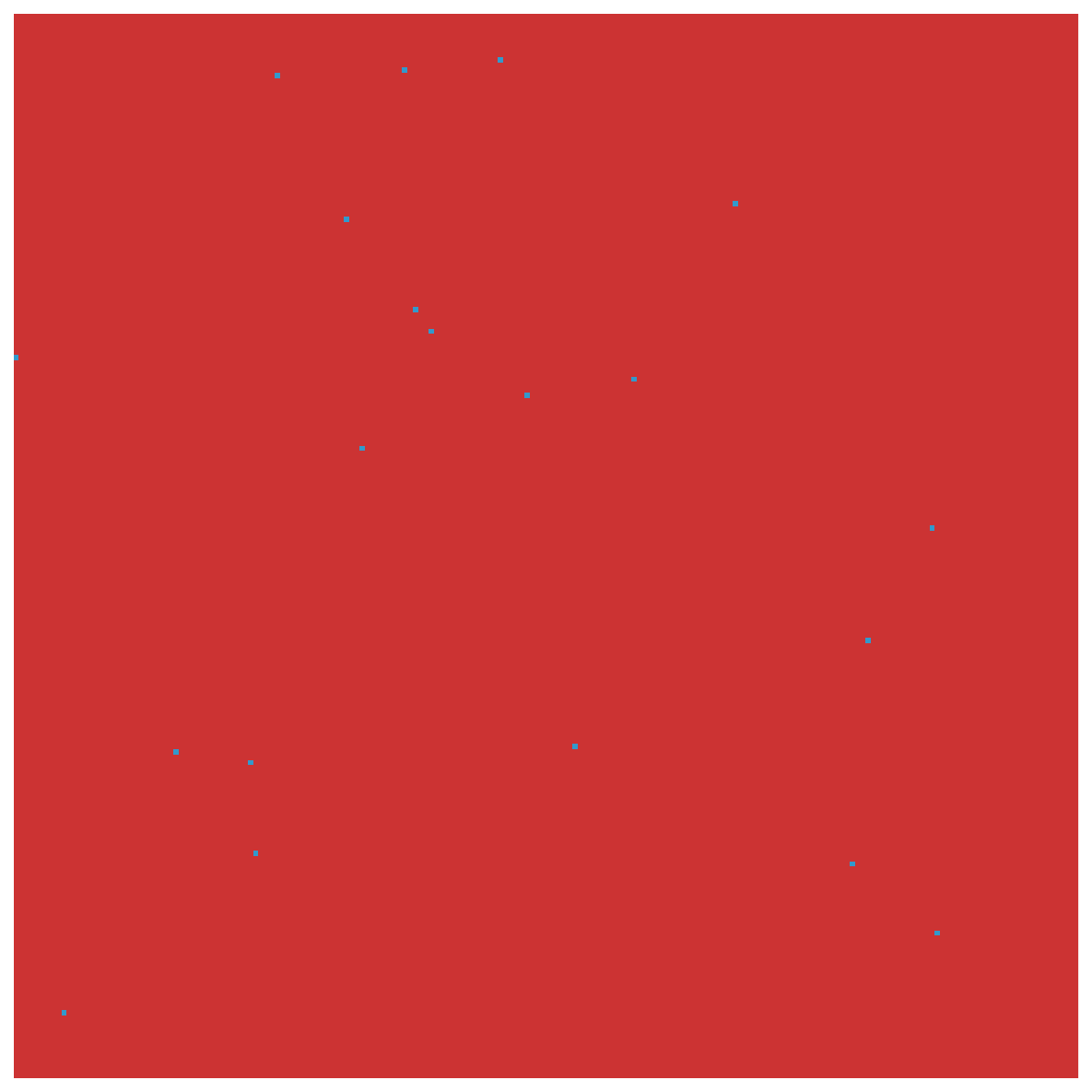}\\
			{\footnotesize t=100}
		\end{minipage}
		\begin{minipage}{0.188\linewidth}
			\centering
			\includegraphics[width=\linewidth]{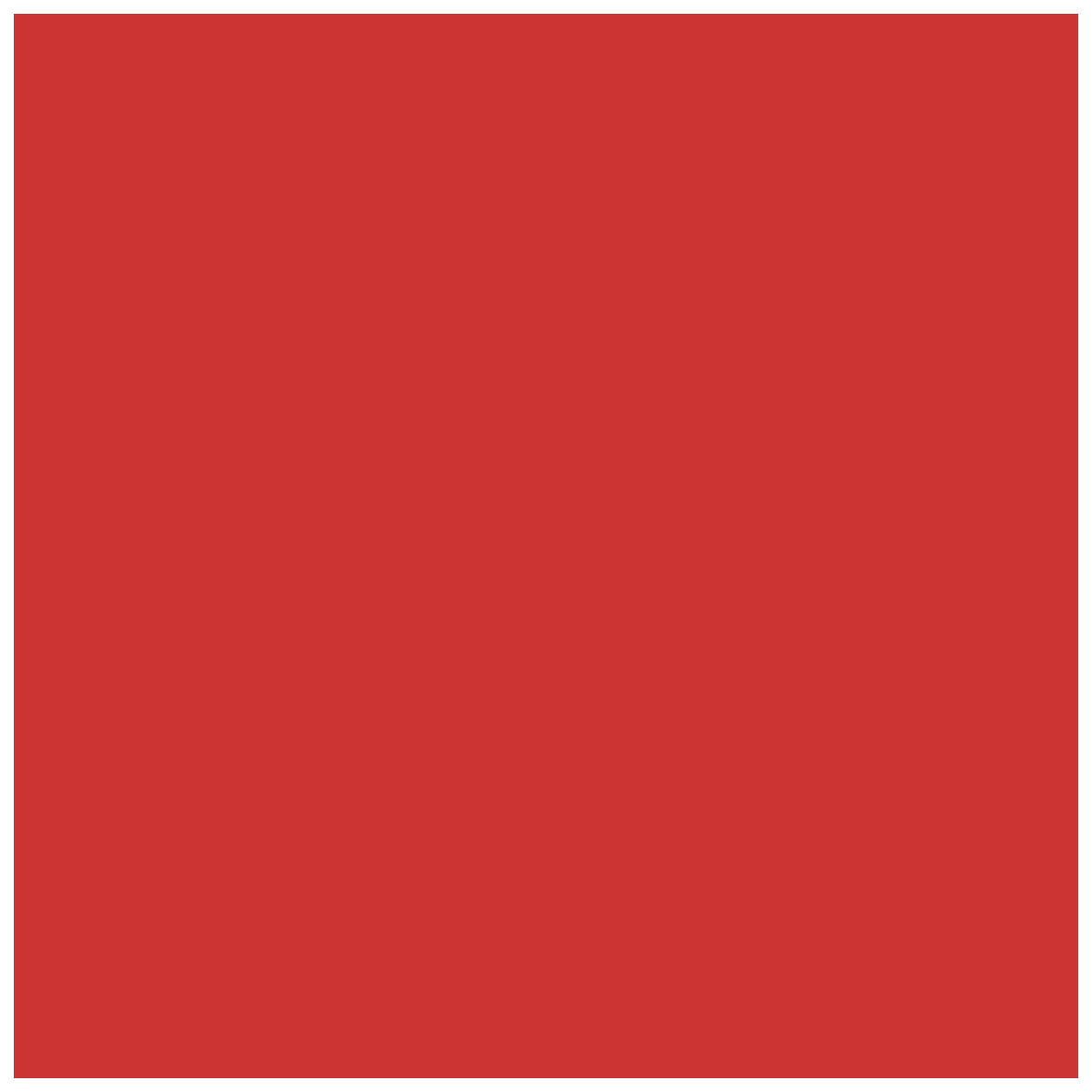}\\
			{\footnotesize t=1000}
		\end{minipage}
		\vspace{2mm}
		\\
		\centering
		{\footnotesize (a) r=4.0}
	\end{minipage}
	\hfill
	\begin{minipage}{0.45\linewidth}
		\begin{minipage}{\linewidth}
			\centering
			\includegraphics[width=\linewidth]{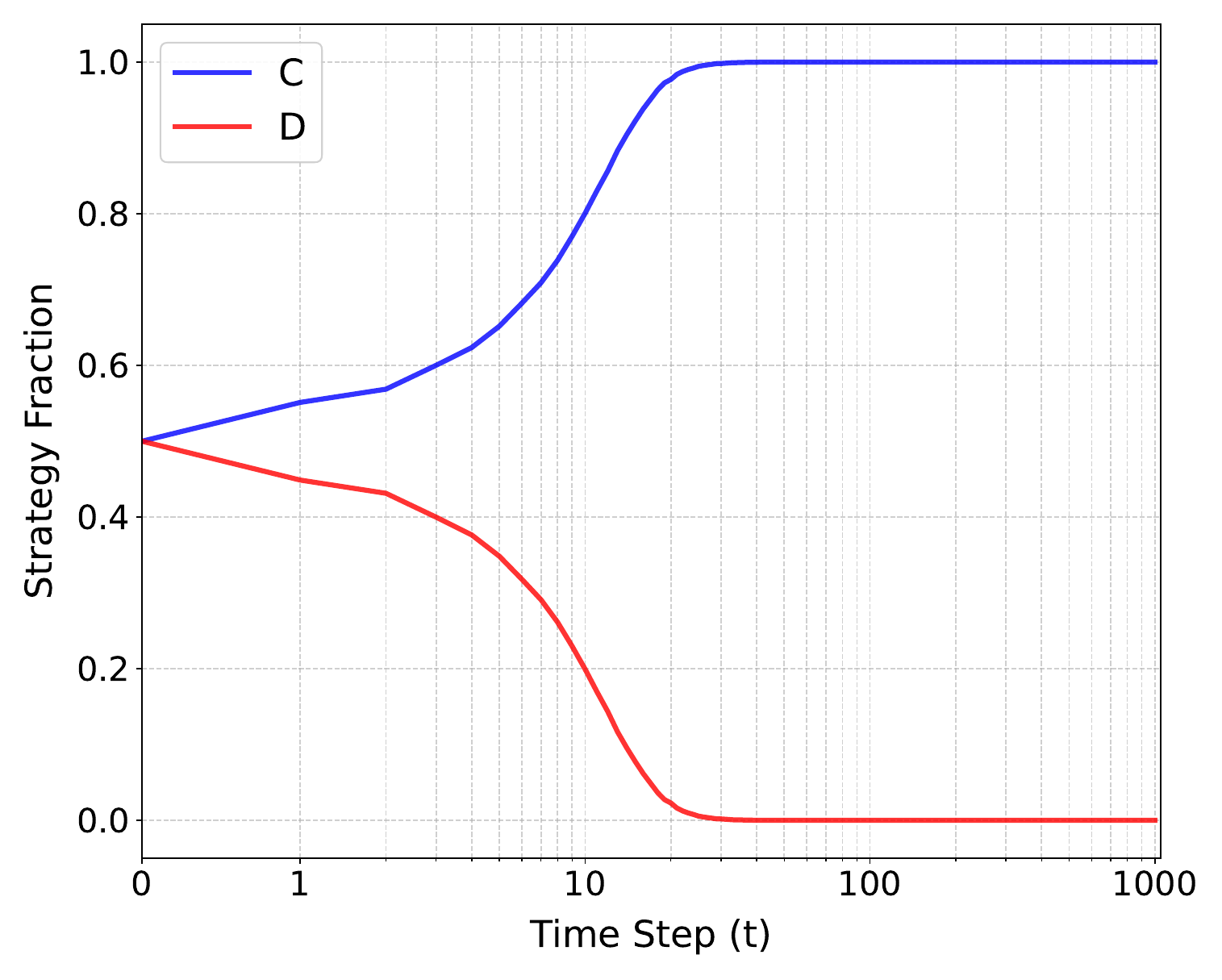}\\
		\end{minipage}
		\vspace{2mm}
		\\
		\begin{minipage}{0.188\linewidth}
			\centering
			\includegraphics[width=\linewidth]{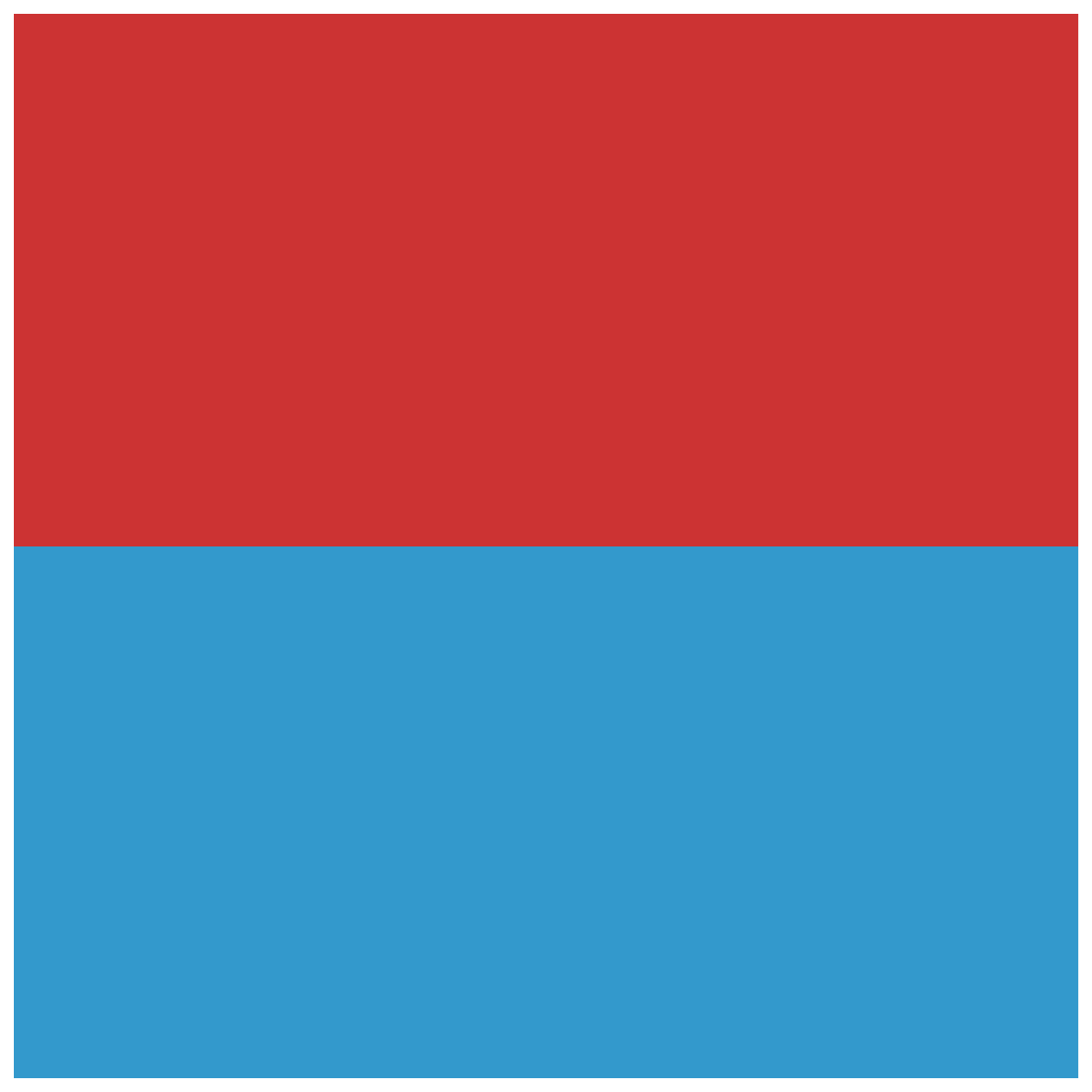}\\
			{\footnotesize t=0}
		\end{minipage}
		\begin{minipage}{0.188\linewidth}
			\centering
			\includegraphics[width=\linewidth]{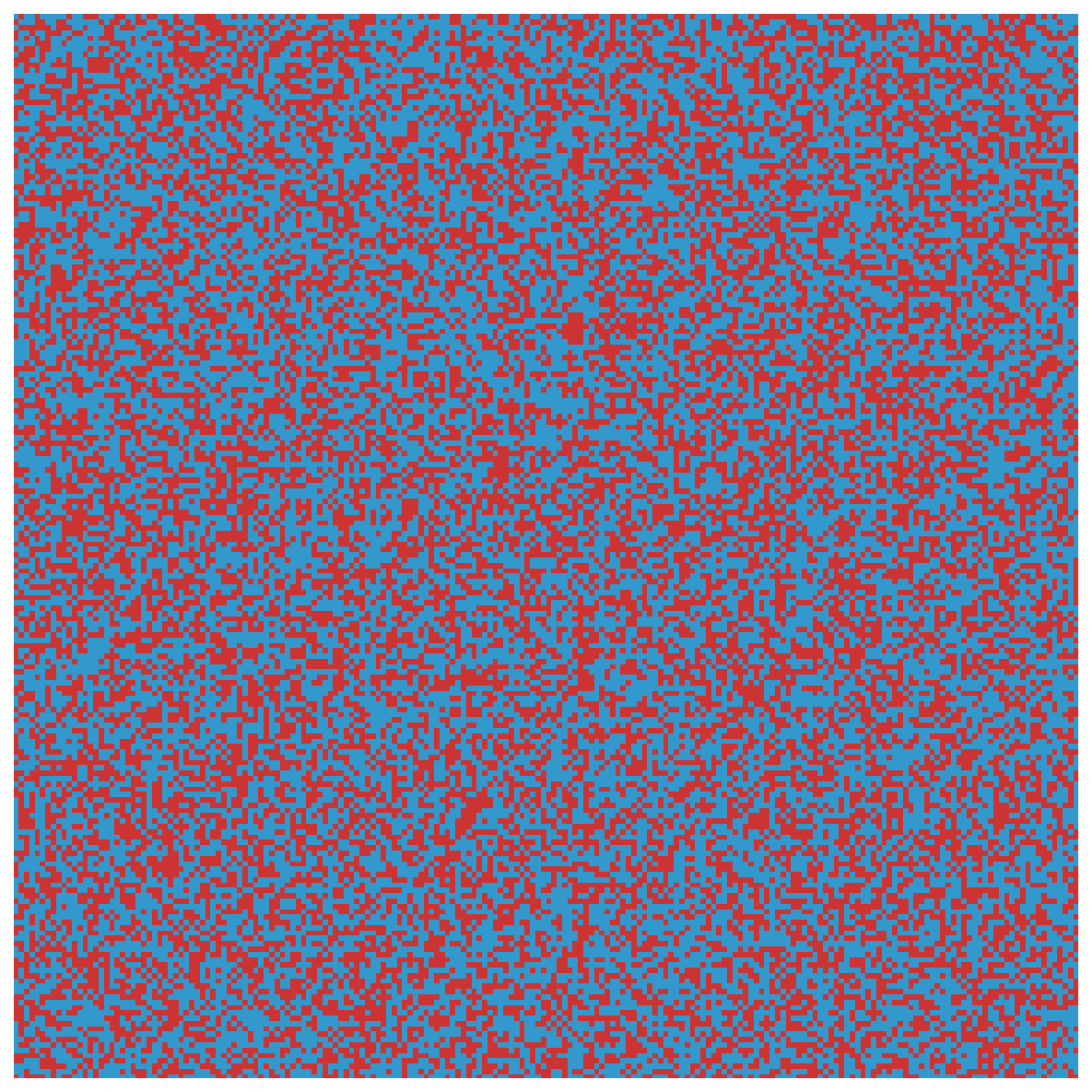}\\
			{\footnotesize t=1}
		\end{minipage}
		\begin{minipage}{0.188\linewidth}
			\centering
			\includegraphics[width=\linewidth]{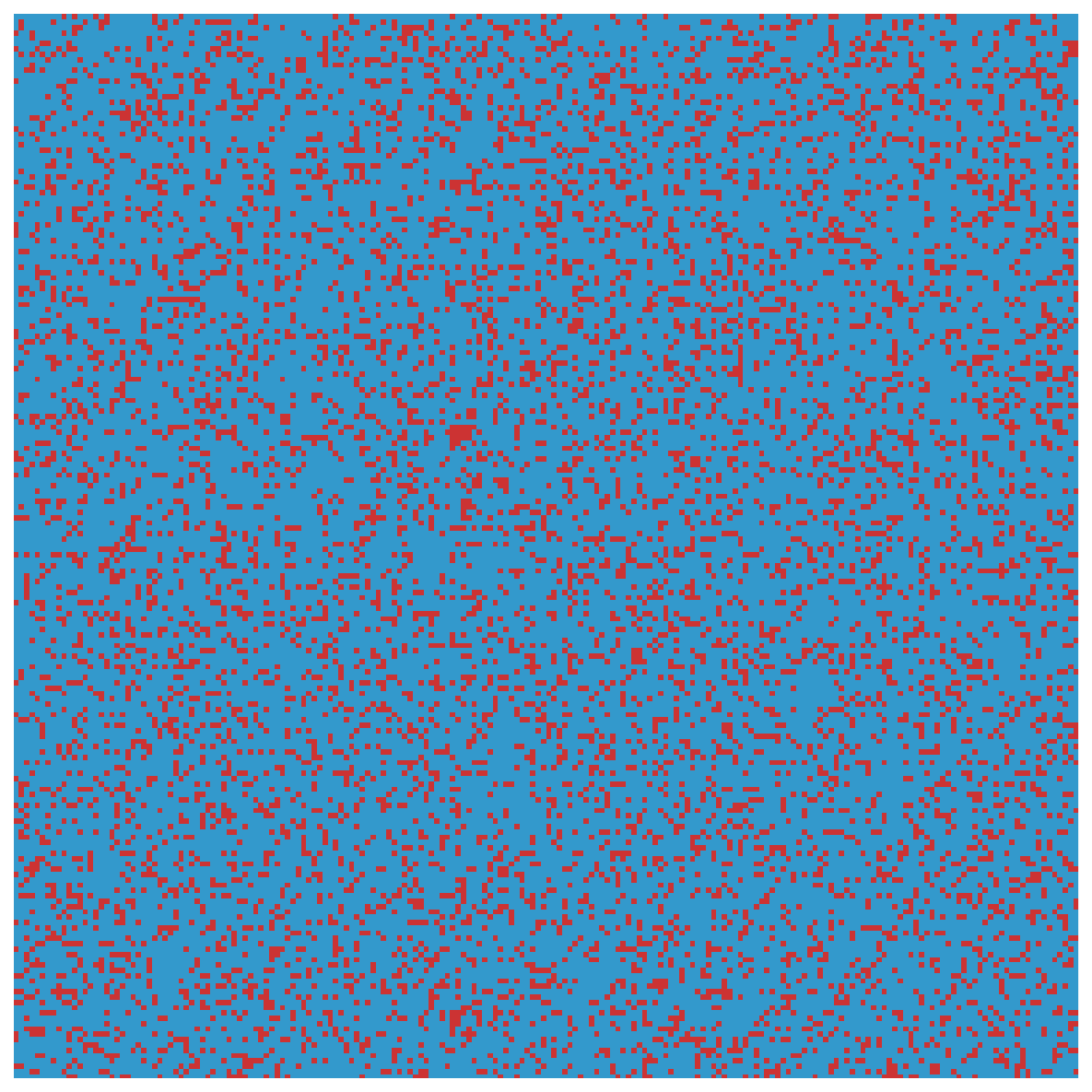}\\
			{\footnotesize t=10}
		\end{minipage}
		\begin{minipage}{0.188\linewidth}
			\centering
			\includegraphics[width=\linewidth]{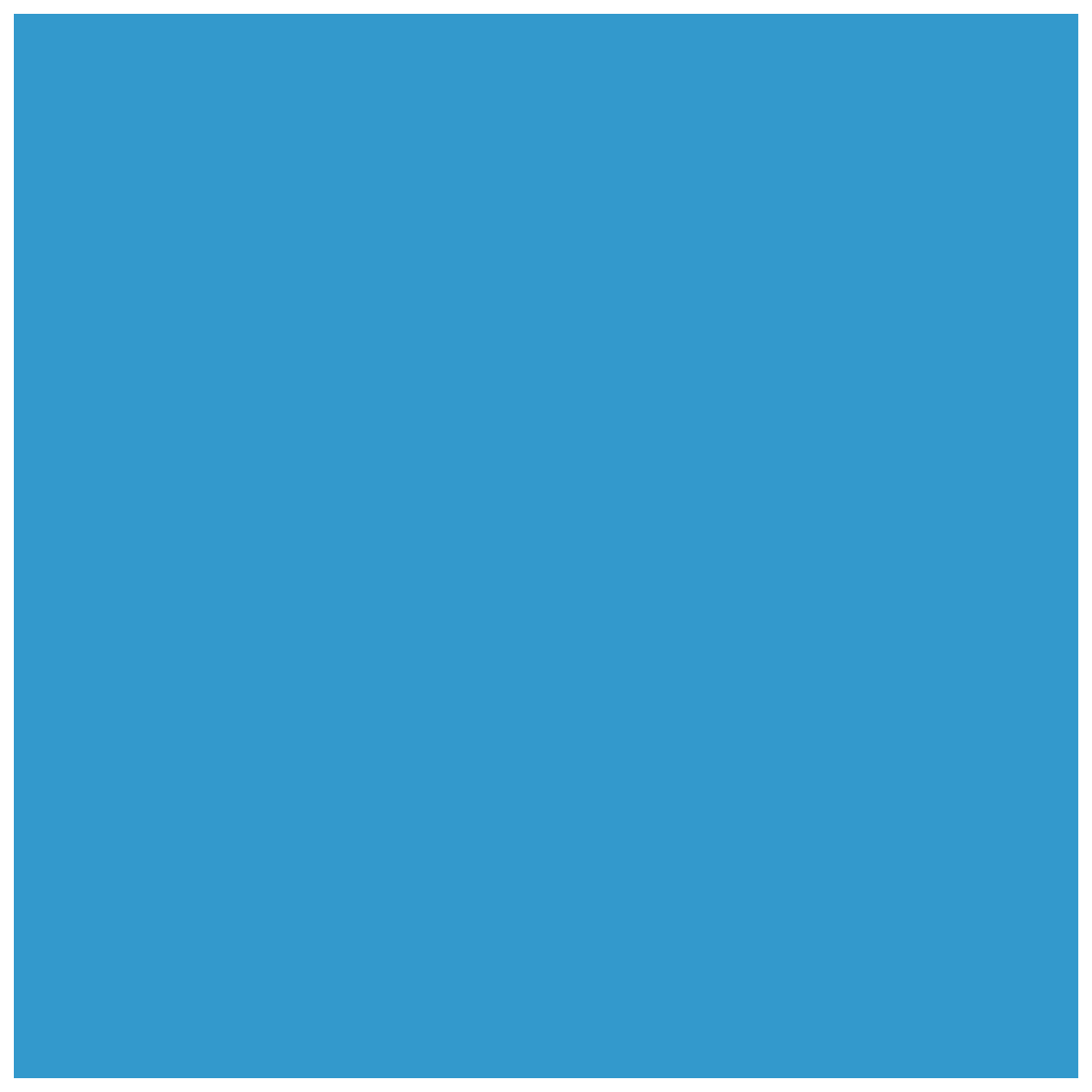}\\
			{\footnotesize t=100}
		\end{minipage}
		\begin{minipage}{0.188\linewidth}
			\centering
			\includegraphics[width=\linewidth]{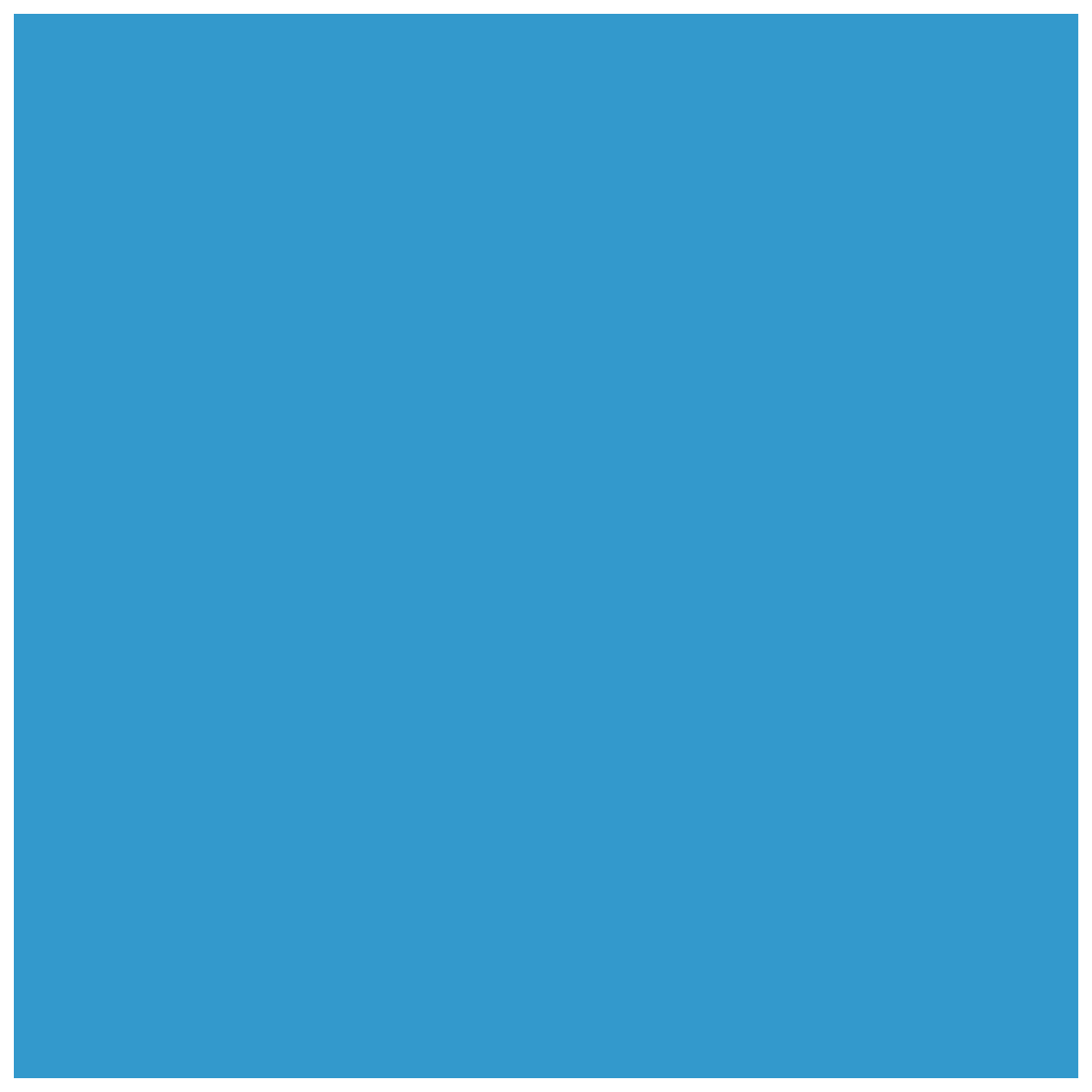}\\
			{\footnotesize t=1000}
		\end{minipage}
		\vspace{2mm}
		\\
		\centering 
		{\footnotesize (b) r=4.5}
	\end{minipage}
	\\
	[2mm]
	\begin{minipage}{\linewidth}
		\begin{minipage}{0.188\linewidth}
			\centering
			\includegraphics[width=\linewidth]{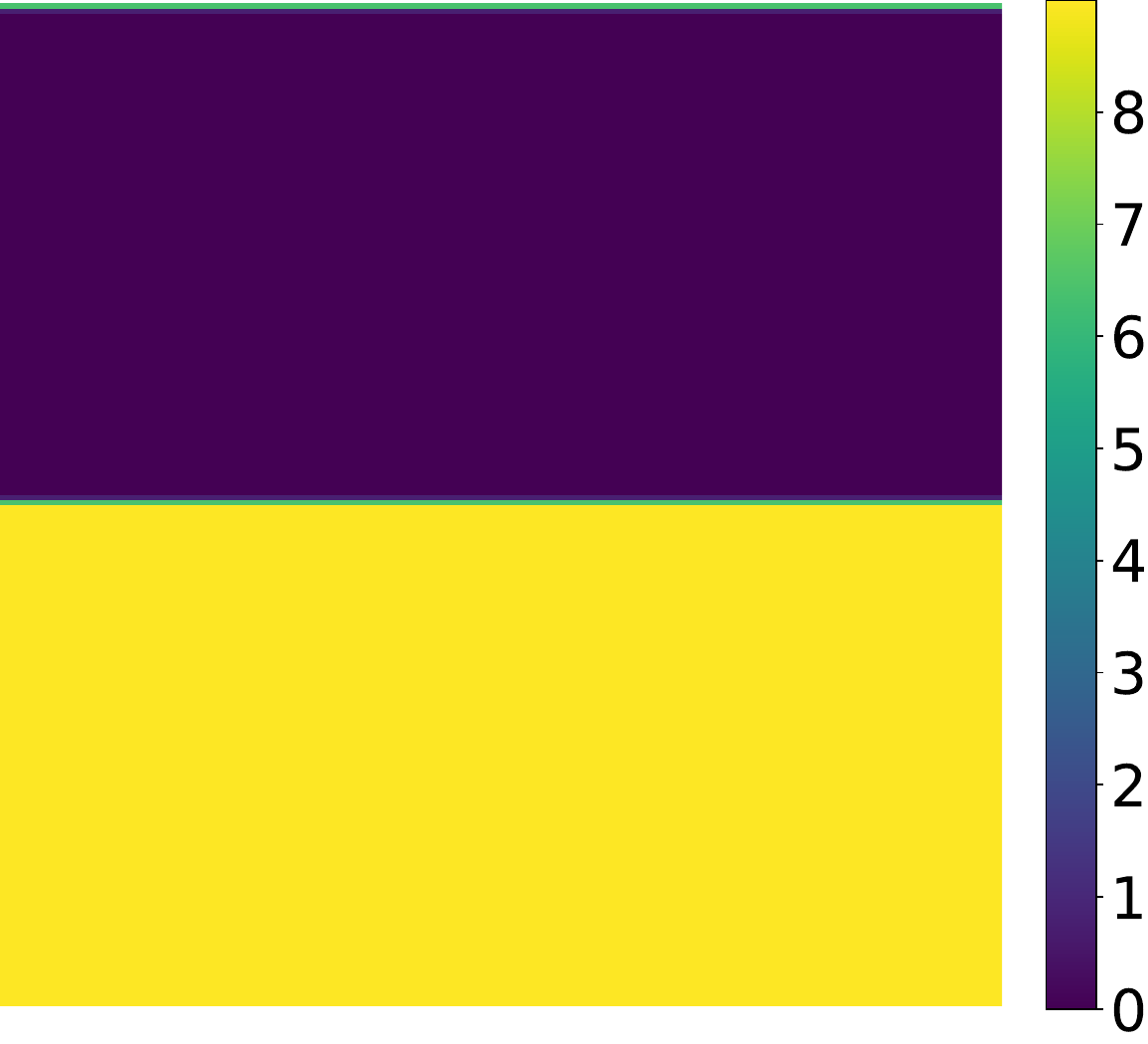}\\
			{\footnotesize t=0}
		\end{minipage}
		\hfill
		\begin{minipage}{0.188\linewidth}
			\centering
			\includegraphics[width=\linewidth]{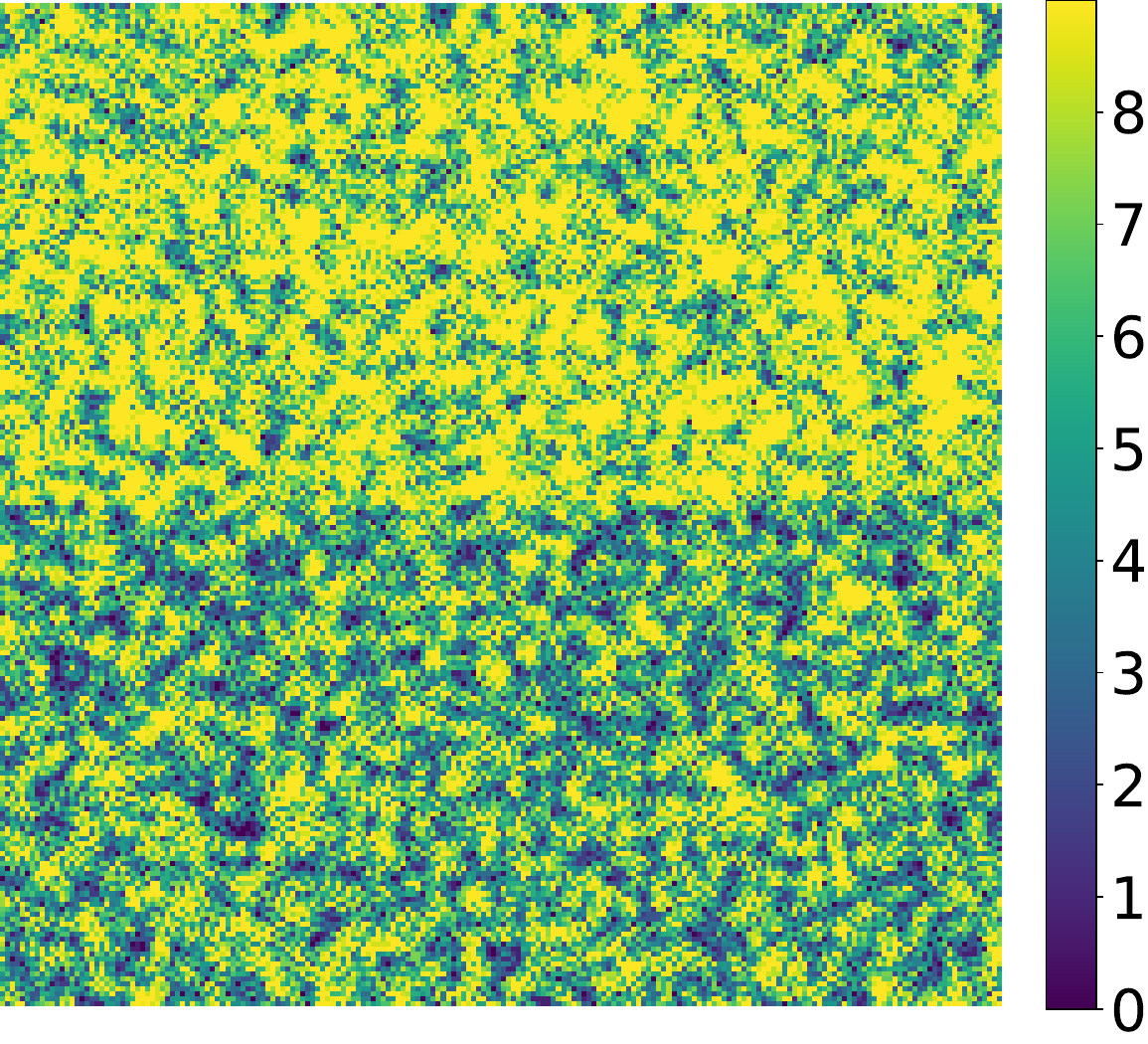}\\
			{\footnotesize t=1}
		\end{minipage}
		\hfill
		\begin{minipage}{0.188\linewidth}
			\centering
			\includegraphics[width=\linewidth]{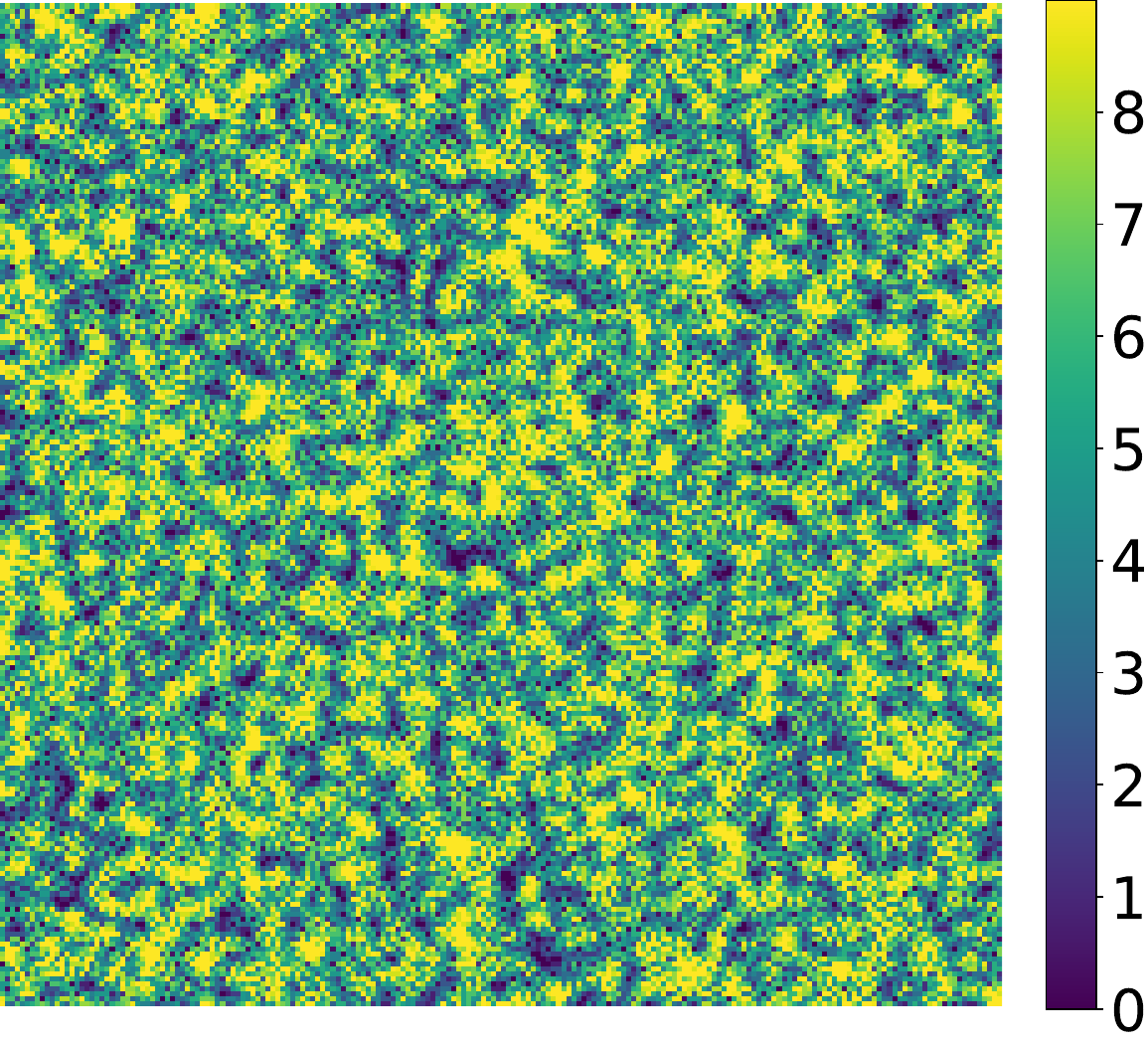}\\
			{\footnotesize t=10}
		\end{minipage}
		\hfill
		\begin{minipage}{0.188\linewidth}
			\centering
			\includegraphics[width=\linewidth]{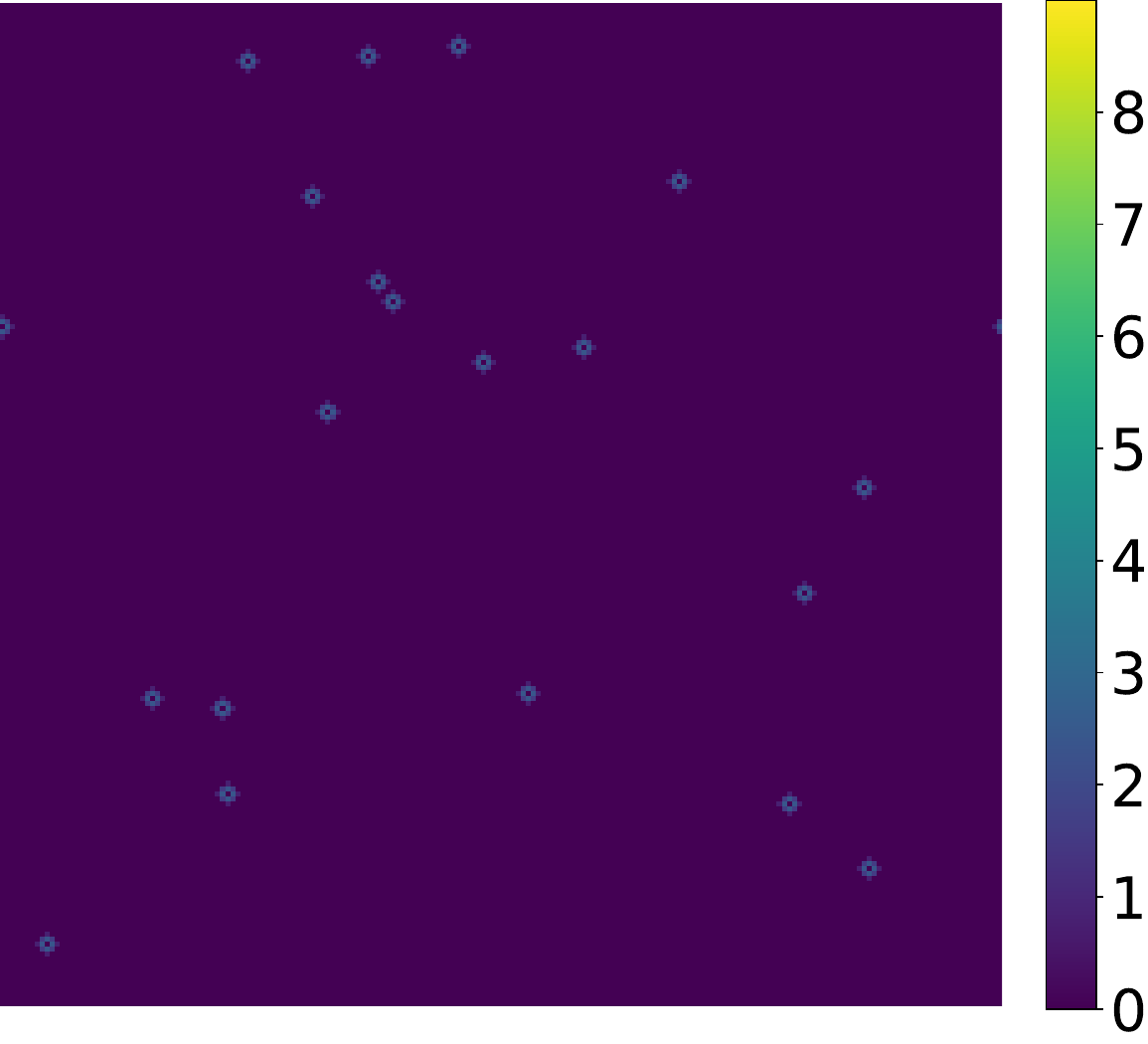}\\
			{\footnotesize t=100}
		\end{minipage}
		\hfill
		\begin{minipage}{0.188\linewidth}
			\centering
			\includegraphics[width=\linewidth]{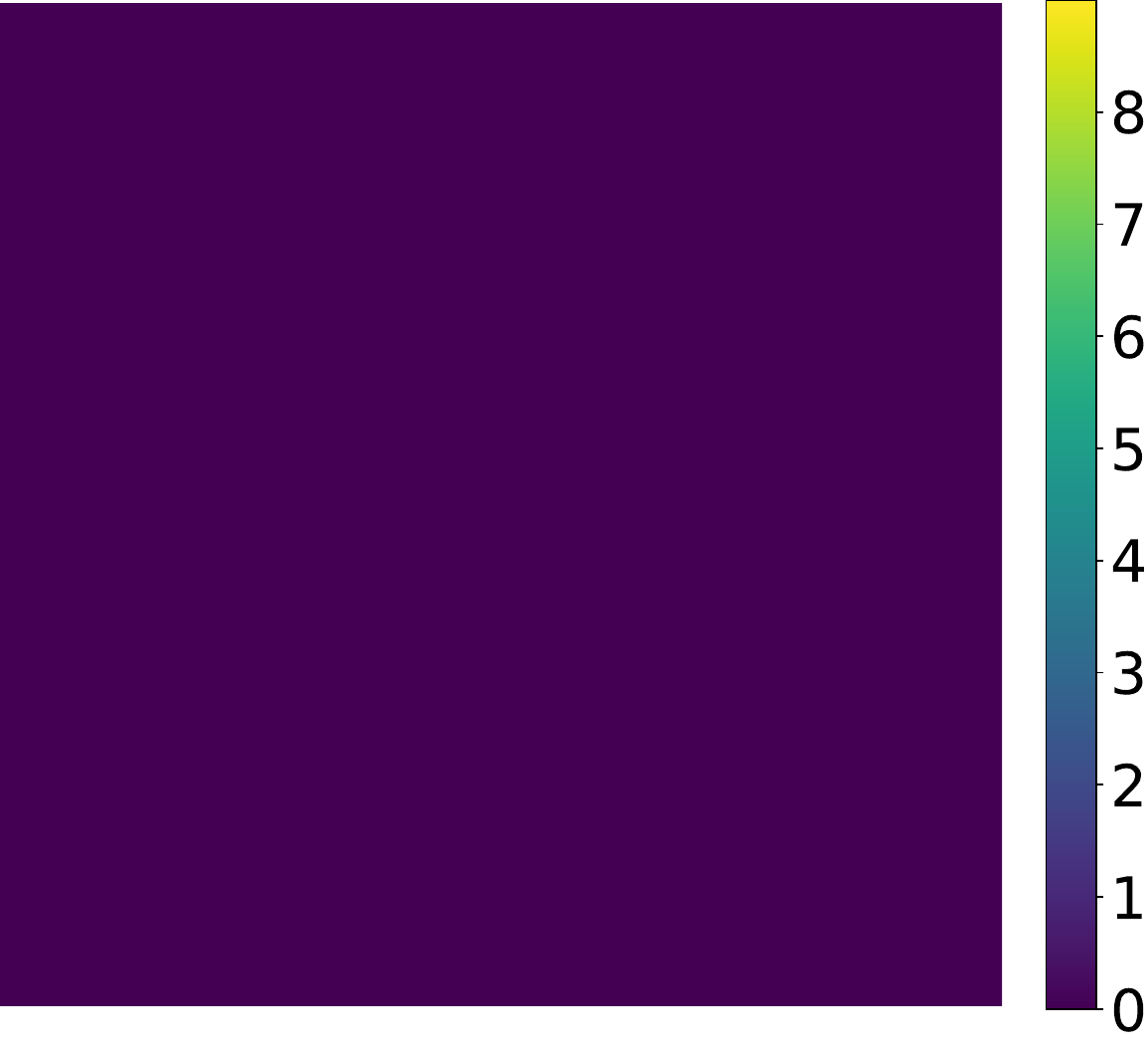}\\
			{\footnotesize t=1000}
		\end{minipage}
		\vspace{2mm}
		\\
		\centering
		{\footnotesize (c) r=4.0 (Payoff heatmaps)}
	\end{minipage}
	\\
	[2mm]
	\begin{minipage}{\linewidth}
		\begin{minipage}{0.188\linewidth}
			\centering
			\includegraphics[width=\linewidth]{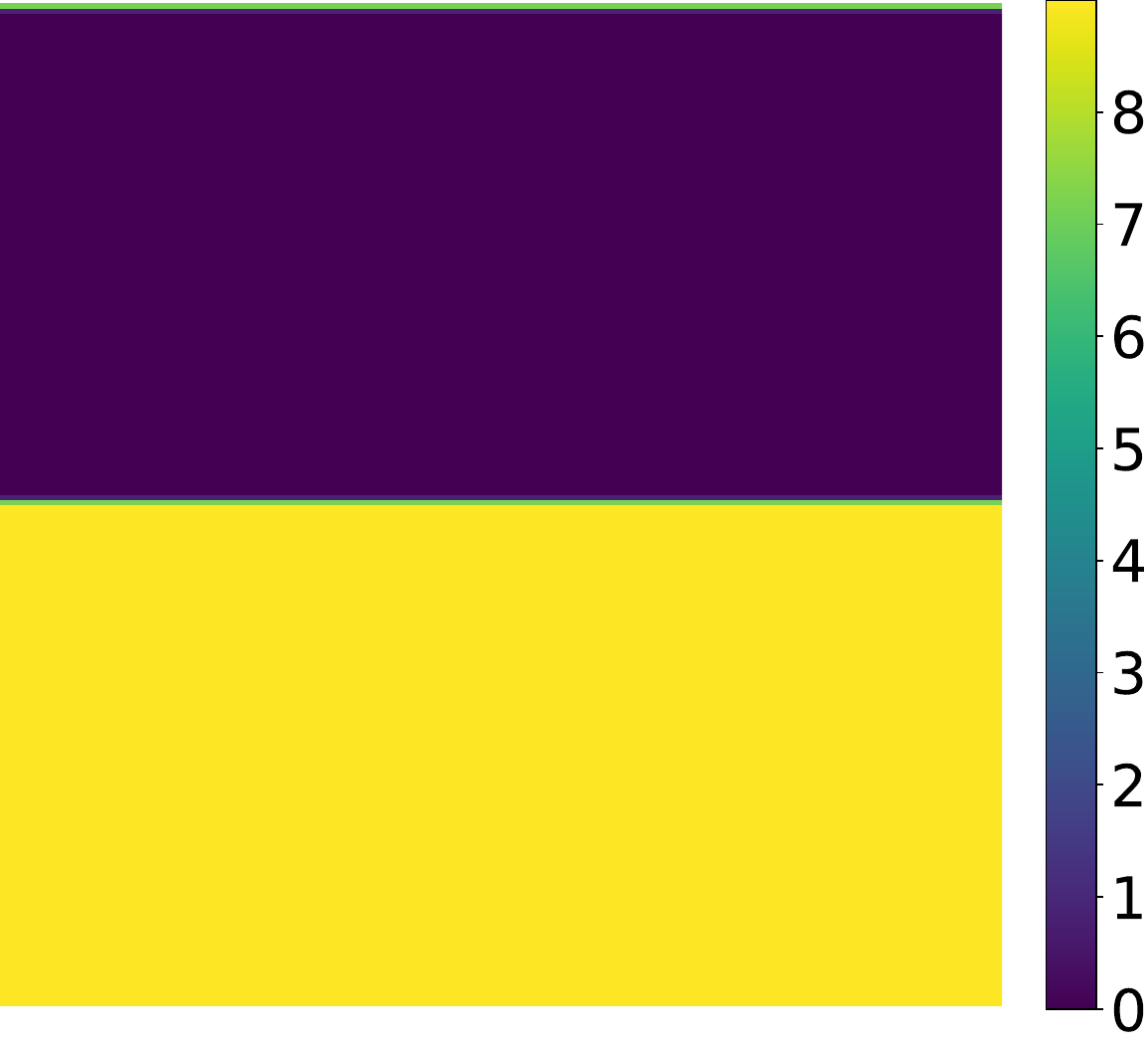}\\
			{\footnotesize t=0}
		\end{minipage}
		\hfill
		\begin{minipage}{0.188\linewidth}
			\centering
			\includegraphics[width=\linewidth]{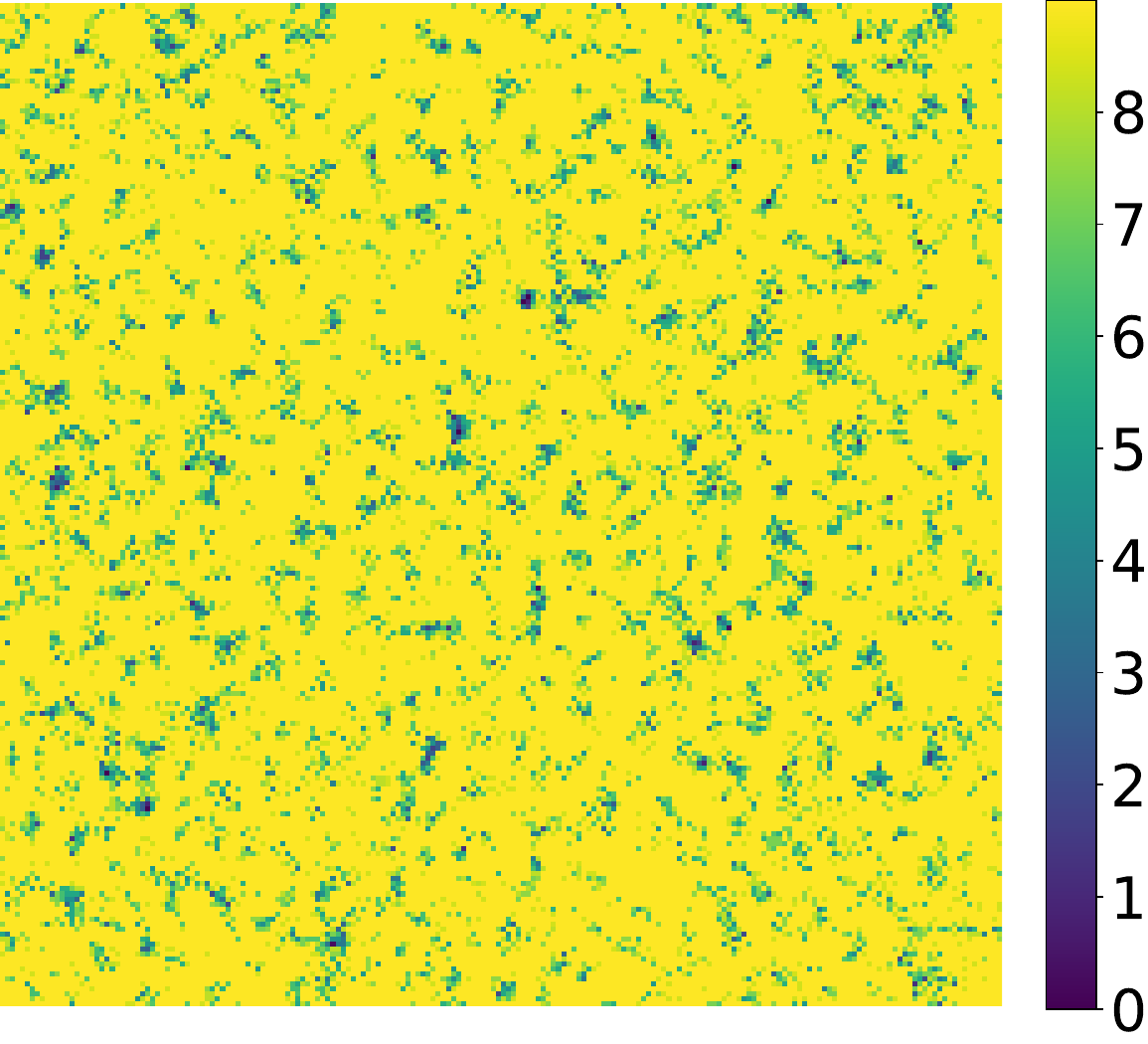}\\
			{\footnotesize t=1}
		\end{minipage}
		\hfill
		\begin{minipage}{0.188\linewidth}
			\centering
			\includegraphics[width=\linewidth]{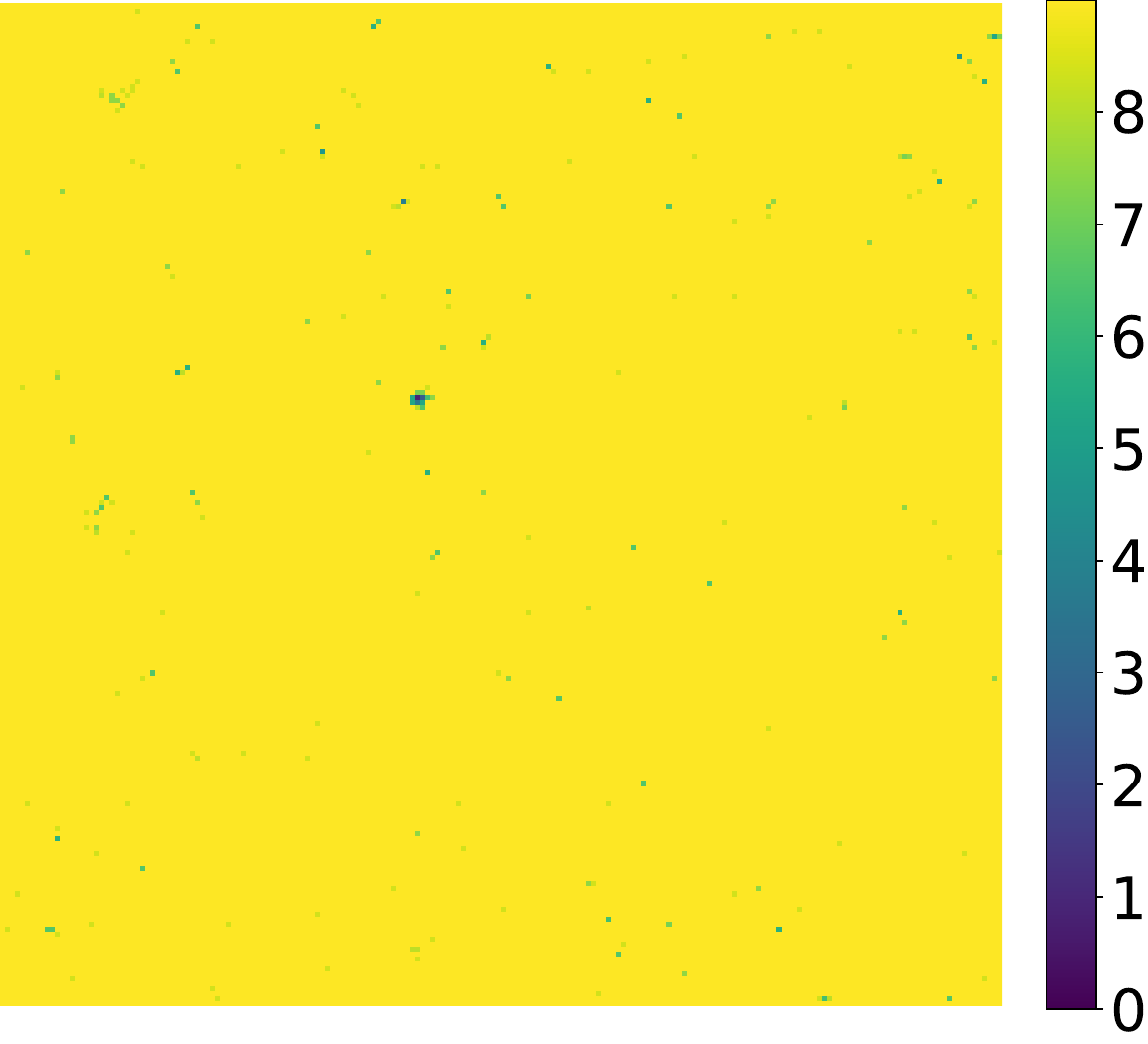}\\
			{\footnotesize t=10}
		\end{minipage}
		\hfill
		\begin{minipage}{0.188\linewidth}
			\centering
			\includegraphics[width=\linewidth]{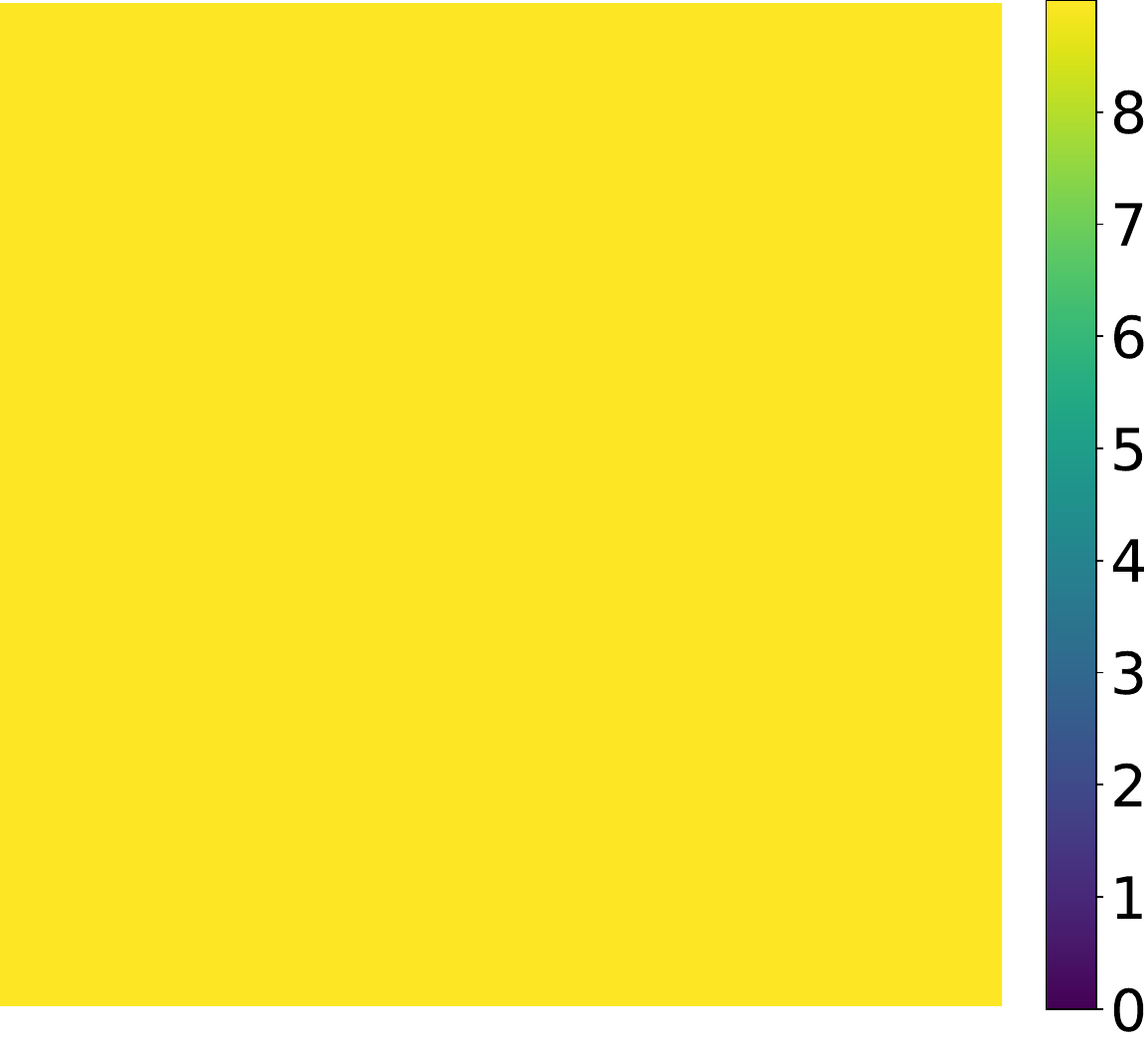}\\
			{\footnotesize t=100}
		\end{minipage}
		\hfill
		\begin{minipage}{0.188\linewidth}
			\centering
			\includegraphics[width=\linewidth]{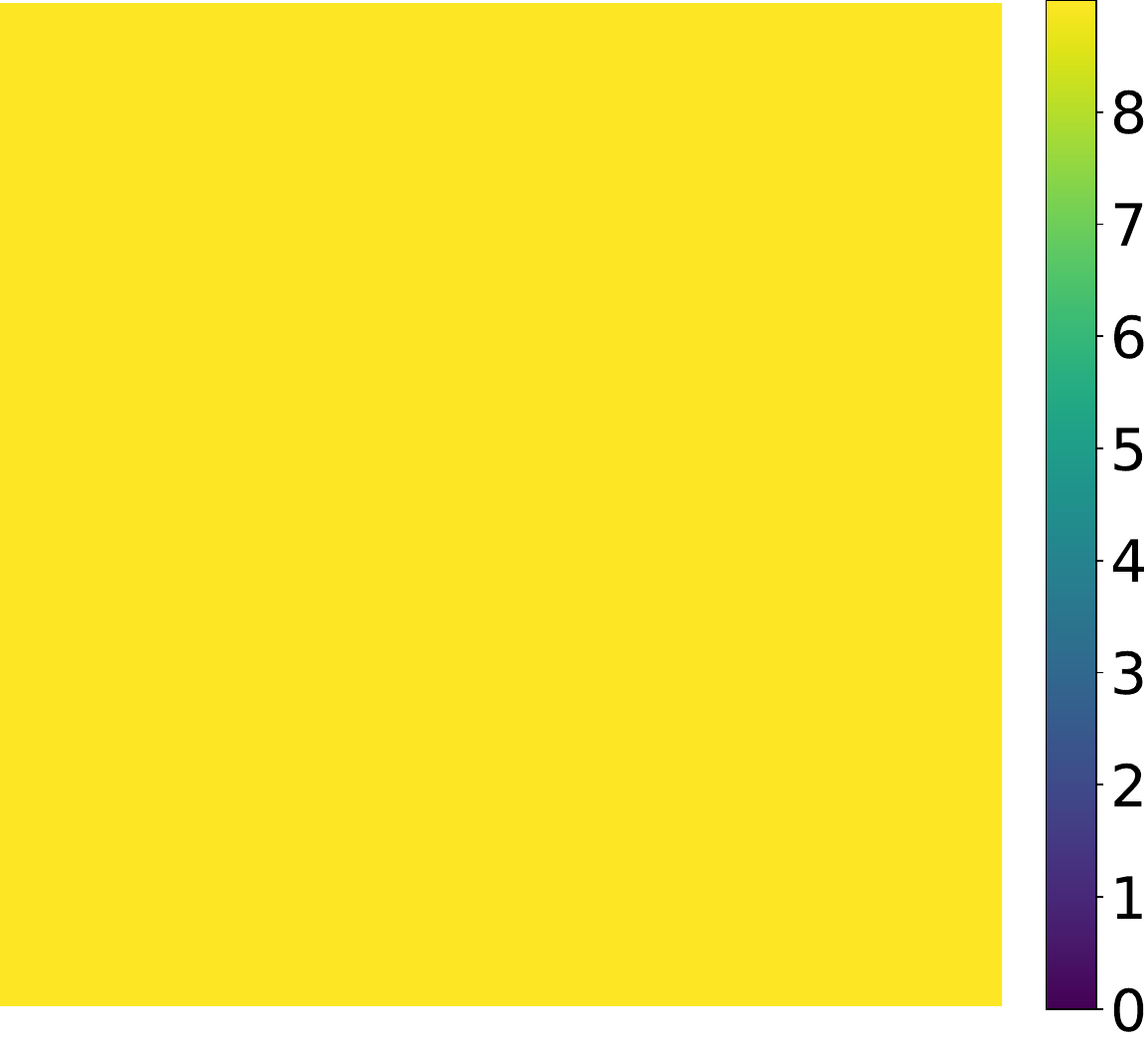}\\
			{\footnotesize t=1000}
		\end{minipage}
		\vspace{2mm}
		\\
		\centering
		{\footnotesize (d) r=4.5 (Payoff heatmaps)}
		
	\end{minipage}
	\caption{Evolutionary dynamics of LMFPPO-UBP under half-and-half initialization. (a,c) At $r=4.0$, cooperation collapses to full defection within 80 iterations, (b,d) At $r=4.5$, rapid global cooperation ($f_C=1.0$) emerges within 30 iterations. Spatial snapshots (top) show strategy propagation, while payoff heatmaps (bottom) reflect collective welfare changes.}
	\label{fig:halfandhalf}
\end{figure*}
The evolutionary dynamics of the LMFPPO‑UBP algorithm were examined under spatially partitioned initial conditions. An $L \times L$ grid was divided equally, with the upper half initialized as defectors $D$ and the lower half as cooperators $C$, creating a well‑defined strategic boundary. As illustrated in Figure~\ref{fig:halfandhalf}, two enhancement factors, $r = 4.0$ and $r = 4.5$, were analyzed to probe the system's sensitivity and the algorithm's capacity to resolve spatial conflict. The results demonstrate a sharp threshold effect governed by the parameter $r$. At $r = 4.0$, the system converges monotonically to an all‑defection equilibrium $f_C = 0$ within 80 iterations. Defectors invade from the initially occupied upper region, ultimately dominating the grid. Corresponding payoff heatmaps reflect a systemic decline in collective welfare. This confirms that defection remains the dominant attractor when the enhancement factor lies below the critical threshold. In contrast, at $r = 4.5$, the system rapidly transitions to full cooperation $f_C = 1.0$ within 30 iterations. Cooperators expand upward from the lower half, forming a uniform cooperative pattern. Payoff heatmaps show a marked improvement in both individual and collective returns. This indicates the emergence of positive feedback under a sufficiently high enhancement factor.

The half‑and‑half initialization provides a stringent test of the algorithm's ability to manage spatial strategic conflict. The successful expansion of cooperation at $r = 4.5$ reveals a key mechanism underpinning the LMFPPO‑UBP framework. Defectors located at the frontier between cooperative and defective regions face maximal punishment because they are surrounded by cooperating neighbors. This creates a localized payoff gradient that systematically disadvantages defection at the boundary. Consequently, cooperator clusters are not only stabilized against invasion but can also actively expand into defector territories. 

\subsection{LMFPPO-UBP with bernoulli random initialization}
\begin{figure*}[htbp!]
	\begin{minipage}{0.45\linewidth}
		\begin{minipage}{\linewidth}
			\centering
			\includegraphics[width=\linewidth]{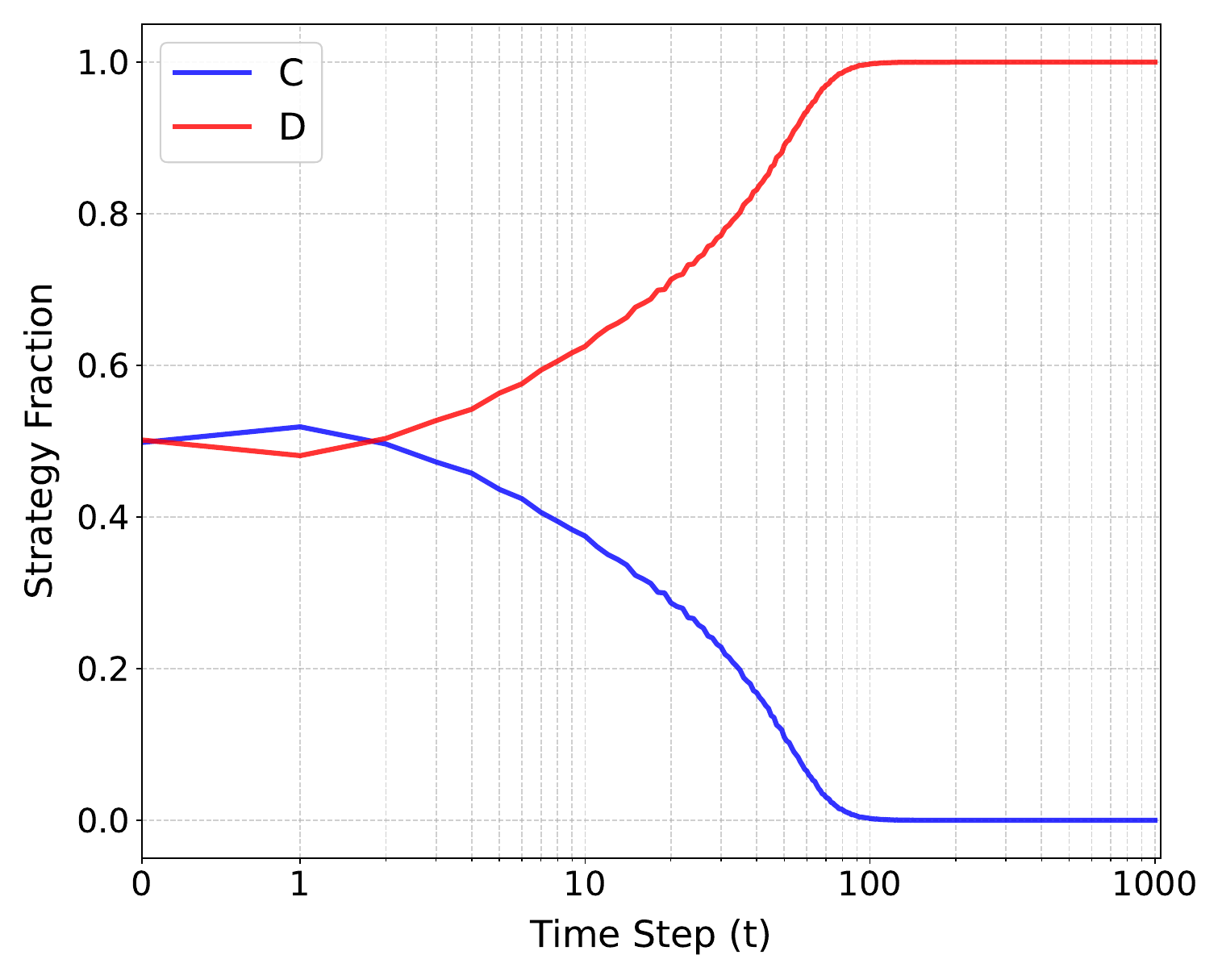}\\
		\end{minipage}
		\vspace{2mm}
		\\
		\begin{minipage}{0.188\linewidth}
			\centering
			\includegraphics[width=\linewidth]{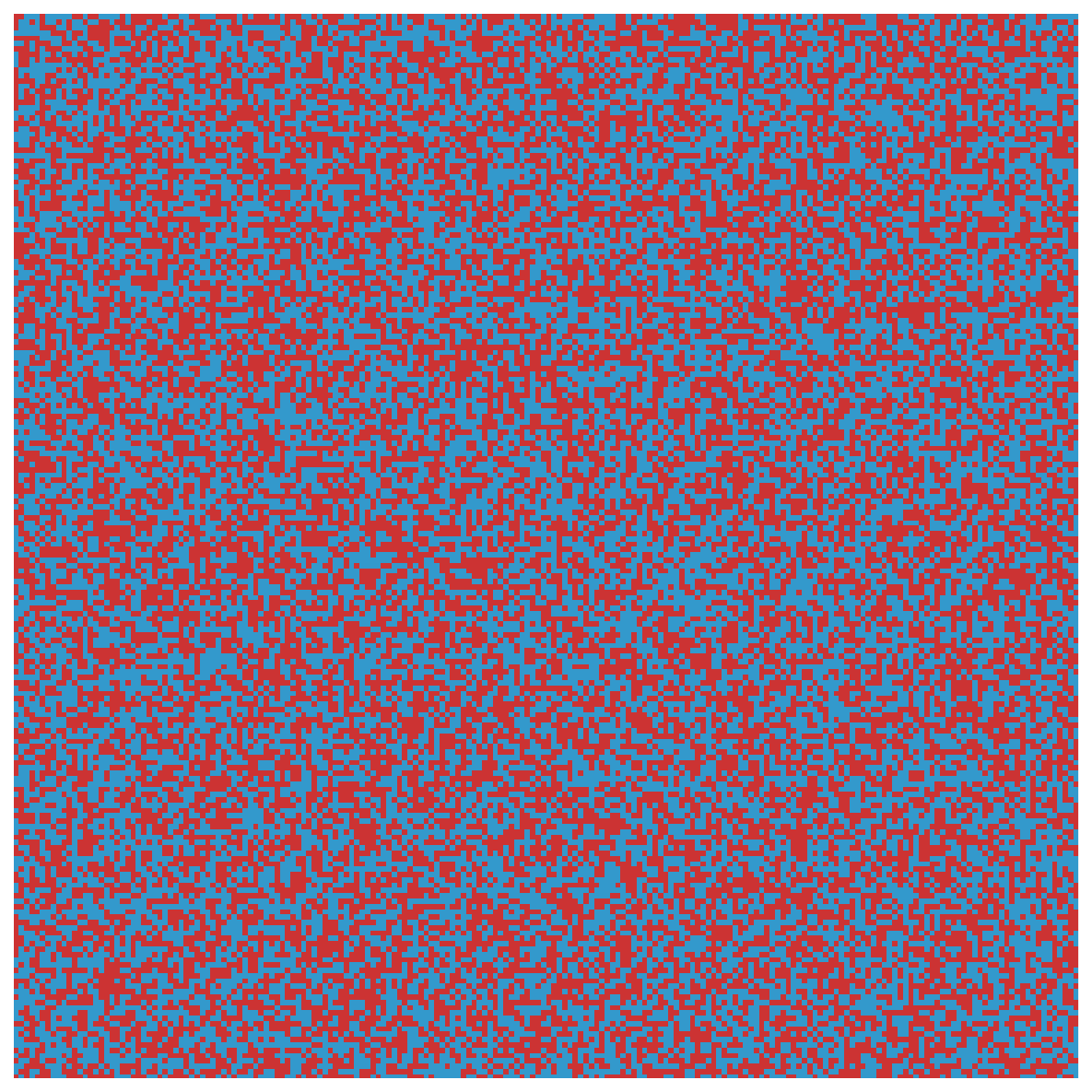}\\
			{\footnotesize t=0}
		\end{minipage}
		\begin{minipage}{0.188\linewidth}
			\centering
			\includegraphics[width=\linewidth]{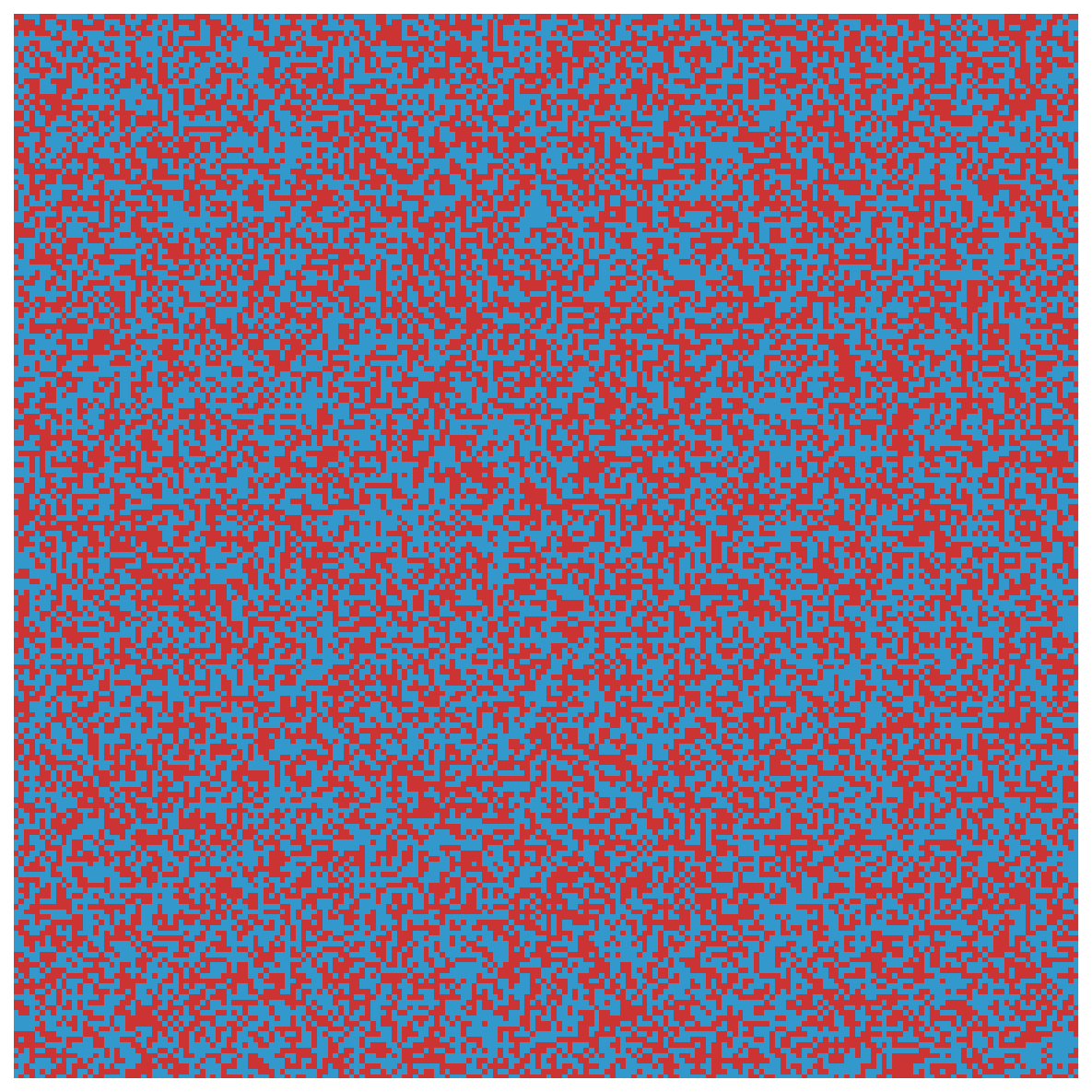}\\
			{\footnotesize t=1}
		\end{minipage}
		\begin{minipage}{0.188\linewidth}
			\centering
			\includegraphics[width=\linewidth]{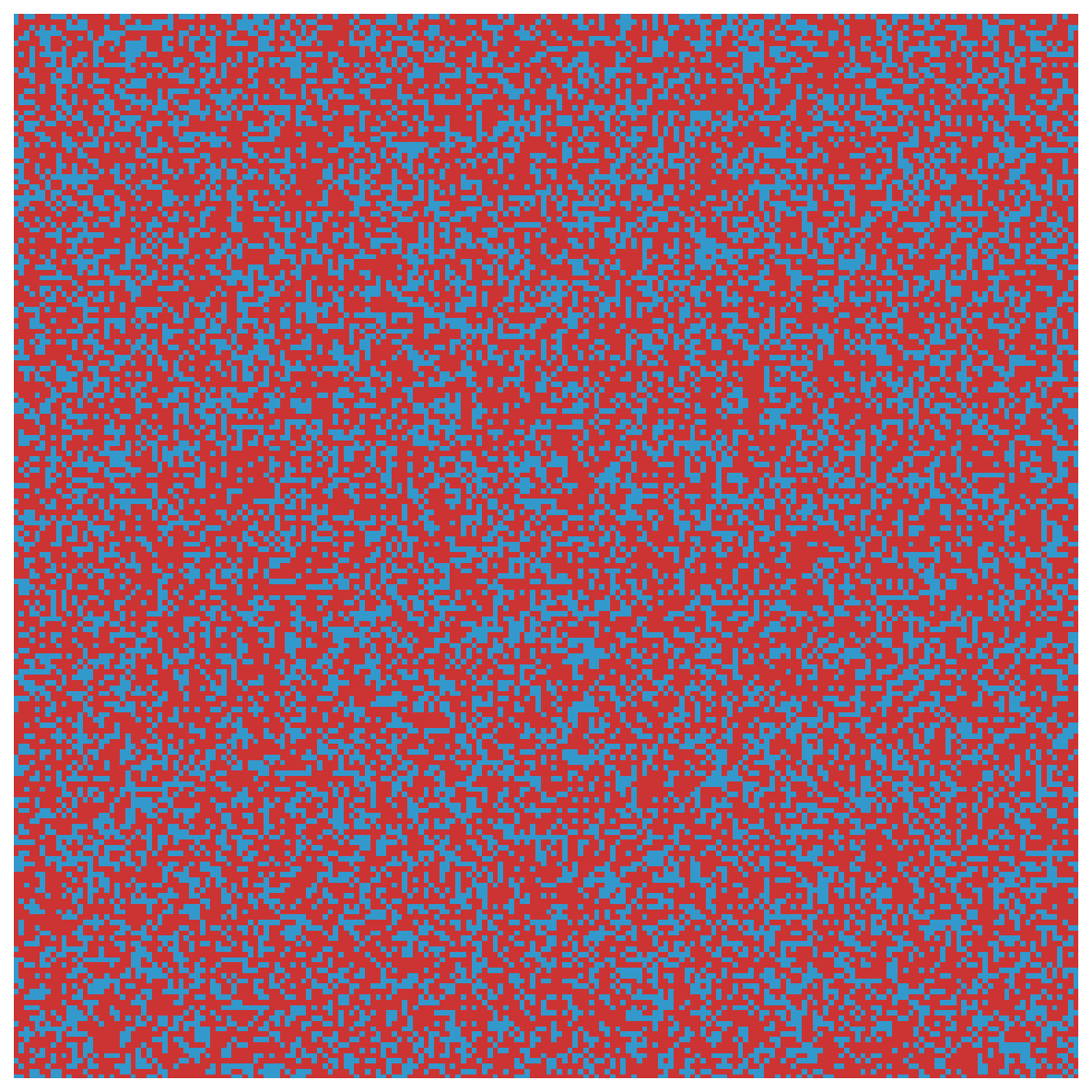}\\
			{\footnotesize t=10}
		\end{minipage}
		\begin{minipage}{0.188\linewidth}
			\centering
			\includegraphics[width=\linewidth]{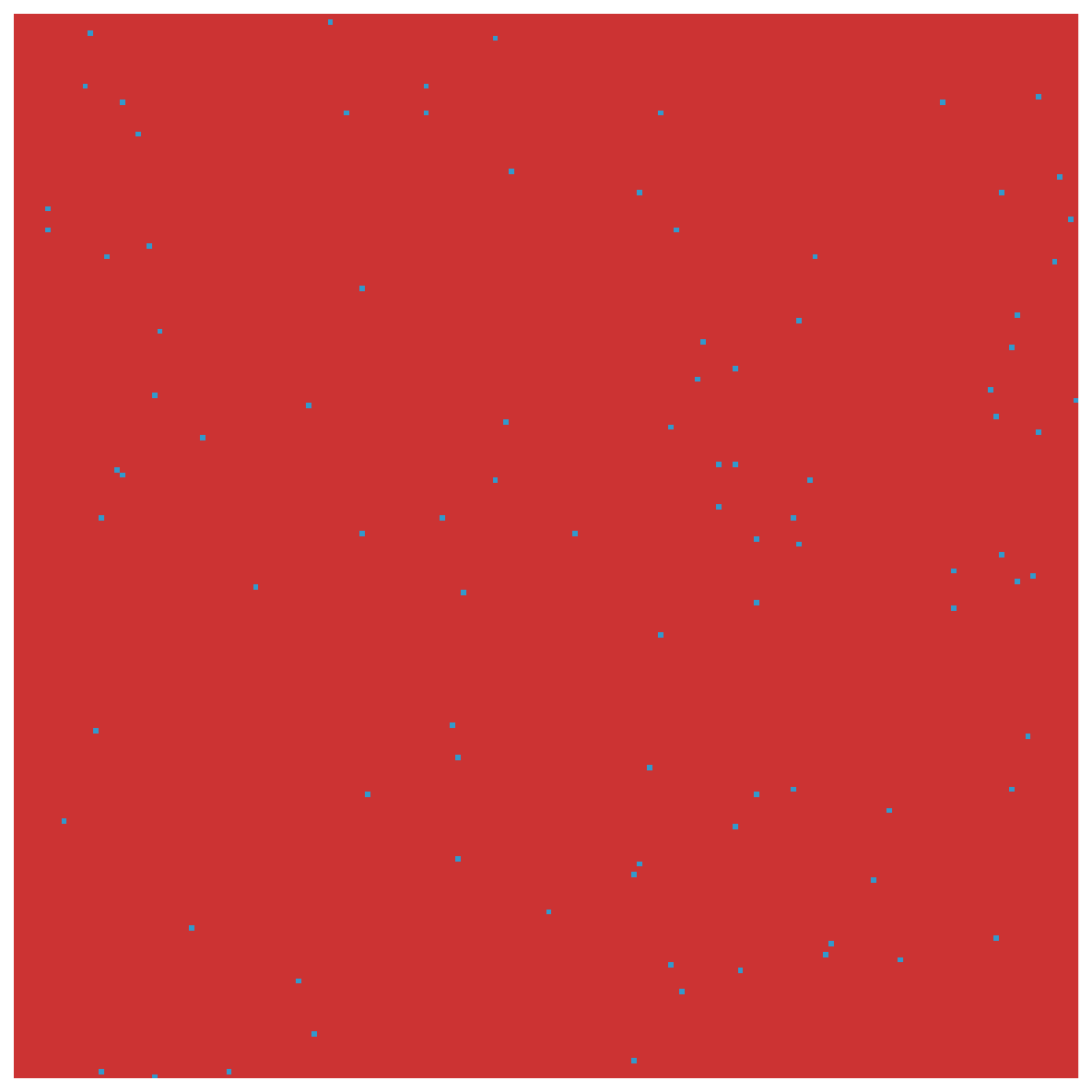}\\
			{\footnotesize t=100}
		\end{minipage}
		\begin{minipage}{0.188\linewidth}
			\centering
			\includegraphics[width=\linewidth]{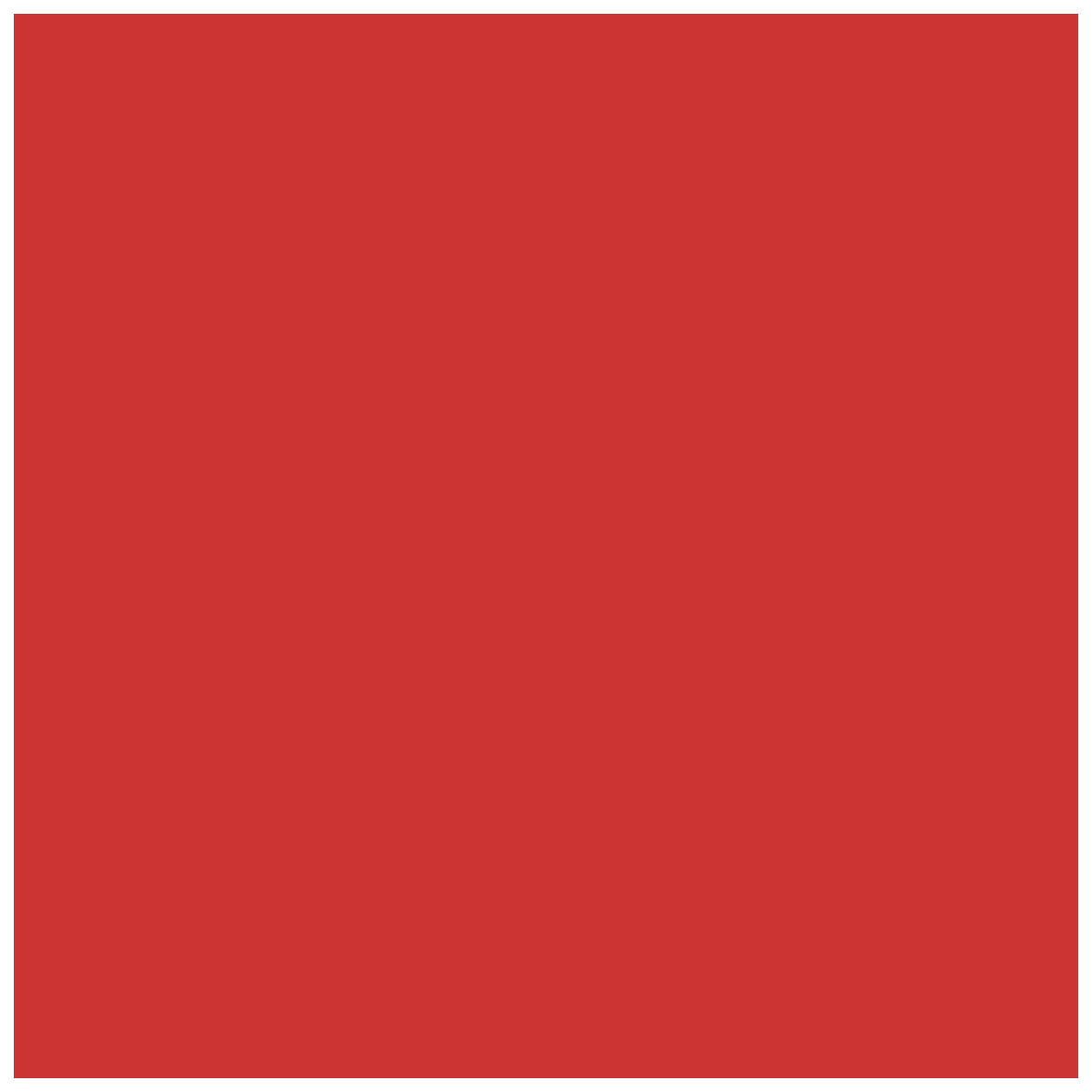}\\
			{\footnotesize t=1000}
		\end{minipage}
		\vspace{2mm}
		\\
		\centering
		{\footnotesize (a) r=4.0}
	\end{minipage}
	\hfill
	\begin{minipage}{0.45\linewidth}
		\begin{minipage}{\linewidth}
			\centering
			\includegraphics[width=\linewidth]{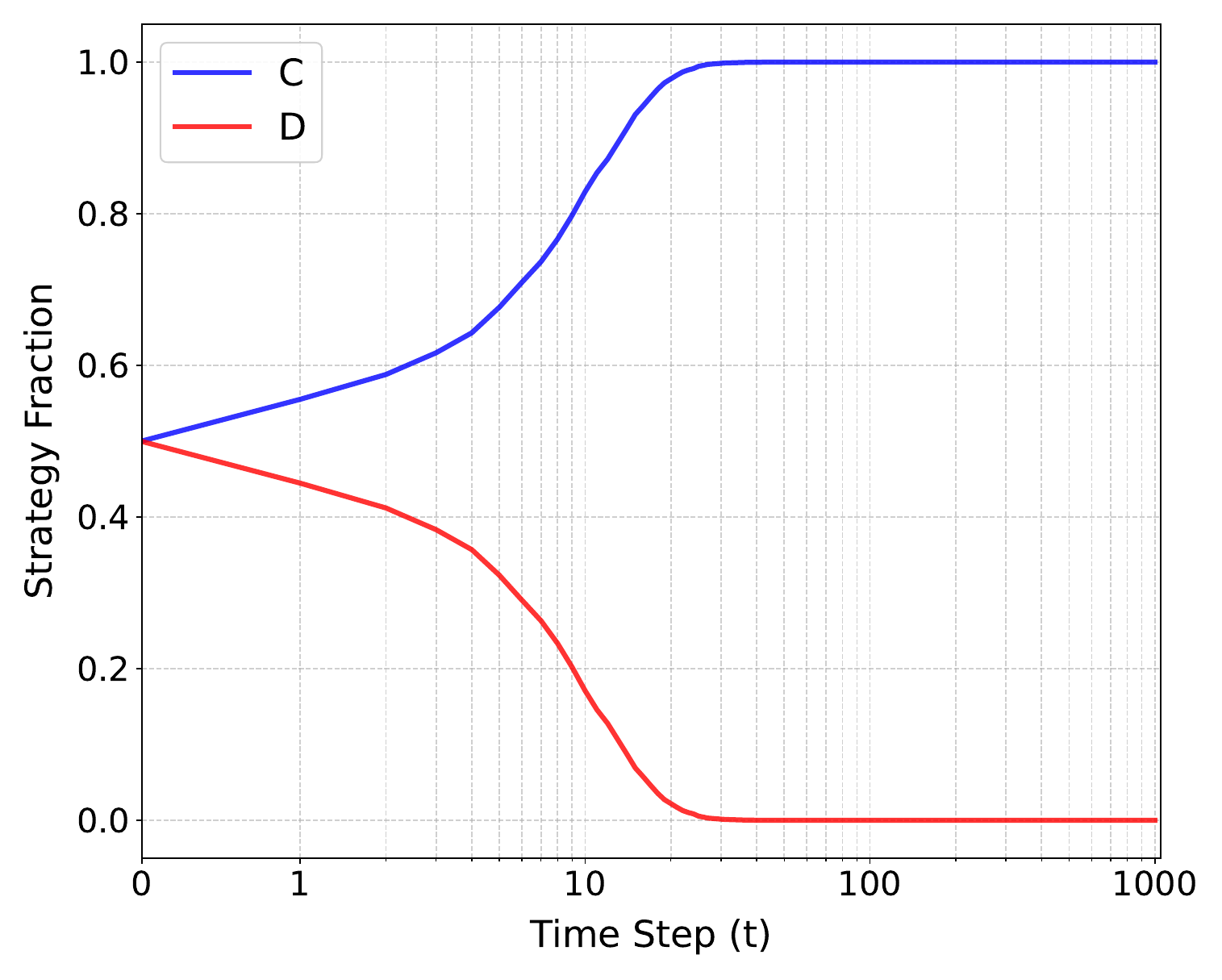}\\
		\end{minipage}
		\vspace{2mm}
		\\
		\begin{minipage}{0.188\linewidth}
			\centering
			\includegraphics[width=\linewidth]{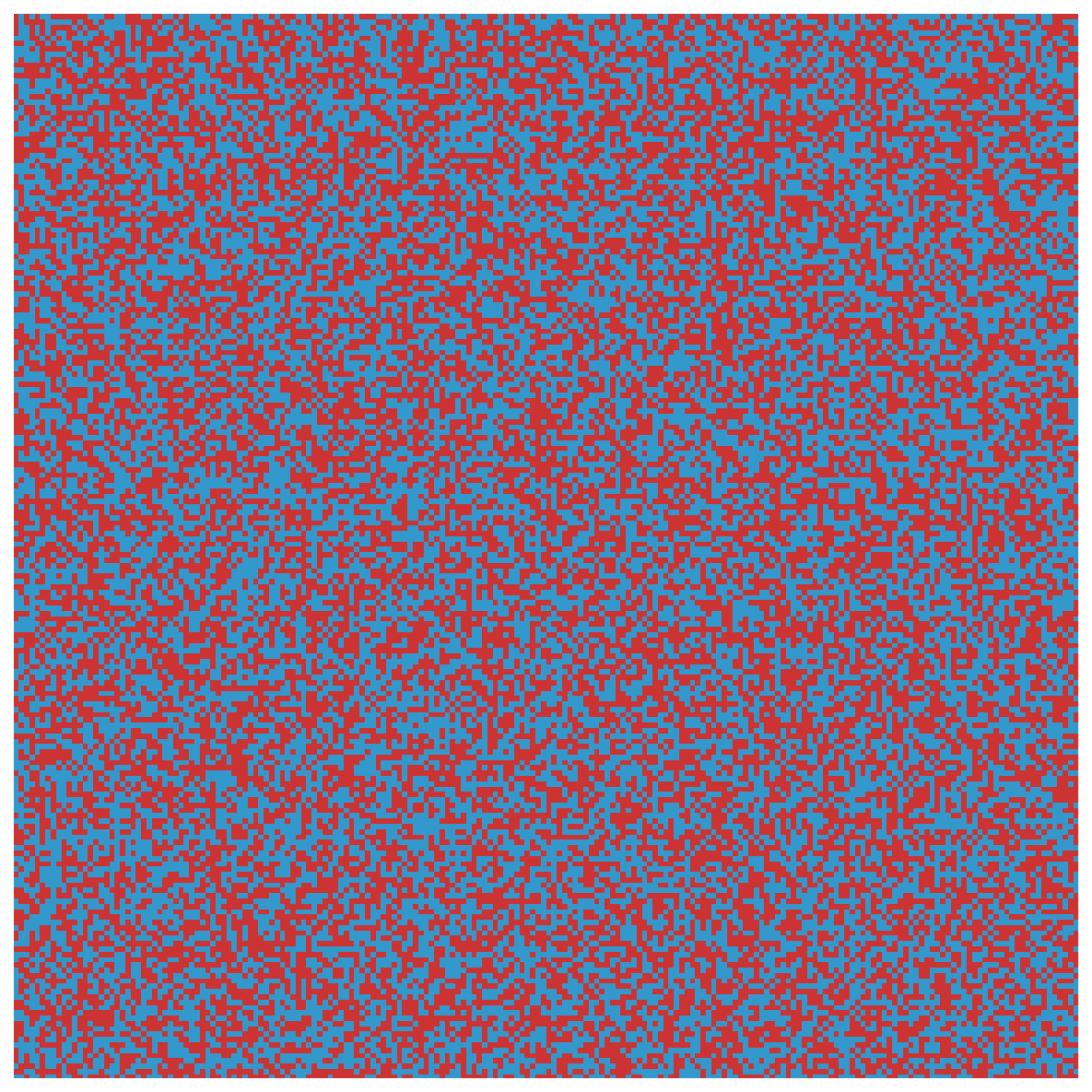}\\
			{\footnotesize t=0}
		\end{minipage}
		\begin{minipage}{0.188\linewidth}
			\centering
			\includegraphics[width=\linewidth]{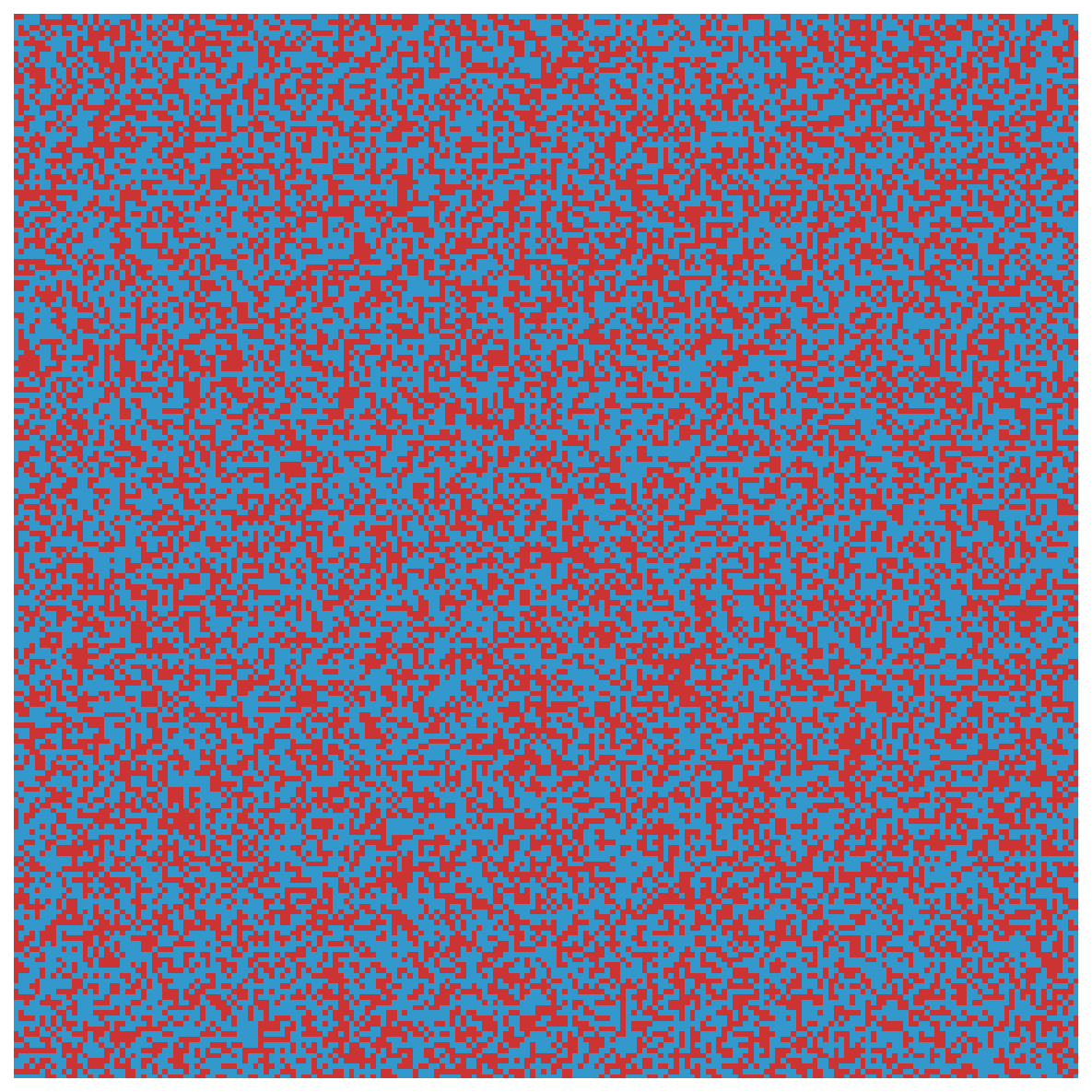}\\
			{\footnotesize t=1}
		\end{minipage}
		\begin{minipage}{0.188\linewidth}
			\centering
			\includegraphics[width=\linewidth]{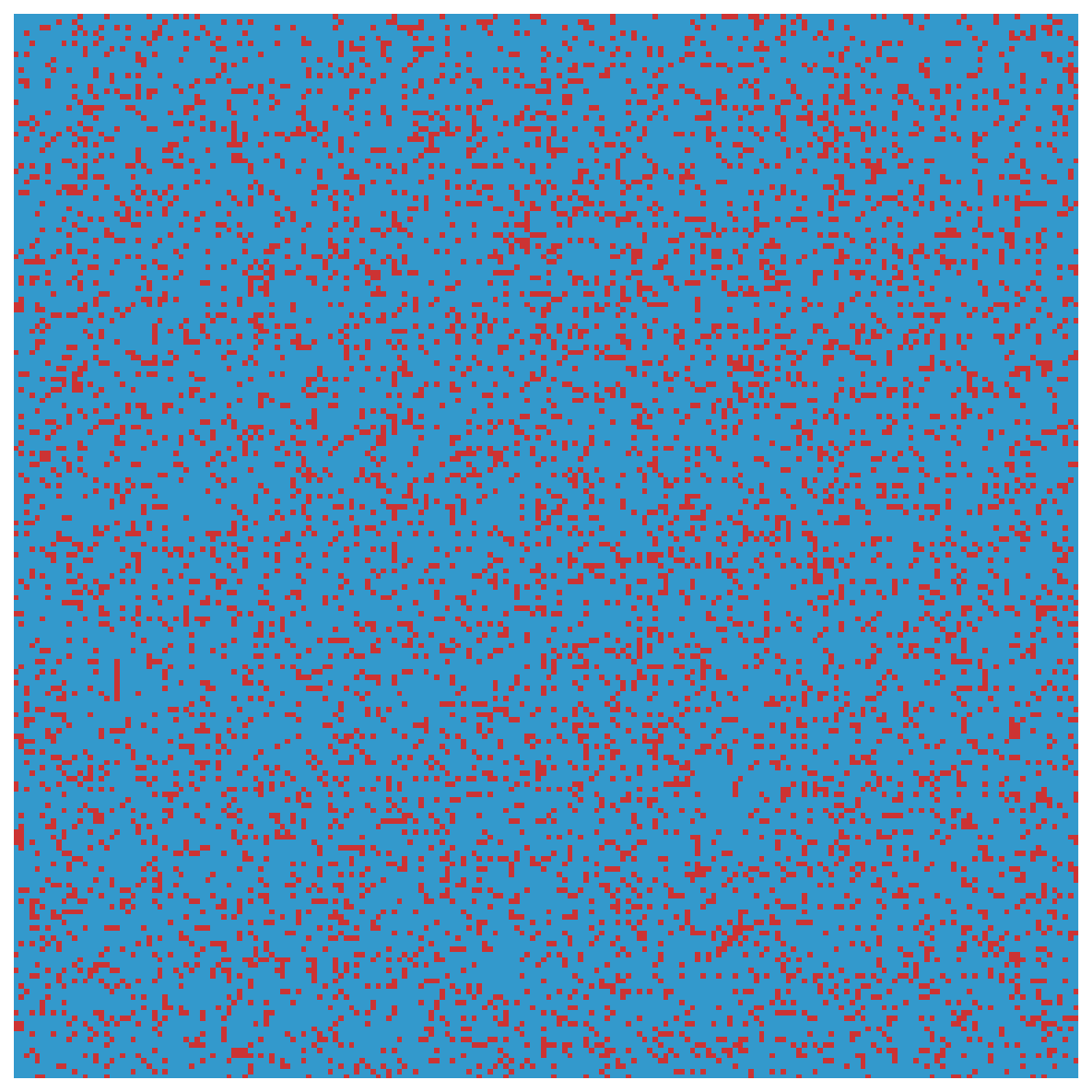}\\
			{\footnotesize t=10}
		\end{minipage}
		\begin{minipage}{0.188\linewidth}
			\centering
			\includegraphics[width=\linewidth]{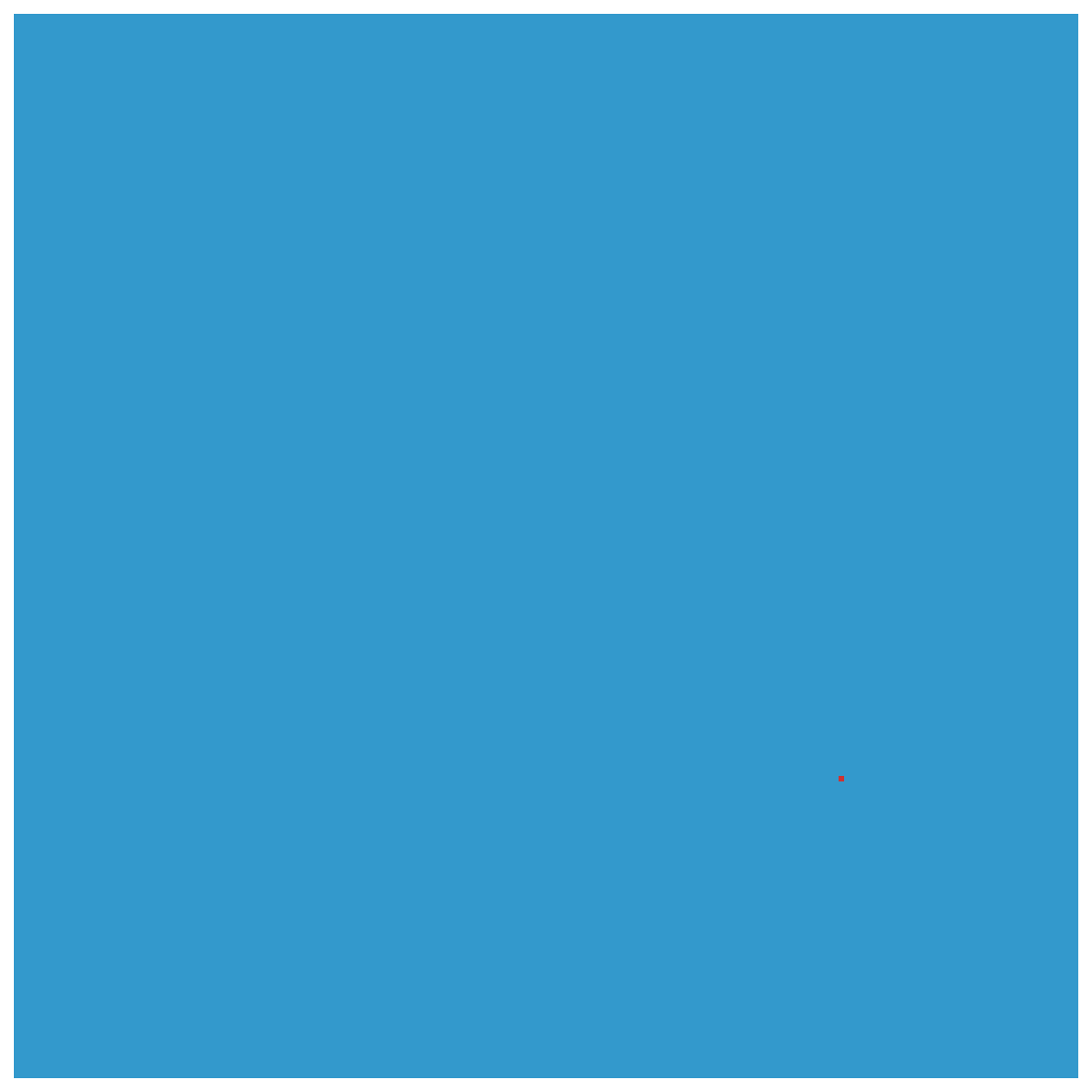}\\
			{\footnotesize t=100}
		\end{minipage}
		\begin{minipage}{0.188\linewidth}
			\centering
			\includegraphics[width=\linewidth]{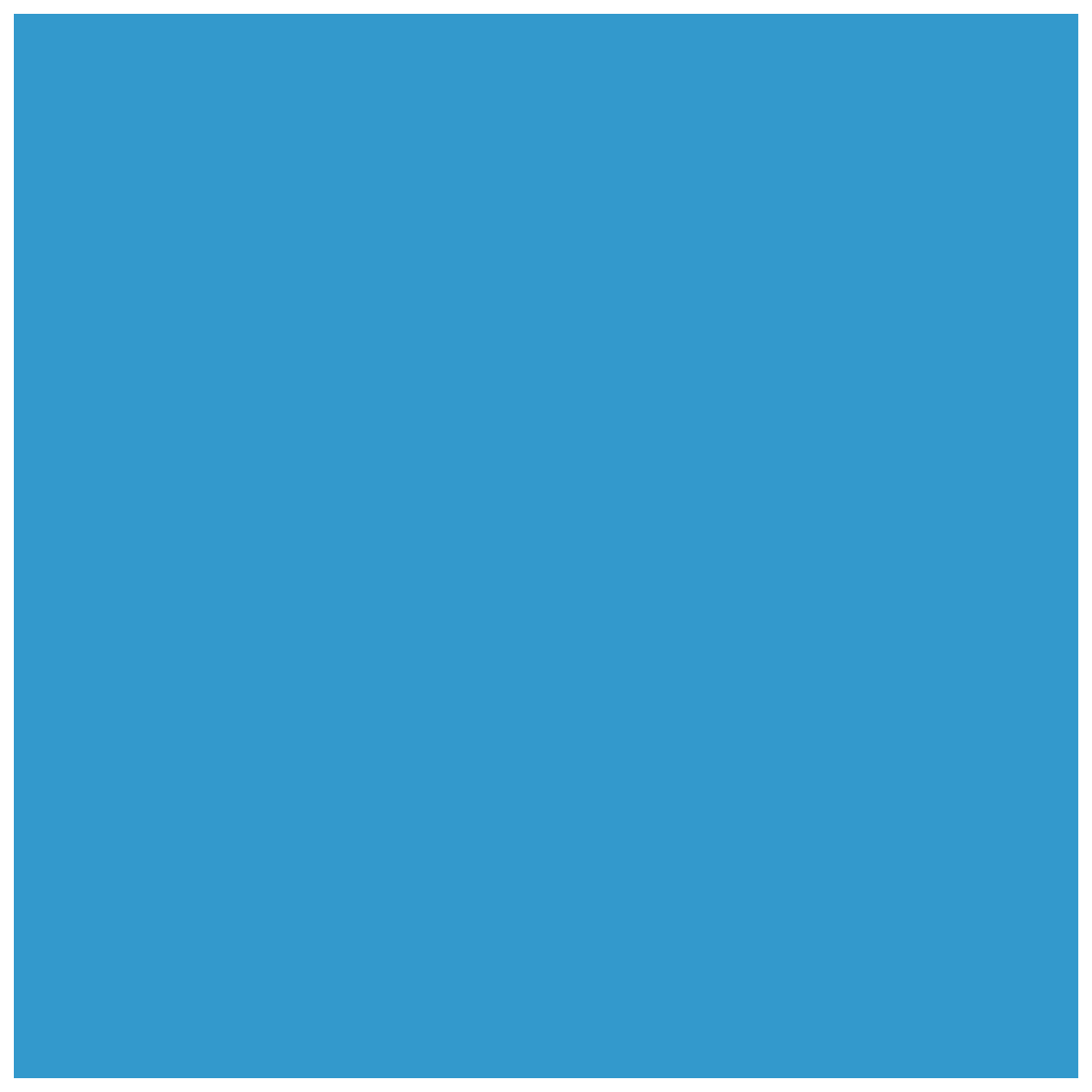}\\
			{\footnotesize t=1000}
		\end{minipage}
		\vspace{2mm}
		\\
		\centering
		{\footnotesize (b) r=4.5}
	\end{minipage}
	\\
	[2mm]
	\begin{minipage}{\linewidth}
		\begin{minipage}{0.188\linewidth}
			\centering
			\includegraphics[width=\linewidth]{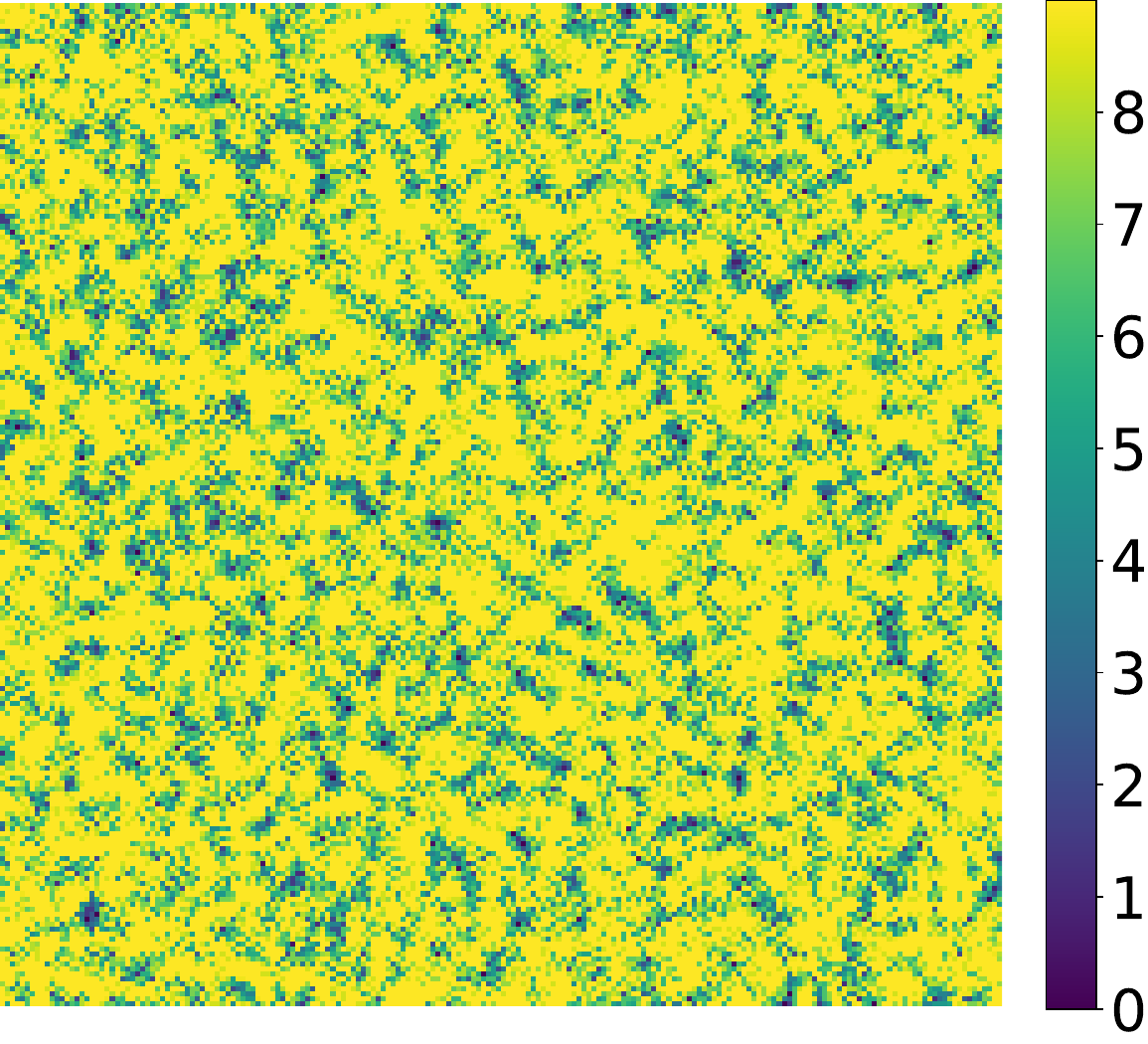}\\
			{\footnotesize t=0}
		\end{minipage}
		\hfill
		\begin{minipage}{0.188\linewidth}
			\centering
			\includegraphics[width=\linewidth]{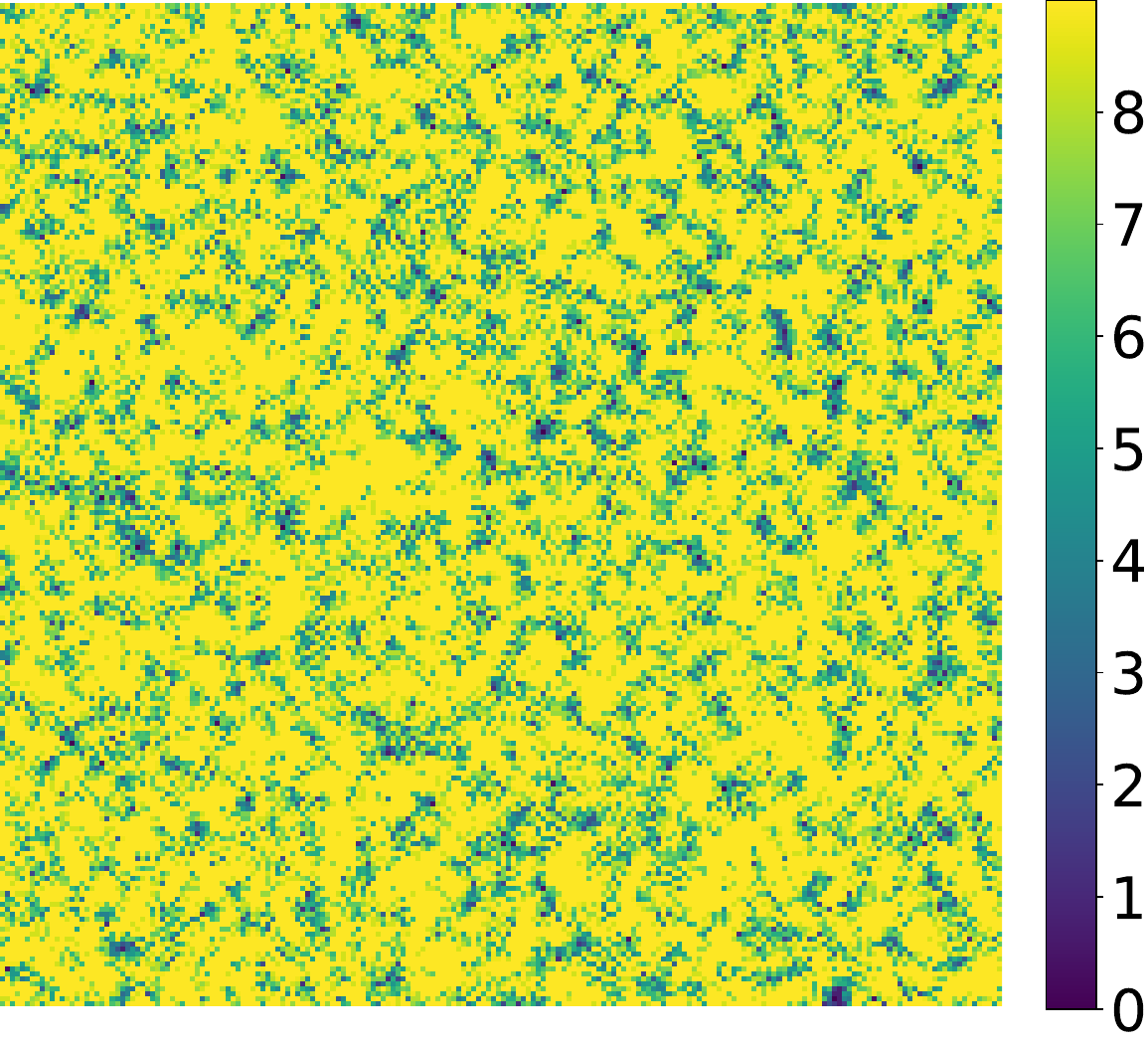}\\
			{\footnotesize t=1}
		\end{minipage}
		\hfill
		\begin{minipage}{0.188\linewidth}
			\centering
			\includegraphics[width=\linewidth]{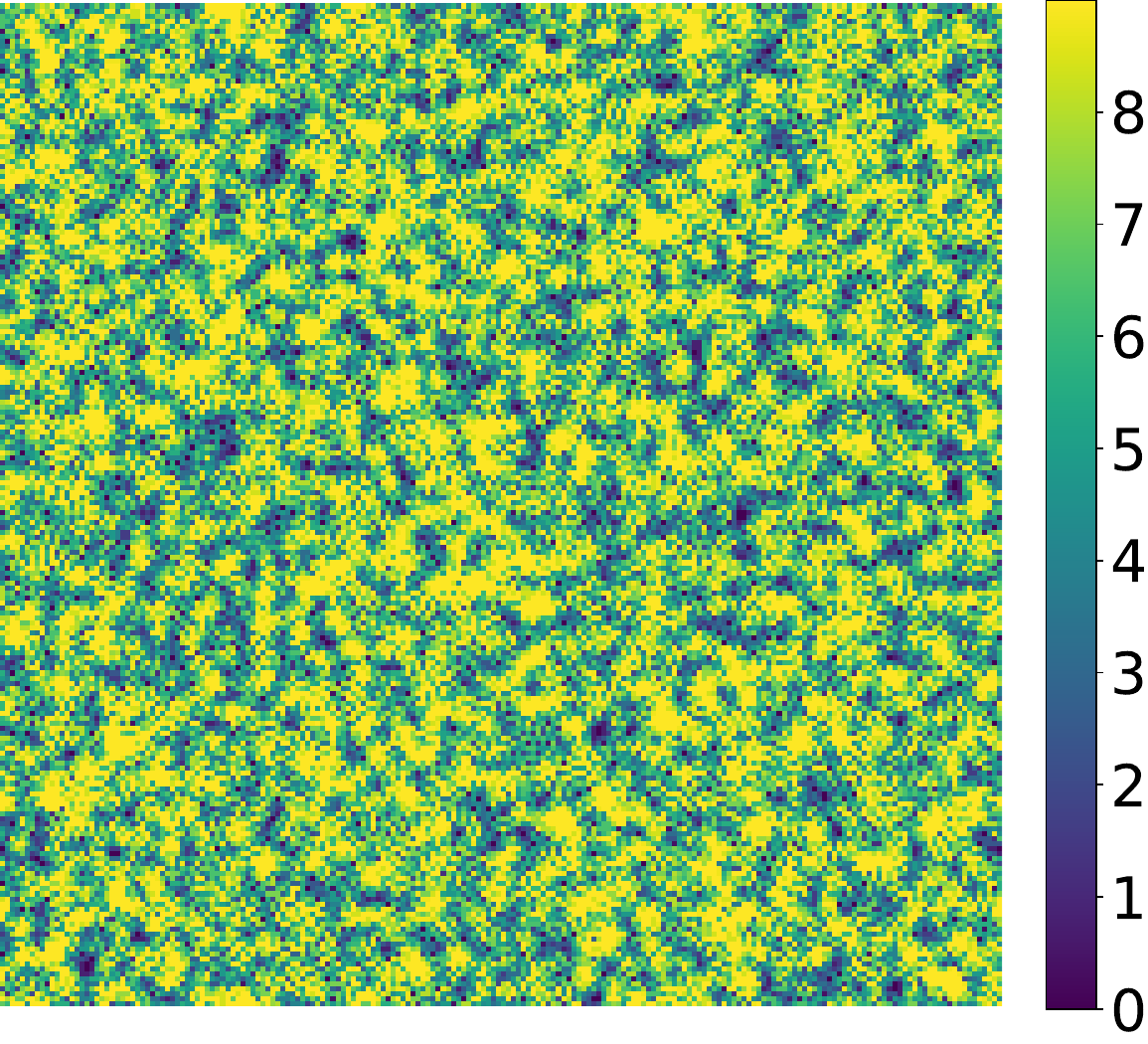}\\
			{\footnotesize t=10}
		\end{minipage}
		\hfill
		\begin{minipage}{0.188\linewidth}
			\centering
			\includegraphics[width=\linewidth]{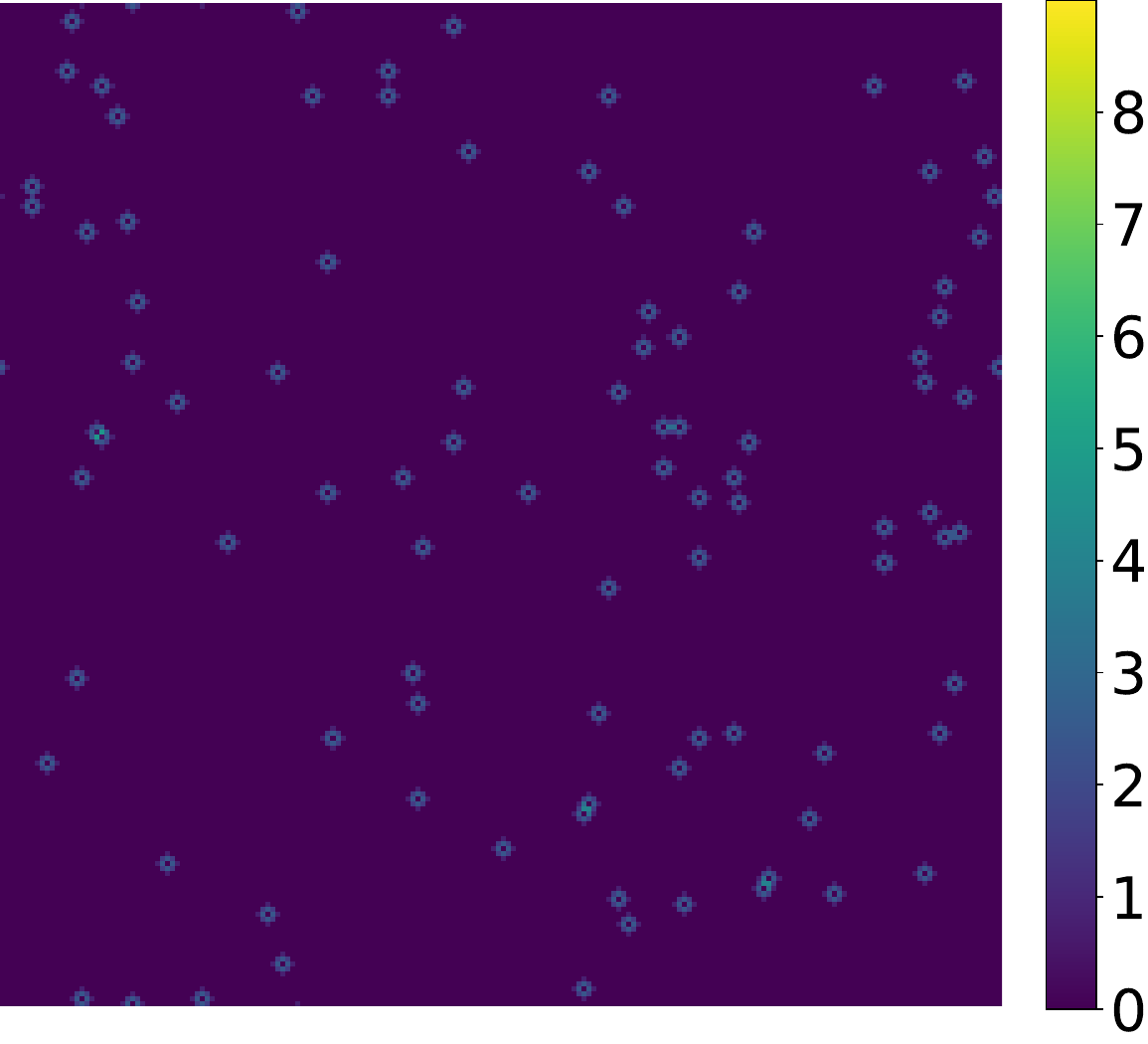}\\
			{\footnotesize t=100}
		\end{minipage}
		\hfill
		\begin{minipage}{0.188\linewidth}
			\centering
			\includegraphics[width=\linewidth]{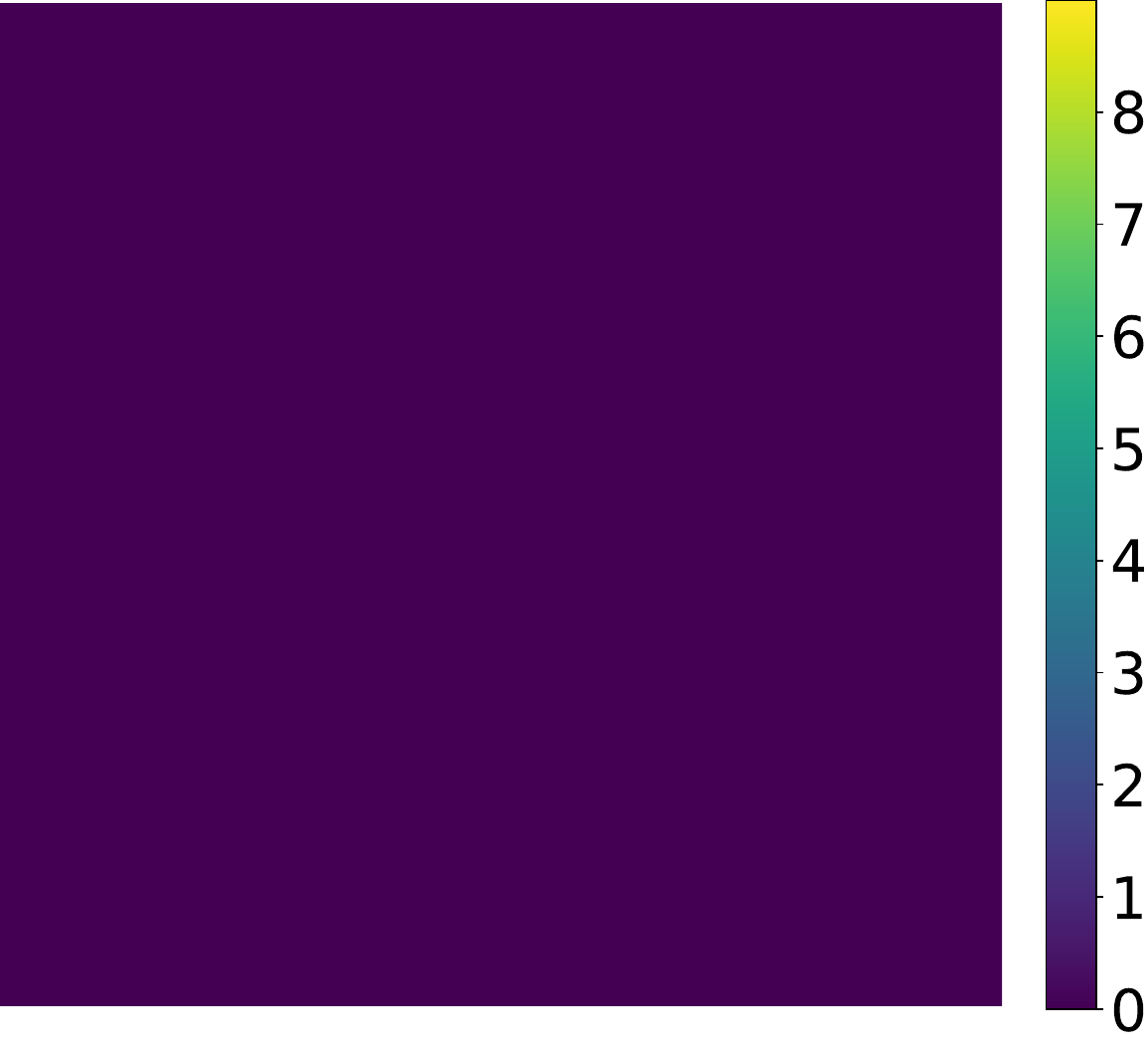}\\
			{\footnotesize t=1000}
		\end{minipage}
		\vspace{2mm}
		\\
		\centering
		{\footnotesize (c) r=4.0 (Payoff heatmaps)}
	\end{minipage}
	\\
	[2mm]
	\begin{minipage}{\linewidth}
		\begin{minipage}{0.188\linewidth}
			\centering
			\includegraphics[width=\linewidth]{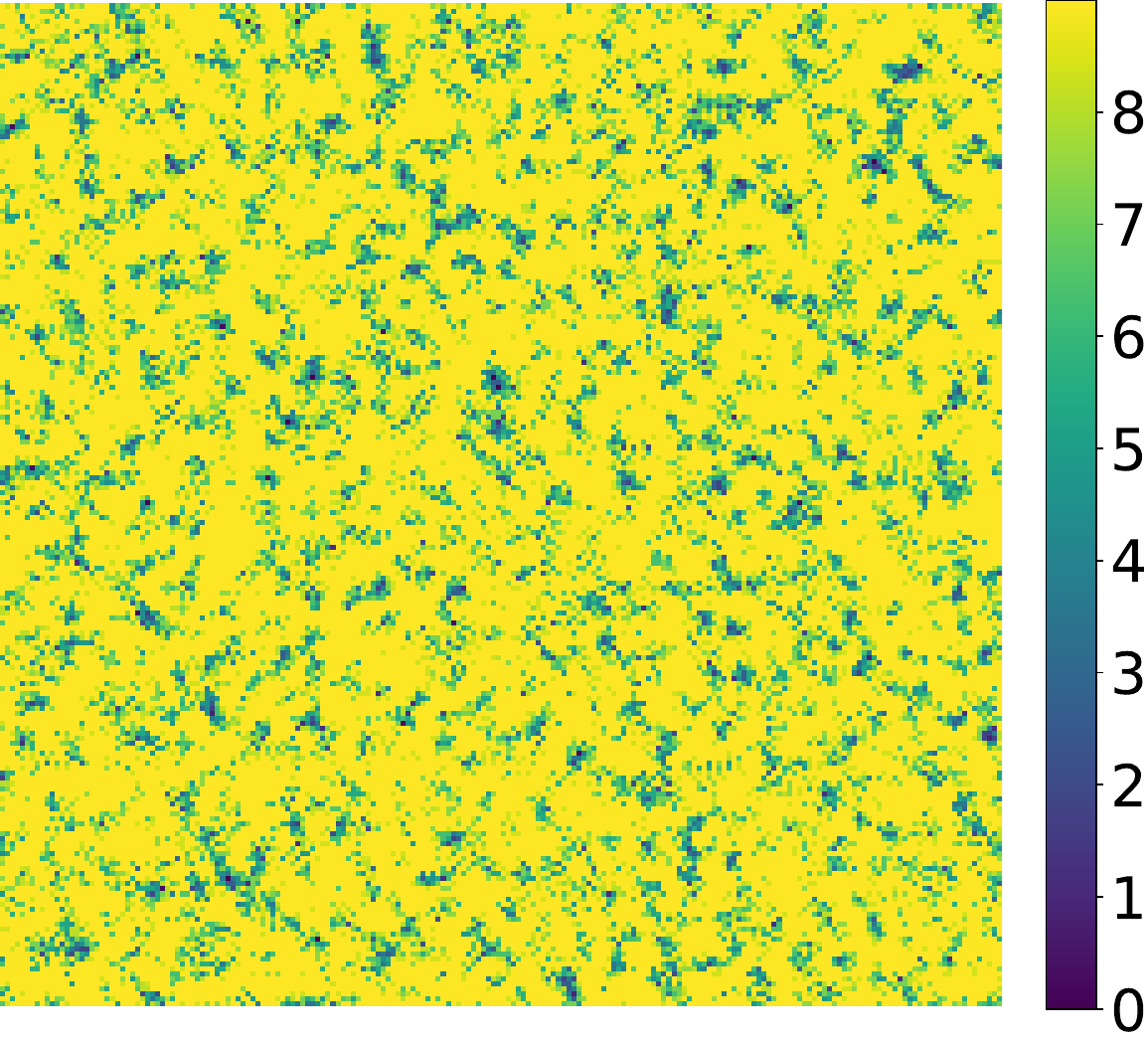}\\
			{\footnotesize t=0}
		\end{minipage}
		\hfill
		\begin{minipage}{0.188\linewidth}
			\centering
			\includegraphics[width=\linewidth]{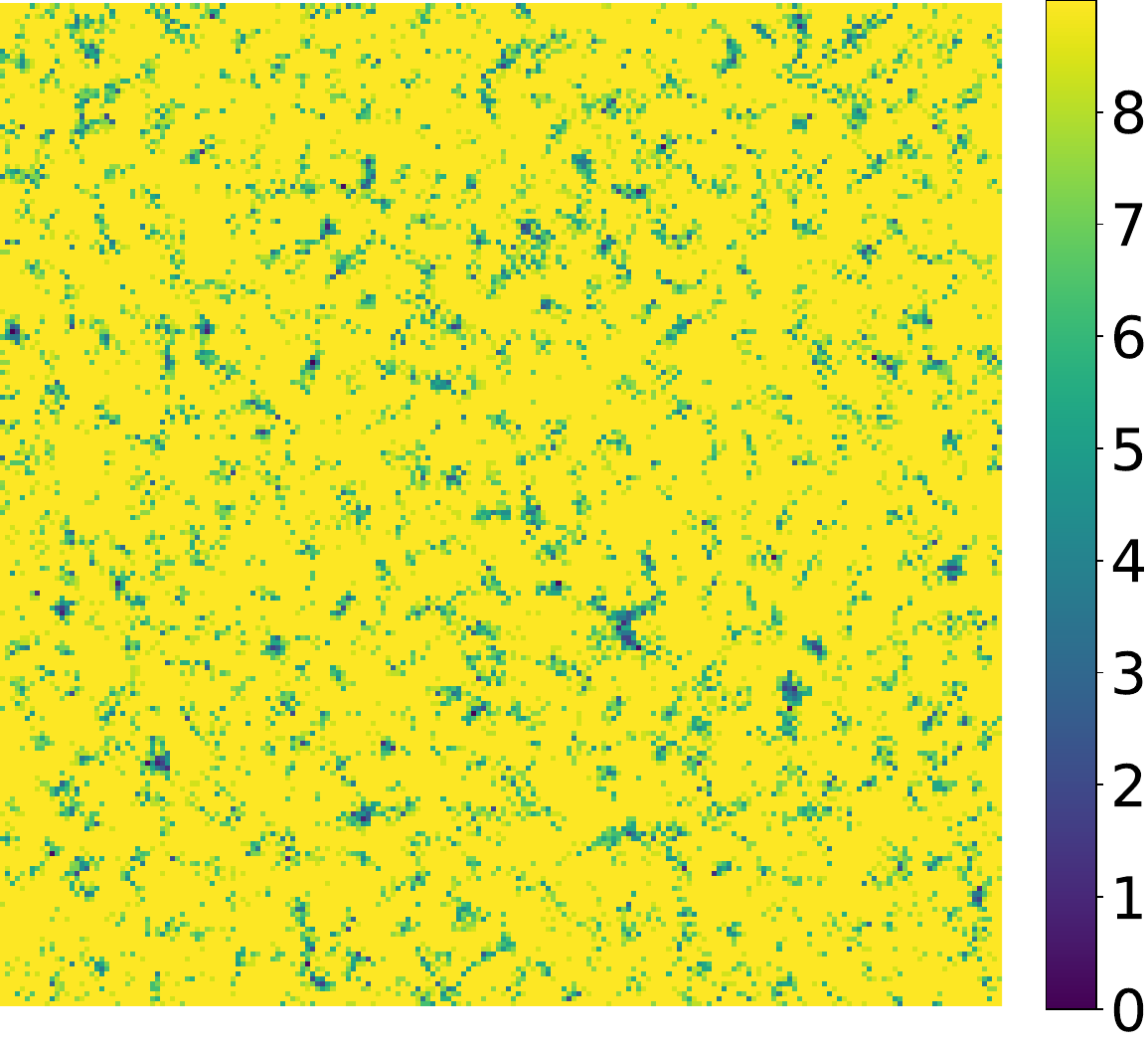}\\
			{\footnotesize t=1}
		\end{minipage}
		\hfill
		\begin{minipage}{0.188\linewidth}
			\centering
			\includegraphics[width=\linewidth]{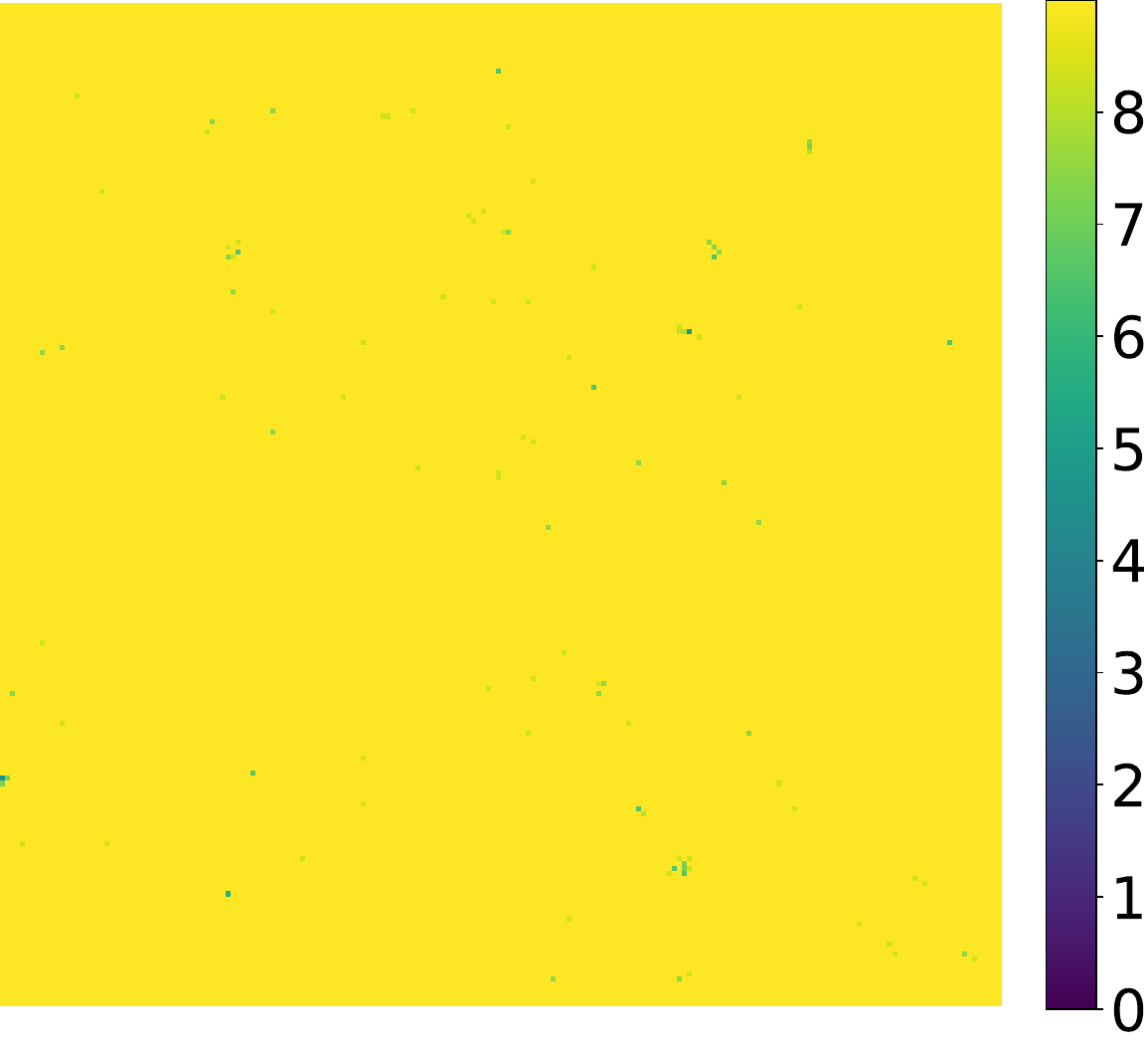}\\
			{\footnotesize t=10}
		\end{minipage}
		\hfill
		\begin{minipage}{0.188\linewidth}
			\centering
			\includegraphics[width=\linewidth]{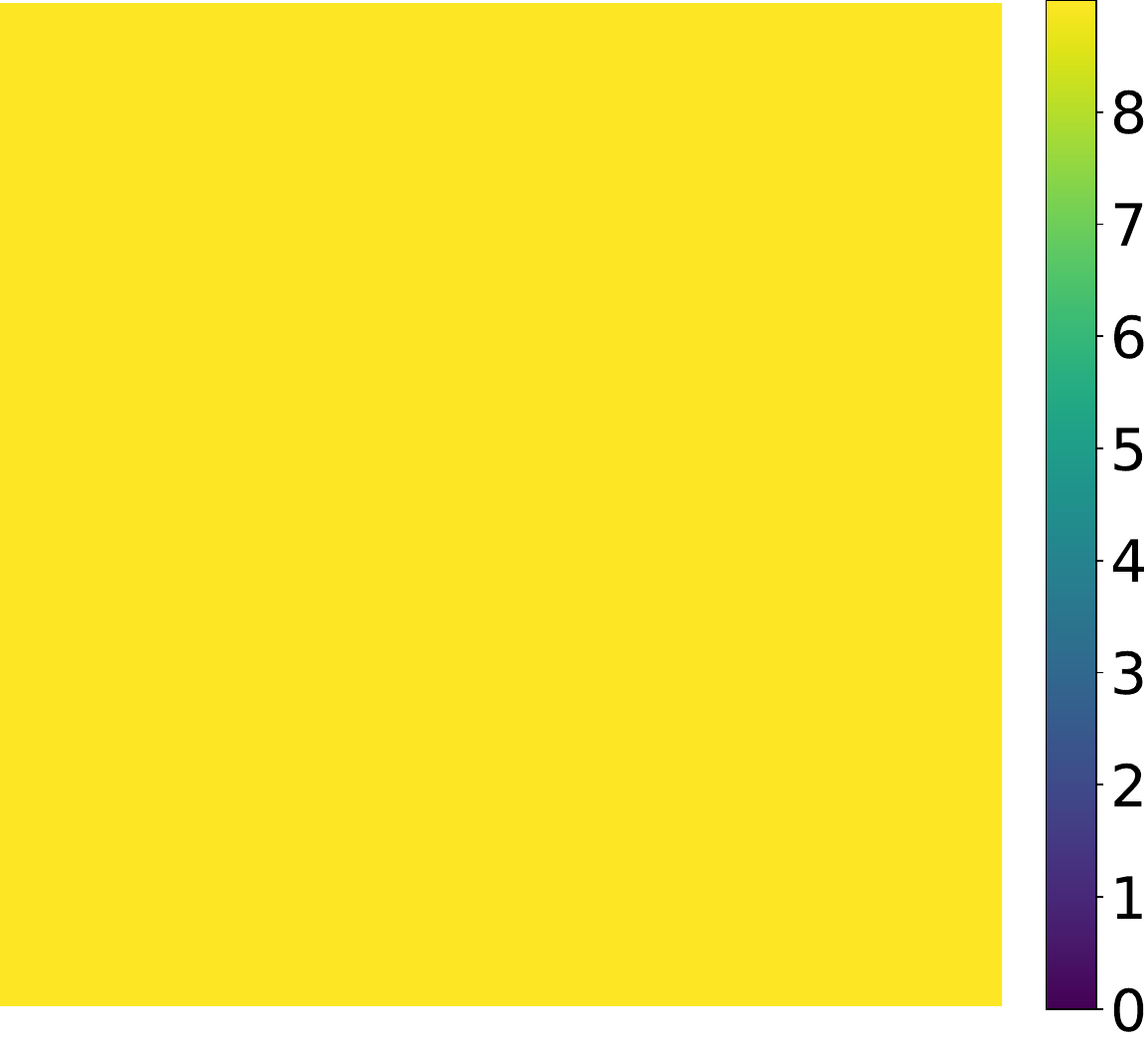}\\
			{\footnotesize t=100}
		\end{minipage}
		\hfill
		\begin{minipage}{0.188\linewidth}
			\centering
			\includegraphics[width=\linewidth]{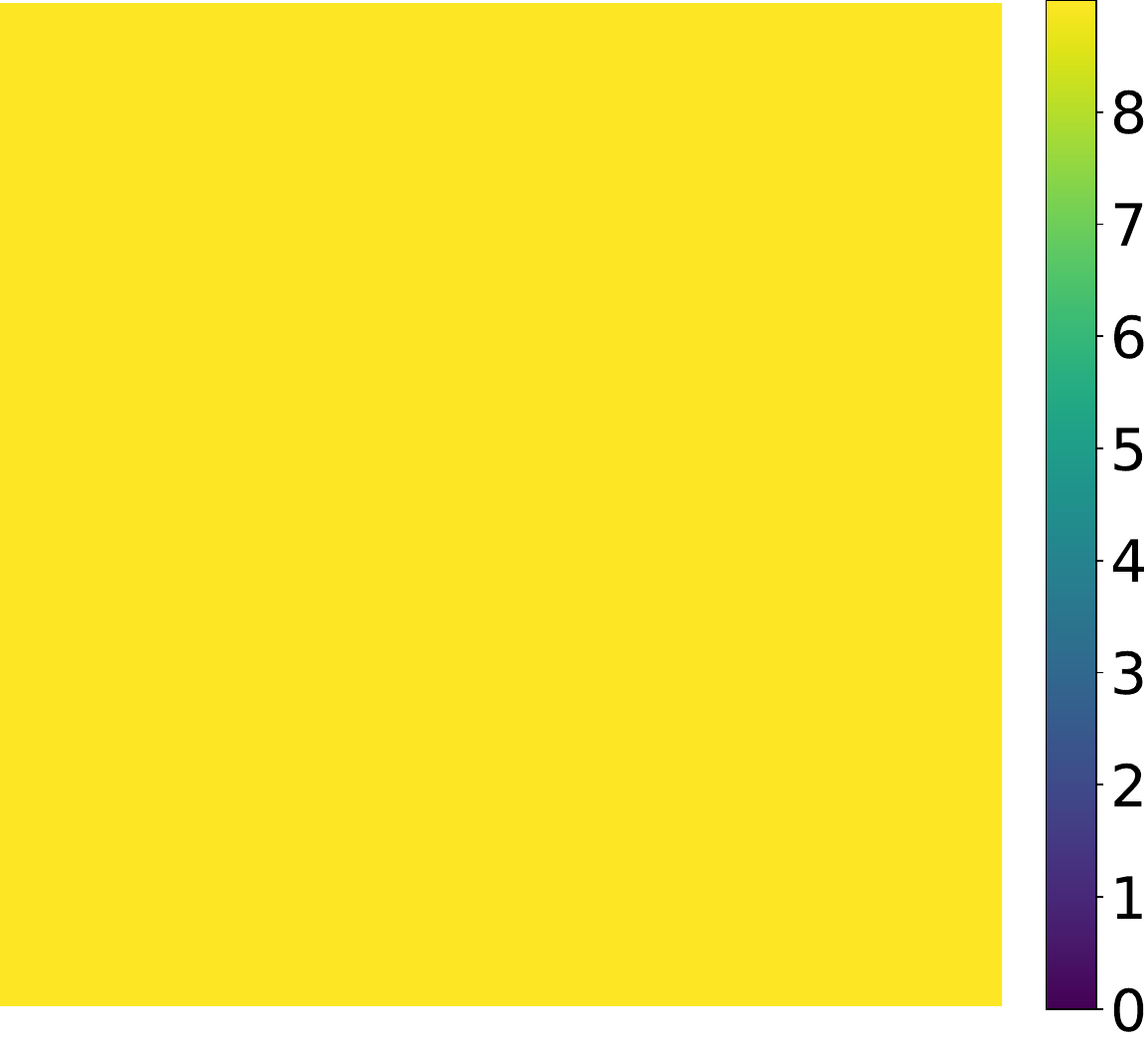}\\
			{\footnotesize t=1000}
		\end{minipage}
		\vspace{2mm}
		\\
		\centering
		{\footnotesize (d) r=4.5 (Payoff heatmaps)}
		
	\end{minipage}
	\caption{LMFPPO-UBP dynamics under Bernoulli random initialization. (a) At $r=4.0$, initial cooperation attempts collapse to full defection by $t=100$, (b) At $r=4.5$, rapid global cooperation ($f_C=1.0$) emerges within 30 iterations despite random starting conditions. Snapshots and payoff heatmaps confirm robust convergence independent of spatial correlations.}
	\label{fig:bernoulli}
\end{figure*}

The robustness of the LMFPPO‑UBP algorithm was evaluated under disordered initial conditions. These conditions were implemented via a Bernoulli random initialization with a success probability of 50\% on a $200 \times 200$ periodic lattice. Evolutionary dynamics were analyzed for two enhancement factors, $r = 4.0$ and $r = 4.5$, as shown in Figure~\ref{fig:bernoulli}.

The results reveal a decisive phase transition driven by $r$. At $r = 4.0$, the system converges to an all‑defection equilibrium $f_C = 0$ within 100 iterations. Despite a brief initial rise, cooperation cannot be sustained, and defectors progressively dominate the disordered landscape. Payoff heatmaps confirm a corresponding decline in collective welfare, illustrating a classic tragedy of the commons outcome. In stark contrast, at $r = 4.5$, the system rapidly achieves full cooperation $f_C = 1.0$ within 30 iterations. Cooperators spontaneously form clusters and assimilate neighboring defectors, leading to a globally cooperative state. Payoff heatmaps reflect a significant enhancement in both individual and collective returns. When $r$ surpasses threshold, the synergistic coupling between LMF perception and the UBP mechanism enables the system to detect and amplify local cooperative advantages. The punishment mechanism creates a selective disadvantage for defectors embedded within or adjacent to cooperative clusters. This effect fosters the nucleation and subsequent growth of cooperation even from a fully disordered initial state. A key theoretical insight from this analysis is that LMFPPO‑UBP does not rely on pre‑existing spatial correlations or strategic coherence to initiate cooperation. Instead, the framework endogenously generates cooperative order through the interplay of policy gradient learning and a spatially‑targeted sanctioning rule. This capability to solve the coordination problem in a disordered strategic landscape underscores the algorithm's potential. It presents a general paradigm for fostering cooperation in systems that lack initial organization or centralized design.

\subsection{LMFPPO-UBP with all-defectors initialization}

\begin{figure*}[htbp!]
	\begin{minipage}{0.45\linewidth}
		\begin{minipage}{\linewidth}
			\centering
			\includegraphics[width=\linewidth]{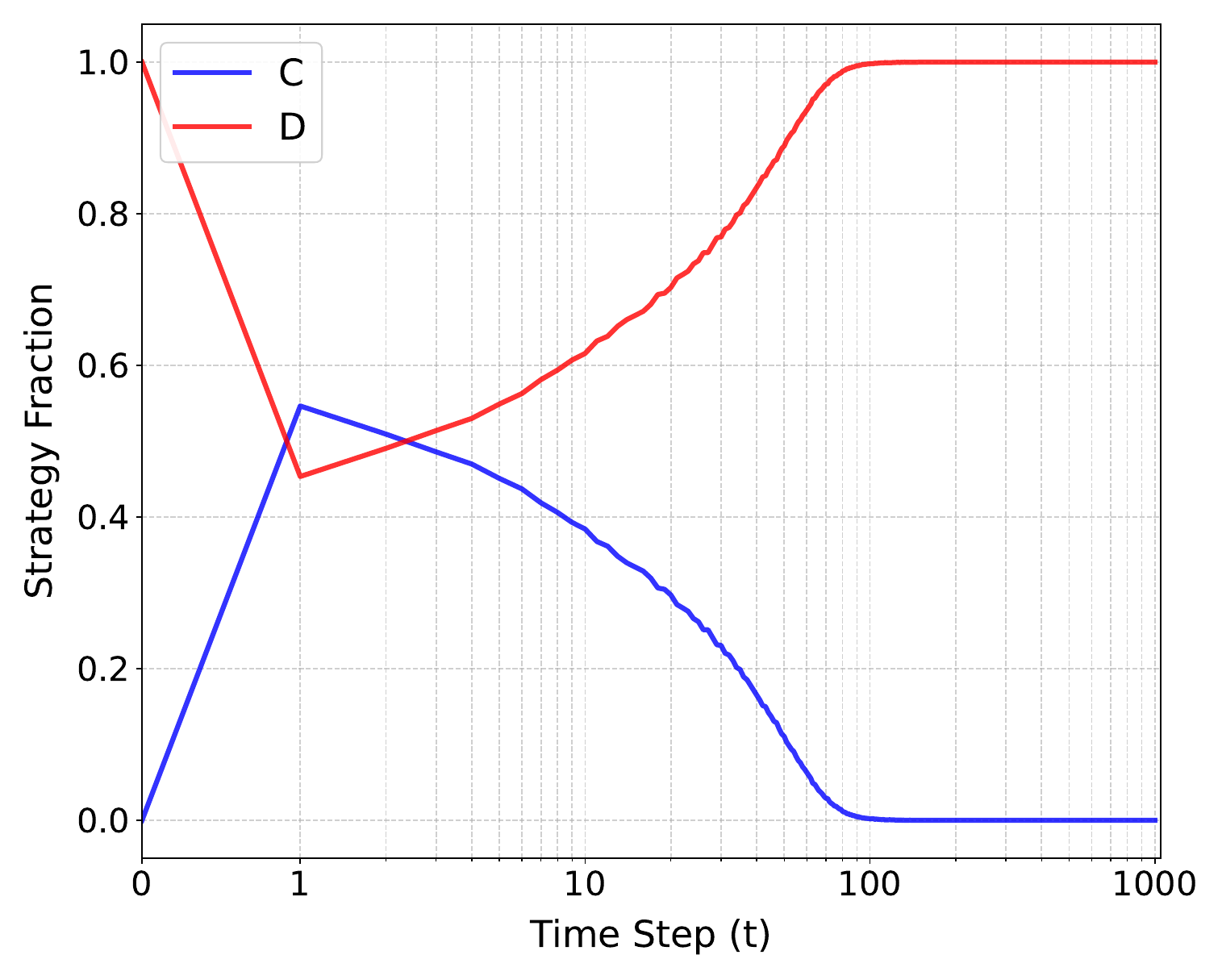}\\
		\end{minipage}
		\vspace{2mm}
		\\
		\begin{minipage}{0.188\linewidth}
			\centering
			\includegraphics[width=\linewidth]{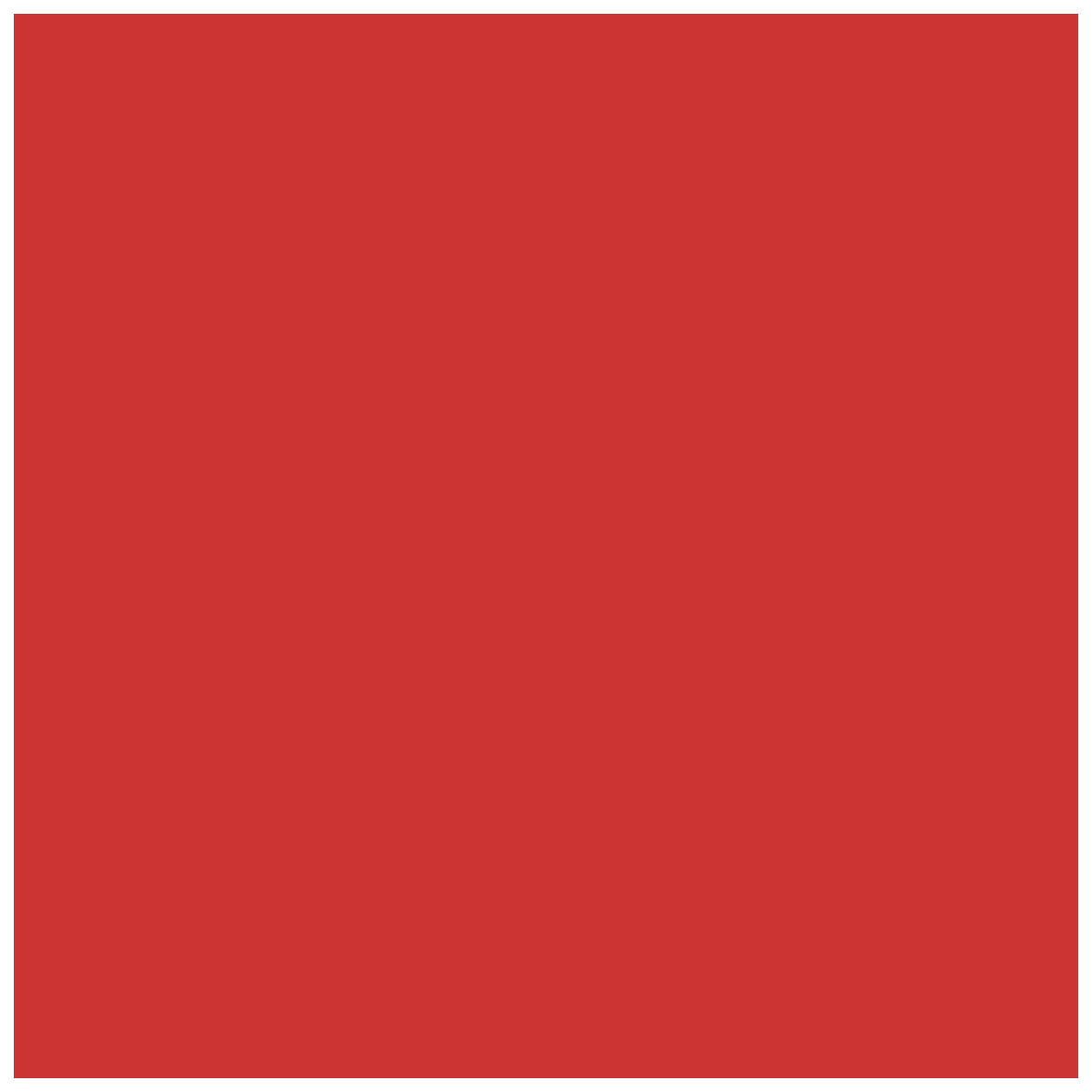}\\
			{\footnotesize t=0}
		\end{minipage}
		\begin{minipage}{0.188\linewidth}
			\centering
			\includegraphics[width=\linewidth]{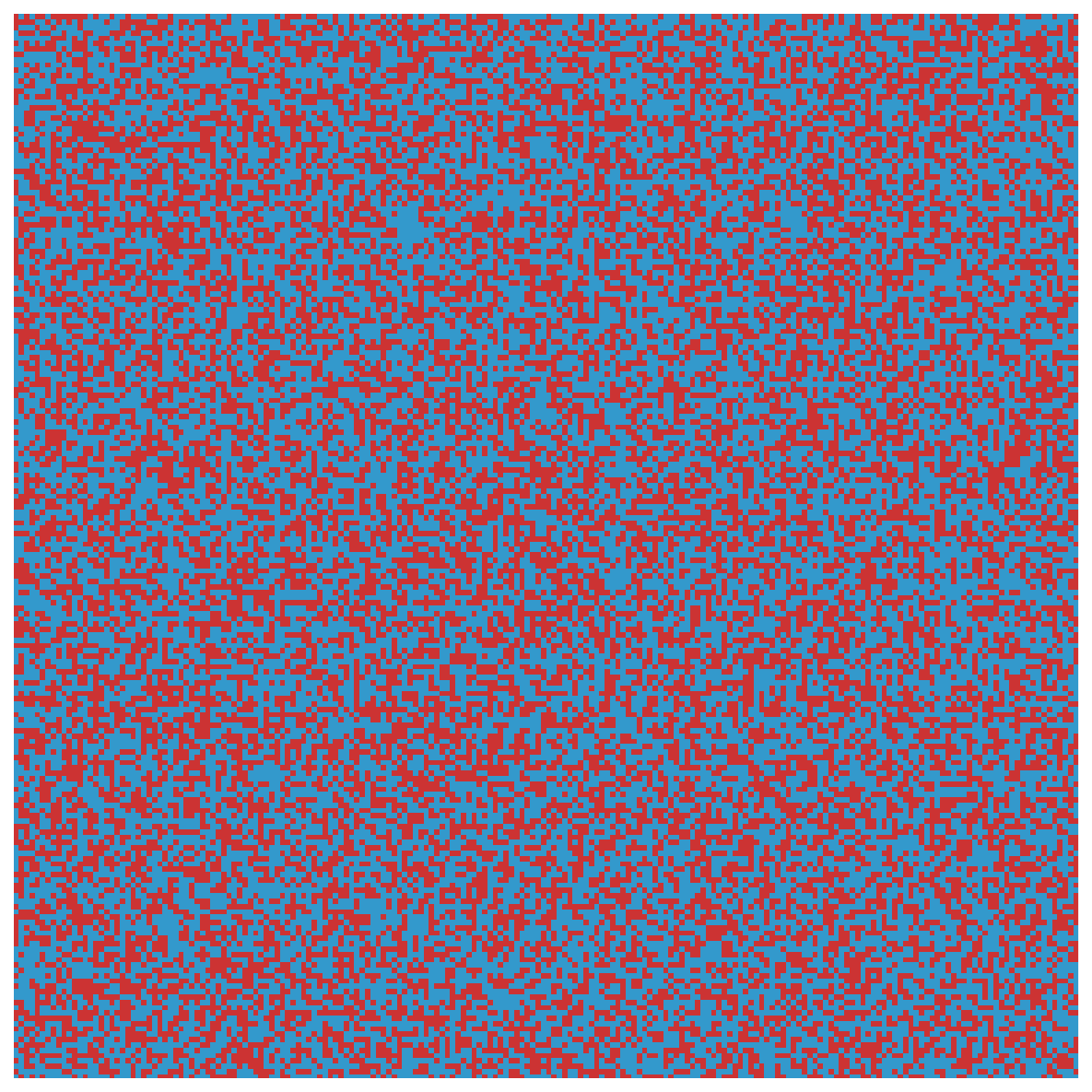}\\
			{\footnotesize t=1}
		\end{minipage}
		\begin{minipage}{0.188\linewidth}
			\centering
			\includegraphics[width=\linewidth]{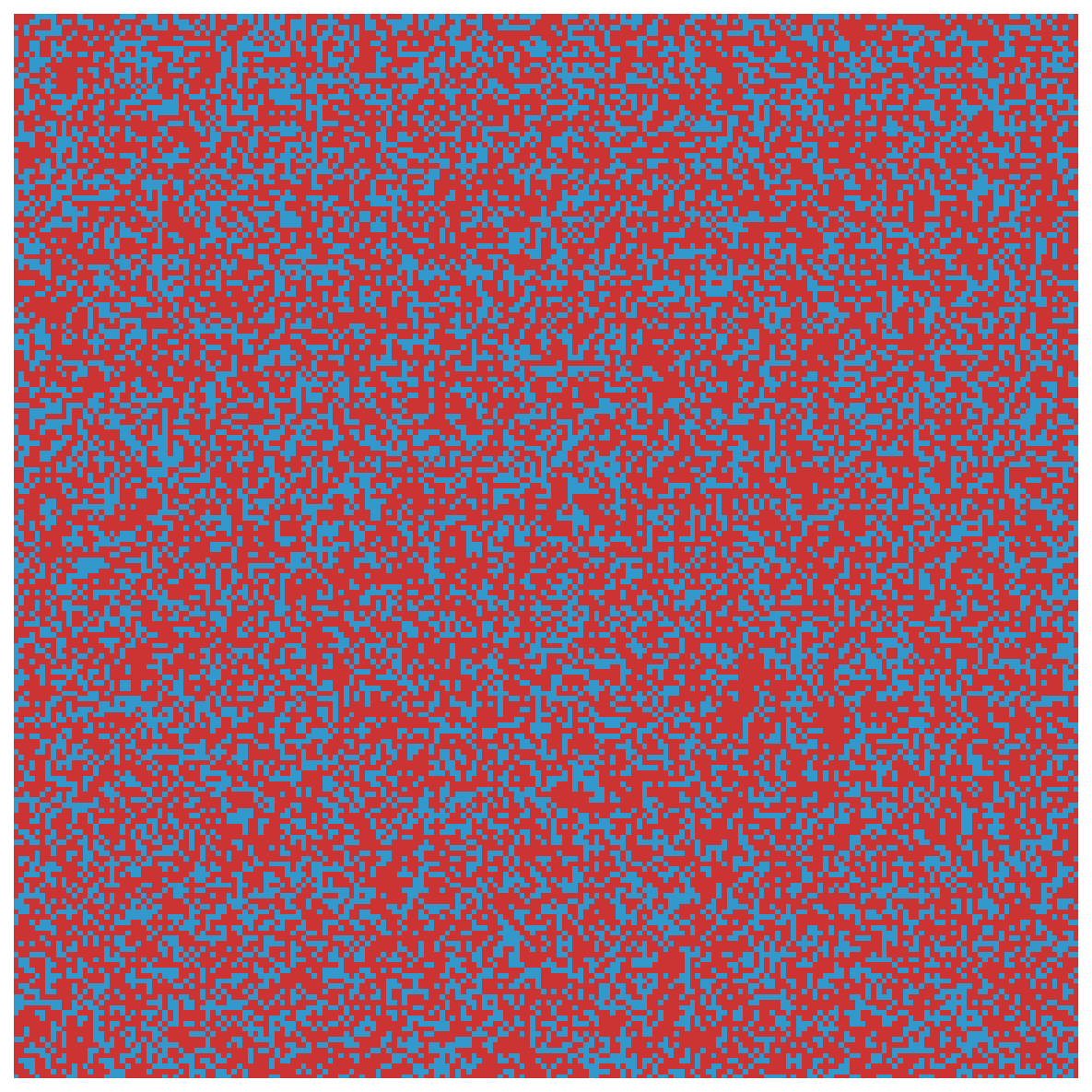}\\
			{\footnotesize t=10}
		\end{minipage}
		\begin{minipage}{0.188\linewidth}
			\centering
			\includegraphics[width=\linewidth]{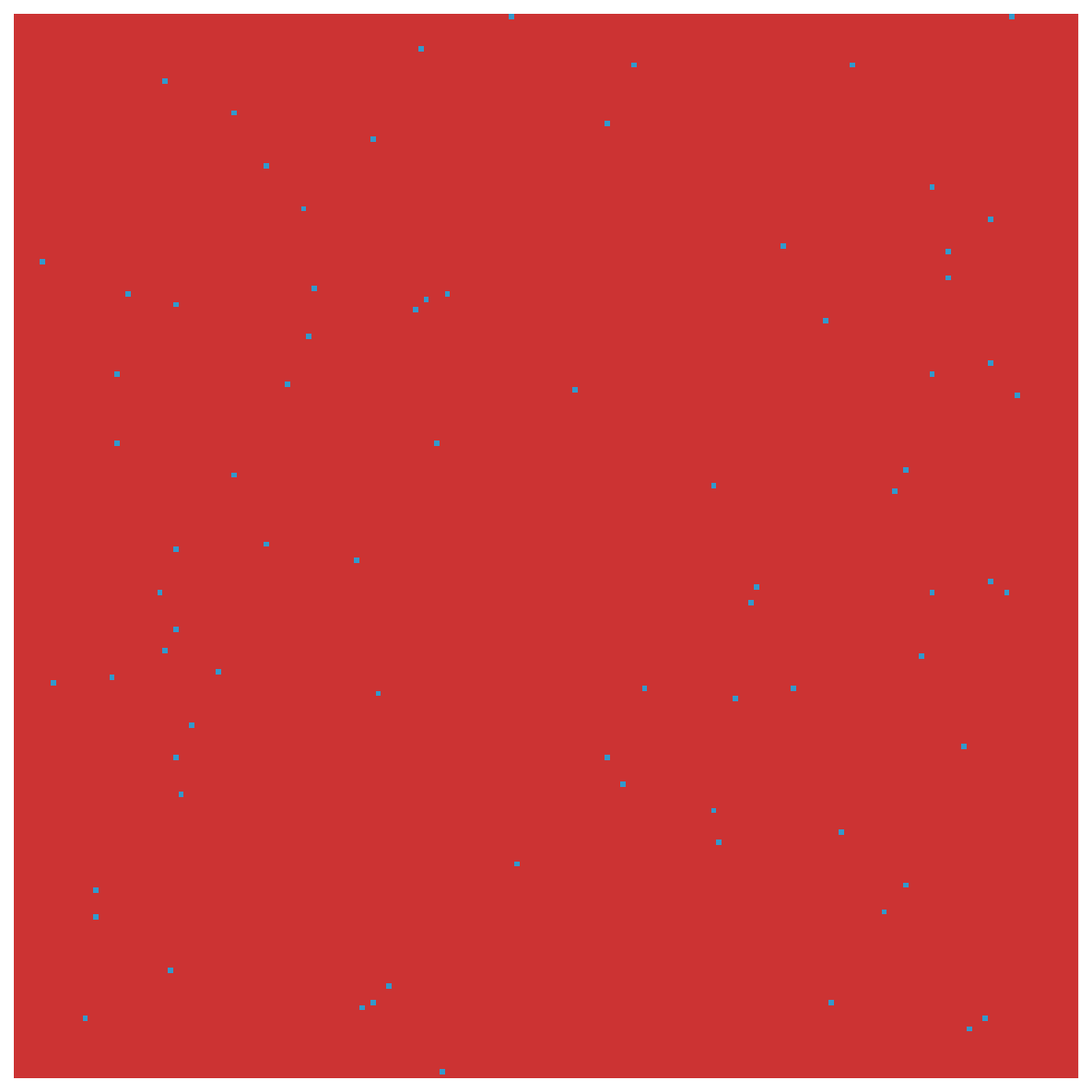}\\
			{\footnotesize t=100}
		\end{minipage}
		\begin{minipage}{0.188\linewidth}
			\centering
			\includegraphics[width=\linewidth]{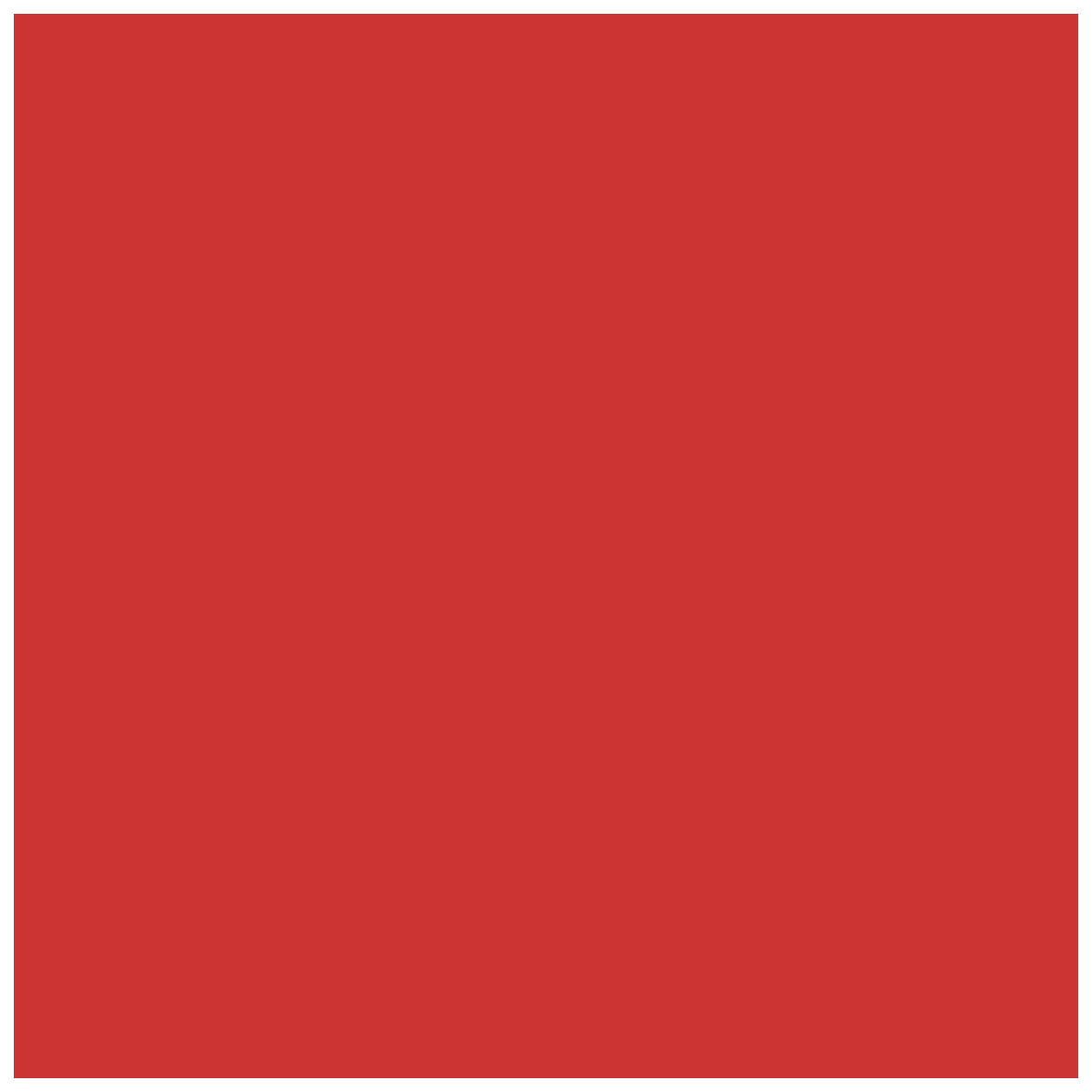}\\
			{\footnotesize t=1000}
		\end{minipage}
		\vspace{2mm}
		\\
		\centering
		{\footnotesize (a) r=4.0}
	\end{minipage}
	\hfill
	\begin{minipage}{0.45\linewidth}
		\begin{minipage}{\linewidth}
			\centering
			\includegraphics[width=\linewidth]{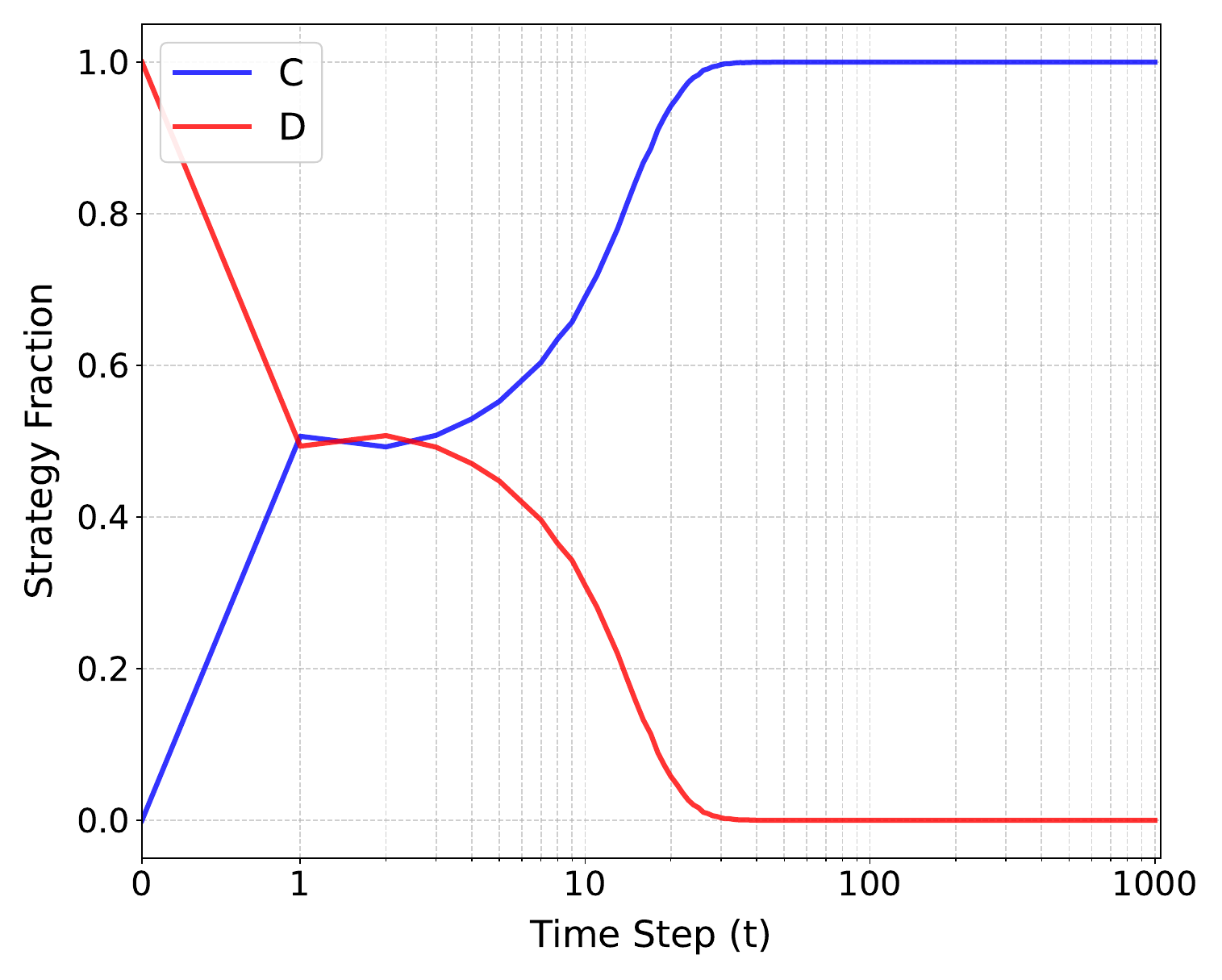}\\
		\end{minipage}
		\vspace{2mm}
		\\
		\begin{minipage}{0.188\linewidth}
			\centering
			\includegraphics[width=\linewidth]{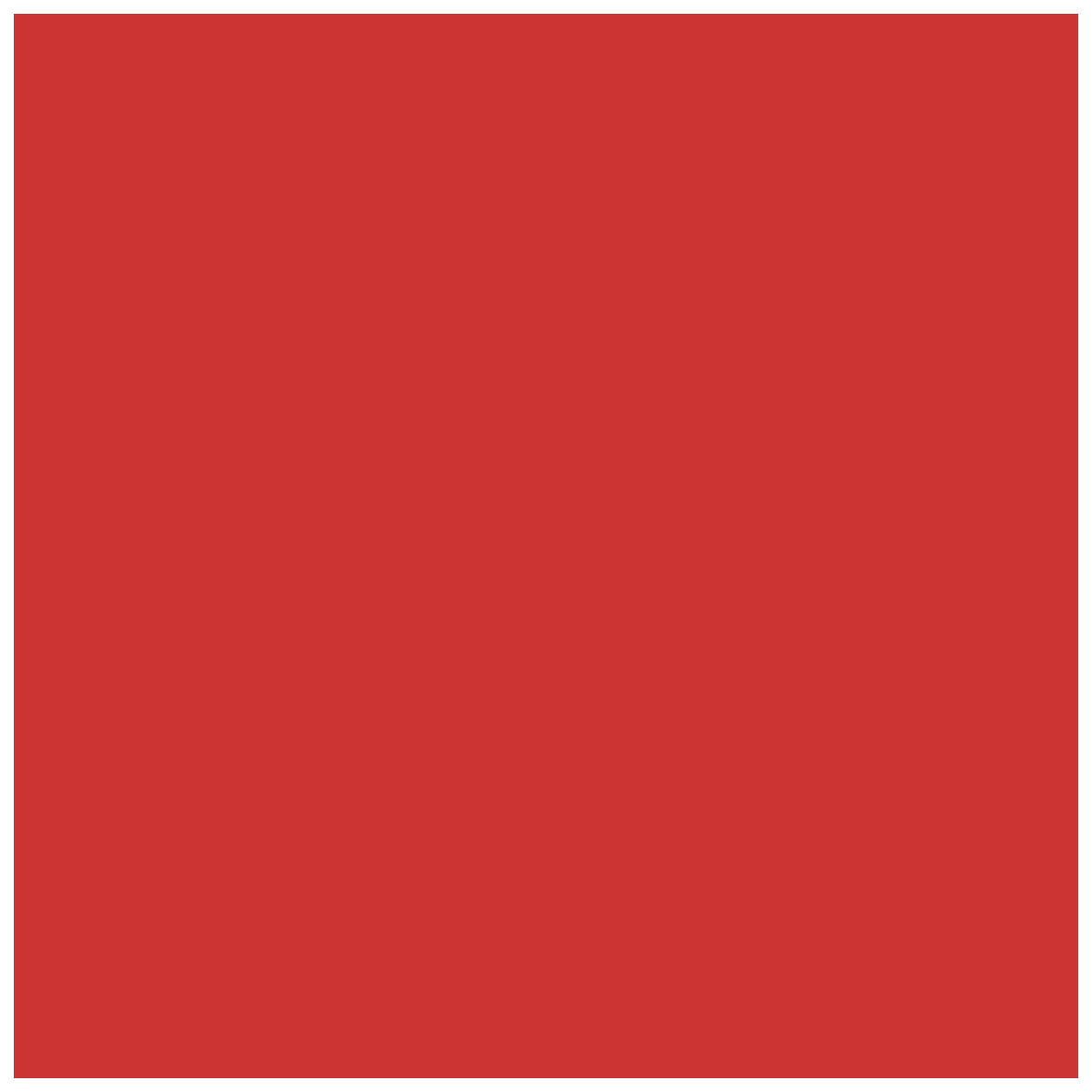}\\
			{\footnotesize t=0}
		\end{minipage}
		\begin{minipage}{0.188\linewidth}
			\centering
			\includegraphics[width=\linewidth]{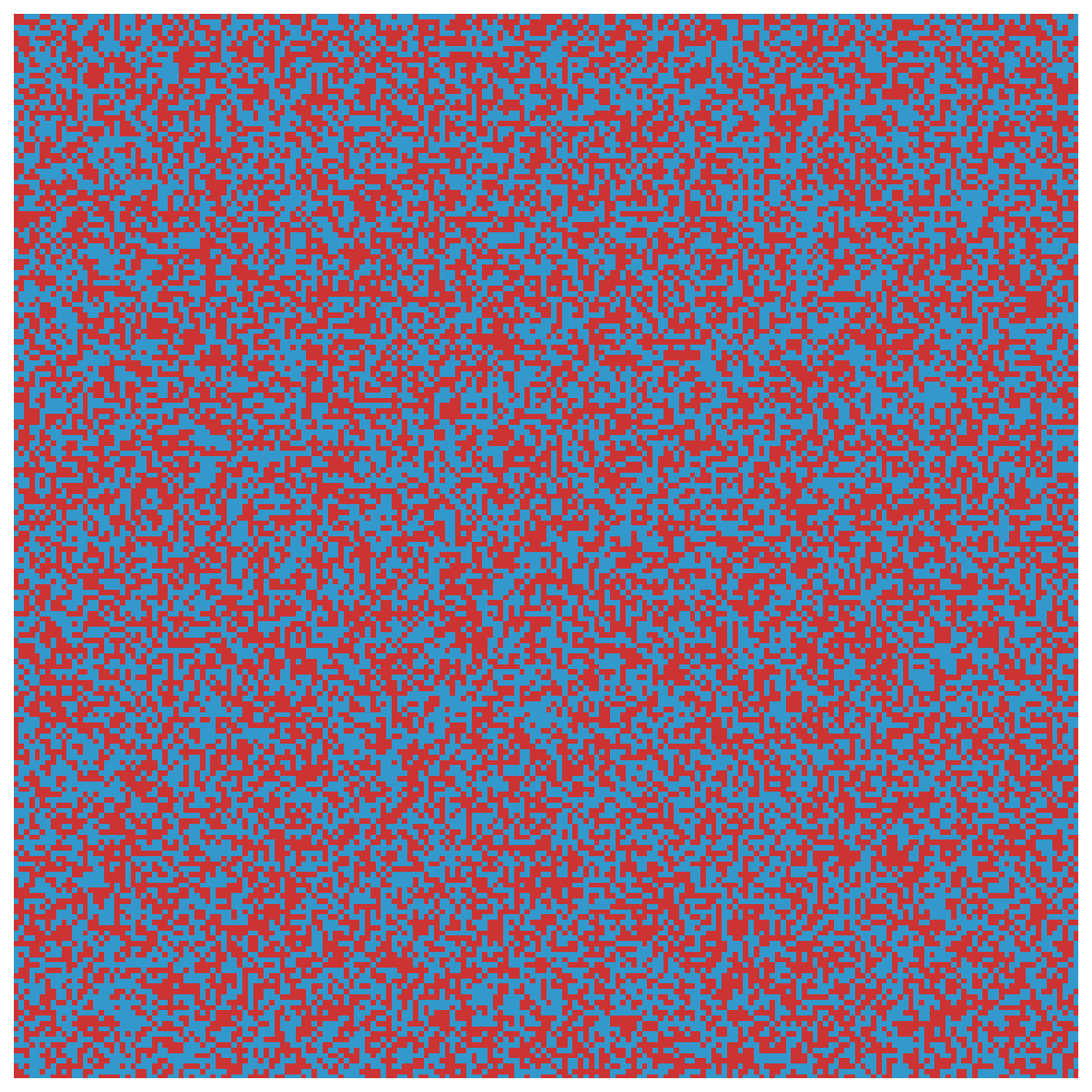}\\
			{\footnotesize t=1}
		\end{minipage}
		\begin{minipage}{0.188\linewidth}
			\centering
			\includegraphics[width=\linewidth]{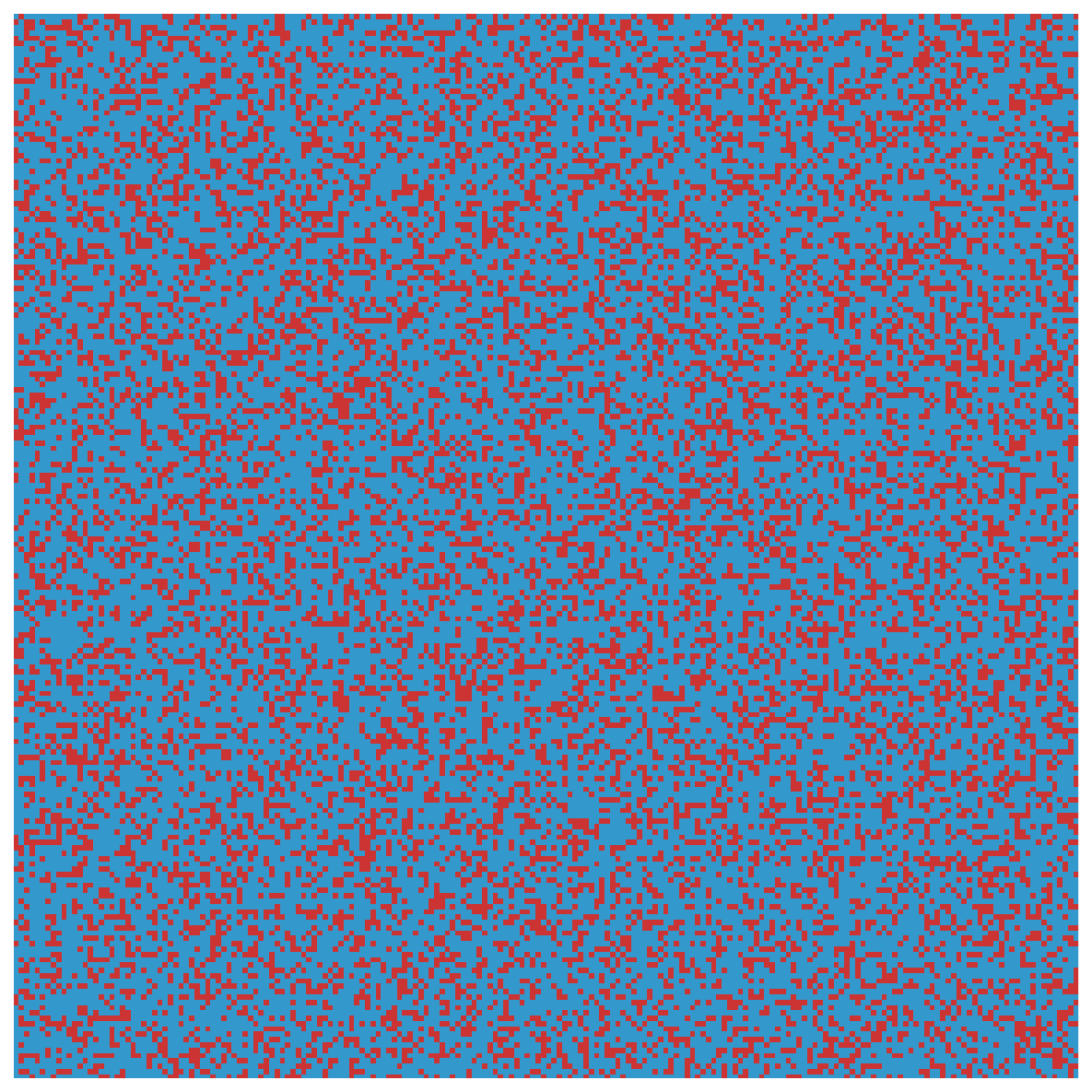}\\
			{\footnotesize t=10}
		\end{minipage}
		\begin{minipage}{0.188\linewidth}
			\centering
			\includegraphics[width=\linewidth]{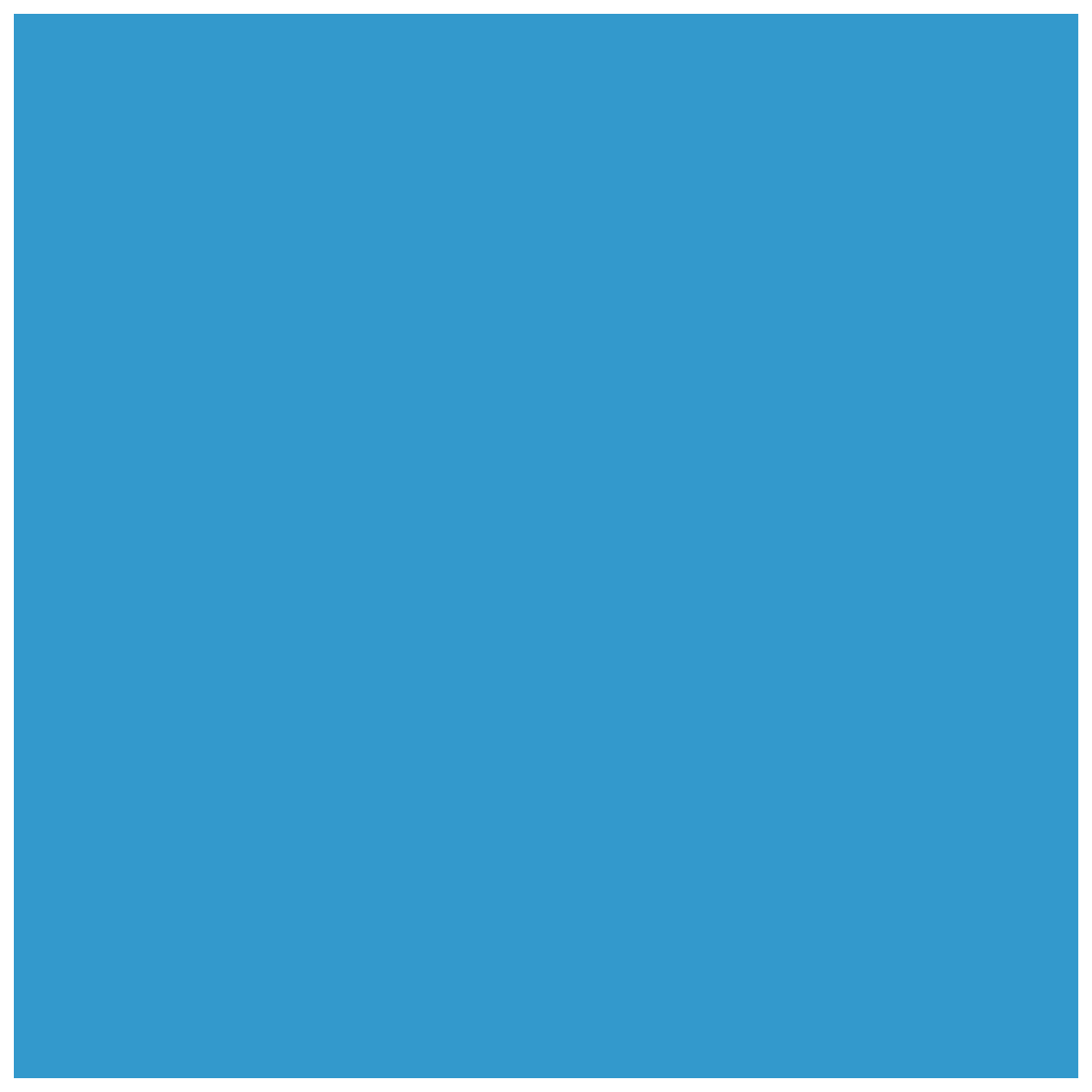}\\
			{\footnotesize t=100}
		\end{minipage}
		\begin{minipage}{0.188\linewidth}
			\centering
			\includegraphics[width=\linewidth]{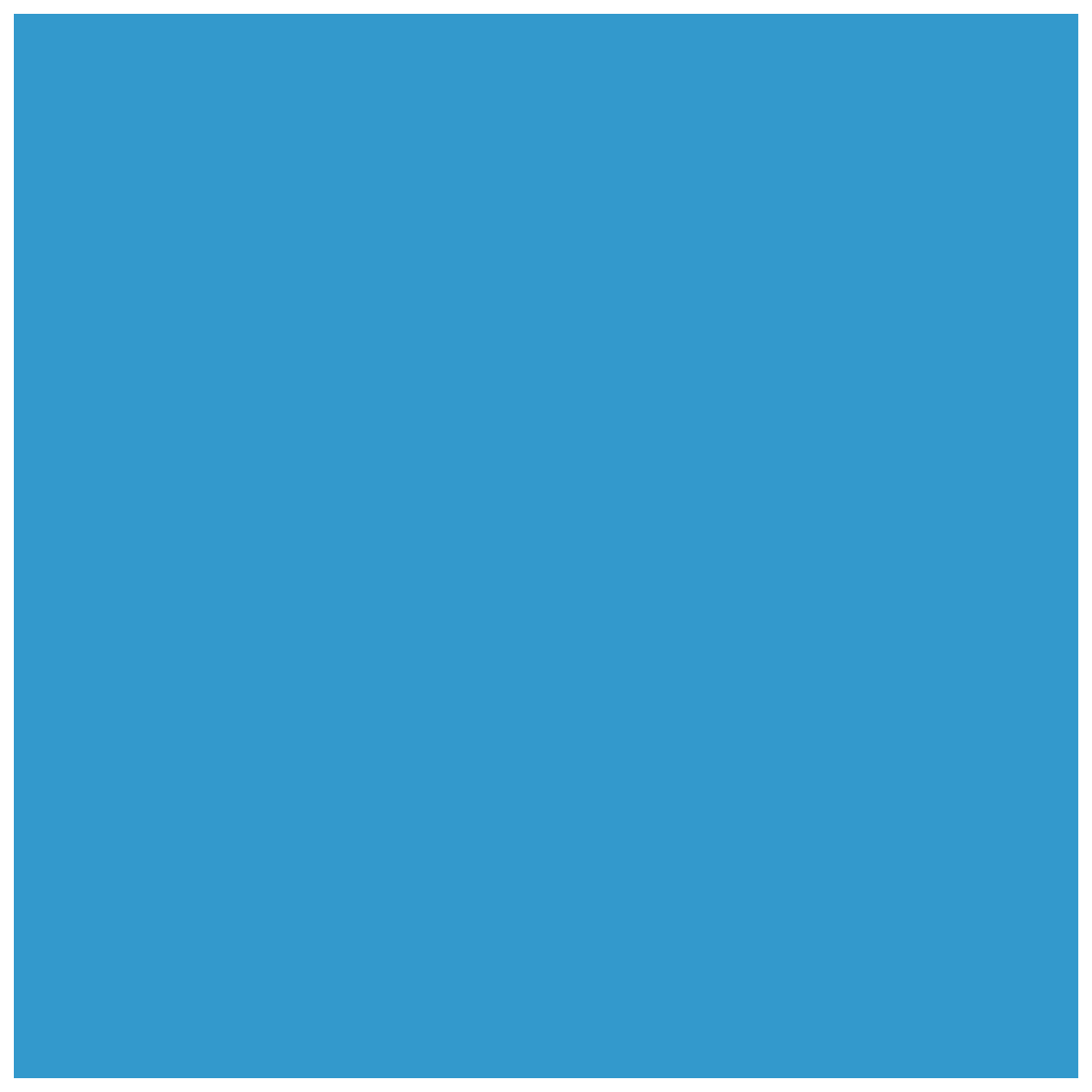}\\
			{\footnotesize t=1000}
		\end{minipage}
		\vspace{2mm}
		\\
		\centering
		{\footnotesize (b) r=4.5}
	\end{minipage}
	\\
	[2mm]
	\begin{minipage}{\linewidth}
		\begin{minipage}{0.188\linewidth}
			\centering
			\includegraphics[width=\linewidth]{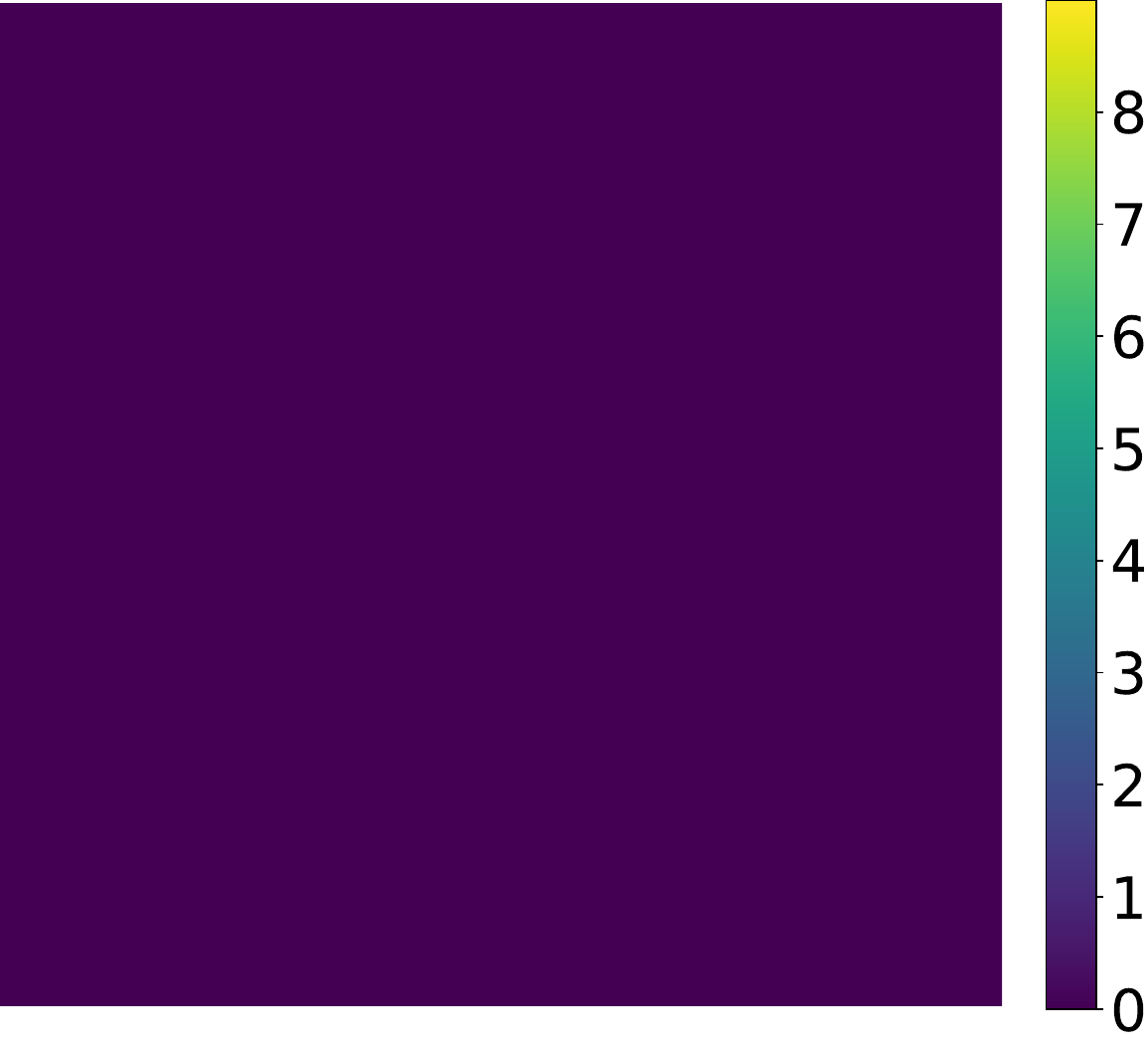}\\
			{\footnotesize t=0}
		\end{minipage}
		\hfill
		\begin{minipage}{0.188\linewidth}
			\centering
			\includegraphics[width=\linewidth]{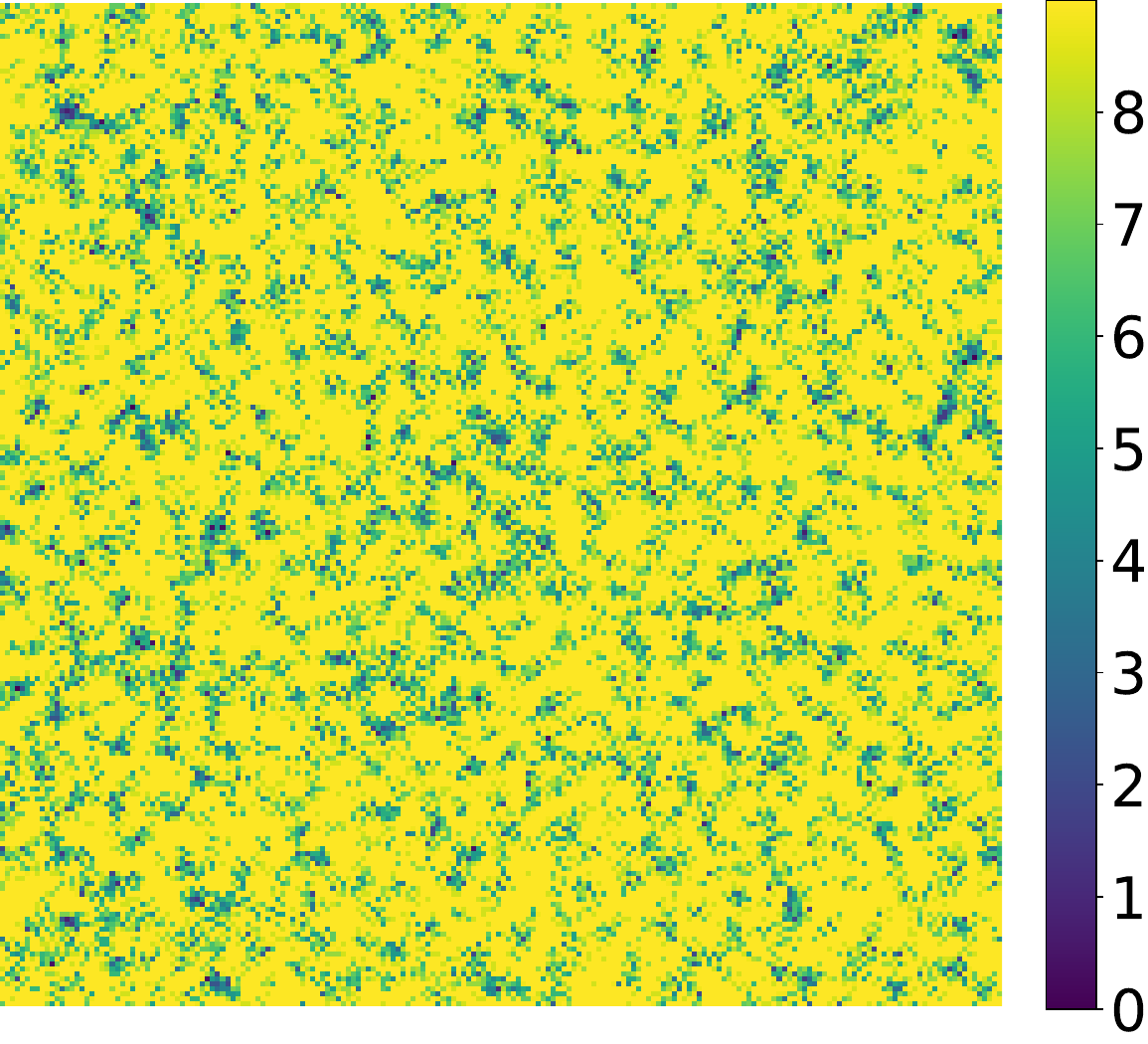}\\
			{\footnotesize t=1}
		\end{minipage}
		\hfill
		\begin{minipage}{0.188\linewidth}
			\centering
			\includegraphics[width=\linewidth]{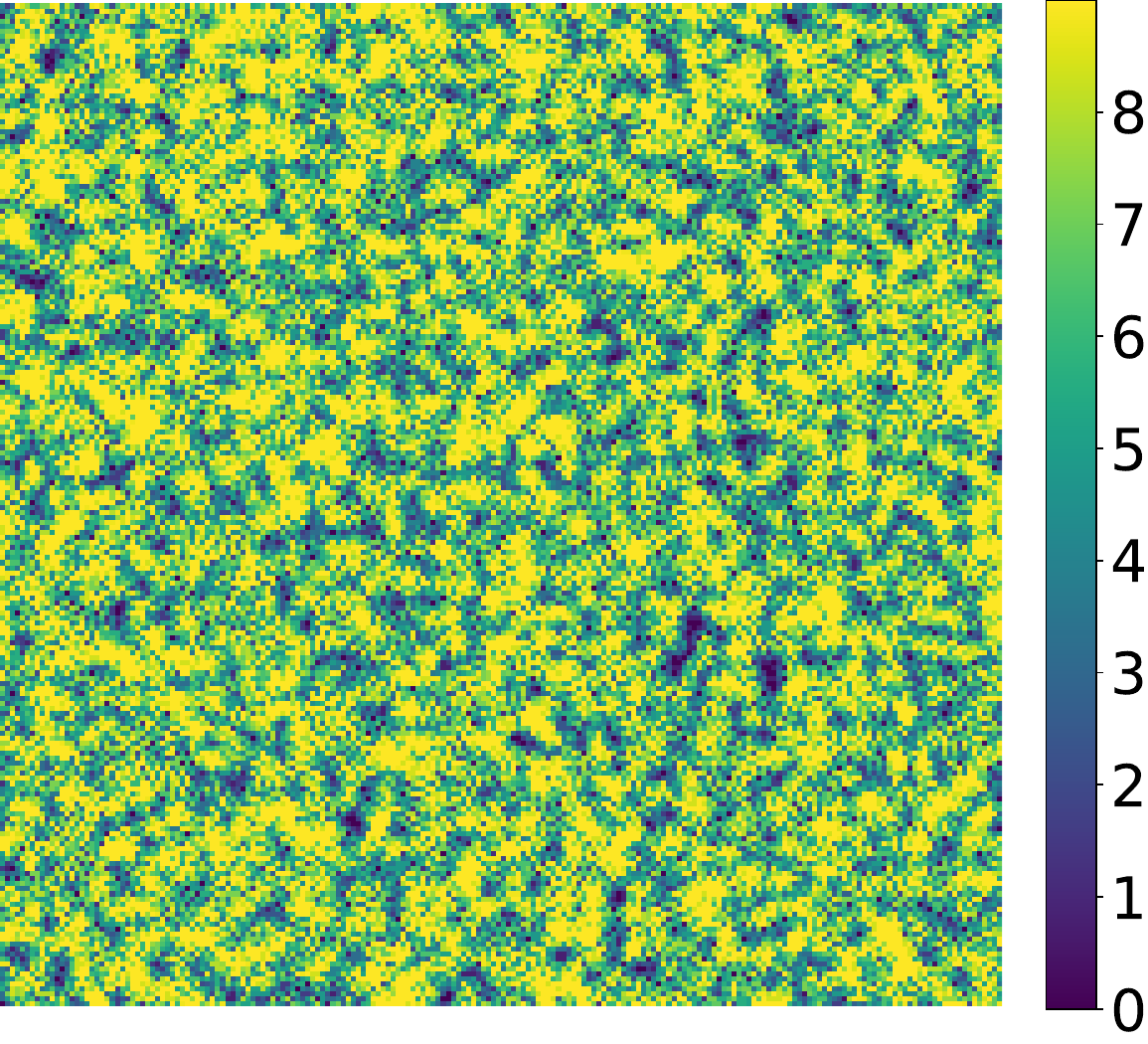}\\
			{\footnotesize t=10}
		\end{minipage}
		\hfill
		\begin{minipage}{0.188\linewidth}
			\centering
			\includegraphics[width=\linewidth]{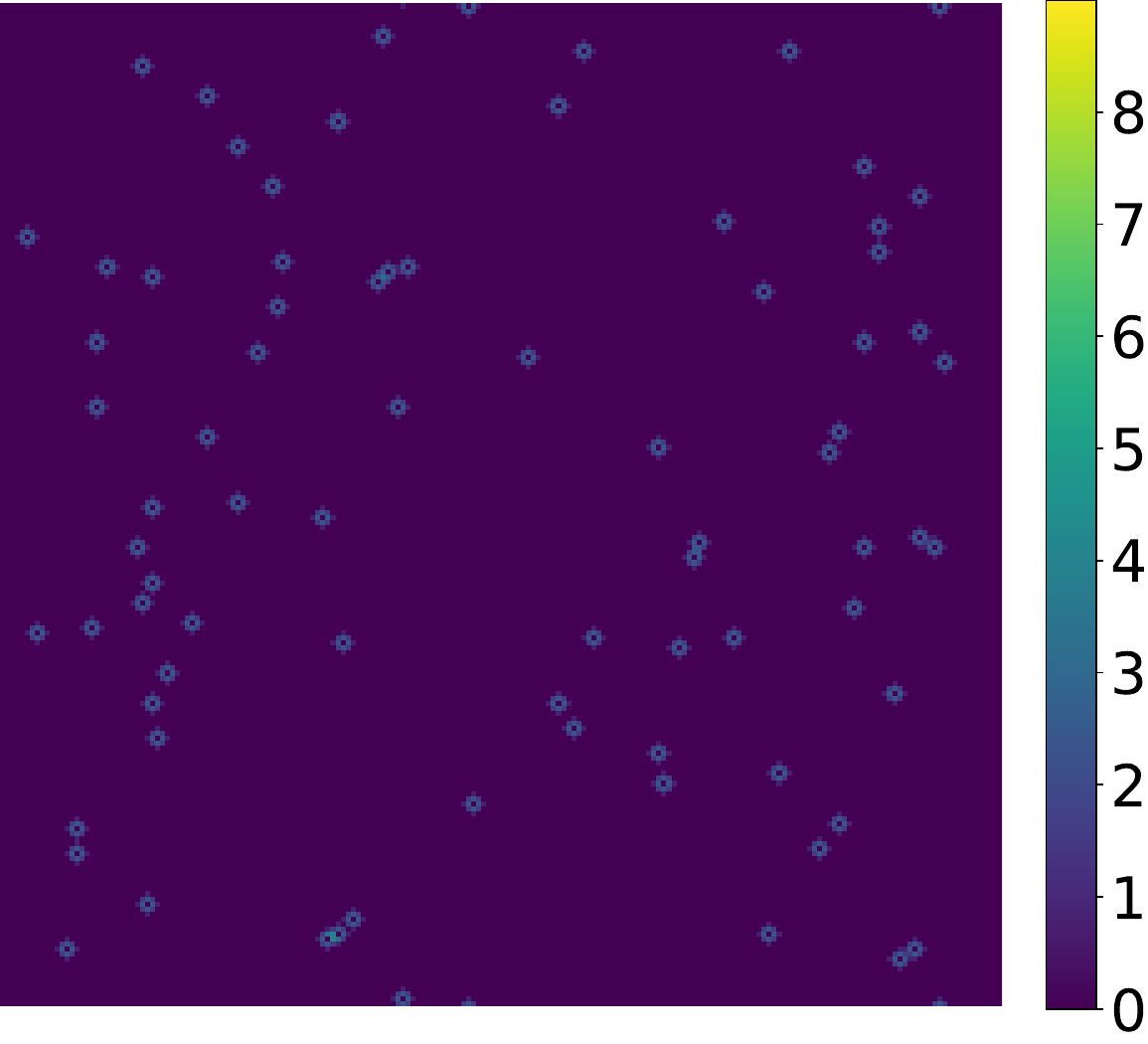}\\
			{\footnotesize t=100}
		\end{minipage}
		\hfill
		\begin{minipage}{0.188\linewidth}
			\centering
			\includegraphics[width=\linewidth]{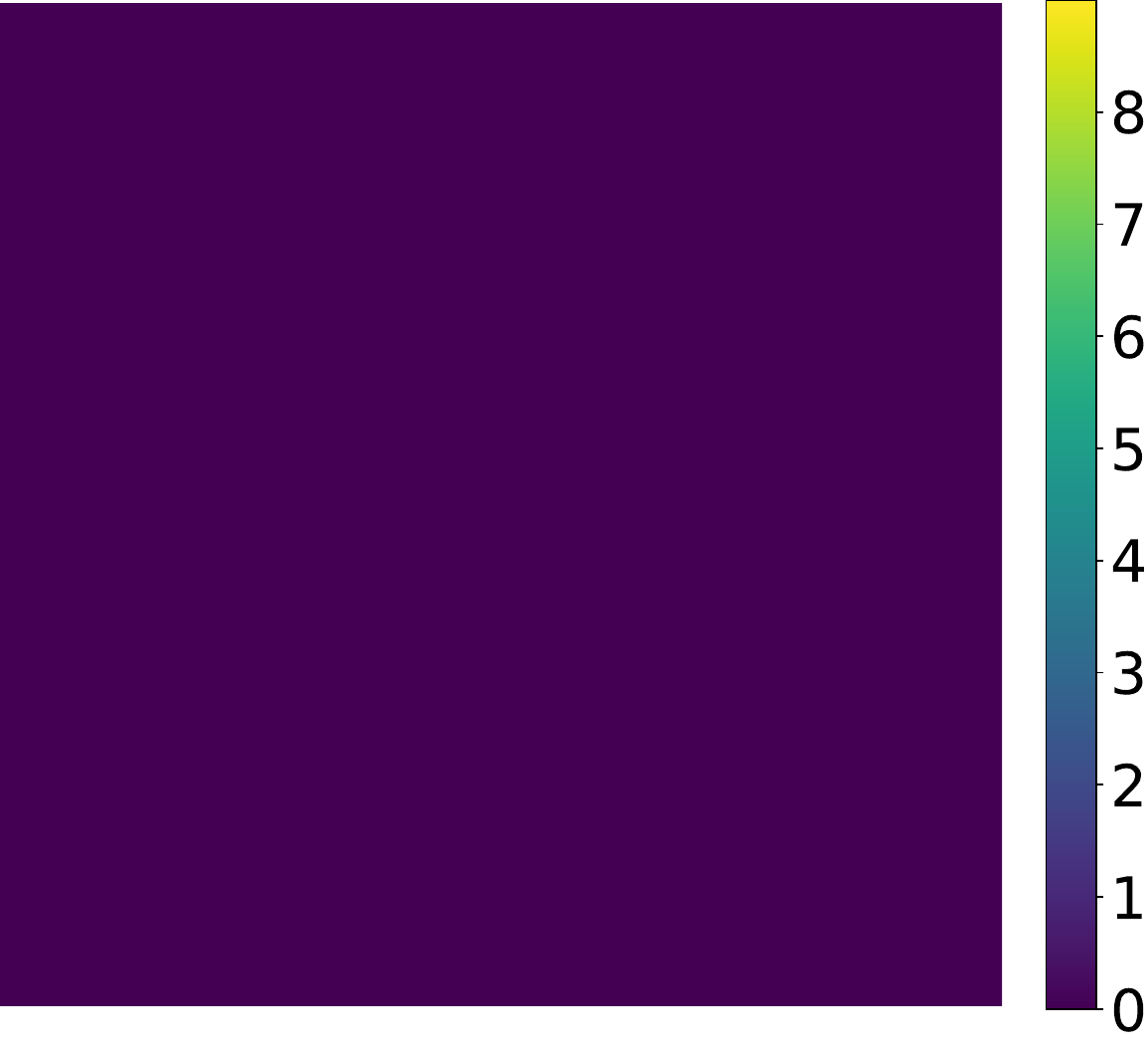}\\
			{\footnotesize t=1000}
		\end{minipage}
		\vspace{2mm}
		\\
		\centering
		{\footnotesize (c) r=4.0 (Payoff heatmaps)}
	\end{minipage}
	\\
	[2mm]
	\begin{minipage}{\linewidth}
		\begin{minipage}{0.188\linewidth}
			\centering
			\includegraphics[width=\linewidth]{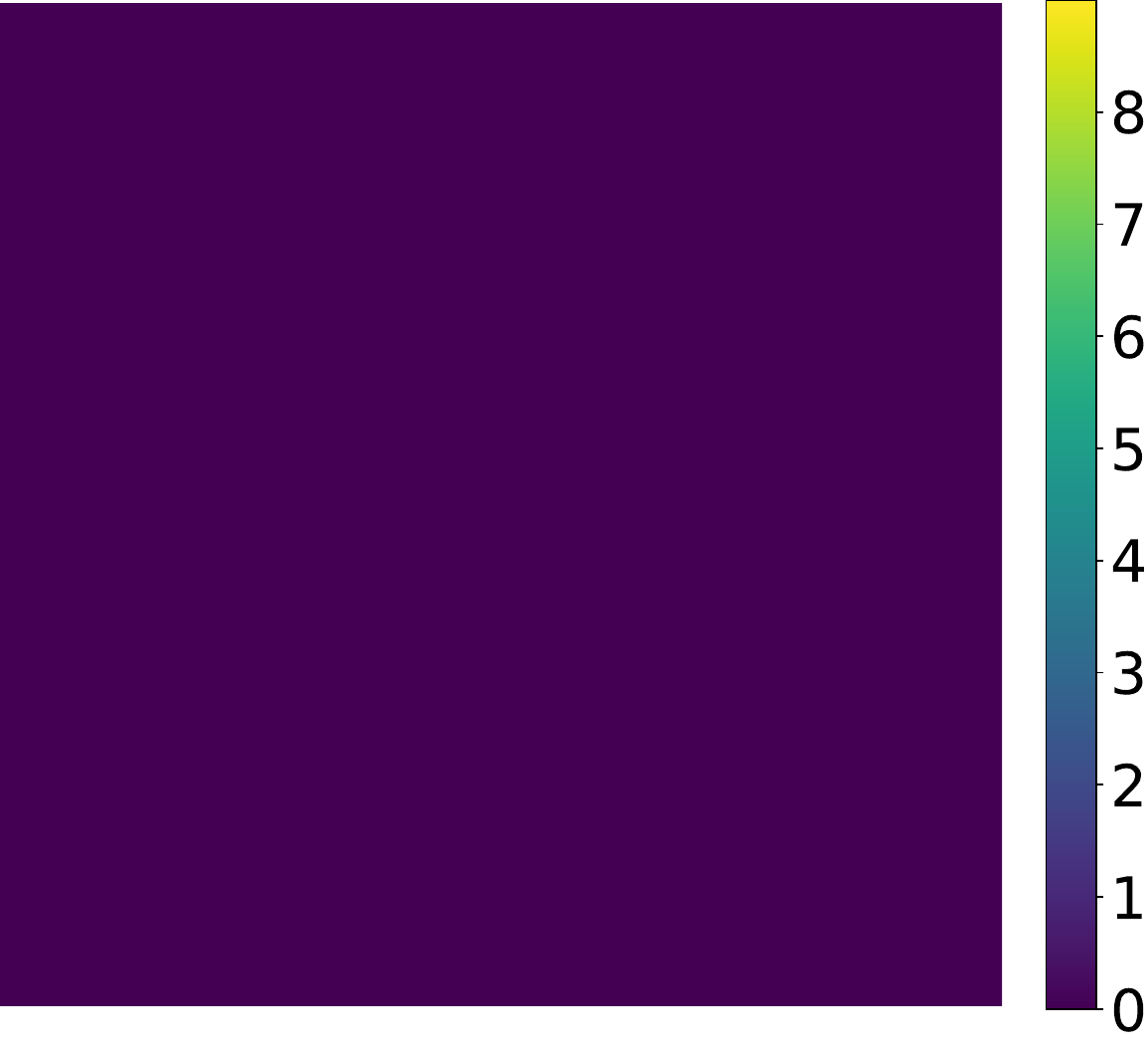}\\
			{\footnotesize t=0}
		\end{minipage}
		\hfill
		\begin{minipage}{0.188\linewidth}
			\centering
			\includegraphics[width=\linewidth]{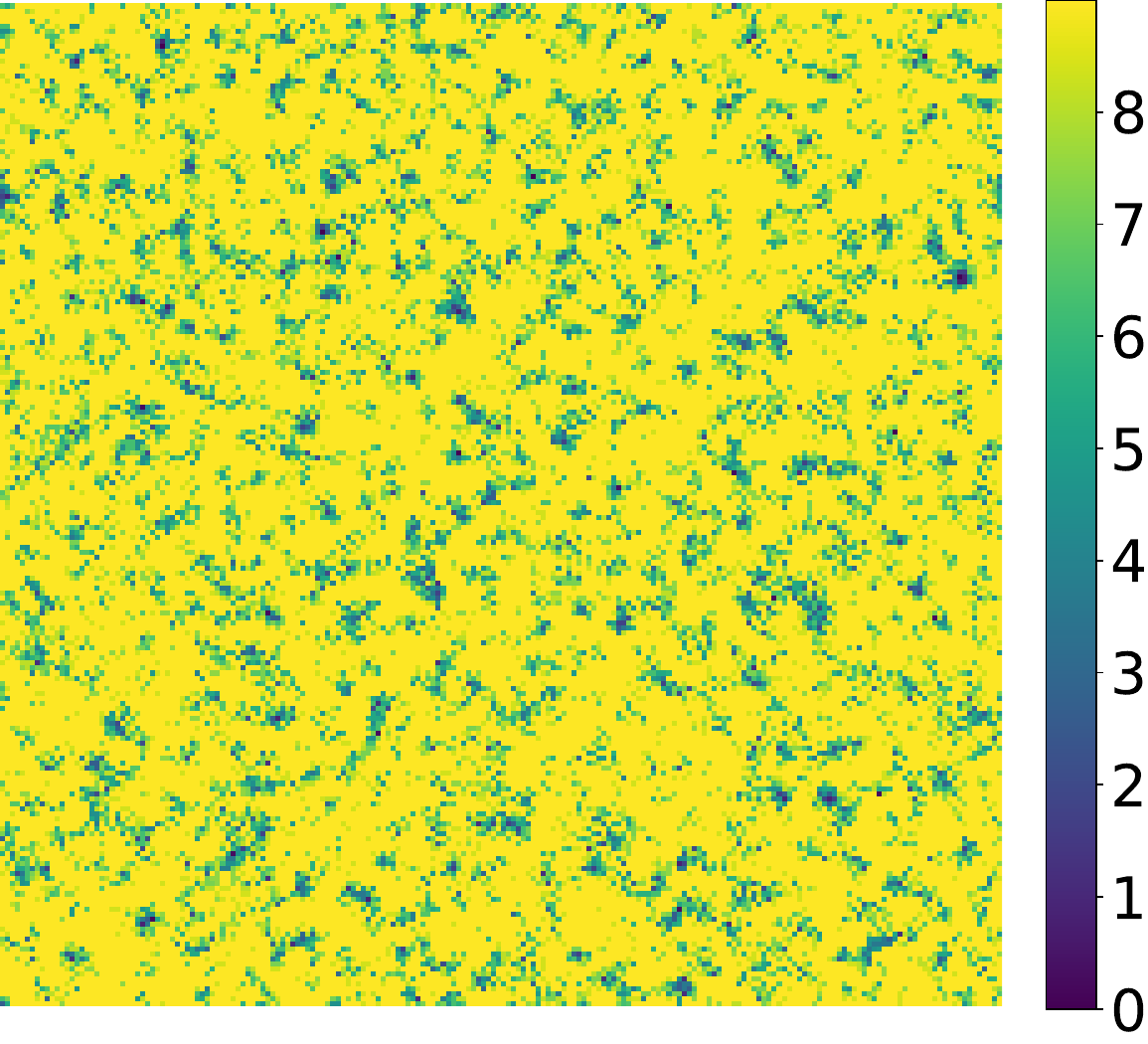}\\
			{\footnotesize t=1}
		\end{minipage}
		\hfill
		\begin{minipage}{0.188\linewidth}
			\centering
			\includegraphics[width=\linewidth]{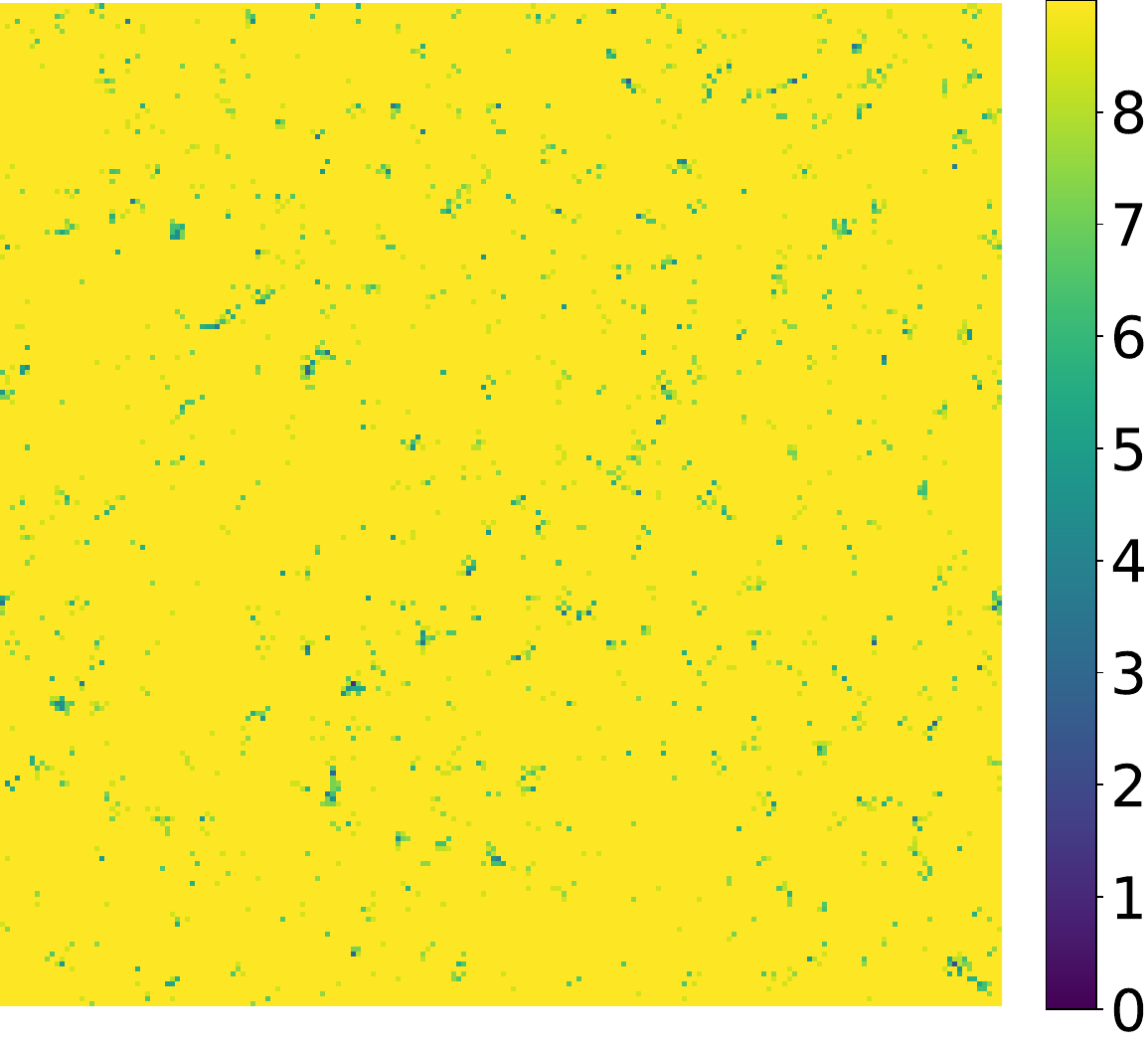}\\
			{\footnotesize t=10}
		\end{minipage}
		\hfill
		\begin{minipage}{0.188\linewidth}
			\centering
			\includegraphics[width=\linewidth]{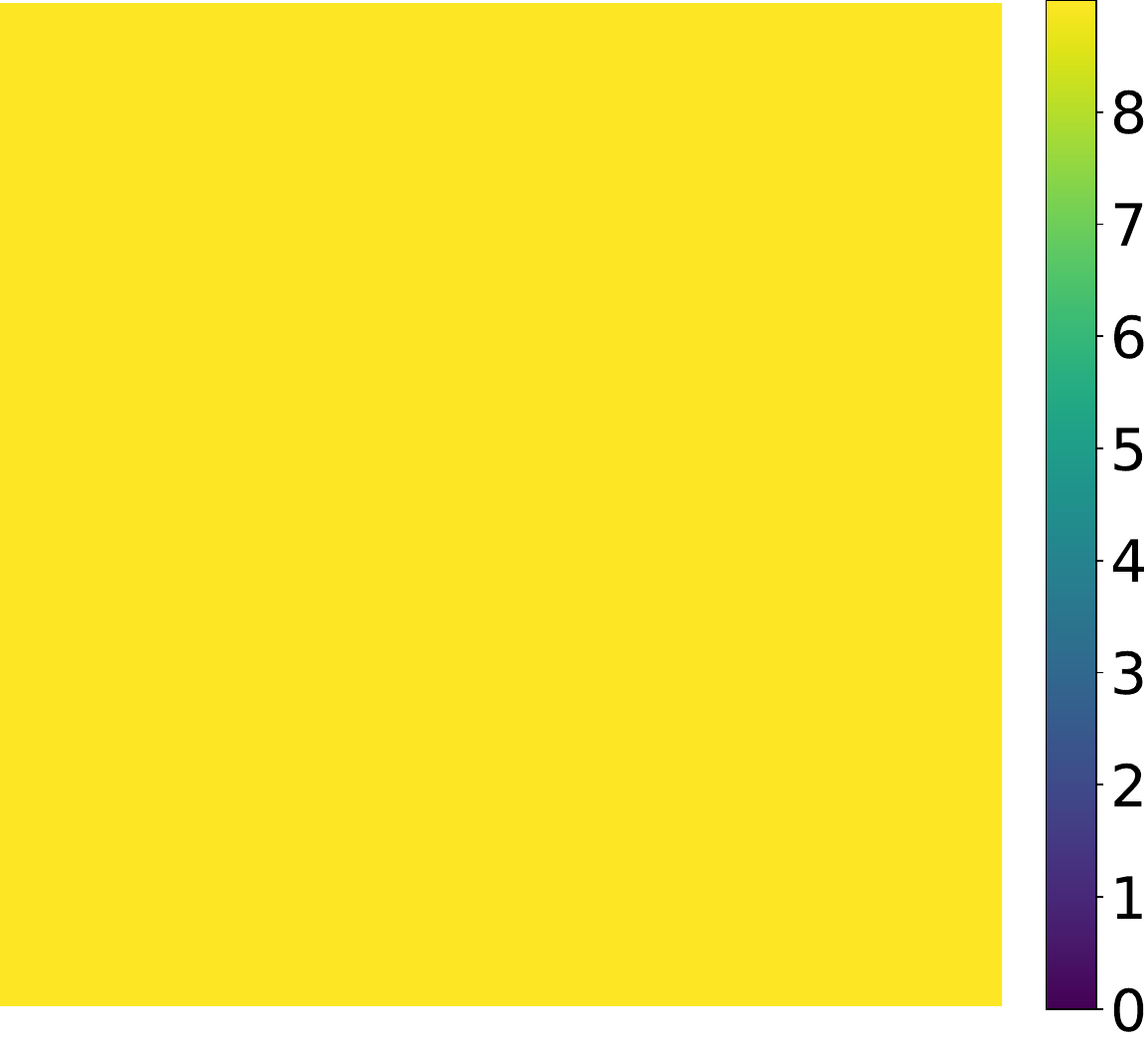}\\
			{\footnotesize t=100}
		\end{minipage}
		\hfill
		\begin{minipage}{0.188\linewidth}
			\centering
			\includegraphics[width=\linewidth]{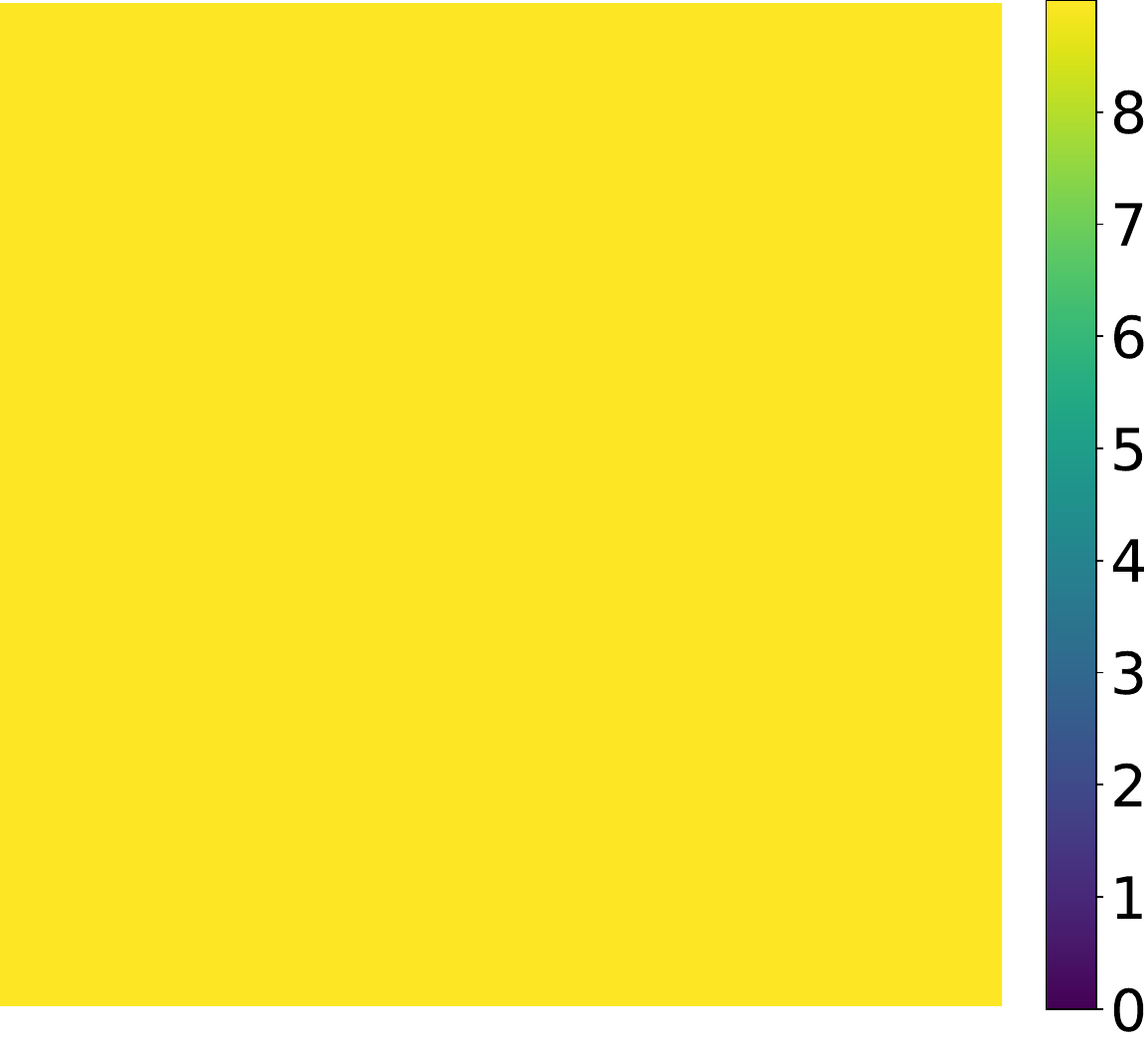}\\
			{\footnotesize t=1000}
		\end{minipage}
		\vspace{2mm}
		\\
		\centering
		{\footnotesize (d) r=4.5 (Payoff heatmaps)}
		
	\end{minipage}
	\caption{LMFPPO-UBP dynamics under all-defectors initialization. (a) At $r=4.0$, transient cooperation peaks at 55\% then collapses to full defection, (b) At $r=4.5$, monotonic growth from zero to full cooperation ($f_C=1.0$) within 30 iterations. The algorithm overcomes traditional limitations by constructing cooperation from homogeneous defection states.}
	\label{fig:all_defectors}
\end{figure*}
The evolutionary dynamics of the LMFPPO‑UBP framework were examined under an extreme initial condition where all agents were initialized as defectors, $f_C(0) = 0$. This scenario tests the algorithm’s capacity to endogenously generate cooperation from a state of universal defection. This poses a significant challenge for traditional imitation‑based rules, as they typically require pre‑existing local strategy diversity. As shown in Figure~\ref{fig:all_defectors}, the results delineate a clear threshold behavior contingent on the enhancement factor $r$. At $r = 4.0$, initial stochastic exploration briefly elevates cooperation to approximately 55\%, but the low collective returns prevent its stabilization. Cooperation subsequently decays to zero within 100 iterations, and spatial analysis reveals transient cooperative clusters being overwhelmed by defectors. The corresponding decline in collective welfare confirms defection as the dominant system attractor under sub‑critical conditions. Conversely, at $r = 4.5$, cooperation grows monotonically from zero to a full‑cooperation equilibrium $f_C = 1.0$ within 30 iterations. Cooperators form nucleating clusters that expand and eventually saturate the grid, accompanied by a systematic improvement in both individual and collective payoffs.

The LMFPPO‑UBP algorithm demonstrates the capacity to construct cooperative behavior de novo, absent any initial cooperative seed within the population. This capability stems from the integrated architecture of policy gradient learning and the UBP mechanism. The policy gradient component enables agents to explore and identify cooperative strategies through trial‑and‑error, even in a uniformly defective environment. Concurrently, the UBP mechanism immediately penalizes any defector surrounded by newly formed cooperators. This action creates a localized incentive structure that actively protects and reinforces the nascent cooperative clusters. This synergistic process allows cooperation to emerge and proliferate from a homogeneous defection state. It demonstrates that structured learning combined with targeted sanctions can overcome the historical path‑dependence often observed in evolutionary systems. Ultimately, this enables the establishment of cooperation as a self‑organizing equilibrium.

\section{Conclusions}
\label{sec:con}
This work redefines cooperation in SPGG through a DRL approach where punishment functions as a structurally embedded incentive to guide the evolutionary process. The proposed LMFPPO-UBP framework models agents as adaptive learners equipped with social perception. These agents collectively optimize their policies by leveraging LMF signals and an integrated UBP mechanism theoretically grounded in inequity aversion. This methodology structurally reshapes payoff landscapes and lowers the critical enhancement factor required for global cooperation while ensuring rapid convergence under diverse initial conditions. By bridging evolutionary game theory with multi-agent reinforcement learning the framework establishes a computational foundation for algorithmic institutional design where agents evolve within adaptive incentive structures. Consequently it enables the development of socially aware and scalable cooperation mechanisms suitable for real-world spatially structured systems. The proposed framework offers a foundation for future research into practical decentralized systems that rely on adaptive local coordination. Potential application domains include traffic networks and sensor arrays as well as energy grids and robotic swarms where robust mechanisms for spatial cooperation are critically needed \citep{2025Asynchronous, Li2025Cooperation, 2025Ajia}.







\printcredits

	\section*{CRediT authorship contribution statement}

\textbf{Jinshuo Yang}: Writing – original draft, Investigation, Writing – review and editing, Methodology, Conceptualization.
\textbf{Zhaoqilin Yang}: Validation, Writing – review and editing, Visualization, Methodology.
\textbf{Xin Wang}: Investigation, Supervision, Writing – original draft.
\textbf{Wenjie Zhou}: Conceptualization, Software, Writing – review and editing.
\textbf{Youliang Tian}: Funding acquisition, Resources, Supervision.

\section*{Declaration of competing interest }

The authors declare that they have no known competing financial interests or personal relationships that could have appeared to influence the work reported in this paper.

\section*{Data availability}

No data was used for the research described in the article.

\section*{Acknowledgments}
This work was supported by the Natural Science Special Project (Special Post) Research Foundation of Guizhou University (No.[2024] 39). Guizhou Provincial Basic Research Program (Natural Science) Youth Guidance Project (No. Qiankehe Foundation QN(2025) 054).
Guizhou Province Basic Research Program General Project (No. Qiankehe Basic MS(2026) 082).
 National Key Research and Development Program of China under Grant 2021YFB3101100; National Natural Science Foundation of China under Grant 62272123; Project of High-level Innovative Talents of Guizhou Province under Grant [2020]6008; Science and Technology Program of Guizhou Province under Grant [2020]5017, [2022]065; Science and Technology Program of Guiyang under Grant [2022]2-4.

\bibliographystyle{cas-model2-names}

\bibliography{cas-refs}



\end{document}